\documentclass[10pt]{article}
\usepackage{colordvi}
\usepackage{epsfig}
\usepackage{axodraw}
\usepackage{epsfig}
\usepackage{graphicx}
\usepackage{rotate}
\usepackage{latexsym}
\usepackage{amssymb}
\usepackage[reqno,tbtags]{amsmath}
\usepackage{multirow}

\allowdisplaybreaks
\newcommand\pubnumber{ PITHA   08/24  \\
                       SFB/CPP-08-73  \\
                       TTP08-42  }
\newcommand\pubdate{\today}

\def\csuma{Institut f\"ur Theoretische Physik E, RWTH Aachen University,\\
           52056 Aachen, Germany}
\def\csumb{Dipartimento di Fisica Teorica, Universit\`a di Torino, Italy\\
           INFN, Sezione di Torino, Italy}
\def\csumc{Physics Department, Brookhaven National Laboratory,\\
           Upton, NY 11973, USA}
\def\csumd{Institut f\"ur Theoretische Teilchenphysik, Universit\"at Karlsruhe,\\
           76128 Karlsruhe, Germany}

\def\Title#1{\begin{center} {\Large\bf #1 } \end{center}}

\def\Author#1{\begin{center}{ \sc #1} \end{center}}
\def\Address#1{\begin{center}{ \it #1} \end{center}}

\newcommand\pubblock{\rightline{\begin{tabular}{l} \pubnumber\\
         \pubdate\\  \end{tabular}}}
\newenvironment{Abstract}{\begin{quotation}  }{\end{quotation}}

\def\Acknowledgments{\bigskip  \bigskip \begin{center}
          \large\bf Acknowledgments\end{center}}
\def\email#1{\footnote{#1}}
\makeatletter
\def\section{\@startsection{section}{0}{\z@}{5.5ex plus .5ex minus
 1.5ex}{2.3ex plus .2ex}{\large\bf}}
\def\subsection{\@startsection{subsection}{1}{\z@}{3.5ex plus .5ex minus
 1.5ex}{1.3ex plus .2ex}{\normalsize\bf}}
\def\subsubsection{\@startsection{subsubsection}{2}{\z@}{-3.5ex plus
-1ex minus  -.2ex}{2.3ex plus .2ex}{\normalsize\sl}}

\renewcommand{\@makecaption}[2]{%
   \vskip 10pt
   \setbox\@tempboxa\hbox{\small #1: #2}
   \ifdim \wd\@tempboxa >\hsize     
       \small #1: #2\par          
     \else                        
       \hbox to\hsize{\hfil\box\@tempboxa\hfil}
   \fi}
 \def\citenum#1{{\def\@cite##1##2{##1}\cite{#1}}}
\def\citea#1{\@cite{#1}{}}
\newcount\@tempcntc
\def\@citex[#1]#2{\if@filesw\immediate\write\@auxout{\string\citation{#2}}\fi
  \@tempcnta\z@\@tempcntb\m@ne\def\@citea{}\@cite{\@for\@citeb:=#2\do
    {\@ifundefined
       {b@\@citeb}{\@citeo\@tempcntb\m@ne\@citea\def\@citea{,}{\bf }\@warning
       {Citation `\@citeb' on page \thepage \space undefined}}%
    {\setbox\z@\hbox{\global\@tempcntc0\csname b@\@citeb\endcsname\relax}%
     \ifnum\@tempcntc=\z@ \@citeo\@tempcntb\m@ne
       \@citea\def\@citea{,}\hbox{\csname b@\@citeb\endcsname}%
     \else
      \advance\@tempcntb\@ne
      \ifnum\@tempcntb=\@tempcntc
      \else\advance\@tempcntb\m@ne\@citeo
      \@tempcnta\@tempcntc\@tempcntb\@tempcntc\fi\fi}}\@citeo}{#1}}
\def\@citeo{\ifnum\@tempcnta>\@tempcntb\else\@citea\def\@citea{,}%
  \ifnum\@tempcnta=\@tempcntb\the\@tempcnta\else
  {\advance\@tempcnta\@ne\ifnum\@tempcnta=\@tempcntb \else\def\@citea{--}\fi
    \advance\@tempcnta\m@ne\the\@tempcnta\@citea\the\@tempcntb}\fi\fi}
\makeatother
\input Input_rosetta.sty
\begin{document}
\begin{titlepage}
\pubblock
\vfill
\def\thefootnote{\fnsymbol{footnote}}
\Title{NNLO Computational Techniques:\\[3mm]
the Cases $H \to \gamma\gamma$ and  $H \to g g$
  \footnote[9]{Work supported by MIUR under contract 2001023713$\_$006, 
               by the European Community's Marie Curie Research 
               Training Network {\it Tools and Precision Calculations for 
               Physics Discoveries at Colliders} under contract 
               MRTN-CT-2006-035505, by the U.S. Department of Energy under 
               contract No. DE-AC02-98CH10886 and by the Deutsche 
               Forschungsgemeinschaft through Sonderforschungsbereich/Transregio 9 
               {\it Computergest\"utzte Theoretische Teilchenphysik}.}}
\vfill
\Author{Stefano Actis       
\email{actis@physik.rwth-aachen.de}}        
\Address{\csuma}
\Author{Giampiero Passarino 
\email{giampiero@to.infn.it}}               
\Address{\csumb}
\Author{Christian Sturm     
\email{sturm@bnl.gov}}            
\Address{\csumc}
\Author{Sandro Uccirati     
\email{uccirati@particle.uni-karlsruhe.de}} 
\Address{\csumd}
\vfill
\vfill
\begin{Abstract}
\noindent 
A large set of techniques needed to compute decay rates at the two-loop level
are derived and systematized. The main emphasis of the paper is on the two 
Standard Model decays $H \to \gamma \gamma$ and $H \to gg$. 
The techniques, however, have a much wider range of application: they give 
practical examples of general rules for two-loop renormalization; they 
introduce simple recipes for handling internal unstable particles in 
two-loop processes; they illustrate simple procedures for the extraction 
of collinear logarithms from the amplitude. 
The latter is particularly relevant to show cancellations, e.g. cancellation 
of collinear divergencies. 
Furthermore, the paper deals with the proper treatment of non-enhanced 
two-loop QCD and electroweak contributions to different physical
(pseudo-)observables, showing how they can be transformed in a way that 
allows for a stable numerical integration. 
Numerical results for the two-loop percentage corrections to 
$H \to \gamma \gamma, gg$ are presented and discussed.
When applied to the process $pp \to gg +X \to H + X$, the results show that 
the electroweak scaling factor for the cross section is between $-4\%$ and 
$+6\%$ in the range $100\,\GeV < \mh < 500\,\GeV$, without incongruent large 
effects around the physical electroweak thresholds, thereby showing that
only a complete implementation of the computational scheme keeps two-loop corrections 
under control. 

\end{Abstract}
\vfill
\begin{center}
\end{center}
\end{titlepage}
\def\thefootnote{\arabic{footnote}}
\setcounter{footnote}{0}
\small
\thispagestyle{empty}
\tableofcontents
\normalsize
\clearpage
\setcounter{page}{1}
\section{Introduction}
In this paper we have collected and systematized a large set of techniques
needed to compute decay rates and production cross sections at the two-loop
level in a spontaneously broken theory characterized by a large number of 
scales. Although the main emphasis will be on the Standard Model (SM) 
Higgs-boson decays to two photons and two gluons and on the related 
production processes, our techniques have a broader range of application. 
Firstly, they represent practical applications of general rules for 
two-loop renormalization and introduce simple recipes for handling internal 
unstable particles. Secondly, they illustrate useful procedures for the 
analytical extraction of collinear logarithms: we explicitly prove that 
the complete $H\to\gamma\gamma$ amplitude is free from collinear logarithms 
in spite of the fact that single Feynman diagrams show a collinear-divergent 
behavior. Finally, our techniques allow to represent the non-enhanced 
two-loop QCD and electroweak contributions to different physical 
(pseudo-)observables in a form suitable for a stable numerical integration.

The lastly mentioned feature represents the important and difficult part 
of our computation. Since we do not rely on any kind of approximation, we 
can safely cover all kinematical regions including different normal thresholds, 
like the $WW$ one, which are peculiar of Higgs- and vector-boson decays. 
Therefore, we are able to place the electroweak component on the same footing 
as the QCD one: a non-trivial but essential task, since in several situations 
the size of the pure electroweak higher order corrections is comparable to 
the theoretical uncertainty related to the parton distribution functions.

Note that two-loop electroweak corrections to $1\to 2$ processes are currently 
investigated by several groups: the authors of Ref.~\cite{Butenschoen:2006ns} 
have derived the dominant contributions to the $H\to b\overline{b}$ decay 
width, and two different collaborations have computed the 
fermionic~\cite{Awramik:2004ge} and bosonic~\cite{Hollik:2005ns,Awramik:2006ar} 
corrections to the effective electroweak mixing angle. 

With the advent of the LHC era, it is clear that the production process 
$gg\to H$ and the decay mode $H\to \gamma\gamma$ are going to play a crucial 
role for a precise comparison of the experimental data with the SM predictions.

Gluon fusion is the main production channel of the SM Higgs boson at hadron 
colliders, and both virtual and real corrections have been thoroughly 
investigated since the beginning of the 90's. The QCD next-to-leading 
order (NLO) radiative corrections to the total Higgs-production cross 
section have been first computed below the $t\overline{t}$ threshold 
in Ref.~\cite{Djouadi:1991tka}, and  using an effective-Lagrangian approach, 
where the top quark is integrated out, in Ref.~\cite{Dawson:1990zj}. Formally, 
the second method defines the heavy-top limit, and it has represented the 
starting point for subsequent higher order improvements. A one-dimensional 
integral representation which proves well-set for numerical evaluation 
has been later derived in Refs.~\cite{Graudenz:1992pv,Spira:1995rr} for the 
entire Higgs-mass range; moreover, the virtual component has been evaluated 
analytically in a closed form by the authors of Ref.~\cite{Harlander:2005rq}. 

Recently, two different groups~\cite{Anastasiou:2006hc,Aglietti:2006tp} have 
provided independent checks of the work of Ref.~\cite{Harlander:2005rq}, 
generalizing the analytic result in a framework where the coupling of the Higgs 
to the external particles is mediated by a scalar field (see also Ref.~\cite{Muhlleitner:2006wx}). 
Since QCD NLO corrections 
increase the cross section by more than $70\,\%$, there was a flurry of activity 
on higher order QCD effects. Partial results for the next-to-next-to-leading order 
(NNLO) corrections to the total cross section in the heavy-top limit have been 
obtained in Ref.~\cite{Harlander:2000mg} for the two-loop virtual corrections to 
the effective heavy-top $Hgg$ vertex, and in Ref.~\cite{Catani:2001ic} for the 
soft components; the complete NNLO result has been finally derived by three 
different groups in Ref.~\cite{Harlander:2002wh}, and later supplemented by an 
all-order resummed calculation of multiple soft-gluon emission at 
next-to-next-to-leading 
logarithmic (NNLL) accuracy in Ref.~\cite{Catani:2003zt} and including the fourth 
logarithmic (N$^3$LL) order in Ref.~\cite{Moch:2005ky} (see also Ref.~\cite{Ahrens:2008qu}). 
In addition, the effect 
of a jet veto on the inclusive cross section has been studied at NNLO in 
Ref.~\cite{Catani:2001cr}, and an improvement with respect to the heavy-top limit 
has been recently obtained by the authors of Ref.~\cite{Marzani:2008az}. Because of 
the cuts on the final states typical for hadron-collider phenomenology, 
fully differential perturbative results have a privileged role; the differential 
cross section for Higgs production has been derived through NNLO in QCD in 
Ref.~\cite{Anastasiou:2004xq}, and later cross-checked through an independent 
subtraction formalism by the authors of Ref.~\cite{Catani:2007vq}. Note that 
the techniques used in Ref.~\cite{Anastasiou:2004xq} have been recently 
extended in Ref.~\cite{Anastasiou:2007mz} to compute the QCD NNLO cross 
section for the $H\to WW\to l\nu l\nu$ signal at the LHC. The Higgs-production 
mechanism in the gluon-fusion channel has clearly a close affinity with the 
Higgs decay to two gluons; the state of the art in QCD is presently 
represented by the two works of Ref.~\cite{Schreck:2007um}, with the full NNLO result and the N$^3$LO 
calculation in the heavy-top limit respectively.

While QCD corrections to Higgs production through gluon fusion at hadron 
colliders are well under control, electroweak effects are less understood. 
The NLO corrections to the total cross section were evaluated in 
Ref.~\cite{Djouadi:1994ge,Djouadi:1997rj} in the heavy-top limit, and turned out 
to be less than $1\%$. The contribution due to the light fermions has been 
calculated analytically in Ref.~\cite{Aglietti:2004nj}, and found to be more 
sizable; the remaining component of the amplitude involving the top quark has been 
computed by means of a Taylor expansion in the Higgs external momentum in 
Ref.~\cite{Degrassi:2004mx}. It is worth stressing that the work of 
Refs.~\cite{Aglietti:2004nj,Degrassi:2004mx} has not been independently 
cross-checked by a different group; moreover, for obvious reasons, the validity 
of the result of Ref.~\cite{Degrassi:2004mx} is restricted to the kinematical
region below the $WW$ threshold. Under the hypothesis of factorization with 
respect to the dominant QCD soft and collinear radiation, the impact of the 
electroweak corrections to Higgs-production in proton-proton collisions has 
been estimated in Ref.~\cite{Aglietti:2006yd}.

The decay of the SM Higgs boson into two photons has also a great phenomenological
interest: on the one hand, it provides a precious information for the discovery
at hadron colliders~\cite{Cranmer:2004uz}; on the other hand, an upgrade option 
at the planned ILC will allow for a high-precision measurement of the partial 
width into two photons~\cite{Monig:2007py}, leading to quantitative tests on 
the existence of new charged particles. QCD corrections to the decay width of 
an intermediate-mass Higgs boson have been computed at NLO in 
Ref.~\cite{Zheng:1990qa} and at NNLO in Ref.~\cite{Steinhauser:1996wy}; the NLO 
result has later been extended in Ref.~\cite{Melnikov:1993tj} to the whole 
Higgs-mass range. 

Electroweak NLO corrections have been computed in Ref.~\cite{Liao:1996td,Djouadi:1997rj} in 
the large top-mass scenario (the so-called ``dominant'' corrections) and in 
Ref.~\cite{Korner:1995xd} in the large Higgs-mass scenario. 
The complexity of this complicated (multi-scale) computation is reflected into 
the fact that seldom an {\em a priori} dominance lives up to its promise;
for instance, $\ord{G_{\ssF} M_t^2}$ corrections in the range 
$80 - 150\,$GeV do not dominate at all.
Recently, the two-loop contributions induced by light fermions have been derived 
in Ref.~\cite{Aglietti:2004nj}, and electroweak corrections due to gauge bosons 
and the top quark have been given in term of an expansion in the Higgs external 
momentum in Ref.~\cite{Degrassi:2005mc}, a result subsequently extended in 
Ref.~\cite{Passarino:2007fp} to cover all kinematical regions including the 
notoriously difficult $WW$ threshold and the NLO QCD corrections.

In this paper we show details about the evaluation carried on for the 
$H\to\gamma\gamma$ decay in Ref.~\cite{Passarino:2007fp} and generalize the 
result to the $gg\to H$ process to the entire Higgs-mass range. Since we are 
not bound to rely on any kind of expansion, neither in the bosonic sector nor 
in the top-bottom one, we can produce extremely accurate results for any value 
of the Higgs-boson mass, including the full dependence on the $W$-, $Z$- and 
Higgs-boson masses and on the top-quark mass. A consistent and gauge-invariant 
treatment of unstable particles allows to produce precise results around the 
$WW$ threshold.

In our approach, Feynman diagrams, up to two loops and including QCD,
are generated by means of an automated procedure which does not rely on any 
external package, and is implemented in the {\sc FORM}~\cite{Vermaseren:2000nd} 
program {$\GS$}~\cite{GraphShot}. After projecting the amplitude for a
given (pseudo-)observable onto form factors and taking traces over
Dirac matrices, three basic simplifications are done recursively:
at first, reducible scalar products are removed; next, tadpole integrals
are reduced using integration-by-parts identities (IBPIs)~\cite{Tkachov:1981wb};
finally, the symmetries of each diagram are exploited in order
to reduce the number of integrals to be evaluated. 

The second step in the calculation of any two-loop process concerns 
renormalization; here we heavily 
rely on the results of Refs.~\cite{Actis:2006ra,Actis:2006rb,Actis:2006rc}. 
After assembling diagrams we perform the usual canonical 
tests to ensure that our result is correct, checking all possible unrenormalized 
Ward-Slavnov-Taylor (WST) identities. The next logical steps consist in removing 
all ultraviolet divergencies, with the due caution to the problem of overlapping 
divergencies, and checking the renormalized WST identities. Then, we express 
renormalized parameters in terms of physical observables belonging to some 
input-parameter set. 

The third step consists in the analytical extraction of collinear logarithms of 
Feynman diagrams. It is worth noting that the amplitude for $H\to\gamma\gamma$ is 
collinear free, and one could adopt the approach where all light fermions are 
considered massless; therefore, the 
collinear behavior of single components is controlled in dimensional regularization 
and collinear poles cancel in the total. We prefer another approach, where 
collinear singularities are controlled by the light-fermion masses; partial 
components of the complete result are divergent, but we check that all 
logarithms of collinear origin cancel in the complete answer. 

Finally, the remaining collinear-free contributions are written in terms of smooth 
integral representations using the methods developed in 
Refs.~\cite{Passarino:2001wv,Passarino:2001jd,Ferroglia:2003yj,Passarino:2006gv,Actis:2004bp}, and then evaluated numerically.

The outline of the paper is as follows: in section~\ref{conv} we summarize our
notation and conventions. Details about the construction of the $H \to 
\gamma \gamma (gg)$ amplitude, including renormalization and choice of the 
input-parameter set, are discussed in section~\ref{Ampl}. In 
section~\ref{sec:RedSym} we explain how we manipulate Feynman integrals in order 
to simplify the amplitude, removing scalar products, reducing tadpole integrals
and symmetrizing Feynman diagrams. The behavior of two-loop diagrams around 
a normal threshold is thoroughly investigated in section~\ref{NormTH}. In 
section~\ref{EoCS} we discuss the extraction of collinear singularities, and in 
section~\ref{EMD} we describe details about the evaluation of massive diagrams. 
Numerical results are shown and discussed in section~\ref{hicsunt}. Finally, we 
summarize our conclusions in section~\ref{sec:conc}. Additional material is 
contained in the appendices.
\section{Notation and conventions}
\label{conv}
\noindent\underline{\emph{Regularization.}} We employ dimensional 
regularization~\cite{'t Hooft:1972fi}, denoting the number of space-time 
dimensions by $n=4-\ep$, and we define short-hand notations for 
regularization-dependent factors,
  \bq \label{defMSB}
    \DUV\,=\,\gamma\,+\,\ln\pi\,+\,\ln\frac{M^2}{\mu^2}, \qquad\qquad
    \DUV(x)\,=\,\DUV\,-\,\ln\frac{M^2}{x},
  \eq
where $\gamma=0.5772156\cdots$ is the Euler constant, $\mu$ is the 't Hooft 
mass unit, $M$ stands for the bare or renormalized $W$-boson mass (in the 
following we will not distinguish unless strictly needed) and $x$ is a 
positive-definite kinematical variable. In our conventions the logarithm 
has a cut along the negative real axis and it is understood that for all 
masses $M^2\!\to\! M^2\! -\! i\, 0$. 

The structure of poles in dimensional regularization is better exploited in 
terms of universal factors. Any one-loop integral $f^1$ can be formally 
decomposed as $f^1\lpar \,\{l\}\,\rpar = \sum_{k=-1}^{1}\, 
f^1\lpar \,\{l\}\,;\,k\,\rpar \, F^1_k(x)$,
where $\{l\}$ stands for a set of arguments representing powers of inverse 
propagators, external kinematical variables and masses of internal particles 
and $x$ denotes a scale which depends on the kind of integral: $M$ for tadpoles 
and a squared external momentum for other configurations. The dependence on the 
dimensional regulator $\ep$ and the regularization-dependent factors introduced 
in \eqn{defMSB} is encapsulated in the universal one-loop 
ultraviolet (hereafter UV) factors
  \bq \label{OLUVF}
    F^1_{-1}(x) = \frac{1}{\ep} - \frac{1}{2}\,\DUV(x) + 
    \frac{1}{8}\,\DUV^2(x)\,\ep, \qquad F^1_0(x) = 1 - \frac{1}{2}\,
    \DUV(x)\,\ep, \qquad F^1_1(x) = \ep.
  \eq
It is worth noting that, because of overlapping divergencies due to UV-divergent 
one-loop sub-diagrams, we include $\ord{\ep}$ terms in all one-loop results. 

A generic two-loop integral $f^2$ can be written as
$f^2\lpar \,\{l\}\,\rpar = \sum_{k=-2}^{0}\,
f^2\lpar \,\{l\}\,;\,k\rpar\, F^2_k(x)$, with two-loop UV factors given by
  \bq \label{UVfactors}
    F^2_{-2}(x) = \frac{1}{\ep^2} - \frac{\DUV(x)}{\ep} + \frac{1}{2}\,\DUV^2(x),
    \qquad F^2_{-1}(x) = \frac{1}{\ep} - \DUV(x), \qquad F^2_0(x) = 1.
  \eq
Note that the product of two one-loop integrals can be written in terms of the 
same UV decomposition of a genuine two-loop integral. 

Finally, let us define our soft/collinear factors; for the electroweak part
of the calculation they are equal to the UV factors of \eqn{UVfactors}.
In appendix~\ref{hereisnab} we will extend their definition to cover non-abelian
QCD configurations.
\vspace{0.2cm}

\noindent\underline{\emph{Classification of two-loop integrals.}}
In this paper we frequently use the notion of scalar, vector and tensor 
two-loop Feynman integrals. An arbitrary two-loop scalar diagram can be cast as
  \bqa \label{Gdiag}
    G^{\alpha\beta\gamma}=\frac{\mu^{2\ep}}{\pi^4}\,
    \intmomsii{n}{q_1}{q_2}\,\prod_{i=1}^{\alpha}\,(k^2_i+m^2_i)^{-1}\,
    \prod_{j=\alpha+1}^{\alpha+\gamma}\,(k^2_j+m^2_j)^{-1}\,
    \prod_{l=\alpha + \gamma+1}^{\alpha+\gamma+\beta}\,(k^2_l+m^2_l)^{-1},
  \eqa
where $\alpha, \beta$ and $\gamma$ give the number of internal lines containing 
the integration momenta $q_1, q_2$ and $q_1-q_2$, and we have introduced
  \bq
  \ba{ll}
    k_i = q_1+\sum_{j=1}^{\ssN}\,\eta^1_{ij}\,p_j, \;&\; i=1,\dots,\alpha,  \\
    k_i = q_1-q_2+\sum_{j=1}^{\ssN}\,\eta^{12}_{ij}\,p_j, \;&\;
    i=\alpha+1,\dots,\alpha+\gamma,  \\
    k_i = q_2+\sum_{j=1}^{\ssN}\,\eta^2_{ij}\,p_j, \;&\;
    i=\alpha+\gamma+1,\dots,
    \alpha+\gamma+\beta.
  \ea
  \eq
Here $N$ is the number of vertices connected to external lines, $\eta^a_{ij} = 
\pm 1, 0$ and $\{p\}$ is the set of external momenta. The capital letter $G$ 
identifies the number of external legs: $T$ for tadpoles, $S$ for self-energies 
and $V$ for vertices. 

The triplet of numbers $\alpha \beta \gamma$ can be represented in an extremely 
compact way: first, we introduce $ \kappa = \gamma_{\rm max}\,[ \alpha_{\rm max}\,
(\beta - 1) + \alpha - 1 ] + \gamma$, where 
for tadpoles $\alpha_{\rm max} = 1$ and $\gamma_{\rm max} = 1$, for self-energies 
$\alpha_{\rm max} = 2$ and $\gamma_{\rm max} = 1$ and for vertices 
$\alpha_{\rm max} 
= 2$ and $\gamma_{\rm max} = 2$; next, we associate a letter of the alphabet to 
each value of $\kappa$ and we derive a set of alpha-numerical correspondences,
  \bqa
    G=T\quad &\Rightarrow& \quad 111 \to A, \qquad
    G=S\quad \Rightarrow \quad 111 \to A, \quad 121 \to C, 
    \quad 131 \to E, \quad 221 \to D,\nl
    G=V\quad &\Rightarrow& \quad 121 \to E, \quad 131 \to I, \quad 141 
    \to M, \quad 221 \to G, \quad 231 \to K, \quad 222 \to H.
  \eqa
All the different groups (sometimes called families or topologies) of 
Feynman diagrams with different numbers of external and internal legs
have been classified for completeness in Figs.~\ref{OL}, \ref{TLselfenrgies}, 
\ref{TLvertices} and \ref{OTLvacuums} of appendix~\ref{app:topos}, where the 
graphical relation between the diagrams and the corresponding integrals can
be found. Note that we follow the notation of Ref.~\cite{Passarino:1979jh} for 
one-loop integrals. At the two-loop level we identify one tadpole (vacuum) 
topology, $T^{\ssA}$, four self-energy topologies, $S^{\ssA},\,S^{\ssC},\,S^{\ssE},
\,S^{\ssD}$, and six vertex topologies, $V^{\aba},\,V^{\aca},\,V^{\ada},\, 
V^{\bba},\,V^{\bca},\,V^{\bbb}$. Note that factorized two-loop topologies, 
associated with the product of two one-loop Feynman integrals, do not receive a 
particular name.  

For a detailed analysis of scalar two-loop self-energies and vertices we refer the 
interested reader to 
Refs.~\cite{Passarino:2001wv,Passarino:2001jd,Ferroglia:2003yj,Passarino:2006gv}. 
The presence of non-trivial structures containing integration momenta in the 
numerators of two-loop integrals requires to introduce tensor structures and form 
factors, as described in sections~7 and~9 of Ref.~\cite{Actis:2004bp} for 
higher rank self-energies and vertices.
\vspace{0.2cm}

\noindent\underline{\emph{Miscellanea.}}
In order to keep our results as compact as possible, we introduce a short-hand
notation for integrals over a simplex of Feynman parameters,
  \bq
    \dsimp{n}(\{x\}) f(x_1\!,\!\cdots\!,\!x_n)\! =\! 
    \prod_{i=1}^{n}\!\int_0^{x_{i-1}} \!\!\!\!\!\!\!\!\!\!
    dx_i f(x_1\!,\!\cdots\!,\!x_n),
    \quad
    \dcub{n}(\{x\}) f(x_1\!,\!\cdots\!,\!x_n) \!=\! \int_0^1\!\!
    \prod_{i=1}^{n}\!dx_i
    f(x_1\!,\!\cdots\!,\!x_n),
  \eq
where $x_0 = 1$. In addition, the so-called $'+'$ and $'+\!+'$ distributions 
will be extensively used,
  \bqa \label{plusdist}
    \intfx{x} \frac{f(x,\{z\})}{x-a}\bmid_+ \! &=&\!
    \intfx{x} \frac{f(x,\{z\}) - f(a,\{z\})}{x-a},\qquad a=0,1,
    \nl
    \intfx{x} \frac{f(x,\{z\})\,\ln^n x}{x}\bmid_+ \!&=&\!
    \intfx{x} \frac{ \left[f(x,\{z\}) - 
      f(0,\{z\}) \right]\,\ln^n x}{x},
  \eqa
  \bq \label{plusplusdist}
    \intsx{x}\,dy\frac{f(x,y,\{z\})}{x\,y}\bmid_{++} =
    \intsx{x}\,dy\,
    \frac{f(x,y,\{z\}) - f(x,0,\{z\}) - f(0,y,\{z\}) + f(0,0,\{z\})}{x\,y}.
  \eq
\section{The amplitude for $H \to \gamma \gamma (gg)$ \label{Ampl}}
We consider the decay of an arbitrary Higgs boson into two photons
$h \left(-P\right)+\gamma\left(p_1\right)+\gamma\left(p_2\right) \to 0$, 
where $P=p_1+p_2$ (all momenta are incoming). The amplitude ${\cal A}$ can be 
written as
\bq
\label{AMP:ffactors1}
{\cal A} \,=\, 
{\cal Z}_\ssA^{-1} \, {\cal Z}_\ssH^{-\frac{1}{2}}
\,\, e^\mu_1 \, e^\nu_2 \,\, 
{\cal A}_{\mu\nu}(0,0,-\mhs),
\eq
where $e_i^\mu=e^\mu(p_i,\lambda_i)$, with $i=1,2$ and $\lambda_i=\pm$, are 
the photon polarization vectors and ${\cal Z}_\ssA$ and ${\cal Z}_\ssH$ are the 
photon and Higgs-boson wave-function renormalization factors.
${\cal A}_{\mu\nu}$ is the amputated Green's function for $h \to\gamma\gamma$, 
whose tensor structure reads
\bq
\label{AMP:ffactors}
{\cal A}_{\mu\nu}(p_1^2,p_2^2,P^2) = \frac{G}{16\pi^2}
\Bigl[                  
                          F_\ssD(p_1^2,p_2^2,P^2) \, \delta_{\mu\nu}    +  
  \sum_{i,j=1}^{2} F^{(ij)}_\ssP(p_1^2,p_2^2,P^2) \, p_{i\mu}\,p_{j\nu} + 
                           F_\ep(p_1^2,p_2^2,P^2) \, 
\ep_{\mu\nu\alpha\beta}\, p_1^\alpha\,p_2^\beta 
\Bigr].
\eq
If $h= H$, the Standard Model CP-even Higgs boson, then $F_{\ep} = 0$
and $G= g^3\,s^2_{\theta}$, where $g$ is the bare $SU(2)\,$-coupling constant and 
$s_{\theta}=\sin \theta$ ($c_{\theta}=\cos \theta$) is the sine (cosine) of the 
bare weak-mixing angle. 
The form factors $F_{\ssD}$, $F_{\ssP}^{(ij)}$ and $F_\epsilon$ are functions of 
the off-shell kinematical invariants $p_1^2$, $p_2^2$ and $P^2$; 
they have to be extracted with suitable projection operators and can be 
subsequently evaluated for on-shell external momenta, $p_1^2=p_2^2=0$ and 
$P^2=-\mhs$, where $\mh$ is the on-shell Higgs-boson mass. 
A general framework for projection operators is discussed in appendix~\ref{proppo}
following the approach of Ref.~\cite{Binoth:2002qh}.

Note that the amplitude for the decay of the SM Higgs boson into two gluons, 
$H\to gg$, is obtained from \eqn{AMP:ffactors1} and \eqn{AMP:ffactors}, 
replacing ${\cal Z}_A$ with ${\cal Z}_g$, the gluon wave-function
renormalization factor, and introducing the appropriate color indices
for the gluon fields; in addition, the overall factor $g^3\, s_\theta^2$
of \eqn{AMP:ffactors} is replaced by $g\, g_\ssS^2$, where $g_\ssS$ is the
$SU(3)\,$-coupling constant of strong interactions.

\begin{figure}[ht]
\vspace{-0.3cm}
$$
\scalebox{0.8}{
\begin{picture}(100,40)(-5,0)
 \SetWidth{1.1}
 \DashLine(-5,0)(25,0){3}                \Text(5,5)[cb]{$H$}
 \Photon(59,-20)(90,-20){2}{6}           \Text(95,-30)[cb]{$\gamma$}
 \Photon(59,20)(90,20){2}{6}             \Text(95,24)[cb]{$\gamma$}
 \ArrowLine(25,0)(59,20)                 \Text(40,15)[cb]{$t$}
 \ArrowLine(59,20)(59,-20)               \Text(67,-3)[cb]{$t$}
 \ArrowLine(59,-20)(25,0)                \Text(40,-22)[cb]{$t$}
\end{picture}
}
\qquad\qquad\qquad
\scalebox{0.8}{
\begin{picture}(100,40)(-5,0)
 \SetWidth{1.1}
 \DashLine(-5,0)(25,0){3}                \Text(5,5)[cb]{$H$}
 \Photon(59,-20)(90,-20){2}{6}           \Text(95,-30)[cb]{$\gamma$}
 \Photon(59,20)(90,20){2}{6}             \Text(95,24)[cb]{$\gamma$}
 \Photon(25,0)(59,20){2}{6}              \Text(40,15)[cb]{$W$}
 \Photon(59,20)(59,-20){2}{6}            \Text(70,-3)[cb]{$W$}
 \Photon(59,-20)(25,0){2}{6}             \Text(40,-24)[cb]{$W$}
\end{picture}
}
\qquad\qquad\qquad
\scalebox{0.8}{
\begin{picture}(100,40)(-5,0)
 \SetWidth{1.1}
 \DashLine(-5,0)(25,0){3}                \Text(5,5)[cb]{$H$}
 \PhotonArc(45,0)(20,0,360){2}{20}       \Text(45,27)[cb]{$W$}
                                         \Text(45,-34)[cb]{$W$}
 \Photon(65,0)(90,20){2}{6}              \Text(95,24)[cb]{$\gamma$}
 \Photon(65,0)(90,-20){2}{6}             \Text(95,-30)[cb]{$\gamma$}
\end{picture}
}
$$
\vspace{0.2cm}
\caption{Examples of one-loop diagrams for the decay $H\to\gamma\gamma$. Other 
bosonic diagrams, not shown here, contain charged unphysical 
Higgs-Kibble scalar and Faddeev-Popov ghost fields.
For $H\to g g$, only fermionic diagrams appear.}
\label{Ste1Loop}
\end{figure}
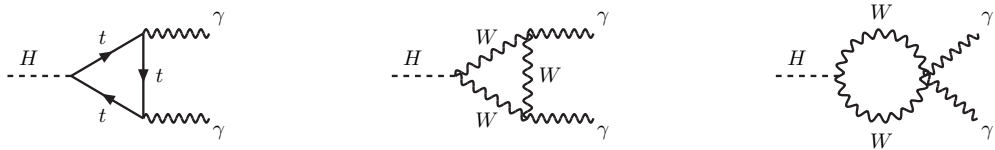

Because of the absence of a tree-level $H \gamma \gamma(gg)$ coupling in the 
SM, the lowest-order amplitude is generated by one-loop fermionic 
and bosonic diagrams, as shown in \fig{Ste1Loop}; 
note that the virtual particles do not decouple also for masses larger 
than the Higgs one, since their coupling to the Higgs boson grows with 
their mass. 

Having singled out the $s_{\theta}$ dependence due to the couplings of the 
external photons, the one-loop form factors do not show any residual dependence 
on $\theta$, because one-loop diagrams containing $Z$ bosons or neutral 
unphysical Higgs-Kibble scalar and Faddeev-Popov ghost fields do not show up. 
Therefore, form factors at one loop contain only the bare masses of the $W$ 
boson, $m_{\ssW}$, the Higgs boson, $m_{\ssH}$, and the fermions, $m_{f}$. 
This simple observation will allow in the following for a straightforward 
implementation of two different renormalization schemes.
On the one hand, the on-shell masses of the $W$ boson, $\mw$, the Higgs boson, $\mh$, and the fermions, $M_f$, will be part of any 
renormalization scheme, and the associated derivatives
of the one-loop form factors will have to be evaluated. On the other hand, $g$ 
and $s_{\theta}$, collected as simple pre-factors in \eqn{AMP:ffactors}, 
will be connected to $G_\ssF$, the Fermi-coupling constant, and $\alpha$, the 
fine-structure constant, or $\mz$, the on-shell $Z$-boson mass.
The choice of the two different input-parameter sets ($G_\ssF, \alpha$) or
($G_\ssF, \mz$) will define our renormalization scheme.
Finally, the $SU(3)$-coupling constant $g_\ssS$ will be related to the 
strong-coupling constant $\alpha_\ssS(\mu_\ssR^2)$ evaluated at the appropriate
renormalization scale $\mu_\ssR$.

\begin{figure}[ht]
\vspace{-0.3cm}
$$
\scalebox{0.8}{
\begin{picture}(140,40)(-5,0)
 \SetWidth{1.1}
 \DashLine(-5,0)(25,0){3}                \Text(5,5)[cb]{$H$}
 \Photon(95,-20)(125,-20){2}{6}          \Text(130,-30)[cb]{$\gamma$}
 \Photon(95,20)(125,20){2}{6}            \Text(130,24)[cb]{$\gamma$}
 \ArrowLine(25,0)(59,20)                 \Text(40,15)[cb]{$t$}
 \ArrowLine(59,20)(59,-20)               \Text(67,-3)[cb]{$b$}
 \ArrowLine(59,-20)(25,0)                \Text(40,-22)[cb]{$t$}
 \Photon(59,-20)(95,-20){2}{7}           \Text(78,-33)[cb]{$W$}
 \Photon(59,20)(95,20){2}{7}             \Text(78,26)[cb]{$W$}
 \Photon(95,-20)(95,20){2}{7}            \Text(105,-3)[cb]{$W$}
\end{picture}
}
\qquad\qquad
\scalebox{0.8}{
\begin{picture}(140,40)(-5,0)
 \SetWidth{1.1}
 \DashLine(-5,0)(25,0){3}                \Text(5,5)[cb]{$H$}
 \Photon(95,-20)(125,-20){2}{6}          \Text(130,-30)[cb]{$\gamma$}
 \Photon(95,20)(125,20){2}{6}            \Text(130,24)[cb]{$\gamma$}
 \ArrowLine(25,0)(59,20)                 \Text(40,15)[cb]{$t$}
 \Photon(59,20)(59,-20){2}{7}            \Text(68,-3)[cb]{$Z$}
 \ArrowLine(59,-20)(25,0)                \Text(40,-22)[cb]{$t$}
 \ArrowLine(95,-20)(59,-20)              \Text(76,-32)[cb]{$t$}
 \ArrowLine(59,20)(95,20)                \Text(76,26)[cb]{$t$}
 \ArrowLine(95,20)(95,-20)               \Text(102,-3)[cb]{$t$}
\end{picture}
}
\qquad\qquad
\scalebox{0.8}{
\begin{picture}(140,40)(-5,0)
 \SetWidth{1.1}
 \DashLine(-5,0)(25,0){3}                \Text(5,5)[cb]{$H$}
 \Photon(95,-20)(125,-20){2}{6}          \Text(130,-30)[cb]{$\gamma$}
 \Photon(95,20)(125,20){2}{6}            \Text(130,24)[cb]{$\gamma$}
 \ArrowLine(25,0)(59,20)                 \Text(40,15)[cb]{$t$}
 \ArrowLine(59,-20)(25,0)                \Text(40,-22)[cb]{$t$}
 \Photon(95,-20)(59,-20){2}{7}           \Text(76,-33)[cb]{$W$}
 \ArrowLine(59,20)(95,20)                \Text(76,26)[cb]{$b$}
 \ArrowLine(95,20)(59,-20)               \Text(94,7)[cb]{$b$}
 \Photon(59,20)(72.5,5){2}{3}            \Text(98,-12)[cb]{$W$}
 \Photon(81.5,-5)(95,-20){2}{3}
\end{picture}
}
$$
\vspace{0.2cm}
\caption{Representative two-loop diagrams giving a non-vanishing contribution 
to the form factor $F_{\ep}$, because of the presence of an axial coupling 
between a massive vector boson and a virtual fermion pair.}
\label{SteGam5}
\end{figure}
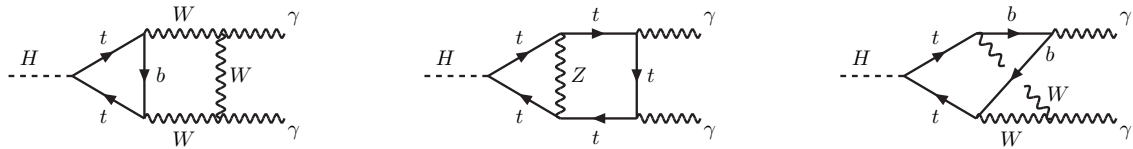

The form factor $F_{\ep}$ is present at two loops in diagrams containing a 
fermion sub-loop, as shown in \fig{SteGam5}, but it does not arise at one 
loop due to the lacking of axial fermion couplings. Due to CP invariance
the contribution of $F_\epsilon$ vanishes in the total. 
Therefore, we can circumvent the notorious problems associated with the 
definition of the $\gamma_5$ matrix in dimensional 
regularization~\cite{'t Hooft:1972fi} (see a detailed discussion
in Ref.~\cite{Jegerlehner:2000dz}) employing a 
completely anticommuting $\gamma_5$.

Finally, we observe that Bose symmetry sets four constraints on the other form 
factors for off-shell external particles,
\bqa
       F_{\ssD}\left(p_1^2,p_2^2,P^2\right) \,&=& 
\,\,\,\, F_{\ssD}\,\,\left(p_2^2,p_1^2,P^2\right),\qquad
F^{(11)}_{\ssP}\left(p_1^2,p_2^2,P^2\right) \,= \,      
F^{(22)}_{\ssP}\left(p_2^2,p_1^2,P^2\right),
\nonumber \\
F^{(12)}_{\ssP}\left(p_1^2,p_2^2,P^2\right)\,&=& \,     
F^{(12)}_{\ssP}\left(p_2^2,p_1^2,P^2\right),\qquad
F^{(21)}_{\ssP}\left(p_1^2,p_2^2,P^2\right)\,= \,       
F^{(21)}_{\ssP}\left(p_2^2,p_1^2,P^2\right).
\eqa
Although $F^{(11)}_{\ssP}$, $F^{(22)}_{\ssP}$ and $F^{(12)}_{\ssP}$ are 
irrelevant for the calculation of the amplitude due to the condition 
$e_i \cdot p_i=0$ for on-shell photons (gluons), they play an essential 
role in computing the relevant Ward-Slavnov-Taylor identities (hereafter WSTIs)
for $H\to\gamma \gamma (g g)$.

Before discussing WSTIs in more detail, let us define what we use for the 
Higgs-boson mass and wave-function renormalization factor.
\subsection{Higgs-boson mass and wave-function renormalization 
\label{Hm&WFR}}
We evaluate the amplitude for $P^2=-\mhs$, where $\mh$ is the 
on-shell Higgs-boson mass, defined as the location of the pole of the 
real part of the associated Dyson-resummed propagator $\overline{\Delta}_\ssH$,
\bq
\label{AMP:mass}
\text{Re}\, \left[\overline{\Delta}_\ssH(p^2)\right]^{-1} 
|_{p^2=-\mhs}\,=\,0 
\qquad\Rightarrow\qquad
\mhs\,=\,m_{\ssH}^2\,-\,\text{Re}\,\Sigma_\ssH(-M_{\ssH}^2).
\eq
Here $m_{\ssH}$ is the bare mass and $\Sigma_\ssH$ is the sum of 
all one-particle irreducible (1PI) diagrams for the Higgs self-energy. For 
definiteness, $\mh$ would be the experimental mass 
extracted from a fit of the hypothetical Higgs-resonance line shape 
obtained by means of a Breit-Wigner function with a running width,
\bq
\overline{\Delta}_\ssH(p^2)\,\propto\,
\frac{1}{p^2\,+\,\mhs\,+\,i\,p^2\, \Gamma_\ssH\,\slash\, \mh}.
\eq

It is well-known that the on-shell definition of mass for an unstable particle 
is gauge-parameter dependent beyond one loop, and it should be replaced by the 
complex pole, $s_\ssH$ in our case, including real and imaginary parts 
of $\overline{\Delta}_\ssH$,
\bq
\left[\overline{\Delta}_\ssH(p^2)\right]^{-1} |_{p^2=-s_\ssH} \,=\,0 
\qquad\Rightarrow\qquad
s_\ssH\,=\,m_{\ssH}^2\, -\, \Sigma_\ssH(-s_\ssH).
\eq
As it is often stated~\cite{Stuart:1991xk}, the complex pole is a 
property of the ${\cal S}$ matrix, and therefore gauge-parameter independent 
at all orders in perturbation theory; the proof of this property relies on the 
Nielsen identities and can be found in Ref.~\cite{Gambino:1999ai}. Therefore, 
fixed-order computations beyond one loop or resummation-improved 
one-loop predictions require to perform mass renormalization 
dropping the concept of on-shell masses and employing complex poles; for
a thorough discussion at one and two loops we refer to Ref.~\cite{Actis:2006rc}.

Concerning $H\to\gamma\gamma$ at two loops, the absence of a tree-level 
amplitude implies that Higgs-mass renormalization has to be performed at one loop: 
this suggests that an on-shell definition of the Higgs-boson mass, being 
gauge-parameter independent at one loop, proves adequate for our purposes. 
Nevertheless, we will show in the following that a naive implementation of 
on-shell mass renormalization for the Higgs boson breaks the standard
two-loop WSTI for $H\to\gamma\gamma$ above the $WW$-production threshold
(light-fermion masses are neglected).
On the one hand, the bare identity develops an imaginary part above the
$WW$ threshold; on the other hand, the Higgs self-energy appearing 
in \eqn{AMP:mass}, $\Sigma_\ssH$, develops an imaginary part at one loop 
for $\mh > 2 \mw$, but on-shell mass renormalization selects its real 
part only. This introduces a mismatch between real and imaginary parts at the 
level of the renormalized WSTI. 
Therefore, the usual statement that renormalization schemes with complex poles or 
on-shell masses are equivalent at one loop should be taken with some caution; for 
$H\to \gamma \gamma$ at two loops, the two schemes are equivalent only below 
the $WW$-production threshold.

A second delicate point concerns the use of a Higgs-boson 
wave-function renormalization (hereafter WFR) factor introduced 
{\emph{{\`a} la}} Lehmann-Symanzik-Zimmermann (LSZ)~\cite{Lehmann:1954rq}. 
The LSZ formalism, in fact, is unambiguously defined for stable particles, but 
it requires some care when external unstable particles appear.

Let us consider first the case of the photon in QED. Here, the LSZ WFR factor 
is fixed by the condition that the transverse part of the photon 
Dyson-resummed propagator, $\overline{\Delta}^\ssT_\ssA$, dressed by its WFR factor 
${\cal Z}_\ssA$, has unity residue at the pole in the $p^2$ plane,
\bq
\label{AMP:wfrA}
{\cal Z}_\ssA\, =\, \frac{\partial}{\partial p^2}\, 
\left[ \overline{\Delta}^\ssT_\ssA(p^2) \right]^{-1}|_{p^2=0} 
\quad \Rightarrow \quad 
{\cal Z}_\ssA\, =\, 1\, -\, \Sigma_{\ssA,p}(0),\quad
\Sigma_{\ssA,p}(p^2)\,=\, \frac{\partial\, \Sigma_\ssA(p^2)}{\partial p^2},
\eq
where $\Sigma_\ssA$ is the sum of all 1PI diagrams for the 
transverse part of the photon self-energy (we will show in the following that 
the presence of a photon$\slash$$Z$-boson mixing in the SM does 
not change the structure of \eqn{AMP:wfrA}). Note that it has been shown long 
ago~\cite{BialynickiBirula:1971sn,'tHooft:1972ue} that the LSZ definition of 
WFR factors leads to gauge-parameter independent ${\cal S}$-matrix elements 
and respects unitarity.

For a stable particle, the pole of the Dyson-resummed propagator is real;
being the Higgs boson an unstable particle, the pole of its Dyson-resummed 
propagator in the $p^2$ plane is complex, and the Higgs WFR factor is 
not unambiguously defined. We define it requiring that the real part of the 
Higgs-boson Dyson-resummed propagator, dressed by its WFR factor 
${\cal Z}_\ssH$, has unity residue at the on-shell Higgs-boson mass,
\bq
\label{AMP:WFRhiggs}
{\cal Z}_\ssH = \text{Re}\, \frac{\partial}{\partial p^2}\, 
\left[ \overline{\Delta}_\ssH(p^2) \right]^{-1}|_{p^2=-\mhs} 
\quad \Rightarrow \quad 
{\cal Z}_\ssH\, =\, 1\, -\, \text{Re}\, \Sigma_{\ssH,p}(-\mhs),\quad 
\Sigma_{\ssH,p}(p^2)\,=\,\frac{\partial\, \Sigma_\ssH(p^2)}{\partial p^2}.
\eq
Other definitions, involving the real part of the derivative of the propagator 
at the complex pole~\cite{Freitas:2002ja}, or the complete derivative, 
including imaginary parts~\cite{Denner:2005fg}, do not look appropriate for 
our computation, which deals with the ideal situation of an asymptotic state 
containing the Higgs boson, whose momentum is required to be real.

Note, however, that the definition of a real WFR factor at the on-shell mass 
is a possible source of inconsistencies. The authors of 
Ref.~\cite{Denner:1997kq}, for example, have shown that unitarity-breaking 
terms can show up in the final form for the amplitude. From a formal 
perspective, indeed, transition amplitudes have to be defined in terms of 
asymptotic states containing stable particles only. To this respect, the 
necessary modifications to the theory have been described in 
Ref.~\cite{Veltman:1963th}, where unitarity and causality of the resulting 
${\cal S}$ matrix are proven.

In the following, we will use a compromise. At first, we will show that, 
concerning $H\to \gamma \gamma (gg)$ at two loops, a naive use of 
\eqn{AMP:WFRhiggs} leads to unphysical singularities at thresholds. 
Then, we will modify our renormalization scheme introducing complex 
poles.
\subsection{Ward-Slavnov-Taylor identities}
Ward-Slavnov-Taylor identities (WSTIs)~\cite{Ward:1950xp} for 
$H \to \gamma \gamma ( g g)$ are essential for organizing our computation. They 
allow to perform severe checks on the algebraic structure of the amplitude, and
set strong constraints on the number of independent form factors.
Although in the following we will focus on the $H\to \gamma \gamma$ process,
it is evident that analogous considerations can be applied to
$H\to g g$.

Let us consider the simplest case, the so-called doubly contracted WSTI 
with an on-shell Higgs boson and both off-shell photons. After 
replacing in \eqn{AMP:ffactors1} the photon polarization vectors with 
their associated four-momenta, we get
\bqa
\label{AMP:WI1}
{\cal W}{\cal I}_{dc}(p_1^2,p_2^2)\,&=&\,
{\cal Z}_\ssH^{-1\slash 2}\,\,p_1^\mu\,\,p_2^\nu\,\,
{\cal A}_{\mu\nu}(p_1^2,p_2^2,-\mhs)
\nonumber\\
&=&- \frac{g^3 s_{\theta}^2}{32 \pi^2}{\cal Z}_\ssH^{-1\slash 2}
\Bigl\{ \left(\mhs+p_1^2+p_2^2\right) \Bigl[ 
                            F_\ssD(p_1^2,p_2^2,-\mhs)+ 
       \sum_{i=1}^2 p_i^2 F^{(ii)}_\ssP(p_1^2,p_2^2,-\mhs)   \nonumber\\ 
&& - \frac{1}{2} \left(\mhs+p_1^2+p_2^2\right)
                     F^{(21)}_\ssP(p_1^2,p_2^2,-\mhs) \Bigr] - 
2 p_1^2 p_2^2    F^{(12)}_\ssP(p_1^2,p_2^2,-\mhs) \Bigr\} = 0.
\eqa
Here the Higgs boson is assumed on its mass shell, $P^2=-\mhs$, and 
provided with its WFR factor ${\cal Z}_\ssH^{- 1\slash 2}$, whereas the form 
factors are evaluated for off-shell photons and photon WFR factors are 
consistently not included. After showing that all form factors are 
regular for on-shell photons, one can set $p_1^2=p_2^2=0$ and get the 
well-known constraint
\bq
\label{AMP:wwi}
{\cal W}{\cal I}_{dc}(0,0)\,=\,
-\,\frac{g^3\,s_{\theta}^2}{32\,\pi^2}\,{\cal Z}_\ssH^{-1\slash 2}\,\mhs\,
\Bigl[\, F_\ssD(0,0,-\mhs)\,-\,
\frac{\mhs}{2}\,F^{(21)}_\ssP(0,0,-\mhs)\,\Bigr]\,=\,0,
\eq
which shows that the $H \to \gamma \gamma$ amplitude can be described through 
a single form factor. Before attempting any evaluation of the amplitude, we 
extract both $F_\ssD$ and $F_\ssP^{(21)}$ employing projection operators for 
off-shell photon momenta. Next, we check that form factors are regular for 
on-shell photons, and we compute the WSTI, verifying that 
\eqn{AMP:wwi} holds  at the algebraic level. This provides a strong test 
on several procedures used throughout the calculation of the amplitude, 
like the reduction of tensor integrals to scalar ones. 
Note, however, that the WSTI, being satisfied at the algebraic level, 
does not provide any information on the analytic structure of the loop 
integrals themselves. Here, a severe  test on the result will be supplemented 
by the extraction of all collinear logarithms: they have to cancel in the full 
amplitude, and indeed we will show that they cancel.

Two additional relations among the form factors can be readily obtained by 
considering simply contracted WSTIs, where one off-shell 
photon leg is contracted with its four-momentum, and the other one is saturated 
with the polarization vector, put on the mass shell and provided with its WFR 
factor,
\bqa
\label{AMP:WI3}
{\cal W}{\cal I}_{sc,1}(p_1^2)\,&=&\,
{\cal Z}_\ssA^{-1\slash 2}\, {\cal Z}_\ssH^{-1\slash 2}\, 
p_1^\mu\, e_2^\nu\, {\cal A}_{\mu\nu}(p_1^2,0,-\mhs)\nonumber\\
&=&\, \frac{g^3\,s_{\theta}^2}{16\, \pi^2}\, (p_1\cdot e_2)\,
{\cal Z}_\ssA^{-1\slash 2}\, {\cal Z}_\ssH^{-1\slash 2}\, \Bigl[\,
F_{\ssD}(p_1^2,0,-\mhs)\,+\, 
p_1^2\,F^{(11)}_{\ssP}(p_1^2,0,-\mhs)\nonumber\\
&&\,-\, \frac{1}{2} (\mhs+p_1^2)\,F^{(21)}_{\ssP}(p_1^2,0,-\mhs)\,\Bigr]
\,=\,0,\\ 
\nonumber\\
\label{AMP:WI3b}
{\cal W}{\cal I}_{sc,2}(p_2^2)\,&=&\,
{\cal Z}_\ssA^{-1\slash 2}\, {\cal Z}_\ssH^{-1\slash 2}\,
e_1^\mu\, p_2^\nu\, {\cal A}_{\mu\nu}(0,p_2^2,-\mhs)\nonumber\\
&=&\,\frac{g^3\, s_{\theta}^2}{16\, \pi^2}\, (p_2 \cdot e_1)\,
{\cal Z}_\ssA^{-1\slash 2}\, {\cal Z}_\ssH^{-1\slash 2}\, \Bigl[\,
F_{\ssD}(0,p_2^2,-\mhs)\,+\, 
p_2^2\,F^{(22)}_{\ssP}(0,p_2^2,-\mhs)\nonumber\\
&&\,-\,\frac{1}{2}  
(\mhs+p_2^2)\, F^{(21)}_{\ssP}(0,p_2^2,-\mhs)\,\Bigr]
\,=\,0.
\eqa

\begin{figure}[ht]
\vspace{-0.3cm}
$$
\raisebox{0.1cm}{\scalebox{0.8}{
\begin{picture}(102,30)(-5,0)
 \SetWidth{1.1}
 \DashLine(-5,0)(25,0){3}                \Text(10,5)[cb]{$H$}
 \GBoxc(-5,0)(7,7){0.9}
 \Photon(56,-14)(90,-14){2}{6}           \Text(80,-27)[cb]{$\gamma$}
 \GCirc(90,-14){4}{0}                
 \Photon(56,14)(90,14){2}{6}             \Text(80,21)[cb]{$\gamma$}
 \GCirc(90,14){4}{0}                
 \GCirc(45,0){18}{0.7}                
\end{picture}
}}
=
\;
0
\qquad\qquad\qquad
\raisebox{0.1cm}{\scalebox{0.8}{
\begin{picture}(102,30)(-5,0)
 \SetWidth{1.1}
 \DashLine(-5,0)(25,0){3}                \Text(10,5)[cb]{$H$}
 \GBoxc(-5,0)(7,7){0.9}
 \Photon(56,-14)(90,-14){2}{6}           \Text(80,-27)[cb]{$\gamma$}
 \GCirc(90,-14){4}{0}                
 \Photon(56,14)(90,14){2}{6}             \Text(80,21)[cb]{$\gamma$}
 \GBoxc(90,14)(7,7){0.9}                
 \GCirc(45,0){18}{0.7}                
\end{picture}
}}
=
\;
0
\qquad\qquad\qquad
\raisebox{0.1cm}{\scalebox{0.8}{
\begin{picture}(102,30)(-5,0)
 \SetWidth{1.1}
 \DashLine(-5,0)(25,0){3}                \Text(10,5)[cb]{$H$}
 \GBoxc(-5,0)(7,7){0.9}
 \Photon(56,-14)(90,-14){2}{6}           \Text(80,-27)[cb]{$\gamma$}
 \GBoxc(90,-14)(7,7){0.9}                
 \Photon(56,14)(90,14){2}{6}             \Text(80,21)[cb]{$\gamma$}
 \GCirc(90,14){4}{0}                
 \GCirc(45,0){18}{0.7}                
\end{picture}
}}
=
\;
0
$$
\vspace{-0.2cm}
\caption{Diagrammatic representation of \eqn{AMP:WI1} (doubly contracted 
WST identity), \eqn{AMP:WI3}  and \eqn{AMP:WI3b} 
(simply contracted 
WST identities). Physical sources are denoted by a gray box, contracted ones, 
for photons, by a  black circle. In the latter case the on-shell 
relation $p_i^2=0$ is relaxed.}
\label{SteWI1}
\end{figure}
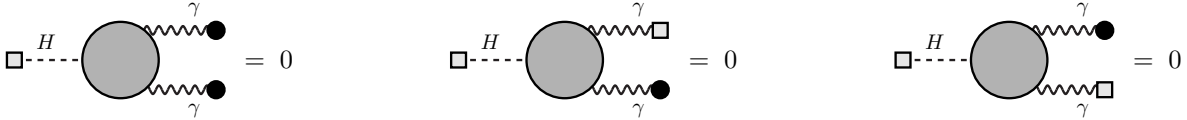

In \fig{SteWI1} we show diagrammatically \eqn{AMP:WI1}, 
\eqn{AMP:WI3} and \eqn{AMP:WI3b}. In addition, we introduce the notions of: 
1) physical source: the external on-shell leg is multiplied by the 
appropriate WFR factor, and, for the case of photons, supplemented by the 
polarization vector; 
2) contracted source: the external off-shell photon leg is 
contracted with its four-momentum.


The computation of a doubly or simply contracted WSTI for an 
on-shell Higgs boson beyond the leading order approximation requires to 
pay special attention to the interplay between the bare and on-shell 
Higgs-boson masses. Let us consider the doubly contracted WSTI of 
\eqn{AMP:wwi} at two loops. The full set of diagrams can be organized in 
three classes: 1)
two-loop diagrams where the Higgs-boson WFR factor is set to unity, 
${\cal Z}_\ssH=1$, and the bare Higgs-boson mass is identified with its 
on-shell experimental value, $m_{\ssH}=\mh$;
2)
one-loop diagrams supplemented with ${\cal Z}_\ssH$ evaluated at one loop 
by means of \eqn{AMP:WFRhiggs}, where we employ again the tree-level 
relation $m_{\ssH}=\mh$;
3)
one-loop diagrams where ${\cal Z}_\ssH=1$, and we replace the tree-level 
identity between the bare and the on-shell Higgs-boson masses with the solution 
of the Higgs-boson mass renormalization equation of \eqn{AMP:mass}, with 
the Higgs-boson self-energy $\Sigma_\ssH$ evaluated at one loop.

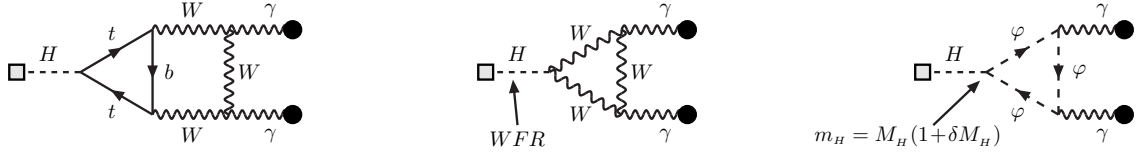
\begin{figure}[ht]
$$
\scalebox{0.8}{
\begin{picture}(140,40)(-5,0)
 \SetWidth{1.1}
 \DashLine(-5,0)(25,0){3}                \Text(10,5)[cb]{$H$}
 \GBoxc(-5,0)(7,7){0.9}
 \Photon(95,-20)(125,-20){2}{6}          \Text(115,-33)[cb]{$\gamma$}
 \GCirc(125,-20){4}{0}                
 \Photon(95,20)(125,20){2}{6}            \Text(115,26)[cb]{$\gamma$}
 \GCirc(125,20){4}{0}                
 \ArrowLine(25,0)(59,20)                 \Text(40,15)[cb]{$t$}
 \ArrowLine(59,20)(59,-20)               \Text(67,-3)[cb]{$b$}
 \ArrowLine(59,-20)(25,0)                \Text(40,-22)[cb]{$t$}
 \Photon(59,-20)(95,-20){2}{7}           \Text(78,-33)[cb]{$W$}
 \Photon(59,20)(95,20){2}{7}             \Text(78,26)[cb]{$W$}
 \Photon(95,-20)(95,20){2}{7}            \Text(105,-3)[cb]{$W$}
\end{picture}
}
\qquad\qquad\qquad
\scalebox{0.8}{
\begin{picture}(100,40)(-5,0)
 \SetWidth{1.1}
 \DashLine(-5,0)(25,0){3}                \Text(10,5)[cb]{$H$}
 \GBoxc(-5,0)(7,7){0.9}
 \Photon(59,-20)(90,-20){2}{6}          \Text(80,-33)[cb]{$\gamma$}
 \GCirc(90,-20){4}{0}                
 \Photon(59,20)(90,20){2}{6}            \Text(80,26)[cb]{$\gamma$}
 \GCirc(90,20){4}{0}                
 \Photon(25,0)(59,20){2}{6}                 \Text(40,15)[cb]{$W$}
 \Photon(59,20)(59,-20){2}{6}               \Text(69,-3)[cb]{$W$}
 \Photon(59,-20)(25,0){2}{6}                \Text(40,-23)[cb]{$W$}
 \Text(10,-35)[cb]{$WFR$}
 \LongArrow(10,-25)(8,-5)
\end{picture}
}
\qquad\qquad\qquad\qquad
\scalebox{0.8}{
\begin{picture}(100,40)(-5,0)
 \SetWidth{1.1}
 \DashLine(-5,0)(25,0){3}               \Text(10,5)[cb]{$H$}
 \GBoxc(-5,0)(7,7){0.9}
 \Photon(59,-20)(90,-20){2}{6}          \Text(80,-33)[cb]{$\gamma$}
 \GCirc(90,-20){4}{0}                
 \Photon(59,20)(90,20){2}{6}            \Text(80,26)[cb]{$\gamma$}
 \GCirc(90,20){4}{0}                
 \DashArrowLine(25,0)(59,20){4}         \Text(40,15)[cb]{$\varphi$}
 \DashArrowLine(59,20)(59,-20){4}       \Text(69,-3)[cb]{$\varphi$}
 \DashArrowLine(59,-20)(25,0){4}        \Text(40,-23)[cb]{$\varphi$}
 \Text(-12,-35)[cb]{$m_\ssH= \mh(1\!+\!\delta \mh)$}
 \LongArrow(0,-25)(21,-5)
\end{picture}
}
$$
\vspace{0.4cm}
\caption{Representative diagrams for the doubly contracted WSTI of 
\eqn{AMP:WI1} at next-to-leading order. As in \fig{SteWI1}, black 
circles correspond to off-shell photons contracted with their four-momenta, 
and a gray box stands for a physical Higgs-boson source. Inclusion of the 
Higgs-boson WFR factor ${\cal Z}_\ssH$ in the second diagram and Higgs-boson 
mass renormalization in the third one are both performed at one loop.}
\label{SteWiReno}
\end{figure}

In \fig{SteWiReno} we show a representative diagram for each class; let 
us consider in detail their expansions in terms of the bare $SU(2)$-coupling 
constant $g$. The first diagram is ${\cal O}(g^5)$; the second one 
is ${\cal O}(g^3)$, and gives an ${\cal O}(g^5)$ contribution once provided 
with the WFR factor ${\cal Z}_\ssH$ at one loop. The most subtle point is 
related to the third diagram, which is ${\cal O}(g^3)$ and is multiplied by a 
tree-level WFR factor, ${\cal Z}_\ssH=1$. However, it entails a vertex where 
one Higgs boson and two unphysical charged Higgs-Kibble scalar fields are 
coupled proportionally to $m_{\ssH}^2$. Since we put the Higgs boson on its 
mass shell, the bare mass $m_{\ssH}$ has to be removed, introducing everywhere
the on-shell value $M_{\ssH}$. Therefore, first we use \eqn{AMP:mass} and 
evaluate $\Sigma_\ssH$ at order ${\cal O}(g^2)$; next, we replace the solution 
in the third diagram and obtain  an ${\cal O}(g^5)$ contribution, which 
proves essential in fulfilling the WSTI.

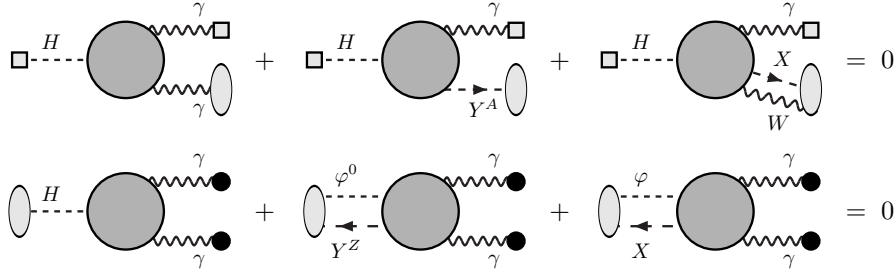
\begin{figure}[ht]
\vspace{-0.3cm}
\bqas
\raisebox{0.1cm}{\scalebox{0.8}{
\begin{picture}(105,35)(-5,0)
 \SetWidth{1.1}
 \DashLine(-5,0)(25,0){3}                \Text(10,5)[cb]{$H$}
 \GBoxc(-5,0)(7,7){0.9}
 \Photon(56,-14)(90,-14){2}{6}           \Text(80,-27)[cb]{$\gamma$}
 \GOval(90,-14)(12,5)(0){0.9}
 \Photon(56,14)(90,14){2}{6}             \Text(80,21)[cb]{$\gamma$}
 \GBoxc(90,14)(7,7){0.9}                
 \GCirc(45,0){18}{0.7}                
\end{picture}
}}
+
\quad
\raisebox{0.1cm}{\scalebox{0.8}{
\begin{picture}(105,35)(-5,0)
 \SetWidth{1.1}
 \DashLine(-5,0)(25,0){3}                \Text(10,5)[cb]{$H$}
 \GBoxc(-5,0)(7,7){0.9}
 \DashArrowLine(56,-14)(90,-14){4}       \Text(75,-27)[cb]{$Y^A$}
 \GOval(90,-14)(12,5)(0){0.9}
 \Photon(56,14)(90,14){2}{6}             \Text(80,21)[cb]{$\gamma$}
 \GBoxc(90,14)(7,7){0.9}                
 \GCirc(45,0){18}{0.7}                
\end{picture}
}}
+
\quad
\raisebox{0.1cm}{\scalebox{0.8}{
\begin{picture}(105,35)(-5,0)
 \SetWidth{1.1}
 \DashLine(-5,0)(25,0){3}                \Text(10,5)[cb]{$H$}
 \GBoxc(-5,0)(7,7){0.9}
 \Photon(56,-14)(93,-24){2}{6}           \Text(75,-33)[cb]{$W$}
 \DashArrowLine(56,-4)(90,-14){4}        \Text(77,-3)[cb]{$X$}
 \GOval(90,-14)(12,5)(0){0.9}
 \Photon(56,14)(90,14){2}{6}             \Text(80,21)[cb]{$\gamma$}
 \GBoxc(90,14)(7,7){0.9}                
 \GCirc(45,0){18}{0.7}                
\end{picture}
}}
&=&
\;
0
\\[0.9cm]
\raisebox{0.1cm}{\scalebox{0.8}{
\begin{picture}(105,30)(-5,0)
 \SetWidth{1.1}
 \DashLine(-5,0)(25,0){3}                \Text(10,5)[cb]{$H$}
 \GOval(-5,0)(12,5)(0){0.9}
 \Photon(56,-14)(90,-14){2}{6}           \Text(80,-27)[cb]{$\gamma$}
 \GCirc(90,-14){4}{0}                
 \Photon(56,14)(90,14){2}{6}             \Text(80,21)[cb]{$\gamma$}
 \GCirc(90,14){4}{0}                
 \GCirc(45,0){18}{0.7}                
\end{picture}
}}
+
\quad
\raisebox{0.1cm}{\scalebox{0.8}{
\begin{picture}(105,30)(-5,0)
 \SetWidth{1.1}
 \DashLine(-5,7)(25,7){3}                \Text(10,12)[cb]{$\varphi^0$}
 \DashArrowLine(25,-7)(-5,-7){4}         \Text(10,-22)[cb]{$Y^Z$}
 \GOval(-5,0)(12,5)(0){0.9}
 \Photon(56,-14)(90,-14){2}{6}           \Text(80,-27)[cb]{$\gamma$}
 \GCirc(90,-14){4}{0}                
 \Photon(56,14)(90,14){2}{6}             \Text(80,21)[cb]{$\gamma$}
 \GCirc(90,14){4}{0}                
 \GCirc(45,0){18}{0.7}                
\end{picture}
}}
+
\quad
\raisebox{0.1cm}{\scalebox{0.8}{
\begin{picture}(105,30)(-5,0)
 \SetWidth{1.1}
 \DashLine(-5,7)(25,7){3}                \Text(10,12)[cb]{$\varphi$}
 \DashArrowLine(25,-7)(-5,-7){4}         \Text(10,-22)[cb]{$X$}
 \GOval(-5,0)(12,5)(0){0.9}
 \Photon(56,-14)(90,-14){2}{6}           \Text(80,-27)[cb]{$\gamma$}
 \GCirc(90,-14){4}{0}                
 \Photon(56,14)(90,14){2}{6}             \Text(80,21)[cb]{$\gamma$}
 \GCirc(90,14){4}{0}                
 \GCirc(45,0){18}{0.7}                
\end{picture}
}}
&=&
\;
0
\eqas
\vspace{-0.2cm}
\caption{Off-shell WSTIs involving special sources, denoted by gray 
ovals. The special source for the photon field can emit a Faddeev-Popov (FP) 
ghost field (dot-line) and two $W$-boson$\slash$charged FP field 
(wavy$\slash$dot-lines) couples. The special source for the Higgs boson can emit 
three neutral and charged Higgs-Kibble$\slash$FP field 
couples. 
See \eqn{AMP:specSources} for the explicit expressions of the special 
sources. As usual, contracted sources are denoted by black circles and 
physical ones by gray boxes.}
\label{SteWiOff}
\end{figure}

Finally, we stress that both doubly 
and simply contracted WSTIs assume that the Higgs boson is emitted 
by a physical source. However, as shown by 't Hooft and Veltman in their 
seminal paper~\cite{'tHooft:1972ue}, a broader class of WSTIs can be 
obtained when we inspect, at the diagrammatic level, the effect of an 
infinitesimal shift of the gauge-fixing functions, ${\cal C}^a \to 
{\cal C}^a + \epsilon \,\, {\cal R}$, with $\epsilon\to 0$. Here $a$ denotes 
one of the gauge fields and ${\cal R}$ stands for an off-shell source 
emitting one or more off-shell particles. As shown in 
Ref.~\cite{'tHooft:1972ue}, off-shell WSTIs connect physical 
or contracted sources to special sources obtained by: 1)
subjecting ${\cal R}$ to an infinitesimal local gauge transformation;
2) replacing the parameters of the transformation with the related Faddeev-Popov 
ghost fields.
Let us consider a simple example one can derive for $H \to \gamma 
\gamma$. After choosing the  gauge-fixing function  for the photon field 
$A_\mu$ in 't Hooft-Feynman gauge, ${\cal C}^{\ssA}=-\partial^\mu A_\mu$, 
we introduce two off-shell sources emitting one photon or one Higgs boson, 
${\cal R}_{\ssA}= J_{\ssA}^\mu A_\mu$ and ${\cal R}_{\ssH}= J_{\ssH} H$. Next, 
we perform a local $SU(2) \times U(1)$ gauge transformation, and replace 
the infinitesimal gauge parameters of the transformation by the appropriate 
Faddeev-Popov ghost fields. We derive two special sources,
\bqa
\label{AMP:specSources}
{\cal R}_{\ssA}\,&\to&\,J_{\ssA}^\mu\,\Bigl[\, A_\mu\,+\,i\, g\, s_{\theta}\, 
\left(\, X^-\, W_\mu^+\, -\, X^+\, W_\mu^-\, \right)\,-\, 
\partial_\mu\, Y^{\ssA}\, \Bigr],
\nonumber\\
{\cal R}_{\ssH}\,&\to&\,J_{\ssH}\,\Bigl[\, H\, +\,
\frac{g}{2\, c_{\theta}}\, Y^{\ssZ}\, \varphi^0\,
+\, \frac{g}{2}\, \left(\, X^-\, \varphi^+\, +\, X^+\, 
\varphi^-\, \right)\, \Bigr],
\eqa
where $Y^{\ssA}$, $Y^{\ssZ}$, $X^+$ and $X^-$ are the Faddeev-Popov ghost 
fields and $\varphi^0$, $\varphi^+$ and $\varphi^-$ are the unphysical 
Higgs-Kibble 
scalar fields. Off-shell WSTIs for $H\to \gamma \gamma$ are then 
readily obtained connecting at least one special source with physical or 
contracted sources. In \fig{SteWiOff} we show two examples: in the first 
case, the Higgs boson and one photon are emitted by physical sources, and the 
second off-shell photon is connected to a special source; in the second case, 
both off-shell photons are contracted with their four-momenta, and the 
off-shell Higgs boson is emitted by a special source. In this case, WFR factor 
and mass renormalization for the Higgs boson have not to be included.

\subsection{One-loop counterterms}
\label{AMP:tad}
In this section we shortly summarize three aspects of the renormalization 
procedure which are needed before performing finite renormalization: 
tadpole renormalization, diagonalization of the neutral sector and 
ultraviolet counterterms. A detailed analysis can be found in 
Refs.~\cite{Actis:2006ra,Actis:2006rb}.

We fix the gauge for the electroweak sector of the SM Lagrangian
introducing six gauge parameters $\xi_i$ ($i=A,Z,AZ,\varphi^{0},W,\varphi$),
\bqa
\label{eq:th:gfun}
{\cal{L}}^{\EW}_{gf}\, &=& 
\, -\,{\cal{C}}^{+}\,{\cal{C}}^{-}\,-\,\frac{1}{2}\,
\Bigl[\, \left({\cal{C}}^{\ssA}\right)^2\, +
\, \left({\cal{C}}^{\ssZ}\right)^2\,\Bigr],
\nl
{\cal{C}}^{\ssA} =
           -\frac{1}{\gpA} \partial_{\mu} A_{\mu} 
           - \gpAZ \,\partial_{\mu} Z_{\mu},
\qquad
{\cal{C}}^{\ssZ} &=&
           -\frac{1}{\gpZ} \partial_{\mu} Z_{\mu} 
           + \xi_{\varphi^{0}}\, \frac{m_\ssW}{c_\theta}\, \varphi^{0},
\qquad
{\cal{C}}^{\pm} =
           -\frac{1}{\gpW} \partial_\mu W^{\pm}_{\mu} 
           + \xi_{\varphi} \, m_{\ssW}\, \varphi^{\pm}.
\eqa
Here $A_\mu$, $Z_\mu$ and $W_\mu^{\pm}$ are the fields for the photon and
the $Z$ and the $W$ bosons and $\varphi^0$ and $\varphi^{\pm}$ are
the fields for the neutral and charged unphysical Higgs-Kibble scalars.
For the QCD sector, we employ the following choice,
\bqa
{\cal{L}}^{\QCD}_{gf}\, =
-\frac{1}{2}\, \sum_a\, {\cal C}^{\ssG,a}\, {\cal C}^{\ssG,a},\qquad
{\cal C}^{\ssG,a} = -\frac{1}{\xi_{\ssG,a}}\, \partial_\mu \, G_\mu^a,
\eqa
which for $\xi_{\ssG}=1$ reduces to the usual 't Hooft-Feynman gauge.
\subsubsection{Higgs tadpoles}
Tadpoles in spontaneously broken theories have been discussed by many authors 
(see Refs.~\cite{Veltman:1994wz,Bardin:1999ak}). The 
tadpole-renormalization prescription used in this paper will be the 
$\beta_t$ scheme, thoroughly discussed in section~2.3 of 
Ref.~\cite{Actis:2006ra}.

Following notation and conventions of Ref.~\cite{Bardin:1999ak}, the minimal
Higgs sector of the SM is provided by the Lagrangian 
\bq
    \cL_S  = -(D_{\mu} K)^\dagger (D_{\mu} K) -\mu^2 K^\dagger K
                  - (\lambda/2) (K^\dagger K)^2,
\label{eq:LS}
\eq
where the covariant derivative is given by
\bq
    D_{\mu} K = \left(\partial_\mu -\frac{i}{2}g B_{\mu}^a \tau^a 
                -\frac{i}{2} g' B_{\mu}^0          \right) K,
\qquad
    K^\dagger = \frac{1}{\sqrt 2} \left( \zeta - i\varphi_0 \,,\,
				  -\varphi_2 - i\varphi_1 \right),
\eq
where $g'/g = -s_\theta/c_\theta$,
$\tau^a$ are the standard Pauli matrices, $B_{\mu}^a$ is a triplet of vector
gauge bosons and $B_{\mu}^0$ a singlet.  For the theory to be stable we must
require $\lambda > 0$. In addition, we choose $\mu^2<0$ in order to have 
spontaneous symmetry breaking. The scalar field belongs to the minimal realization 
of the SM, where $\zeta$ and the Higgs-Kibble fields $\varphi_0$, $\varphi_1$ and
$\varphi_2$ are real. In
particular, we choose $\zeta +i\varphi_0$ to be the component of $K$ to develop
the non-zero vacuum expectation value, and we set 
$\langle \varphi_0 \rangle_0 = 0$ and $\langle \zeta\rangle_0 \neq 0$. We then 
introduce the (physical) Higgs field as $H =\zeta - v$.
The parameter $v$ is not a new parameter of the model and its value must
be fixed by the requirement that $\langle H \rangle_0 = 0$ (i.e.~$\langle K
\rangle_0 = (1/\sqrt{2})(v,0)$), so that the vacuum does not absorb or create
Higgs particles.

It is then convenient to define the bare parameters $m_\ssW$ (the $W$ boson 
mass), $m_\ssH$ (the mass of the physical Higgs particle) and $\beta_t$ (the 
tadpole constant) according to the following ``$\beta_t$ scheme'',
\renewcommand{\arraystretch}{1.5}
\bq
\left\{
  \begin{array}{lll}
     m_\ssW (1+\beta_t) \!\! &=& g v/2 \\
     m_\ssH^2 &=& \lambda \left( 2m_\ssW/g \right)^2 \\
     0 &=& \mu^2  +\frac{\lambda}{2}\left(2 m_\ssW/g\right)^2
  \end{array}
\right.
\!\! \Longrightarrow~
\left\{
  \begin{array}{lll}
    v &=& 2m_\ssW(1+\beta_t)/g \\
    \lambda &=& \left(g m_\ssH/2 m_\ssW\right)^2 \\
    \mu^2 &=& -\frac{1}{2} m_\ssH^2
  \end{array}
\right. .
\label{eq:betat}
\eq
\renewcommand{\arraystretch}{1}

Next, we define $\beta_t = \beta_{t_0} +\beta_{t_1} g^2 + \beta_{t_2} g^4 + 
\cdots $ and we fix the parameter $\beta_t$ such that the vacuum-expectation 
value of the Higgs field remains zero order by order in perturbation theory.
At the lowest order we can simply set $\beta_t=0$, i.e. $\beta_{t_0} = 0$.
At one loop we take into account all tadpole diagrams and we get
\bqa
\beta_{t_1} &=& \frac{m_\ssW^2}{16\, \pi^2 m^2_{\ssH}}
\Bigl\{  3 \sum_{i=1}^3 
  \Bigl[ \frac{m_{d,i}^4}{m_\ssW^4} a_0(m_{d,i}) + 
       \frac{m_{u,i}^4}{m_\ssW^4} a_0(m_{u,i}) \Bigr]
  + \sum_{i=1}^3 \frac{m_{l,i}^4}{m_\ssW^4} a_0(m_{l,i})
\nl
{}&-& \frac{1}{4 \ctw^2} \Bigl( 
\frac{n-1}{\ctw^2}+ \frac{m^2_{\ssH}}{2 m_\ssW^2} \Bigr)
  a_0(m_\ssZ)  -  \frac{1}{2} \Bigl( n - 1 +\frac{m^2_{\ssH}}{2 m_\ssW^2} \Bigr)
  a_0(m_\ssW)
- \frac{3}{8} \frac{m_{\ssH}^4}{m_\ssW^4} a_0(m_\ssH)\Bigr\},
\label{eq:one:beta1res}
\eqa
where $m_\ssZ= m_\ssW \slash c_\theta$ is the bare mass of the $Z$ boson.
The tadpole integral $a_0$ reads 
\bq
a_0(m) = \sum_{k=-1}^{1}\,a_0(m\,;\,k)\,F^1_k(m_\ssW^2),
\label{cUV}
\eq
with universal one-loop UV factors given by \eqn{OLUVF} and coefficients
\bq
a_0(m;-1) = -2, \qquad
a_0(m;0) = - 1 + \ln\frac{m^2}{m_\ssW^2},
\qquad
a_0(m;1)  = \frac{1}{2} \Bigl[ - 1 - \frac{1}{2} \zeta(2)
          + \Bigl( 1 - 
         \frac{1}{2} \ln \frac{m^2}{m_\ssW^2}\Bigr) 
\ln \frac{m^2}{m_\ssW^2} \Bigr],
\label{defAfun}
\eq
where $\zeta(x)$ is the Riemann's zeta function.
The full list of $\beta_t$-dependent Feynman rules needed for our computation
is given in appendix~B of Ref.~\cite{Actis:2006ra}.
\subsubsection{Diagonalization of the neutral sector}
\label{diagon}
The $Z$--$\gamma$ transition in the SM does not vanish at zero squared
momentum transfer. Although this fact does not pose any serious problem, not
even for the renormalization of the electric charge, it is preferable to use
an alternative strategy. We will follow the treatment of
Ref.~\cite{Passarino:1990xx}; consider the new $SU(2)$-coupling
constant $\bar{g}$, the new mixing angle $\bar{\theta}$ and the new $W$ mass
$\overline{m}_\ssW$ in the $\beta_t$ scheme,
\renewcommand{\arraystretch}{1.5}
\bq
g = \bar{g} \left(1+\Gamma\right),  
\qquad
g'= -\,\frac{\sin\bar{\theta}}{\cos\bar{\theta}}\,\bar{g}, 
\qquad
v = \frac{2\,\overline{m}_\ssW}{\bar{g}}(1+\beta_t), 
\qquad
\lambda = \left( \frac{\bar{g}\,m_\ssH}{2\,\overline{m}_\ssW} \right)^2, 
\qquad
\mu^2 = -\frac{1}{2}m_\ssH^2,
\label{eq:gbinbetat}
\eq
where $\Gamma = \Gamma_1 \, \bar{g}^2 + \Gamma_2 \,\bar{g}^4 + \cdots~$ is a 
new parameter yet to be specified. This change of parameters entails new 
$A_{\mu}$ and $Z_{\mu}$ fields related to $B^3_\mu$ and $B^0_\mu$ by
\renewcommand{\arraystretch}{1}
\bq
    \left(\begin{array}{c} 
      Z_\mu \\ A_\mu \end{array} \right) =
    \left(\begin{array}{cc} 
      \cos\bar{\theta}& -\sin\bar{\theta} \\ 
      \sin\bar{\theta}& \cos\bar{\theta}\end{array} \right)
    \left(\begin{array}{c} B^3_\mu \\ B^0_\mu \end{array} \right).
\label{eq:newAZ}
\eq
In our approach $\Gamma$ is fixed, order by order, requiring that
the $Z$--$\gamma$ transition is zero at $p^2 = 0$ in 't Hooft-Feynman gauge.
The explicit result for $\Gamma_1$ is
\bq
\Gamma_1 = \frac{n-2}{16\, \pi^2} \, a_0(m_\ssW),
\label{eq:one:gamma1res}
\eq
with $a_0(m)$ given in \eqn{defAfun}.
The full list of $\Gamma$-dependent Feynman rules needed for our computation
is given in appendix~C of Ref.~\cite{Actis:2006ra}. In addition,
for $H\to \gamma \gamma$ we need also a three-leg vertex with one
Higgs boson and two photons given by $-g^5 s_\theta^2 m_\ssW \Gamma_1^2 
\delta_{\mu\nu}$.
\subsubsection{$\overline{\text{MS}}$ counterterms}
We relate bare quantities to renormalized ones introducing multiplicative 
renormalization constants $Z_i$ and (if not otherwise stated) we 
expand them through the renormalized $SU(2)$-coupling constant $g_\ssR$,
\bq
Z_{i} = 1 + \sum_{n=1}^{\infty}\,\lpar \frac{g^2_{\ssR}}{16\,\pi^2}
\rpar^n\,\delta Z^{(n)}_{i},
\label{exppar}
\eq
where $\delta Z^{(n)}_{i}$ are counterterms and the subscript $i$ refers to 
masses, couplings, gauge parameters and fields. 

In particular, for all bare masses we write $m=Z^{1/2}_{m}\,m_{\ssR}$ and
for $g$, $g_\ssS$ and $c_\theta$ ($s_\theta$) we define
$p=Z_{p}\,p_{\ssR}$, where $p=g, g_{\ssS}, c_\theta,s_\theta$.
For a given bare field $\phi$ we find convenient to write $\phi = 
Z^{1/2}_{\phi}\,\phi_{\ssR}$, where $\phi_{\ssR}$ is a renormalized field, and 
we expand $Z_{\phi}$ through \eqn{exppar}. The bare photon field $A^\mu$ 
represents an exception, and here we use
\bq
A^\mu = Z^{1/2}_{\ssA\ssA}\,A^\mu_{\ssR} +
Z^{1/2}_{\ssA\ssZ}\,Z^\mu_{\ssR},
\qquad
Z^{1/2}_{\ssA\ssZ} =
\sum_{n=1}^{\infty}\,\lpar \frac{g^2_{\ssR}}{16\,\pi^2}
\rpar^n\,\delta Z^{(n)}_{\ssA\ssZ},
\eq
where $A^\mu_\ssR$ and $Z^\mu_\ssR$ are the renormalized fields for the photon
and the $Z$ boson. Note that $Z_{\ssA\ssA}$ is expanded through \eqn{exppar}.
In addition, bare fermion fields $\psi$ (we omit flavor labels) are written 
by means of bare left-handed and right-handed chiral fields $\psi^\ssL$ and 
$\psi^\ssR$. The latter are traded for renormalized fields  
$\psi^\ssL_{\ssR}$ and $\psi^\ssR_{\ssR}$ expanding the renormalization 
constants through \eqn{exppar},
\bq
\psi^{\ssL,\ssR} = \frac{1}{2}\,(1 \pm \gamma^5)\psi, \qquad
\psi^{\ssL,\ssR} = Z^{1/2}_{\psi_{_{\ssL,\ssR}}}\,\psi_{\ssR}^{\ssL,\ssR}.
\label{fermRENchi}
\eq
Faddeev-Popov ghost fields are not renormalized.
For the bare gauge parameters introduced in \eqn{eq:th:gfun}
we use $\xi = Z_{\xi}\,\xi_{\ssR}$, where $\xi$ is one of the bare
gauge parameters and $\xi_{\ssR}$ is the associated renormalized quantity.
$Z_\xi$ is expanded by means of \eqn{exppar} except for the case 
$\xi=\xi_{\ssA\ssZ}$, where we use
\bq
Z_{\xi_{\ssA\ssZ}} = \sum_{n=1}^{\infty}\,\lpar \frac{g^2_{\ssR}}{16\,\pi^2}
\rpar^n\,\delta Z^{(n)}_{\xi_{\ssA\ssZ}}.
\eq
After the expansions, we use the freedom in choosing the values for
the renormalized gauge parameters and set $\xi=1$.
Next, we define a minimal $\MSB$ subtraction scheme,
\bq
\delta Z^{(1)}_i =
\lpar - \frac{2}{\ep} + \DUV \rpar\,\Delta\,Z^{(1)}_i,
\label{eq:one:pole1ctaaaDUV}
\eq
and fix the counterterms in order to remove the poles at $\ep=0$ for any 
one-loop Green's function.
The full list of one-loop counterterms in the $\MSB$ scheme can be found
in section~5 of Ref.~\cite{Actis:2006rb}. For completeness, we list
here all the needed results, using short-hand notations for sums over 
fermions ($l\to$ charged leptons, $u,d\to$ quarks), 
\bq
X_{l}^{j} = \sum_{i=1}^3 x_{l,i}^{j}, \qquad \quad
X_{u}^{j} = \sum_{i=1}^3 x_{u,i}^{j}, \qquad \quad
X_{d}^{j} = \sum_{i=1}^3 x_{d,i}^{j},
\label{eq:one:SUMMA}
\eq
and we introduce scaled masses, $x_i=m_{i,\ssR}^2\slash m_{\ssW,\ssR}^2$.
In the following we drop everywhere the subscript $R$, since all quantities
are renormalized ones.
\vspace{0.2 cm}

\noindent
\underline{\emph{Gauge parameters.}}
\bqa
\Delta Z_{\gpA}^{(1)} &=&
\frac{1}{2}\, \Delta Z_{\ssA\ssA}^{(1)},
\qquad\qquad
\Delta Z_{\gpAZ}^{(1)} = - \,\Delta Z_{\ssA\ssZ}^{(1)},
\nl
\Delta Z_{\gpZ}^{(1)} &=& \frac{1}{2}\, \Delta Z_{\ssZ}^{(1)},
\qquad\qquad
\Delta Z_{\xi_{\varphi^0}}^{(1)} =  -\, \frac{1}{2}\,
\Bigl( \, \Delta Z_{\varphi^0}^{(1)} \,+\, \Delta Z_{m_\ssW}^{(1)}\, \Bigr)\,
+\, \Delta Z_{\ctw}^{(1)},
\nl
\Delta Z_{\gpW}^{(1)} &=& \frac{1}{2} \,\Delta Z_{\ssW}^{(1)},
\qquad\qquad
\Delta Z_{\xi_{\varphi}}^{(1)} = -\, \,\frac{1}{2}\,\Bigl(\,
\Delta Z_{\varphi}^{(1)}\, +\, \Delta Z_{m_\ssW}^{(1)}\, \Bigr).
\label{eq:one:oneCOUNTER}
\eqa
\vspace{0.2cm}

\noindent
\underline{\emph{Gauge-boson and Higgs-Kibble fields.}}
\bqa
\Delta Z^{(1)}_{\ssA\ssA} &=& 
\frac{23}{3} \stw^2,\qquad\qquad\qquad\qquad\ \ \ \
\Delta Z_{\ssA\ssZ}^{(1)} = - \frac{\stw}{3}
   \Bigl( \frac{41}{2}\frac{1}{ \ctw} - 23 \, \ctw \Bigr),
\nl
\Delta Z_{\ssZ}^{(1)} &=& \frac{1}{3}
   \Bigl( \frac{41}{2}\frac{1}{ \ctw^2} - 41 + 23 \, \ctw^2 \Bigr),\qquad
\Delta Z_{\varphi^0}^{(1)} =
   - 1 - \frac{1}{2} 
   \Bigl[ \frac{1}{\ctw^2} - X_{l} - 3 \, \Bigl( X_{d} + X_{u} \Bigr) \Bigl],
\nl
\Delta Z^{(1)}_{\ssW} &=& \frac{5}{6},\qquad\qquad\qquad\qquad\qquad\ \ 
\Delta Z^{(1)}_{\varphi} = \Delta Z^{(1)}_{\varphi^0}.
\label{csymI}
\eqa
\noindent
\underline{\emph{Masses and couplings.}}
\bqa
\Delta Z^{(1)}_{m_\ssW} &=&
    - \frac{3}{4}\frac{1}{ \ctw^2}
    - \frac{7}{3}
    + \frac{1}{x_{\ssH}}
    \Bigl[ 
       \frac{3}{2}\frac{1}{ \ctw^4} + 3  - 2 \, X_{l}^2 - 6 \,\Bigl( X_{d}^2 
+ X_{u}^2 \Bigr)
    \Bigr] + \frac{1}{2} \Bigl[ \frac{3}{2} x_{\ssH} + X_{l}
    + 3 \, \Bigl( X_{d} + X_{u} \Bigr)\Bigr],
\nl
\Delta Z^{(1)}_{\ctw} &=& \frac{1}{2}
  \Bigl(   \frac{41}{6}\frac{1}{ \ctw^2} - \frac{29}{2} + \frac{23}{3} \,
\ctw^2 \Bigr),\qquad
\Delta Z^{(1)}_{\stw} = - \frac{\ctw^2}{\stw^2} \, \Delta Z^{(1)}_{\ctw},\qquad
\Delta Z_g^{(1)} = - \Delta Z_{\stw}^{(1)} - 
\frac{1}{2}\, \Delta Z_{\ssA\ssA}^{(1)}.
\label{eq:one:orrendo}
\eqa
\vspace{0.2cm}

\noindent
\underline{\emph{Higgs-boson field and mass.}}
\bqa
\Delta Z^{(1)}_{\ssH} &=&  - 1 - \frac{1}{2}
  \Bigl[ \frac{1}{\ctw^2}  - X_{l}  - 3 \,\Bigl(  X_{d} + X_{u}  \Bigr)
  \Bigr],
\quad
\Delta Z^{(1)}_{\ssM_{\ssH}} = \frac{3}{2}
  \Bigl[ \frac{1}{2}\frac{1}{\ctw^2} + 1 - \frac{1}{2}\, x_{\ssH}
  - \frac{1}{3}\, X_{l}  - \Bigl( X_{d} + X_{u} \Bigr)  \Bigr].
\label{eq:one:oneCOUNTERstop}
\eqa
\noindent
\underline{\emph{Fermion fields and masses.}}
\bqa
\Delta Z_{\nu_R}^{(1)}\! &=& 0, 
\quad
\Delta Z_{l_R}^{(1)}\!\! = \frac{1}{\ctws} - 1 + \frac{x_l}{2},
\quad
\Delta Z_{u_R}^{(1)}\! =
  \frac{4}{9\ctws}  - \frac{4}{9}
  + \frac{x_u}{2} + \frac{4}{3}\frac{g^2_s}{g^2},
\quad
\Delta Z_{d_R}^{(1)}\! =
  \frac{1}{9\ctws}
  - \frac{1}{9} + \frac{x_d}{2} + \frac{4}{3}\frac{g^2_s}{g^2},
\nl
\Delta Z_{\nu_L}^{(1)}\! &=&
\Delta Z_{l_L}^{(1)}\!\! = 
 \frac{1}{4}\Bigl( \frac{1}{\ctws} + 2 + x_{l} \Bigr),
\qquad
\Delta Z_{u_L}^{(1)}\! =
\Delta Z_{d_L}^{(1)}\! =
  \frac{1}{4} \Bigl(
  \frac{1}{9\ctws} + \frac{26}{9}
  + x_u + x_d + \frac{16}{3}\frac{g^2_s}{g^2} \Bigr),
\label{eq:one:FER1}
\eqa
\bqa
\Delta Z_{m_l}^{(1)} &=&
  3 \, \frac{\stw^2}{\ctw^2}
  + \frac{1}{x_{\ssH}}
  \Bigl[ \frac{3}{2}\frac{1}{ \ctw^4} + 3  - 2\, X_{l}^2
  - 6 \,\Bigl( X_{u}^2 + X_{d}^2 \Bigr)
  \Bigr]
  + \frac{3}{4} \Bigl( x_{\ssH} - x_{l} \Bigr),
\nl
\Delta Z_{m_u^{(1)}} &=&
  \frac{2}{3}\,\frac{\stw^2}{\ctw^2}
  + \frac{1}{x_{\ssH}} \Bigl[
  \frac{3}{2}\frac{1}{ \ctw^4} + 3 
  - 2\, X_{l}^2 - 6\, \Bigl( X_{u}^2 + X_{d}^2 \Bigr)
  \Bigr]
  + \frac{3}{4} \Bigl(
  x_{\ssH} - x_{u} + x_{d}
  \Bigr)  + 8\,\frac{g^2_s}{g^2},
\nl
\Delta Z_{m_d}^{(1)} &=&
  - \frac{1}{3}\,\frac{s^2_{\theta}}{\ctw^2}
  + \frac{1}{x_{\ssH}} \Bigl[
  \frac{3}{2}\frac{1}{ \ctw^4}+ 3    - 2\, X_{l}^2
  -6 \,\Bigl( X_{u}^2 + X_{d}^2 \Bigr)
  \Bigr]
  + \frac{3}{4} \Bigl( x_{\ssH} + x_{u} - x_{d} \Bigr)
  + 8\,\frac{g^2_s}{g^2}.
\label{eq:one:FER2}
\eqa

\noindent
\underline{\emph{QCD Counterterms.}}
\bqa
\Delta Z_{\xi_{\ssG}}^{(1)} &=&
\frac{1}{2}\, \Delta Z_{\ssG}^{(1)},
\qquad
\Delta Z_{\ssG}^{(1)}= -\frac{g_{\ssS}^2}{g^2},
\qquad
\Delta Z_{g_S}^{(1)}= \frac{7}{2}\, \frac{g_{\ssS}^2}{g^2}.
\eqa
\subsection{Finite renormalization}
We devote this section to discuss: 1)
wave-function renormalization (WFR): the one-loop photon (gluon) and 
Higgs-boson WFR factors, 
${\cal Z}_\ssA$ (${\cal Z}_\ssG$) and 
${\cal Z}_\ssH$, defined in \eqn{AMP:wfrA} and \eqn{AMP:WFRhiggs}, 
are inserted in the amplitude of \eqn{AMP:ffactors1};
2) finite renormalization: all renormalized parameters showing 
up in the UV-finite amplitude are related to two different 
experimental input-parameter sets (IPSs) through the one-loop solutions of 
the SM renormalization equations.
In addition, we show through a detailed analysis of finite 
renormalization that on-shell mass renormalization clashes with the simplest 
WST identity at hand, the doubly contracted relation for on-shell photons of 
\eqn{AMP:wwi}.

In order to deal with compact expressions, we 
use the relations $e_i\cdot p_i\,=\,0$ for on-shell photons and set
$F_\ssP\,=\,F_\ssP^{(21)}$; the amplitude for $H\to \gamma \gamma$ reads
\bq\label{AMP:AMP}
{\cal A}\,=\,\frac{g^3\,s_\theta^2}{16\,\pi^2}\,
{\cal Z}_\ssA^{-1}\,{\cal Z}_\ssH^{-1\slash 2}\,{\cal M},
\qquad\;
{\cal M}\,=\,
\left(\,e_1\cdot e_2\,\right)\,F_\ssD(0,0,-\mhs)
\,+\,
\left(\,e_1\cdot p_2\,\right)\,
\left(\,e_2\cdot p_1\,\right)\,F_\ssP(0,0,-M_{\ssH}^2).
\eq
The expression for the $H\to gg$ amplitude is obtained
introducing color indices and replacing $g^2\,s^2_\theta$ with $g_\ssS^2$
and ${\cal Z}_\ssA$ with ${\cal Z}_\ssG$.

Furthermore, after expanding the form factors $F_\ssD$ and $F_\ssP$ and the 
function ${\cal M}$ at two loops, we split pure ${\cal O}(g^2)$ electroweak 
corrections and ${\cal O}(g_\ssS^2)$ QCD components, where 
$g_{\ssS}$ is the renormalized SU(3)-coupling constant of the QCD Lagrangian,
\bq\label{AMP:Mexp}
F\,=\,F^{(1)}\,+\,\frac{g^2}{16\,\pi^2}\,F^{(2,\EW)}\,+\,
\frac{g_\ssS^2}{16\,\pi^2}\,F^{(2,\QCD)},
\qquad F\,=\,F_\ssD\,,\,F_\ssP\,,\,{\cal M}.
\eq
\subsubsection{Wave-function renormalization}
In the context of the $\overline{\text{MS}}$-renormalization prescription 
used in this paper, field-renormalization constants are not chosen in 
order to compensate virtual corrections induced by WFR factors 
{\emph{{\`a} la}} LSZ. 
Therefore, at variance with the on-mass-shell prescription, external 
legs have to be properly dressed through the formalism introduced 
in \eqn{AMP:wfrA} and \eqn{AMP:WFRhiggs}.

Before expanding the WFR factors, we briefly recall our notation. Let 
$\Sigma_{i}^{\mu\nu}$ be the sum of all 1PI diagrams for 
a vector-boson self-energy ($i=A,Z,W$) or the transition
between the photon and the $Z$ boson ($i=AZ$). We isolate tensor 
structures and introduce perturbative expansions according to
\bq\label{AMP:PIaa}
\Sigma_i^{\mu\nu}(p^2)\,=\,\Sigma_i(p^2)\,
t^{\mu\nu}\,+\,P_i(p^2)\,l^{\mu\nu},\qquad
\Sigma_i(p^2)\,=\,\sum_{n=1}^{\infty}\,
\frac{g^{2n}}{(16\,\pi^2)^n}\,\Sigma_i^{(n)}(p^2),
\eq
where $t^{\mu\nu}\,=\,\delta^{\mu\nu}\,-\,l^{\mu\nu}$ and 
$l^{\mu\nu}\,=\,p^\mu p^\nu\slash p^2$. The same expansion in $g$ will be 
used for the Higgs boson or a fermion self-energy, both denoted 
by $\Sigma_i$ ($i=H,f$).

The LSZ WFR factor for a photon, ${\cal Z}_{\ssA}$, is fixed 
through \eqn{AMP:wfrA}. The transverse part of the photon Dyson-resummed 
propagator $\overline{\Delta}^{\mu\nu}_\ssA$ can be expressed using 
Eq.~(105) of Ref.~\cite{Actis:2006ra}; it involves the transverse parts of the 
photon and $Z$-boson self-energies and the photon$\slash$$Z$-boson 
transition,
\bq
\overline{\Delta}_\ssA^{\mu\nu}(p^2)\,=\, t^{\mu\nu}\, 
\overline{\Delta}^\ssT_\ssA(p^2)\,+\,
l^{\mu\nu}\, \overline{\Delta}^\ssL_\ssA(p^2),\qquad
\left[\overline{\Delta}_\ssA^\ssT(p^2)\right]^{-1}\,=\, 
p^2-\Sigma_\ssA(p^2)-
\frac{\left[\Sigma_{\ssA\ssZ}(p^2)\right]^2}{p^2+m_\ssZ^2-\Sigma_\ssZ(p^2)},
\eq
where $m_\ssZ$ is the renormalized $Z$-boson mass. For a one-loop accuracy, 
self-energies and transitions are computed at order ${\cal O}(g^2)$, and 
factors involving the weak-mixing angle $\theta$ are singled out in the same 
spirit of the LQ-basis formalism, thoroughly discussed in section~6 of 
Ref.~\cite{Actis:2006ra},
\bq\label{AMP:LQ}
\Sigma_\ssA^{(1)}(p^2)\,=\, p^2\, s_\theta^2\, \Pi^{(1)}_\ssA(p^2),\qquad
\Sigma_{\ssA\ssZ}^{(1)}(p^2)\,=\, p^2\, 
\frac{s_\theta}{c_\theta}\, \Pi^{(1)}_{\ssA\ssZ}(p^2),\qquad
\Sigma_\ssZ^{(1)}(p^2)\,=\, \frac{1}{c^2_\theta}\, \Pi^{(1)}_\ssZ(p^2).
\eq
The one-loop photon vacuum-polarization function $\Pi_\ssA^{(1)}$ and the 
residual function $\Pi_\ssZ^{(1)}$ are regular at $p^2=0$. In  
addition, because of the diagonalization procedure summarized in 
section~\ref{diagon}, also $\Pi_{\ssA\ssZ}^{(1)}$ is regular at $p^2=0$. 
Therefore, the canonical LSZ condition of \eqn{AMP:wfrA} for the photon 
WFR factor allows to express ${\cal Z}_{\ssA}$ also in the full SM 
by means of $\Pi_\ssA^{(1)}$ evaluated at zero-momentum  transfer, as in QED,
\bq\label{AMP:wfr1}
{\cal Z}_{\ssA}\,=\, 
1\, - \frac{g^2\, s_\theta^2}{16\, \pi^2}\, \Pi_\ssA^{(1)}(0).
\eq
Note that an analogous relation holds at the two-loop level, see section~5 of 
Ref.~\cite{Actis:2006rc} and Ref.~\cite{Degrassi:2003rw}, where the same 
result was obtained through a background-field method analysis.

The WFR factor for the Higgs boson, ${\cal Z}_{\ssH}$, is fixed in 
\eqn{AMP:WFRhiggs}; 
respect to the photon case, here mixings are 
not present because of CP conservation. After expanding the Higgs-boson 
self-energy at one loop and introducing the real part of its derivative, 
$\Sigma_{\ssH,p}^{(1)}$, we get
\bq\label{AMP:wfr2}
{\cal Z}_{\ssH}\,=\, 1\, - \frac{g^2}{16\, \pi^2}\,
\text{Re} \Sigma_{\ssH,p}^{(1)}(-\mhs),\qquad
\Sigma_{\ssH,p}^{(1)}(p^2)\,=\, 
\frac{\partial\Sigma_{\ssH}^{(1)}(p^2)}{\partial\, p^2}.
\eq
Including the  WFR factors of \eqn{AMP:wfr1} and \eqn{AMP:wfr2}  
in the amplitude of \eqn{AMP:AMP}, and expanding ${\cal M}$ through 
\eqn{AMP:Mexp}, we get finally
\bq
\label{AMP:ampSplit}
{\cal A}\,=\,\frac{g^3\,s_\theta^2}{16\,\pi^2}\,
\Bigl\{\, {\cal M}^{(1)}\, +\, \frac{g^2}{16\, \pi^2}\, 
\Bigl[\,{\cal M}^{(2,\EW)}\,
+\, {\cal M}^{(1)}\, \Bigl(\,
\frac{1}{2}\,\text{Re}\Sigma_{\ssH,p}^{(1)}(-\mhs)\,
+\,s_\theta^2\, \Pi_{\ssA}^{(1)}(0)\,\Bigl)\,\Bigr]
+\, \frac{g_{\ssS}^2}{16\, \pi^2}\, {\cal M}^{(2,\QCD)}\,\Bigr\}.
\eq
Of course, an on-mass-shell prescription for the counterterms would 
shift $\Sigma_{\ssH,p}^{(1)}$ and $\Pi_{\ssA}^{(1)}$ in the one-loop 
expressions for the counterterms themselves; the associated
contribution, in our notation, would be hidden in ${\cal M}^{(2,\EW)}$.
\subsubsection{Finite renormalization for masses and couplings}
The second step in building the amplitude consists in performing finite 
renormalization; the residual dependence of the UV-finite amplitude on the 
renormalized parameters is removed through the solutions of the SM 
renormalization equations, truncated at the appropriate order. For a 
NLO accuracy, we need tree-level solutions for all 
${\cal O}(g^5)$ and ${\cal O}(g^3\, g_\ssS^2)$ terms of 
\eqn{AMP:ampSplit} and one-loop ones for the ${\cal O}(g^3)$ component. 
As a result, the final expression of the amplitude will contain only the 
selected experimental IPS and will be ready for 
numerical evaluation.

Note that the dependence of the amplitude on the renormalized parameters 
appears at two different levels:
1) explicitly, by means of the pre-factors containing the coupling 
constants $g$, $s_\theta$ and $g_{\ssS}$;
2) implicitly, through the functions ${\cal M}^{(1)}$, 
${\cal M}^{(2,\EW)}$, ${\cal M}^{(2,\QCD)}$,
          $\Sigma_{\ssH,p}^{(1)}$ and $\Pi_{\ssA}^{(1)}$, which 
a priori depend on the renormalized $W$-boson, 
          Higgs-boson and fermion masses $m_\ssW$, $m_\ssH$ and $m_f$, 
and the weak-mixing angle $\theta$ (or, 
          equivalently, on the renormalized $Z$-boson mass 
$m_\ssZ=m_\ssW\slash c_\theta$). Here the 
          Cabibbo-Kobayashi-Maskawa matrix is identified with the unit matrix.

Finite renormalization for ${\cal M}^{(2,\EW)}$, 
${\cal M}^{(2,\QCD)}$, $\Sigma_{\ssH,p}^{(1)}$ and 
$\Pi_{\ssA}^{(1)}$ is trivially achieved by identifying the renormalized 
parameters showing up in their explicit expressions with the tree-level 
solutions of the renormalization equations, which will depend on the
chosen IPS. 

The case of ${\cal M}^{(1)}$, instead, is more subtle. Let us 
introduce an appropriate suffix, ${\cal M}^{(1)}_{r}$, to indicate that the 
renormalized parameters $\{p_{i,r}\}$ appear in ${\cal M}^{(1)}$. 
Solving at one-loop the renormalization equations amounts to replacing 
$\{p_{i,r}\}\to\{p_{i,e}+\delta p_i\}$, 
where $\{p_{i,e}\}$ are taken from experiment and belong to the chosen IPS, 
whereas $\{\delta p_i\}$ summarize the one-loop corrections. 
Finite renormalization is obtained by 
\bq\label{AMP:finite}
{\cal M}^{(1)}_{r}\,\to\, 
{\cal M}^{(1)}_{e}\, +\, \sum_{j} {\cal M}_{p_{j,e}}^{(1)} \delta p_j,\qquad
{\cal M}_{p_{j,e}}^{(1)}=
\frac{\partial {\cal M}^{(1)}_{r}}{\partial\, p_{j,r}} |_{\{p_{i,r}=p_{i,e}\}}.
\eq
For $H \to \gamma \gamma$, the dependence of ${\cal M}^{(1)}$ on $g$ and $\theta$ 
is encapsulated in the overall pre-factors of the amplitude of 
\eqn{AMP:ampSplit}, and the renormalized mass of the $Z$ boson $m_\ssZ$ 
does not show up at one loop. Therefore, it is convenient to choose the 
on-shell masses of the $W$ boson, $M_{\ssW}$, the Higgs boson, $M_{\ssH}$, and 
the fermions, $M_f$, to be part of any IPS, and to evaluate derivatives respect 
to the associated renormalized masses $m_\ssW$, $m_\ssH$ and $m_f$.

We employ on-shell mass renormalization, as already discussed for the 
Higgs boson in \eqn{AMP:mass}.
We use the one-loop solution of the renormalization equations for the Higgs-, 
$W$-boson and fermion masses,
\bq
m_\ssB^2\,=\, M_\ssB^2\, + \frac{g^2}{16\, \pi^2}\,
\text{Re}\Sigma_\ssB^{(1)}(-M_\ssB^2),\qquad B=H,W,\qquad
m_f\,=\, M_f \, + \frac{g^2}{16\,\pi^2}\, \text{Re}\Sigma_f^{(1)}(-M_f^2),
\eq
and write the amplitude of \eqn{AMP:ampSplit} as
\bq
\label{AMP:ampSplit3}
{\cal A}\,=\,\frac{g^3\,s_\theta^2}{16\,\pi^2}\,
\Bigl\{\, {\cal M}^{(1)}\, +\, \frac{g^2}{16\, \pi^2}\, 
\overline{{\cal M}}^{(2,\EW)}\,
+\, \frac{g^2\, s_\theta^2}{16\, \pi^2}\, {\cal M}^{(1)}\,\Pi_{\ssA}^{(1)}(0)\,
+\, \frac{g_{\ssS}^2}{16\, \pi^2}\, {\cal M}^{(2,\QCD)}\,\Bigr\},
\eq
where we have factorized the residual dependence on the renormalized coupling 
constants $g$, $\theta$ and $g_\ssS$, defining
\bq
\label{AMP:ampSplitdef}
\overline{{\cal M}}^{(2,\EW)}=
{\cal M}^{(2,\EW)}
+\frac{1}{2} {\cal M}^{(1)} \text{Re}\Sigma_{\ssH,p}^{(1)}(-\mhs)
+\sum_{i=W,H} {\cal M}^{(1)}_{M^2_{i}} \text{Re} \Sigma_{i}^{(1)}(-M_{i}^2)
+ 2 \sum_f M_f {\cal M}^{(1)}_{M^2_f} \text{Re} \Sigma_{f}^{(1)}(-M_{f}^2).
\eq
The last term in \eqn{AMP:ampSplitdef} contains a dependence
on $g_\ssS$ coming from QCD corrections to finite top-quark mass renormalization.
At this stage, we can safely identify all renormalized parameters showing up 
in the explicit expressions for ${\cal M}^{(1)}$, 
$\overline{{\cal M}}^{(2,\EW)}$, ${\cal M}^{(2,\QCD)}$ and 
$\Pi_{\ssA}^{(1)}$ with the tree-level solutions of the chosen renormalization 
equations.

For electroweak corrections to $H\to gg$ a similar decomposition holds,
\bq
\label{AMP:ampSplit3Glu}
{\cal A}\,=\,\frac{g\,g_\ssS^2}{16\,\pi^2}\,
\Bigl\{\, {\cal M}^{(1)}\, +\, \frac{g^2}{16\, \pi^2}\, 
\overline{{\cal M}}^{(2,\EW)}\,
\Bigr\},
\eq
provided with color indices.
Note that, concerning $H\to gg$, the derivative with respect to the Higgs-boson
mass in \eqn{AMP:ampSplitdef} vanishes, since there is no dependence on 
$m_\ssH$ at LO.

We complete finite renormalization by relating the renormalized coupling 
constants $g$, $\theta$ and $g_\ssS$, collected as simple pre-factors, to three 
additional experimental input data. For $g_\ssS$, we write 
$g_\ssS^2= 4\pi\alpha_\ssS(\mu_\ssR^2)$,
where $\alpha_\ssS$ is the strong-coupling constant and $\mu_\ssR$ the appropriate 
renormalization scale. 
For $g$ and $\theta$, instead, we can select two data among $G_\ssF$, 
the Fermi-coupling constant, $\alpha$, the fine-structure constant, and 
$\mz$, the on-shell $Z$-boson mass.
\vspace{0.2cm}

\noindent\underline{\emph{Finite renormalization for $H\to \gamma \gamma$: IPS 1.}}
The first IPS is $\{G_\ssF,\alpha\}$. We start writing 
$g^2 s_\theta^2= e^2$, where $e$ is the renormalized electromagnetic-coupling 
constant, and use the relation for one-loop electric-charge renormalization,
\bq
e^2= \frac{4\, \pi\, \alpha}{1 + \frac{\alpha}{4\pi} \, \Pi_{\ssA}^{(1)}(0)},
\eq
where $\Pi_{\ssA}^{(1)}$, the one-loop photon vacuum-polarization function, is 
defined in \eqn{AMP:LQ}. After using charge renormalization in the 
amplitude of \eqn{AMP:ampSplit3}, we observe that all terms containing 
$\Pi_{\ssA}^{(1)}$ cancel out and we obtain
\bq
\label{AMP:ampSplit3b}
{\cal A}\,=\,\frac{g\, \alpha}{4\,\pi}\,
\Bigl\{\, {\cal M}^{(1)}\, +\, \frac{g^2}{16\, \pi^2}\, 
\overline{{\cal M}}^{(2,\EW)}\,
+\, \frac{\alpha_{\ssS}(\mu_\ssR^2)}{4 \pi}\, 
{\cal M}^{(2,\QCD)}\,\Bigr\}.
\eq
For $g$, we employ the relation following from muon decay~\cite{Green:1980bd},
\bq\label{AMP:gf}
g\, =\, 2\, \mw\, \Bigl(\,\sqrt{2}\, G_\ssF\, \Bigr)^{1\slash 2}\, \Bigl\{
\,1\, +\, \frac{G_\ssF}{4\,\sqrt{2}\,\pi^2}\,
\Bigl[\,
\text{Re} \Sigma_{\ssW}^{(1)}(-M_{\ssW}^2)\,
-\,\Sigma_{\ssW}^{(1)}(0) \,-\, \mw^2 \delta_\ssG\,
\Bigr]\,
\Bigr\},
\eq
where, following Ref.~\cite{Passarino:1989ta}, we split
universal hard corrections to the muon lifetime, encapsulated in the $W-$boson 
self-energy $\Sigma^{(1)}_{\ssW}$, and process-dependent components, 
summarized by the quantity $\delta_\ssG$, whose explicit expression reads
\bq\label{AMP:deltaG}
\delta_\ssG= 6 + \frac{7-4 s_\theta^2}{2 s^2_\theta} \ln (c_\theta^2).
\eq
Note that $s_\theta$ and $c_\theta$ are consistently fixed at the lowest 
order in perturbation theory by
\bq
c^2_\theta=1-s_\theta^2,\qquad 
s^2_\theta=\frac{e^2}{g^2}=\frac{\pi\alpha}{\sqrt{2}G_\ssF \mw^2},
\eq
and we are avoiding any reference to the on-shell $Z$-boson mass $\mz$.
The amplitude will finally read
\bq
\label{AMP:ampSplit3bb}
{\cal A}\,=\,\frac{\alpha\, \mw}{2\,\pi}\,
\Bigl( \sqrt{2}\, G_\ssF\Bigr)^{1\slash 2}
\Bigl\{\, {\cal M}^{(1)}\, +\, 
\frac{G_\ssF \, \mw^2}{2\, \sqrt{2}\, \pi^2}\, 
\Bigl(\, {\cal M}^{(2,\EW)}\, + 
{\cal M}^{(2,\EW)}_{\ssI\ssP\ssS 1}\,
\,\Bigr)\, + \, \frac{\alpha_{\ssS}(\mu_\ssR^2)}{4 \pi}\, 
{\cal M}^{(2,\QCD)}\,\Bigr\},
\eq
where the total contribution from finite renormalization reads 
\bqa
{\cal M}^{(2,\EW)}_{\ssI\ssP\ssS 1}\,&=&\, \frac{1}{2}\, {\cal M}^{(1)}\,
\bigg[\,
\frac{\text{Re} \Sigma_{\ssW}^{(1)}(-M_{\ssW}^2)\,
-\,\Sigma_{\ssW}^{(1)}(0)}{\mw^2}\, - \, \delta_\ssG\,
\,+ \text{Re}\Sigma_{\ssH,p}^{(1)}(-\mhs)
\bigg]
\nl
&+&\sum_{i=W,H} {\cal M}^{(1)}_{M^2_{i}} \text{Re} \Sigma_{i}^{(1)}(-M_{i}^2)
+ 2 \sum_f M_f {\cal M}^{(1)}_{M^2_f} \text{Re} \Sigma_{f}^{(1)}(-M_{f}^2).
\eqa
\vspace{0.2cm}

\noindent\underline{\emph{Finite renormalization for $H\to \gamma \gamma$: IPS 2.}}
The second IPS is $\{G_\ssF,\mz\}$. We start replacing 
\eqn{AMP:gf} in \eqn{AMP:ampSplit3}, removing $g$. We get
\bqa
\label{AMP:ampSplit3c}
{\cal A}\,&=&\,s_\theta^2\,\frac{\mw^3}{2\,\pi^2}\, \left(
\sqrt{2}\,G_\ssF
\right)^{3\slash 2}
\bigg\{\, 
{\cal M}^{(1)}\, +\, 
\frac{G_\ssF \, \mw^2}{2\, \sqrt{2}\, \pi^2}\, 
\bigg[\,
\overline{{\cal M}}^{(2,\EW)}\, + \frac{3}{2}{\cal M}^{(1)}\,
\Bigl(
\frac{\text{Re} \Sigma_{\ssW}^{(1)}(-M_{\ssW}^2)
-\Sigma_{\ssW}^{(1)}(0)}{\mw^2} - \delta_\ssG
\Bigr)
\bigg]
\nl
&& \quad \qquad\quad \qquad\qquad\quad +\,  s_\theta^2 
\frac{G_\ssF \, \mw^2}{2\, \sqrt{2}\, \pi^2}\,  
{\cal M}^{(1)}\,\Pi_{\ssA}^{(1)}(0)\,
+\, \frac{\alpha_{\ssS}(\mu_\ssR^2)}{4 \pi}\, 
{\cal M}^{(2,\QCD)}
\,\bigg\}.
\eqa
Next, we write $s_\theta^2=1-c_\theta^2$ and use the relation between 
renormalized masses  $c_\theta=m_\ssW \slash m_\ssZ$. After using vector-boson 
on-shell mass renormalization it follows that
\bq\label{AMP:last}
s_\theta^2\,=\, \left(1\,-\,\frac{\mw^2}{\mz^2}\right)
\Bigl\{1
\,-\,\frac{G_\ssF\, \mw^4}{2\, \sqrt{2} \pi^2 (\mz^2-\mw^2)}\,
\Bigl[\,\frac{\text{Re}\Sigma_{\ssW\ssW}(-\mw^2)}{\mw^2}\,-\,
\frac{\text{Re}\Sigma_{\ssZ\ssZ}(-\mz^2)}{\mz^2}\, \Bigr]\Bigr\}.
\eq
After employing \eqn{AMP:last} in \eqn{AMP:ampSplit3c}, the 
amplitude will then be expressed through $\gf$, $\mw$
and $\mz$;
$\delta_\ssG$ is given in \eqn{AMP:deltaG}, $s_\theta$ and $c_\theta$ 
are fixed at the lowest order in perturbation theory by
$s^2_\theta=1-c_\theta^2$ and $c^2_\theta= \mw^2/\mz^2$,
and any reference to the fine-structure constant $\alpha$ is avoided.
\vspace{0.2cm}

\noindent\underline{\emph{Finite renormalization for $H\to gg$.}}
Concerning finite renormalization for $H\to gg$, we start from
\bq
\label{AMP:ampSplit4Glu}
{\cal A}\,=\,\frac{g\,\alpha_\ssS(\mu_\ssR^2)}{4\,\pi}\,
\Bigl\{\, {\cal M}^{(1)}\, +\, \frac{g^2}{16\, \pi^2}\, 
\overline{{\cal M}}^{(2,\EW)}\,
\Bigr\},
\eq
and we use the result of \eqn{AMP:gf} for muon decay to express $g$ 
through $G_\ssF$.
\subsubsection{The doubly contracted WST identity for $H\to \gamma\gamma$ at 
two loops}
\label{sec:aaa}
Having introduced WFR factors and finite renormalization, we can now have a 
closer look at the doubly contracted WST identity with two on-shell photons 
of \eqn{AMP:wwi}. As previously stressed, the computation of the identity 
requires Higgs-boson mass renormalization as a key ingredient, because bosonic 
couplings proportional to the $\overline{\text{MS}}$-renormalized Higgs-boson 
mass, $m_\ssH$, appear at the one-loop level. Since the Higgs boson is emitted 
from a physical source, the associated momentum is on the mass shell, 
$P^2=-\mhs$, and $m_\ssH$ has to be consistently traded everywhere with 
$\mh$, including radiative corrections.

We write ${\cal W}{\cal I}_{dc}$ of \eqn{AMP:wwi} through the same 
decomposition already used for the amplitude in \eqn{AMP:AMP},
\bq
{\cal W}{\cal I}_{dc}(0,0)\,=\,
\frac{g^3\,s_\theta^2}{16\,\pi^2}\, {\cal Z}_\ssH^{-1\slash 2}\, {\cal X},
\qquad
{\cal X}\,=\,
-\,\frac{\mhs}{2}\,\Bigl[\,
F_\ssD(0,0,-\mhs)\,-\,\frac{\mhs}{2}\,F_\ssP(0,0,-\mhs) 
\,\Bigr],
\eq
where $F_\ssP=F_\ssP^{(21)}$. Next, we expand at one loop the Higgs-boson 
WFR factor ${\cal Z}_\ssH$ employing \eqn{AMP:wfr2}, and the function 
${\cal X}$ through \eqn{AMP:Mexp}, for $F={\cal X}$,
\bq
{\cal W}{\cal I}_{dc}(0,0)\,=\,
\frac{g^3\,s_\theta^2}{16\,\pi^2}\,\Bigl\{\,  
{\cal X}^{(1)}\,+\,\frac{g^2}{16\, \pi^2}\,
\Bigl[\,{\cal X}^{(2,\EW)}\,+\, \frac{1}{2}\, 
{\cal X}^{(1)}\,\text{Re}\,\Sigma_{\ssH,p}^{(1)}(-\mhs)\,\Bigr]\,
+\,\frac{g_\ssS^2}{16\, \pi^2}\,{\cal X}^{(2,\QCD)} \, \Bigr\}.
\eq
Since the identity holds for an on-shell Higgs boson, we perform finite 
renormalization employing \eqn{AMP:finite} and obtain
\bq
\!{\cal W}{\cal I}_{dc}(0,0) =
\frac{g^3 s_\theta^2}{16\pi^2}\bigg\{\!
  {\cal X}^{(1)}\!
+ \frac{g^2}{16\pi^2}\bigg[\!
    {\cal X}^{(2,\EW)}\!
  + \,\frac{{\cal X}^{(1)}}{2}\text{Re}\,\Sigma_{\ssH,p}^{(1)}(-\mhs)\,
  + {\cal X}^{(1)}_{\mhs}\text{Re}\Sigma_{\ssH}^{(1)}(-\mhs)
  \bigg]
+ \frac{g_\ssS^2}{16\pi^2}{\cal X}^{(2,\QCD)}
\!\bigg\}.
\eq
Note that we do not need to employ other renormalization equations, since the 
identity has to be proven both for renormalized and experimental masses and 
couplings.
We compute the identity at one loop and we prove that ${\cal X}^{(1)}=0$. The 
final form for the identity at two loops reads
\bq
{\cal W}{\cal I}_{dc}(0,0)\,=\,
\frac{g^3\,s_\theta^2}{16\,\pi^2}\,\Bigl\{\, \frac{g^2}{16\, \pi^2}\,
\Bigl[\,{\cal X}^{(2,\EW)}\,+\, {\cal X}^{(1)}_{\mhs} \, 
\text{Re}\Sigma_{\ssH}^{(1)}(-\mhs)\, 
\Bigr]\,
+\,\frac{g_\ssS^2}{16\, \pi^2}\,{\cal X}^{(2,\QCD)} \, \Bigr\}.
\eq
Finally, it is harmless to replace renormalized parameters with the tree-level 
solutions of the renormalization equations,
\bq
{\cal W}{\cal I}_{dc}(0,0)\,=\,
\frac{\alpha\,\mw}{2\,\pi}\,\left( \sqrt{2} G_\ssF\right)^{1\slash 2} 
\Bigl\{\, \frac{G_\ssF \, \mw^2}{2\,\sqrt{2} \pi^2}\,
\Bigl[\,{\cal X}^{(2,\EW)}\,+\, {\cal X}^{(1)}_{\mhs} \, 
\text{Re}\Sigma_{\ssH}^{(1)}(-\mhs)\, 
\Bigr]\,
+\,\frac{\alpha_\ssS(\mu_\ssR^2)}{4\, \pi}\,{\cal X}^{(2,\QCD)} \, 
\Bigr\}.
\eq
We prove that ${\cal W}{\cal I}_{dc} \not=0$, explicitly we obtain
\bq
{\cal W}{\cal I}_{dc}(0,0)\,=\,
\left( \sqrt{2} G_\ssF\right)^{3\slash 2}\,M_{\ssW}^4\,
\frac{\alpha}{16\,\pi^3}\,C_0\lpar-\mhs,0,0\,;\,\mw,\mw,\mw\rpar\,
\text{Im}\,\Sigma^{(1)}_{\ssH}(-\mhs), 
\label{para}
\eq
where $C_0$ is the scalar three-point function. The analysis of this paradox 
-- violation of WSTI -- will be postponed till section~\ref{subsec:complexmass}.
\section{Manipulating Feynman integrals after generation of 
Feynman diagrams}
\label{sec:RedSym}
In this section we summarize the techniques used after the generation
of Feynman diagrams, performed with the {\sc{FORM}}~\cite{Vermaseren:2000nd} 
program {$\GS$}~\cite{GraphShot}, the projection of the amplitude onto the
form factors of~\eqn{AMP:ffactors} with the projectors discussed in 
appendix~\ref{proppo} and the standard operations concerning the Dirac algebra.
Before attempting the semi-analytical or numerical evaluation of 
(pseudo-)observables, we perform three kinds of simplifications and symbolic 
manipulations: firstly, we remove reducible scalar products; secondly, we employ 
integration-by-parts (IBP) identities~\cite{Tkachov:1981wb} (see also 
Ref.~\cite{'t Hooft:1972fi}) for simplifying tadpole diagrams; finally, 
we achieve an optimal level of symmetrization for loop integrals. 
\subsection{Reduction of scalar products}
\label{sec:ScalarReduct}
Given a generic loop integral, it is possible to assign a one-to-one
correspondence between a specific reducible scalar product containing 
at least one integration momentum and a particular propagator. 
This simple observation allows to recursively write loop-momentum 
dependent scalar products in terms of propagators, and employ a basic 
cancellation mechanism, when non-trivial numerator structures appear, 
through the algebraic master relation
  \bq \label{reduc}
    \frac{2\,q\cdot p}{(q^2+m_1^2)[(q+p)^2+m_2^2]}=
    \frac{1}{q^2+m_1^2}
    -\frac{1}{(q+p)^2+m_2^2}
    -\frac{p^2-m_1^2+m_2^2}{(q^2+m_1^2)[(q+p)^2+m_2^2]}\,.
  \eq
Note that the cancellation of a scalar product is associated with the
disappearance of lines in the corresponding diagram; as a result, each diagram 
generates a set of child diagrams with a smaller number of propagators.

The number of independent scalar products involving loop momenta can
exceed the number of propagators in a given diagram; therefore, some 
irreducible scalar products cannot be removed from the numerator functions.
This is the well-known obstacle in achieving a full reduction for two-loop
diagrams; for the special case of two-point functions, it can be by-passed 
through a judicious sub-loop reduction, as shown by the authors of 
Ref.~\cite{Weiglein:1993hd}. Obviously, for a fixed number of loops and 
external legs, diagrams with a large number of internal lines, and thus 
more propagators, exhibit less irreducible scalar products than diagrams 
with a small number of internal lines. 
Note also that the choice of the scalar products which are considered as 
reducible and irreducible is to a large extent arbitrary.

For two-loop three-point functions we have two independent external momenta
and two integration momenta; therefore, we have to deal with $7-I$ irreducible 
scalar products, where $I$, with $4\leq I\leq 6$, is the number of internal lines,
and a full reducibility using \eqn{reduc} is clearly not at hand.

As an example, let us consider the $V^{\bbb}$-family diagram shown on the
left-hand side of \eqn{ex:VH}, where the Higgs boson couples to photons 
through a couple of $W$ bosons and a top-bottom loop. 
After acting on the diagram with the projector $P^{\mu\nu}_{\ssD}$ as 
described in appendix~\ref{proppo} and saturating all free Lorentz 
indices, we remove all possible reducible scalar products and obtain:
\vspace{-0.2cm}
  \bq \label{ex:VH}
    \raisebox{0.1cm}{\scalebox{0.7}{
    \begin{picture}(150,75)(0,0)
    \SetWidth{1.2}                        
    \DashLine(0,0)(40,0){3}                \Text(5,7)[cb]{$H$}
    \Photon(128,-53)(100,-35){2}{5}        \Text(130,-66)[cb]{$\gamma$}
    \Photon(128,53)(100,35){2}{5}          \Text(130,58)[cb]{$\gamma$} 
    \ArrowLine(70,-17.5)(100,35)           \Text(100,10)[cb]{$t$}
    \ArrowLine(100,35)(70,17.5)            \Text(82,32)[cb]{$t$}
    \Photon(70,-17.5)(40,0){2}{5}          \Text(53,-23)[cb]{$W$}
    \Photon(70,17.5)(40,0){2}{5}           \Text(53,16)[cb]{$W$}
    \ArrowLine(82,-3.5)(100,-35)
    \Line(70,17.5)(78,3.5)            \Text(100,-16)[cb]{$b$}
    \ArrowLine(100,-35)(70,-17.5)          \Text(82,-41)[cb]{$b$}
    \end{picture}
    }}
    \!\!\!\!\!\!\!\!\!\!\!\!\!\!\!\!\!\!\!\!\!\!
    \otimes\,P^{\mu\nu}_{\ssD}\,
    =\!\!\!
    \,\,
    \tilde{C}_{\!\ssH}
    \Bigl[\!M_W^2-\! M^2_t\!+\!2\spro{p_1}{q_1}\!\Bigl(1\!-\!
    \frac{\spro{p_1}{q_1}}{\spro{p_1}{p_2}}\Bigr)\!\Bigr]\,
    \raisebox{0.1cm}{\scalebox{0.7}{
    \begin{picture}(140,75)(0,0)
    \SetWidth{1.2}
    \DashLine(0,0)(40,0){3}                
    \Photon(128,-53)(100,-35){2}{5}        
    \Photon(128,53)(100,35){2}{5}          
    \Line(70,-17.5)(100,35)       \Text(100,10)[cb]{$M_{t}$}
    \Line(100,35)(70,17.5)        \Text(80,31)[cb]{$M_{t}$}
    \Line(70,-17.5)(40,0)                  \Text(53,-23)[cb]{$M_W$}
    \Line(70,17.5)(40,0)                   \Text(53,16)[cb]{$M_W$}
    \DashLine(82,-3.5)(100,-35){1.4}
    \DashLine(70,17.5)(78,3.5){1.4}        \Text(100,-16)[cb]{$M_{b}$}
    \DashLine(100,-35)(70,-17.5){1.4}      \Text(82,-39)[cb]{$M_{b}$}
    \LongArrow(4,8)(24,8)                  \Text(9,13)[cb]{$-P$}
    \LongArrow(118,-56)(104,-47)           \Text(125,-65)[cb]{$p_1$}
    \LongArrow(118,56)(104,47)             \Text(125,57)[cb]{$p_2$}
    \end{picture}
    }}
    \!\!\!\!\!\!\!\!\!\!\!\!\!\!\!\!
    +
    \left(\!
    \ba{l}
    \hbox{\footnotesize{reduced}}\\[-0.1cm]
    \hbox{\footnotesize{diagrams}}
    \ea
    \!\!
    \right)\!.\!\!
  \vspace{1.7cm}
  \eq
Here the non-planar diagram on the right-hand side corresponds  to the 
$V^{\bbb}$ configuration defined in \fig{TLvertices} (the dot-line 
denotes a light fermion), where the tensor structures have been stripped 
and explicitly collected as an overall factor in square brackets; 
the symbol $\tilde{C}_{\!\ssH}$ is proportional to 
$\tilde{C}_{\!\ssH}\propto \spro{p_1}{p_2} + M_W^2 - M^2_t\,$ 
and we have denoted by ``reduced diagrams'' all diagrams with at least 
one internal line less. 
The scalar product $p_1\cdot q_1$ survives as an irreducible one, showing the 
presence of scalar, vector and tensor $V^{\ssH}$-type integrals, which appear in 
the combination of \eqn{ex:VH}. 

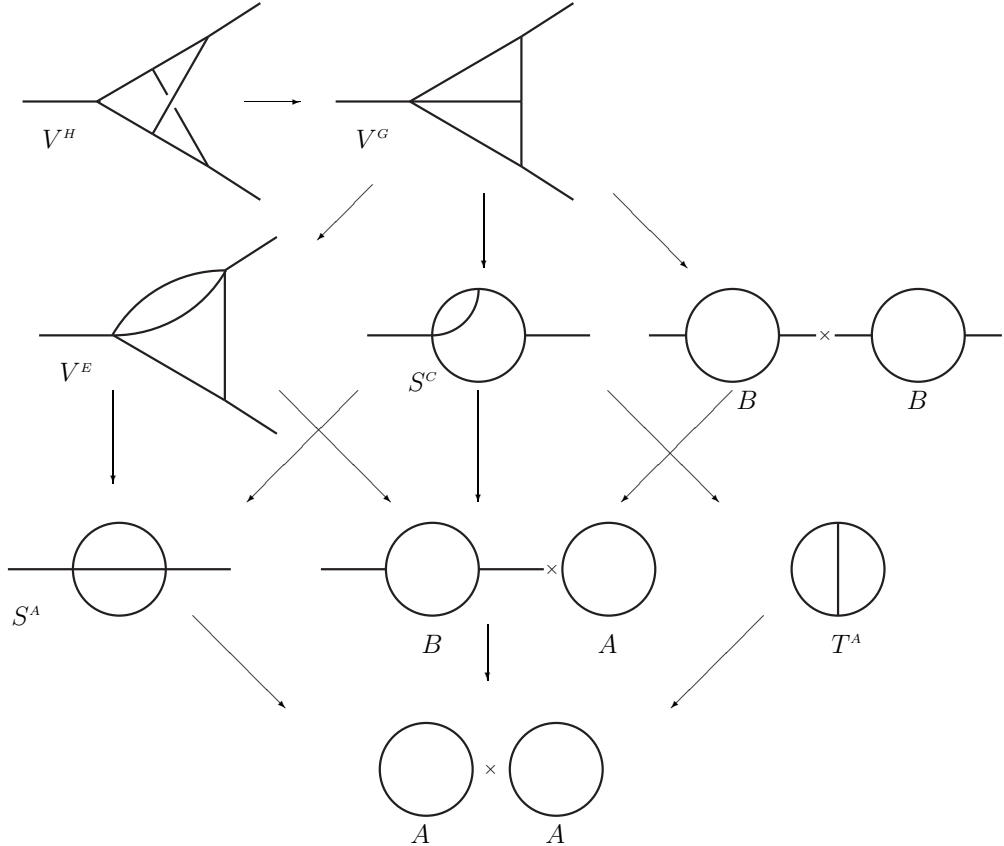
\begin{figure}[!b]
$$
\raisebox{0.1cm}{\scalebox{0.7}{
\begin{picture}(130,75)(0,0)
 \SetWidth{1.2}
 \Line(0,0)(42,0)         
 \Line(128,-53)(100,-35)  
 \Line(128,53)(100,35)    
 \Line(70,-17.5)(100,35)                            
 \Line(100,35)(70,17.5)                             
 \Line(70,-17.5)(40,0)                              
 \Line(70,17.5)(40,0)                               
 \Line(100,-35)(82,-3.5)\Line(78,3.5)(70,17.5)      
 \Line(100,-35)(70,-17.5)                           
 \Text(20,-25)[cb]{\Large $V^{\bbb}$}
\put(120,0){\vector(1,0){30}}
\end{picture}
}}
\hspace{0.8cm}
\raisebox{0.1cm}{\scalebox{0.7}{
\begin{picture}(130,75)(0,0)
 \SetWidth{1.2}
 \Line(0,0)(42,0)         
 \Line(128,-53)(100,-35)  
 \Line(128,53)(100,35)    
 \Line(100,-35)(40,0)           
 \Line(100,0)(100,35)           
 \Line(100,0)(40,0)             
 \Line(100,-35)(100,0)          
 \Line(100,35)(40,0)            
 \Text(20,-25)[cb]{\Large $V^{\bba}$}
\put(20,-45){\vector(-1,-1){30}}
\put(80,-50){\vector(0,-1){40}}
\put(150,-50){\vector(1,-1){40}}
\end{picture}
}}
\hspace{5cm}
$$
\vspace{0.5cm}
$$
\raisebox{0.1cm}{\scalebox{0.7}{
\begin{picture}(130,75)(0,0)
 \SetWidth{1.2}
 \Line(0,0)(40,0)         
 \Line(128,-53)(100,-35)  
 \Line(128,53)(100,35)    
 \CArc(100,-35)(70,90,150)      
 \CArc(40,70)(70,270,330)       
 \Line(100,-35)(40,0)           
 \Line(100,-35)(100,35)         
 \Text(20,-25)[cb]{\Large $V^{\aba}$}
\put(40,-30){\vector(0,-1){50}}
\put(130,-30){\vector(1,-1){60}}
\end{picture}
}}
\hspace{1cm}
\raisebox{0.1cm}{\scalebox{0.7}{
\begin{picture}(130,75)(0,0)
 \SetWidth{1.2}
 \Line(0,0)(35,0)         
 \CArc(60,0)(25,0,360)          
 \CArc(35,25)(25,-90,0)         
 \Line(85,0)(120,0)         
 \Text(30,-30)[cb]{\Large $S^{\ssC}$}
\put(-5,-30){\vector(-1,-1){60}}
\put(60,-30){\vector(0,-1){60}}
\put(130,-30){\vector(1,-1){60}}
\end{picture}
}}
\hspace{1cm}
\raisebox{0.1cm}{\scalebox{0.7}{
\begin{picture}(130,75)(0,0)
 \SetWidth{1.2}
 \Line(-25,0)(-5,0)         
 \CArc(20,0)(25,0,360)          
 \Line(45,0)(65,0)
 \Text(70,0)[c]{$\times$}
 \Line(75,0)(95,0)
 \CArc(120,0)(25,0,360)         
 \Line(145,0)(165,0)         
 \Text(28,-40)[cb]{\Large $B$}
 \Text(120,-40)[cb]{\Large $B$}
\put(20,-30){\vector(-1,-1){60}}
\end{picture}
}}
$$
\vspace{0.5cm}
$$
\raisebox{0.1cm}{\scalebox{0.7}{
\begin{picture}(130,75)(0,0)
 \SetWidth{1.2}
 \Line(0,0)(35,0)         
 \CArc(60,0)(25,0,360)          
 \Line(35,0)(85,0)              
 \Line(85,0)(120,0)         
 \Text(10,-30)[cb]{\Large $S^{\ssA}$}
\put(100,-25){\vector(1,-1){50}}
\end{picture}
}}
\hspace{0.8cm}
\raisebox{0.1cm}{\scalebox{0.7}{
\begin{picture}(180,75)(0,0)
 \SetWidth{1.2}
 \Line(0,0)(35,0)         
 \CArc(60,0)(25,0,360)          
 \Line(85,0)(120,0)         
 \CArc(155,0)(25,0,360)      
 \Text(125,0)[c]{$\times$}
 \Text(60,-45)[cb]{\Large $B$}
 \Text(155,-45)[cb]{\Large $A$}
\put(90,-30){\vector(0,-1){30}}
\end{picture}
}}
\hspace{0.8cm}
\raisebox{0.1cm}{\scalebox{0.7}{
\begin{picture}(130,75)(0,0)
 \SetWidth{1.2}
 \CArc(60,0)(25,0,360)          
 \Line(60,25)(60,-25)           
 \Text(65,-45)[cb]{\Large $T^{\ssA}$}
\put(20,-25){\vector(-1,-1){50}}
\end{picture}
}}
$$
\vspace{0.05cm}
$$
\hspace*{-0.9cm}
\raisebox{0.1cm}{\scalebox{0.7}{
\begin{picture}(130,75)(0,0)
 \SetWidth{1.2}
 \CArc(50,0)(25,0,360)          
 \Text(85,0)[c]{$\times$}
 \CArc(120,0)(25,0,360)
 \Text(47,-40)[cb]{\Large $A$}
 \Text(120,-40)[cb]{\Large $A$}
\end{picture}
}}
$$
\vspace{0.1cm}
\caption[]{\label{fig:child} Generic child topologies of the
  $V^{\ssH}$ parent topology. The five-line $V^{\ssG}$ diagram is
  obtained by removing one line of the $V^{\ssH}$ diagram; the second
  line contains the child topologies of $V^{\ssG}$ ($V^{\aba}$, $S^{\ssC}$ 
  and $B\times B$). The third line contains the topologies $S^{\ssA}$, 
  $B\times A$ and $T^A$, obtained by removing one line from the diagrams above.
  The arrows indicate the correspondences between parent and child topologies.}
\end{figure}

The reduction procedure 
is repeated 
iteratively on the reduced diagrams; as a matter of fact, the $V^{\ssH}$ 
topology generates a set of eight sub-topologies, including simple 
factorized topologies (products of one-loop functions) and two-loop 
vertices, self-energies and vacuum diagrams, as illustrated 
in~\fig{fig:child}.

During the reduction procedure, we devote special care to preserve the 
canonical routing of loop momenta defined in appendix~\ref{app:topos}. 
In \fig{fig:examplereduction} we show an explicit example for reducing 
the scalar products of a $V^{\ssG}$-family diagram.
Note that this configuration can appear after performing the reduction of 
the $V^{\ssH}$ diagram using \eqn{ex:VH}. In step ({\rm 1a}), the 
scalar product $q_2\cdot p_1$ is removed; this operation leads to the 
appearance of an additional integral belonging to the $V^{\ssE}$ family 
which does not possess the standard routing of momenta as defined in 
\fig{TLvertices}. 
The canonical routing of momenta is recovered in step ({\rm 1b}) through 
a shift in the loop momenta which acts also on the remaining scalar product 
$q_1\cdot p_2$ and generates a new reducible scalar product $q_2\cdot p_2$. 
The latter is then removed in step ({\rm 2}), leading to an additional 
integral of the $S^{\ssA}$ family.
\begin{figure}[ht]
$$
{}_{ 4[(q_2\!\cdot p_1)(q_1\!\cdot p_2)] }
\scalebox{0.5}{
\begin{picture}(140,75)(0,0)
 \SetWidth{1.4}
 \Line(0,0)(40,0)                    \Text(10,5)[cb]{$-P$}
 \Line(128,-53)(100,-35)             \Text(118,-65)[cb]{$p_1$}
 \Line(128,53)(100,35)               \Text(118,57)[cb]{$p_2$}
 \Line(100,35)(40,0)                 
 \Line(100,-35)(40,0)                
 \Line(100,-35)(100,35)              
\Line(40,0)(100,0)                   
\end{picture}
}
$$
\vspace{0.5cm}
$$
{}_{ = \;\; 2[q_1\!\cdot p_2] }
\scalebox{0.5}{
\begin{picture}(130,75)(6,0)
 \SetWidth{1.4}
 \Text(-60,9)[cb]{\mbox{\bf{\Large{(1a)}}}}
 \Line(5,0)(40,0)                    \Text(15,5)[cb]{$-P$}
 \Line(128,-53)(100,-35)             \Text(118,-65)[cb]{$p_1$}
 \Line(128,53)(100,35)               \Text(118,57)[cb]{$p_2$}
 \CArc(100,-35)(70,90,150)           
 \CArc(40,70)(70,270,330)            
 \Line(100,-35)(40,0)                
 \Line(100,-35)(100,35)              
\end{picture}
}
\hspace{-0.4cm}
{}_{ - \;2[(q_2^2\!+m_4^2)q_1\!\cdot p_2] }
\scalebox{0.5}{
\begin{picture}(130,75)(6,0)
 \SetWidth{1.4}
 \Line(5,0)(40,0)                    \Text(15,5)[cb]{$-P$}
 \Line(128,-53)(100,-35)             \Text(118,-65)[cb]{$p_1$}
 \Line(128,53)(100,35)               \Text(118,57)[cb]{$p_2$}
 \Line(100,35)(40,0)                 
 \Line(100,-35)(40,0)                
 \Line(100,-35)(100,35)              
\Line(40,0)(100,0)                   
\end{picture}
}
\hspace{-0.1cm}
{}_{ = \;\; - [ 2 q_2\!\cdot p_2 \!+ P^2 ] }
\scalebox{0.5}{
\begin{picture}(130,75)(6,0)
 \SetWidth{1.4}
 \Text(-107,9)[cb]{\mbox{\bf{\Large{(1b)}}}}
 \Line(5,0)(40,0)                    \Text(15,5)[cb]{$-P$}
 \Line(128,-53)(100,-35)             \Text(118,-65)[cb]{$p_1$}
 \Line(128,53)(100,35)               \Text(118,57)[cb]{$p_2$}
 \CArc(100,-35)(70,90,150)           
 \CArc(40,70)(70,270,330)            
 \Line(100,-35)(40,0)                
 \Line(100,-35)(100,35)              
\end{picture}
}
\hspace{-0.4cm}
{}_{ - \;[(q_2^2\!+m_4^2)q_1\!\cdot p_2]  }
\scalebox{0.5}{
\begin{picture}(130,75)(6,0)
 \SetWidth{1.4}
 \Line(5,0)(40,0)                    \Text(15,5)[cb]{$-P$}
 \Line(128,-53)(100,-35)             \Text(118,-65)[cb]{$p_1$}
 \Line(128,53)(100,35)               \Text(118,57)[cb]{$p_2$}
 \Line(100,35)(40,0)                 
 \Line(100,-35)(40,0)                
 \Line(100,-35)(100,35)              
\Line(40,0)(100,0)                   
\end{picture}
}
$$
\vspace{0.5cm}
$$
{}_{ = \quad -\;}
\scalebox{0.5}{
\begin{picture}(130,75)(6,0)
 \SetWidth{1.4}
 \Text(-43,9)[cb]{\mbox{\bf{\Large{(2)}}}}
 \Line(5,0)(40,0)                    \Text(15,5)[cb]{$-P$}
 \Line(128,-53)(86.5,0)              \Text(118,-65)[cb]{$p_1$}
 \Line(128,53)(86.5,0)               \Text(118,57)[cb]{$p_2$}
 \Line(40,0)(86.5,0)                 
 \CArc(63.5,0)(23,0,360)             
\end{picture}
}
\hspace{-0.2cm}
{}_{ + \;\;[m_3^2+q_2^2-P^2] }
\scalebox{0.5}{
\begin{picture}(130,75)(6,0)
 \SetWidth{1.4}
 \Line(5,0)(40,0)                    \Text(15,5)[cb]{$-P$}
 \Line(128,-53)(100,-35)             \Text(118,-65)[cb]{$p_1$}
 \Line(128,53)(100,35)               \Text(118,57)[cb]{$p_2$}
 \CArc(100,-35)(70,90,150)           
 \CArc(40,70)(70,270,330)            
 \Line(100,-35)(40,0)                
 \Line(100,-35)(100,35)              
\end{picture}
}
\hspace{-0.2cm}
{}_{ - \;\;[(q_2^2+m_4^2)q_1\!\cdot p_2]}
\scalebox{0.5}{
\begin{picture}(130,75)(6,0)
 \SetWidth{1.4}
 \Line(5,0)(40,0)                    \Text(15,5)[cb]{$-P$}
 \Line(128,-53)(100,-35)             \Text(118,-65)[cb]{$p_1$}
 \Line(128,53)(100,35)               \Text(118,57)[cb]{$p_2$}
 \Line(100,35)(40,0)                 
 \Line(100,-35)(40,0)                
 \Line(100,-35)(100,35)              
\Line(40,0)(100,0)                   
\end{picture}
}
$$
\vspace{0.5cm}
\caption[]{\label{fig:examplereduction}Reduction of scalar products for
  an example of the $V^{\ssG}$ family. The equations are valid for $p_1^2=p_2^2=0$.}  
\end{figure}

In some cases, only the scalar configuration survives after the projection
procedure.  An example is the $V^{\ssK}\,$ configuration in the left-hand side
of \eqn{ex:VK}, where the Higgs boson couples to two photons through a
couple of $Z$ bosons and a top-quark loop. After applying the projector $P^{\mu\nu}_{\ssD}$
and removing reducible scalar products, one obtains the decomposition
\vspace{-0.4cm}
\bq
\raisebox{0.1cm}{\scalebox{0.7}{
\begin{picture}(150,75)(0,0)
 \SetWidth{1.2}                       
 \DashLine(0,0)(40,0){3}                \Text(5,7)[cb]{$H$}
 \Photon(128,-53)(100,-35){2}{5}        \Text(130,-66)[cb]{$\gamma$}
 \Photon(128,53)(100,35){2}{5}          \Text(130,58)[cb]{$\gamma$} 
 \Photon(70,-17.5)(40,0){2}{5}          \Text(53,-23)[cb]{$Z$}
 \Photon(70,17.5)(40,0){2}{5}           \Text(53,16)[cb]{$Z$}
 \ArrowLine(70,17.5)(70,-17.5)          \Text(81,-3)[cb]{$t$}
 \ArrowLine(70,-17.5)(100,-35)          \Text(82,-41)[cb]{$t$}
 \ArrowLine(100,-35)(100,35)            \Text(110,-3)[cb]{$t$}
 \ArrowLine(100,35)(70,17.5)            \Text(82,32)[cb]{$t$}
\end{picture}
}}
\!\!\!\!\!\!\!\!\!\!\!\!
\otimes\,P^{\mu\nu}_{\ssD}\,\,
=
\,\,
\tilde{C}_{\!\ssK}
\,
\raisebox{0.1cm}{\scalebox{0.7}{
\begin{picture}(140,75)(0,0)
 \SetWidth{1.2}
 \DashLine(0,0)(40,0){3}            
 \Photon(128,-53)(100,-35){2}{5}    
 \Photon(128,53)(100,35){2}{5}      
 \Line(70,-17.5)(40,0)                \Text(53,-23)[cb]{$M_Z$}
 \Line(70,17.5)(40,0)                 \Text(53,16)[cb]{$M_Z$}
 \Line(70,17.5)(70,-17.5)             \Text(82,-3)[cb]{$M_{t}$}
 \Line(70,-17.5)(100,-35)    \Text(82,-39)[cb]{$M_{t}$}
 \Line(100,-35)(100,35)      \Text(110,-3)[cb]{$M_{t}$}
 \Line(100,35)(70,17.5)      \Text(82,31)[cb]{$M_{t}$}
\LongArrow(4,8)(24,8)       \Text(9,13)[cb]{$-P$}
\LongArrow(118,-56)(104,-47)\Text(125,-65)[cb]{$p_1$}
\LongArrow(118,56)(104,47)  \Text(125,57)[cb]{$p_2$}
\end{picture}
}}
\!\!\!\!\!\!\!\!\!\!\!
+\;
\left(\!
\ba{l}
 \hbox{\footnotesize{reduced}}\\[-0.1cm]
    \hbox{\footnotesize{diagrams}}
\ea
\!\!
\right),\!\!
\label{ex:VK}
\vspace{2cm}
\eq
where the symbol $\tilde{C}_{\!\ssK}$ is proportional to $
\tilde{C}_{\!\ssK}\propto 32\,(v_+^2+v_-^2)\,M_t^2\,( \spro{p_1}{p_2} - M_Z^2 +
2\,M_t^2) - 128\,v_+ v_-\,M_t^2\,( \spro{p_1}{p_2} + 2\,M_t^2)$.
Here, $v_{\pm}$ are the $V \pm A$ couplings $Z {\bar t} t$.
Clearly, here no irreducible scalar products for the $V^{\ssK}$ topology
remain, and only the scalar configuration survives.
\subsection{Vacuum diagrams}
The removal of scalar products illustrated in \fig{fig:child} shows that the
vacuum diagram $T^A$ appears after the reduction of the $S^{\ssC}$ self-energy. 
The explicit evaluation of $T^A$ has been carried out in Ref.~\cite{Caffo:1998du}.
In addition, if the external momentum of the $S^{\ssC}$ integral corresponds 
to one of the photon momenta, the $S^{\ssC}$ integral is a vacuum
integral with an increased power of one propagator.  Furthermore, if additional 
loop-momentum dependent scalar products appear in the numerator of the 
$S^{\ssC}$ diagram, the integral has to be expanded in powers of the external 
momentum around zero; therefore, higher increased powers of the propagators can 
arise.  

Vacuum integrals with increased powers of the propagators can be related to the 
$T^A$ tadpole integral using the traditional IBP method. 
In particular, the relation between $T^B$, the tadpole diagram with one increased
power of the third propagator (shown diagrammatically through
a dot on the corresponding line) and $T^A$ is given in Eq.~\eqref{eq:Tad} by
\vspace{-1.2cm}
\bqa
\label{eq:Tad}
\raisebox{0.1cm}{\scalebox{0.7}{
\begin{picture}(130,75)(0,0)
 \SetWidth{1.2}
 \CArc(60,0)(25,0,360)          \Text(43,-3)[cb]{$m_1$}
 \Vertex(85,0){4}               \Text(73,10)[cb]{$m_3$}
                                \Text(73,-15)[cb]{$m_3$}
 \Line(60,25)(60,-25)           \Text(67,-3)[cb]{$m_2$}
\end{picture}
\hspace*{-1.2cm}
}}
&=&(n-3)\frac{m_1^2+m_2^2-m_3^2}{\lambda(m_1^2,m_2^2,m_3^2)}
\raisebox{0.1cm}{\scalebox{0.7}{
\hspace*{-1.2cm}
\begin{picture}(130,75)(0,0)
 \SetWidth{1.2}
 \CArc(60,0)(25,0,360)          \Text(43,-3)[cb]{$m_1$}
                                \Text(73,-15)[cb]{$m_3$}
 \Line(60,25)(60,-25)           \Text(67,-3)[cb]{$m_2$}
\end{picture}
\hspace*{-1.7cm}
}}
+ \frac{(2-n)}{\lambda(m_1^2,m_2^2,m_3^2)}\*
\raisebox{0.1cm}{\scalebox{0.7}{
\hspace{-1.7cm}
\begin{picture}(130,75)(0,0)
 \SetWidth{1.2}
 \CArc(60,0)(15,0,360)          \Text(60,-3)[cb]{$m_1$}
 \CArc(90,0)(15,0,360)          \Text(90,-3)[cb]{$m_2$}
\end{picture}
\hspace*{-1.2cm}
}} \nonumber \\[-1cm]
&+& (2-n)\frac{m_1^2-m_2^2-m_3^2}{2\,m_3^2\,\lambda(m_1^2,m_2^2,m_3^2)}
\raisebox{0.1cm}{\scalebox{0.7}{
\hspace{-1.7cm}
\begin{picture}(130,75)(0,0)
 \SetWidth{1.2}
 \CArc(60,0)(15,0,360)          \Text(60,-3)[cb]{$m_1$}
 \CArc(90,0)(15,0,360)          \Text(90,-3)[cb]{$m_3$}
\end{picture}
\hspace*{-1.0cm}
}}
+ (n-2)\frac{m_1^2-m_2^2+m_3^2}{2\,m_3^2\,\lambda(m_1^2,m_2^2,m_3^2)}
\raisebox{0.1cm}{\scalebox{0.7}{
\hspace{-1.7cm}
\begin{picture}(135,75)(0,0)
 \SetWidth{1.2}
 \CArc(60,0)(15,0,360)          \Text(60,-3)[cb]{$m_2$}
 \CArc(90,0)(15,0,360)          \Text(90,-3)[cb]{$m_3$}
\end{picture}
\hspace*{-1.2cm}
}}
,
\qquad
\eqa
with $\lambda(x,y,z)=x^2+y^2+z^2-2\*x\*y-2\*x\*z-2\*y\*z$. 

After projecting the amplitude on the relevant form factors, it turns out that only 
the tadpole diagrams $T^A$ and $T^B$, related to each other through \eqn{eq:Tad},
contribute to the amplitudes $H\rightarrow\gamma\gamma$ and $H\rightarrow gg$. 
\subsection{Symmetrization of loop integrals}
It is essential to exploit the symmetries of each diagram in order to reduce the
number of integrals to be calculated and to identify equal configurations.
We have taken into account the symmetries of the appearing one-loop and
two-loop topologies, summarizing them in \tabn{tab:symmetries1loop}
and \tabn{tab:symmetries2loop}. In both tables, the first column denotes the 
topology and the second column enumerates the different symmetry transformations 
for a given topology; the third column contains the transformation of the loop 
momenta $q_1$ and $q_2$ and the fourth column the corresponding interchange of 
masses and external momenta. The identity transformation and
the total reflection of all external momenta ($p_1 \to -p_1$, $p_2 \to -p_2$ and
$P\to -P$, corresponding to the loop-momentum transformation $q_1\to -q_1$
and $q_2\to -q_2$), which leaves the loop integral unchanged, 
are not listed in \tabn{tab:symmetries1loop} and \tabn{tab:symmetries2loop}.
The largest number of symmetries can be observed for the $V^{\ssH}$ family. 
\begin{table}[!ht]
\begin{center}
\begin{tabular}{|c||c||ll||l|}
\hline
\multirow{1}{*}{$B$}
&(I)&$q\rightarrow -p-q'$&&
     $m_1\leftrightarrow m_2$\\\hline 
\multirow{7}{*}{$C$}
&(I)&$q\rightarrow  -p_1-q'$&&
     $m_1\leftrightarrow m_2$, 
     $-P\leftrightarrow p_2$\\
&(II)&$q\rightarrow  -P-q'$&&
     $m_1\leftrightarrow m_3$, 
     $p_1\leftrightarrow p_2$\\
&(III)&$q\rightarrow -q'$&&
     $m_2\leftrightarrow m_3$, 
     $-P\leftrightarrow p_1$\\
&(IV)&$q\rightarrow -P+q'$&&
     $m_1\rightarrow m_3$, 
     $m_2\rightarrow m_1$, 
     $m_3\rightarrow m_2$,\\
&&&& $p_1\rightarrow -P$, 
     $p_2\rightarrow p_1$, 
     $-P \rightarrow -p_2$\\
&(V)&$q\rightarrow  -p_1+q'$&&
     $m_1\rightarrow m_2$, 
     $m_2\rightarrow m_3$, 
     $m_3\rightarrow m_1$,\\
&&&& $p_1\rightarrow p_2$, 
     $p_2\rightarrow -P$, 
     $-P \rightarrow p_1$\\\hline 
\end{tabular}
\end{center}
\vspace*{-0.65cm}
\caption[]{\label{tab:symmetries1loop}Symmetry transformations for
  one-loop topologies.}
\end{table}
\begin{table}[!ht]
\begin{center}
\begin{tabular}{|c||c||ll||l|}
\hline
\multirow{6}{*}{$T^{\ssA}$}
&(I)&$q_1\rightarrow q'_1$,& $q_2\rightarrow q'_1 - q'_2$&
     $m_2\leftrightarrow m_3$\\
&(II)&$q_1\rightarrow  -q'_1 + q'_2$,& $q_2\rightarrow q'_2$&
     $m_1\leftrightarrow m_2$\\
&(III)&$q_1\rightarrow -q'_2 $,& $q_2\rightarrow -q'_1$&
     $m_1\leftrightarrow m_3$\\
&(IV)&$q_1\rightarrow -q'_1 + q'_2$,& $q_2\rightarrow -q'_1$&
     $m_1\rightarrow m_3$, 
     $m_2\rightarrow m_1$, 
     $m_3\rightarrow m_2$\\
&(V)&$q_1\rightarrow -q'_2$,& $q_2\rightarrow q'_1 - q'_2$&
     $m_1\rightarrow m_2$, 
     $m_2\rightarrow m_3$, 
     $m_3\rightarrow m_1$\\\hline 
\multirow{5}{*}{$S^{\ssA}$}
&(I)&$q_1\rightarrow -q'_2$,& $q_2\rightarrow -q'_1$&
     $m_1\leftrightarrow m_3$\\
&(II)&$q_1\rightarrow -p-q'_1+q'_2$,& $q_2\rightarrow q'_2$&
     $m_1\leftrightarrow m_2$\\
&(III)&$q_1\rightarrow q'_1$,& $q_2\rightarrow p+q'_1-q'_2$&
     $m_2\leftrightarrow m_3$\\
&(IV)&$q_1\rightarrow -q'_2 $,& $q_2\rightarrow p+q'_1-q'_2$&
     $m_1\rightarrow m_2$, 
     $m_2\rightarrow m_3$, 
     $m_3\rightarrow m_1$\\
&(V)&$q_1\rightarrow -p-q'_1+q'_2$,& $q_2\rightarrow -q'_1$&
     $m_1\rightarrow m_3$, 
     $m_2\rightarrow m_1$, 
     $m_3\rightarrow m_2$\\\hline 
%
\multirow{1}{*}{$S^{\ssC}$}
&(I)&$q_1\rightarrow -q'_1+q'_2$,& $q_2\rightarrow q'_2$&
     $m_1\leftrightarrow m_2$\\\hline 
%
\multirow{1}{*}{$S^{\ssD}$}
&(I)&$q_1\rightarrow -p-q'_1$,& $q_2\rightarrow -p-q'_2$&
     $m_1\leftrightarrow m_2$, 
     $m_4\leftrightarrow m_5$\\\hline 
%
\multirow{3}{*}{$S^{\ssE}$}
&(I)&$q_1\rightarrow -q'_1+q'_2$,& $q_2\rightarrow q'_2$&
     $m_1 \leftrightarrow m_2$\\
&(II)&$q_1\rightarrow q'_1$,& $q_2\rightarrow q'_2$&
     $m_3 \leftrightarrow m_5$\\
&(III)&$q_1\rightarrow -q'_1+q'_2$,& $q_2\rightarrow q'_2$&
     $m_1 \leftrightarrow m_2$, 
     $m_3\leftrightarrow m_5$\\\hline 
%
\multirow{3}{*}{$V^{\ssE}$}
&(I)&$q_1\rightarrow -q'_1+q'_2 $,& $q_2\rightarrow q'_2$&
     $m_1\leftrightarrow m_2$\\
&(II)&$q_1\rightarrow -q'_1$,& $q_2\rightarrow -q'_2$&
     $m_3\leftrightarrow m_4$, 
     $p_2\leftrightarrow -P$\\
&(III)&$q_1\rightarrow q'_1-q'_2$,& $q_2\rightarrow -q'_2$&
     $m_1\leftrightarrow m_2$, 
     $m_3\leftrightarrow m_4$, 
     $-P\leftrightarrow p_2$\\\hline 
\multirow{3}{*}{$V^{\ssI}$}
&(I)&$q_1\rightarrow -q'_1+q'_2$,& $q_2\rightarrow q'_2$&
     $m_1\leftrightarrow m_2$\\
&(II)&$q_1\rightarrow-q'_1$,& $q_2\rightarrow -q'_2$& 
     $m_4\leftrightarrow m_5$, 
     $-P\leftrightarrow p_1$\\
&(III)& $q_1\rightarrow q'_1-q'_2$,& $q_2\rightarrow -q'_2$&
     $m_1\leftrightarrow m_2$, 
     $m_4\leftrightarrow m_5$, 
     $-P\leftrightarrow p_1$\\\hline
\multirow{3}{*}{$V^{\ssM}$}
&(I)&$q_1\rightarrow -q'_1 + q'_2$,& $q_2\rightarrow q'_2$&
     $m_1 \leftrightarrow m_2$\\
&(II)&$q_1\rightarrow -q'_1$,& $q_2\rightarrow -q'_2$&
     $m_4\leftrightarrow m_5$, 
     $-P\leftrightarrow p_1$\\
&(III)&$q_1\rightarrow q'_1-q'_2$,& $q_2\rightarrow -q'_2$&
     $m_1 \leftrightarrow m_2$, 
     $m_4\leftrightarrow m_5$, 
     $-P\leftrightarrow p_1$\\\hline 
\multirow{1}{*}{$V^{\ssG}$}
&(I)&$q_1\rightarrow -P-q'_2$,& $q_2\rightarrow -P-q'_1$&
     $m_1\leftrightarrow m_5$, 
     $m_2\leftrightarrow m_4$, 
     $p_1\leftrightarrow p_2$\\\hline 
\multirow{1}{*}{$V^{\ssK}$}
&(I)&$q_1\rightarrow -P-q'_1$,& $q_2\rightarrow -P-q'_2$&
     $m_1\leftrightarrow m_2$, 
     $m_4\leftrightarrow m_6$, 
     $p_1\leftrightarrow p_2$\\\hline 
\multirow{19}{*}{$V^{\ssH}$}
&(I)&$q_1\rightarrow p_2 - q'_1$,& $q_2\rightarrow p_1 - q'_2$&
     $m_1\leftrightarrow m_2$, 
     $m_3\leftrightarrow m_4$, 
     $m_5\leftrightarrow m_6$\\
&(II)&$q_1\rightarrow -p_1 - q'_1 + q'_2$,& $q_2\rightarrow q'_2$&
     $m_1\leftrightarrow m_3$, 
     $m_2\leftrightarrow m_4$, 
     $-P\leftrightarrow p_2$\\
&(III)&$q_1\rightarrow q'_1$,& $q_2\rightarrow -p_2 + q'_1 - q'_2$&
     $m_3\leftrightarrow m_6$, 
     $m_4\leftrightarrow m_5$, 
     $-P\leftrightarrow p_1$\\
&(IV)&$q_1\rightarrow p_2 - q'_2$,& $q_2\rightarrow  p_1 - q'_1$&
     $m_1\leftrightarrow m_6$, 
     $m_2\leftrightarrow m_5$, 
     $p_1\leftrightarrow p_2$\\
&(V)&$q_1\rightarrow P+q'_1-q'_2$,& $q_2\rightarrow p_1-q'_2$&
     $m_1\leftrightarrow m_4$, 
     $m_2\leftrightarrow m_3$, 
     $m_5\leftrightarrow m_6$, 
     $p_2\leftrightarrow -P$\\
&(VI)&$q_1\rightarrow p_2-q'_1$,& $q_2\rightarrow P-q'_1+q'_2$&
     $m_1\leftrightarrow m_2$, 
     $m_3\leftrightarrow m_5$, 
     $m_4\leftrightarrow m_6$, 
     $p_1\leftrightarrow -P$\\
&(VII)&$q_1\rightarrow q'_2$,& $q_2\rightarrow q'_1$&
     $m_1\leftrightarrow m_5$, 
     $m_2\leftrightarrow m_6$, 
     $m_3\leftrightarrow m_4$, 
     $p_1\leftrightarrow p_2$\\
&(VIII)&$q_1\rightarrow q'_2$,& $q_2\rightarrow -p_2-q'_1+q'_2$&
     $m_1\rightarrow m_4$, 
     $m_2\rightarrow m_3$, 
     $m_3\rightarrow m_5$,\\
&&&& $m_4\rightarrow m_6$, 
     $m_5\rightarrow m_1$, 
     $m_6\rightarrow m_2$,\\
&&&& $p_1\rightarrow p_2$, 
     $p_2\rightarrow -P$, 
     $-P\rightarrow p_1$\\
&(IX)&$q_1\rightarrow  p_2 - q'_2$,& $q_2\rightarrow P+q'_1-q'_2$&
     $m_1\rightarrow m_3$, 
     $m_2\rightarrow m_4$, 
     $m_3\rightarrow m_6$,\\
&&&& $m_4\rightarrow m_5$, 
     $m_5\rightarrow m_2$, 
     $m_6\rightarrow m_1$,\\
&&&& $p_1\rightarrow p_2$, 
     $p_2\rightarrow -P$, 
     $-P \rightarrow p_1$\\
&(X)&$q_1\rightarrow P - q'_1 + q'_2$,& $q_2\rightarrow p_1-q'_1$&
     $m_1\rightarrow m_6$, 
     $m_2\rightarrow m_5$, 
     $m_3\rightarrow m_1$,\\
&&&& $m_4\rightarrow m_2$, 
     $m_5\rightarrow m_4$, 
     $m_6\rightarrow m_3$,\\
&&&& $p_2\rightarrow p_1$, 
     $p_1\rightarrow -P$, 
     $-P \rightarrow p_2$\\
&(XI)&$q_1\rightarrow -p_1+q'_1-q'_2$,& $q_2\rightarrow q'_1$&
     $m_1\rightarrow m_5$, 
     $m_2\rightarrow m_6$, 
     $m_3\rightarrow m_2$,\\
&&&& $m_4\rightarrow m_1$, 
     $m_5\rightarrow m_3$, 
     $m_6\rightarrow m_4$,\\
&&&& $p_1\rightarrow -P$, 
     $p_2\rightarrow p_1$, 
     $-P\rightarrow p_2$\\\hline 
\end{tabular}
\end{center}
\vspace*{-0.65cm}
\caption[]{\label{tab:symmetries2loop}Symmetry transformations for
  two-loop topologies.}
\end{table}

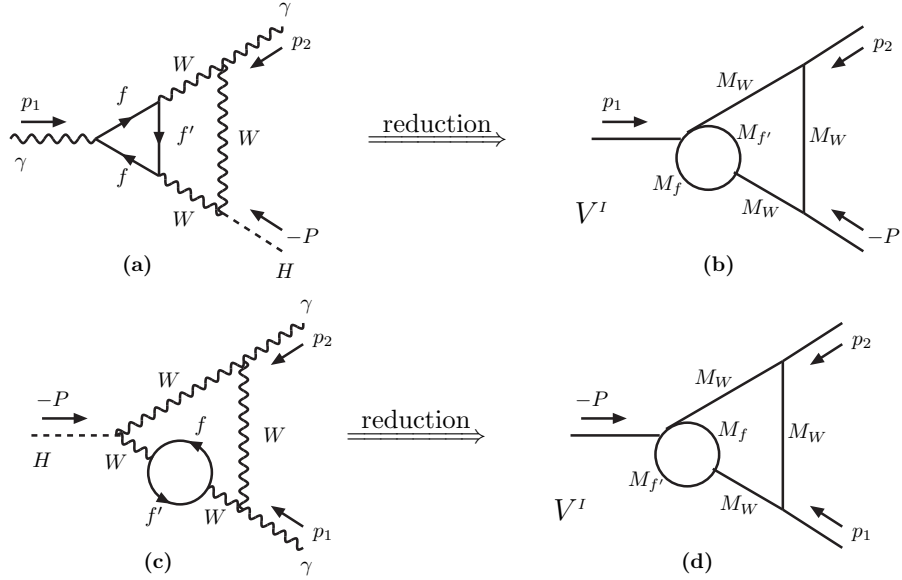
\begin{figure}
\begin{center}
\raisebox{0.1cm}{\scalebox{0.8}{
\begin{picture}(150,75)(0,0)
 \SetWidth{1.2}                       
 \Photon(0,0)(40,0){2}{5}             \Text(5,-15)[cb]{$\gamma$}
 \DashLine(128,-53)(100,-35){3}       \Text(130,-66)[cb]{$H$} 
 \Photon(128,53)(100,35){2}{5}        \Text(130,58)[cb]{$\gamma$}

 \ArrowLine(70,-17.5)(40,0)           \Text(53,-23)[cb]{$f$}
 \ArrowLine(40,0)(70,17.5)            \Text(53,16)[cb]{$f$}

 \ArrowLine(70,17.5)(70,-17.5)        \Text(83,-3)[cb]{$f'$}
 \Photon(70,-17.5)(100,-35){2}{5}     \Text(82,-41)[cb]{$W$}
 \Photon(100,-35)(100,35){2}{10}      \Text(113,-3)[cb]{$W$}
 \Photon(100,35)(70,17.5){2}{5}       \Text(82,32)[cb]{$W$}
 \Text(60,-65)[cb]{$\mbox{\bf{(a)}}$}
 \LongArrow(5,8)(25,8)          \Text(10,13)[cb]{$p_{1}$}
 \LongArrow(128,-43)(114,-34)   \Text(138,-50)[cb]{$-P$}
 \LongArrow(128,43)(114,34)     \Text(138,42)[cb]{$p_{2}$}
\end{picture}
}}
\hspace{0.2cm}
$\stackrel{\mbox{reduction}}{=\!=\!=\!=\!=\!=\!\Longrightarrow}$
\hspace{0.8cm}
\raisebox{0.1cm}{\scalebox{0.8}{
\begin{picture}(130,75)(0,0)
 \SetWidth{1.2}
 \Line(0,0)(42,0)         
 \LongArrow(5,8)(25,8)          \Text(10,13)[cb]{$p_{1}$}
 \Line(128,-53)(100,-35)  
 \LongArrow(128,-43)(114,-34)   \Text(138,-50)[cb]{$-P$}
 \Line(128,53)(100,35)    
 \LongArrow(128,43)(114,34)     \Text(138,41)[cb]{$p_{2}$}
 \CArc(55,-9)(15,0,360)         \Text(77,-3)[cb]{$M_{f'}$}
                                \Text(36,-27)[cb]{$M_f$}
 \Line(100,-35)(67,-15.75)      \Text(80,-37)[cb]{$M_W$}
 \Line(100,-35)(100,35)         \Text(112,-3)[cb]{$M_W$}
 \Line(100,35)(45,3)            \Text(68,23)[cb]{$M_W$}
 \Text(0,-40)[cb]{\Large $V^{\aca}$}
 \Text(60,-65)[cb]{$\mbox{\bf{(b)}}$}
\end{picture}
}}\\[1.7cm]
\raisebox{0.1cm}{\scalebox{0.8}{
\begin{picture}(130,75)(0,0)
 \SetWidth{1.2}
 \DashLine(0,0)(42,0){3}        \Text(10,13)[cb]{$-P$} 
 \LongArrow(5,8)(25,8)          
 \Photon(128,-53)(100,-35){2}{5}  
 \LongArrow(128,-43)(114,-34)   \Text(138,-50)[cb]{$p_{1}$}
 \Photon(128,53)(100,35){2}{5}    
 \LongArrow(128,43)(114,34)     \Text(138,42)[cb]{$p_{2}$}
 \Photon(57,-10)(40,0){2}{3}    \Text(40,-16)[cb]{$W$}
 \ArrowArc(70,-17.5)(15,-28,155)\Text(80,0)[cb]{$f$}
 \ArrowArc(70,-17.5)(15,-210,-28)
                                \Text(58,-43)[cb]{$f'$}
 \Photon(100,-35)(83,-25){2}{3} \Text(87,-42)[cb]{$W$}
 \Photon(100,-35)(100,35){2}{10} \Text(115,-3)[cb]{$W$}
 \Photon(100,35)(40,0){2}{10}    \Text(65,23)[cb]{$W$}
 \Text(60,-65)[cb]{$\mbox{\bf{(c)}}$}
 \Text(5,-15)[cb]{$H$} 
 \Text(130,-66)[cb]{$\gamma$} 	 
 \Text(130,58)[cb]{$\gamma$}
\end{picture}
}}
\hspace{0.2cm}
$\stackrel{\mbox{reduction}}{=\!=\!=\!=\!=\!=\!\Longrightarrow}$
\hspace{0.8cm}
\raisebox{0.1cm}{\scalebox{0.8}{
\begin{picture}(130,75)(0,0)
 \SetWidth{1.2}
 \Line(0,0)(42,0)         
 \LongArrow(5,8)(25,8)          \Text(10,13)[cb]{$-P$}
 \Line(128,-53)(100,-35)  
 \LongArrow(128,-43)(114,-34)   \Text(138,-50)[cb]{$p_{1}$}
 \Line(128,53)(100,35)    
 \LongArrow(128,43)(114,34)     \Text(138,41)[cb]{$p_{2}$}
 \CArc(55,-9)(15,0,360)         \Text(77,-3)[cb]{$M_f$}
                                \Text(36,-27)[cb]{$M_{f'}$}
 \Line(100,-35)(67,-15.75)      \Text(80,-37)[cb]{$M_W$}
 \Line(100,-35)(100,35)         \Text(112,-3)[cb]{$M_W$}
 \Line(100,35)(45,3)            \Text(68,23)[cb]{$M_W$}
 \Text(0,-40)[cb]{\Large $V^{\aca}$}
 \Text(60,-65)[cb]{$\mbox{\bf{(d)}}$}
\end{picture}
}}
\end{center}
\vspace{1.1cm}
\caption[]{\label{ex:VIsym} The $V^{\ssK}$- and $V^{\ssM}$-type Feynman diagrams (a) and (c), 
  where $f$ and $f'$ stand for fermions of the same doublet, 
  lead to the $V^{\ssI}$ integrals (b) and (d) after reduction of scalar products.}
\end{figure}

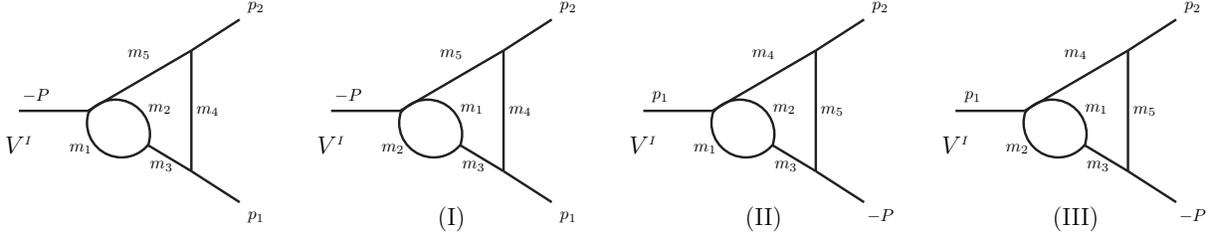
\begin{figure}[!ht]
$$
\scalebox{0.65}{
\begin{picture}(140,75)(0,0)
 \SetWidth{1.4}
 \Line(0,0)(40,0)                    \Text(10,5)[cb]{$-P$}
 \Line(128,-53)(100,-35)             \Text(138,-65)[cb]{$p_1$}
 \Line(128,53)(100,35)               \Text(138,57)[cb]{$p_2$}
 \CArc(56,-13.5)(20,-21,140)         \Text(36,-25)[cb]{$m_1$}
 \CArc(59.5,-7)(20,160,320)          \Text(82,-2)[cb]{$m_2$}
 \Line(40,0)(100,35)                 \Text(70,30)[cb]{$m_5$}
 \Line(100,-35)(100,35)              \Text(110,-3)[cb]{$m_4$}
 \Line(100,-35)(75,-20)              \Text(83,-35)[cb]{$m_3$}
                                     \Text(0,-25)[cb]{\Large  $V^{\ssI}$}
\end{picture}
}
\hspace{0.8cm}
\scalebox{0.65}{
\begin{picture}(140,75)(0,0)
 \SetWidth{1.4}
 \Line(0,0)(40,0)                    \Text(10,5)[cb]{$-P$}
 \Line(128,-53)(100,-35)             \Text(138,-65)[cb]{$p_1$}
 \Line(128,53)(100,35)               \Text(138,57)[cb]{$p_2$}
 \CArc(56,-13.5)(20,-21,140)         \Text(36,-25)[cb]{$m_2$}
 \CArc(59.5,-7)(20,160,320)          \Text(82,-2)[cb]{$m_1$}
 \Line(40,0)(100,35)                 \Text(70,30)[cb]{$m_5$}
 \Line(100,-35)(100,35)              \Text(110,-3)[cb]{$m_4$}
 \Line(100,-35)(75,-20)              \Text(83,-35)[cb]{$m_3$}
                                     \Text(0,-25)[cb]{\Large  $V^{\ssI}$}
                                     \Text(70,-70)[cb]{\Large{(I)}}
\end{picture}
}
\hspace{0.8cm}
\scalebox{0.65}{
\begin{picture}(140,75)(0,0)
 \SetWidth{1.4}
 \Line(0,0)(40,0)                    \Text(10,5)[cb]{$p_1$}
 \Line(128,-53)(100,-35)             \Text(138,-65)[cb]{$-P$}
 \Line(128,53)(100,35)               \Text(138,57)[cb]{$p_2$}
 \CArc(56,-13.5)(20,-21,140)         \Text(36,-25)[cb]{$m_1$}
 \CArc(59.5,-7)(20,160,320)          \Text(82,-2)[cb]{$m_2$}
 \Line(40,0)(100,35)                 \Text(70,30)[cb]{$m_4$}
 \Line(100,-35)(100,35)              \Text(110,-3)[cb]{$m_5$}
 \Line(100,-35)(75,-20)              \Text(83,-35)[cb]{$m_3$}
                                     \Text(0,-25)[cb]{\Large  $V^{\ssI}$}
                                     \Text(70,-70)[cb]{\Large{(II)}}
\end{picture}
}
\hspace{0.8cm}
\scalebox{0.65}{
\begin{picture}(140,75)(0,0)
 \SetWidth{1.4}
 \Line(0,0)(40,0)                    \Text(10,5)[cb]{$p_1$}
 \Line(128,-53)(100,-35)             \Text(138,-65)[cb]{$-P$}
 \Line(128,53)(100,35)               \Text(138,57)[cb]{$p_2$}
 \CArc(56,-13.5)(20,-21,140)         \Text(36,-25)[cb]{$m_2$}
 \CArc(59.5,-7)(20,160,320)          \Text(82,-2)[cb]{$m_1$}
 \Line(40,0)(100,35)                 \Text(70,30)[cb]{$m_4$}
 \Line(100,-35)(100,35)              \Text(110,-3)[cb]{$m_5$}
 \Line(100,-35)(75,-20)              \Text(83,-35)[cb]{$m_3$}
                                     \Text(0,-25)[cb]{\Large  $V^{\ssI}$}
                                     \Text(70,-70)[cb]{\Large{(III)}}
\end{picture}
}
$$
\vspace{0.5cm}
\caption[]{\label{fig:VIsymmetries} The symmetries of the
  $V^{\ssI}$ family enumerated in \tabn{tab:symmetries2loop}.} 
\end{figure}

As an explicit simple example concerning the application of symmetry
transformations, let us consider the case of the $V^{\ssI}$-family
integrals arising after the reduction of the $V^{\ssK}$- and  $V^{\ssM}$-type 
Feynman diagrams shown in \fig{ex:VIsym} (a) and (c). Note that the momenta 
routings of the two child $V^{\ssI}$-family diagrams in \fig{ex:VIsym} (b) and (d) 
are different; the application of 
symmetries aims to map the various integral representations onto a single one
in order to allow for cancellations.
The $V^I$ family has three basic symmetry transformations, summarized in
\tabn{tab:symmetries2loop} and graphically illustrated in
\fig{fig:VIsymmetries}: here the first diagram corresponds to the 
standard integral representation of $V^{\ssI}$ as defined in
appendix~\ref{app:diagrams}; the symmetry transformation (I) amounts to exchanging 
the first and the second line of the self-energy insertion in the diagram; 
(II) corresponds to an exchange of the fourth and fifth lines
(in this case also the external momenta are interchanged, $p_1
\leftrightarrow -P$); symmetry (III) is a combination of symmetries (I) 
and (II).
The two $V^{\ssI}$ child topologies in \fig{ex:VIsym}(b) and (d) are
related through the symmetry transformation (III). 

Finally, we stress that the procedures of reducing scalar products and
symmetrizing loop integrals have been recursively performed in order to 
achieve a maximal simplification of the amplitude.
Next, the remaining integrals are classified as scalar-, vector- and 
tensor-type integrals according to the number of irreducible scalar 
products in the numerators. 
The Lorentz structure of the loop integrals are expressed through the
external momenta $p_j$, with $j=1,2$, and the metric tensor introducing 
suitable form factors. 
A {\sc FORTRAN} code is generated and the form factors are then 
evaluated numerically employing the {\sc NAG} library~\cite{naglib}.
\section{Behavior of two-loop diagrams around a normal threshold}
\label{NormTH}
In this section we will discuss one of the main results of the paper:
by carefully studying all singularities of Feynman diagrams we propose the
two-loop implementation of a renormalization scheme which cures anomalous
behaviors of the amplitude.

Feynman diagrams have a complicated analytical structure as functions
of the external Mandelstam invariants and the internal masses. A 
frequently encountered singular behavior is associated with the so-called
normal thresholds: the leading Landau singularities~\cite{Landau:1959fi} of 
self-energy-like diagrams which can appear, in more complicated diagrams, 
as sub-leading singularities. In this section we discuss how the amplitudes 
for $H\to\gamma\gamma$ and $H \to gg$ behave around a normal threshold,
with special emphasis to the problem of possible square-root and logarithmic
singularities.

\begin{figure}[ht]
\vspace{-0.3cm}
$$
\scalebox{0.8}{
\begin{picture}(140,40)(0,0)
 \DashLine(82,40)(82,-40){3}
 \SetWidth{1.1}
 \DashLine(0,0)(25,0){3}                 \Text(10,5)[cb]{$H$}
 \Photon(95,-20)(125,-20){2}{6}          \Text(130,-30)[cb]{$\gamma$}
 \Photon(95,20)(125,20){2}{6}            \Text(130,24)[cb]{$\gamma$}
 \Photon(25,0)(59,20){2}{6}              \Text(40,15)[cb]{$W$}
 \Photon(59,-20)(25,0){2}{6}             \Text(40,-22)[cb]{$W$}
 \ArrowLine(59,-20)(59,20)               \Text(68,-3)[cb]{$d$}
 \ArrowLine(95,-20)(59,-20)              \Text(72,-32)[cb]{$u$}
 \ArrowLine(59,20)(95,20)                \Text(72,26)[cb]{$u$}
 \ArrowLine(95,20)(95,-20)               \Text(102,-3)[cb]{$u$}
\end{picture}
}
\scalebox{0.8}{
\begin{picture}(140,40)(0,0)
 \DashLine(100,40)(20,-40){3}
 \SetWidth{1.1}
 \DashLine(0,0)(25,0){3}                 \Text(10,5)[cb]{$H$}
 \Photon(95,-20)(125,-20){2}{6}          \Text(130,-30)[cb]{$\gamma$}
 \Photon(95,20)(125,20){2}{6}            \Text(130,24)[cb]{$\gamma$}
 \Photon(25,0)(59,20){2}{6}              \Text(40,15)[cb]{$W$}
 \Photon(59,-20)(25,0){2}{6}             \Text(31,-19)[cb]{$W$}
 \ArrowLine(59,-20)(59,20)               \Text(68,-5)[cb]{$d$}
 \ArrowLine(95,-20)(59,-20)              \Text(76,-32)[cb]{$u$}
 \ArrowLine(59,20)(95,20)                \Text(76,26)[cb]{$u$}
 \ArrowLine(95,20)(95,-20)               \Text(102,-3)[cb]{$u$}
\end{picture}
}
\scalebox{0.8}{
\begin{picture}(140,40)(0,0)
 \DashLine(100,40)(20,-40){3}
 \SetWidth{1.1}
 \DashLine(0,0)(25,0){3}                 \Text(10,5)[cb]{$H$}
 \Photon(95,-20)(125,-20){2}{6}          \Text(130,-30)[cb]{$\gamma$}
 \Photon(95,20)(125,20){2}{6}            \Text(130,24)[cb]{$\gamma$}
 \Photon(25,0)(59,20){2}{6}              \Text(40,15)[cb]{$Z$}
 \Photon(59,-20)(25,0){2}{6}             \Text(31,-19)[cb]{$Z$}
 \ArrowLine(59,-20)(59,20)               \Text(68,-5)[cb]{$u$}
 \ArrowLine(95,-20)(59,-20)              \Text(76,-32)[cb]{$u$}
 \ArrowLine(59,20)(95,20)                \Text(76,26)[cb]{$u$}
 \ArrowLine(95,20)(95,-20)               \Text(102,-3)[cb]{$u$}
\end{picture}
}
\scalebox{0.8}{
\begin{picture}(140,40)(0,0)
 \DashLine(47,40)(47,-40){3}
 \SetWidth{1.1}
 \DashLine(0,0)(25,0){3}                 \Text(10,5)[cb]{$H$}
 \Photon(95,-20)(125,-20){2}{6}          \Text(130,-30)[cb]{$\gamma$}
 \Photon(95,20)(125,20){2}{6}            \Text(130,24)[cb]{$\gamma$}
 \Photon(25,0)(59,20){2}{6}              \Text(36,14)[cb]{$W$}
 \Photon(59,-20)(25,0){2}{6}             \Text(36,-21)[cb]{$W$}
 \ArrowLine(59,-20)(59,20)               \Text(68,-3)[cb]{$d$}
 \ArrowLine(95,-20)(59,-20)              \Text(76,-32)[cb]{$u$}
 \ArrowLine(59,20)(95,20)                \Text(76,26)[cb]{$u$}
 \ArrowLine(95,20)(95,-20)               \Text(102,-3)[cb]{$u$}
\end{picture}
}
$$
\vspace{0.2cm}
\caption[]{Sample two- and three-particle cut diagrams for 
$H \to \gamma\gamma$. The mass of the up and down quarks is neglected.}
\label{noVKzerocut}
\end{figure}
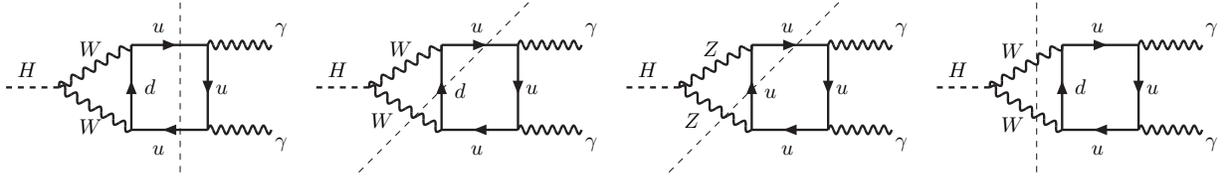

Without loss of generality, let us consider the case of the $H \to \gamma\gamma$ 
decay in the setup where all light fermions are taken to be massless. 
As far as normal thresholds are concerned, we have the possibilities 
$\mh = \mw\,,\,\mz\,,\,2\,\mw\,,\,2\,\mz\,,\,2 M_t$, as illustrated in 
terms of cut diagrams in \fig{noVKzerocut}. 
Note that, as observed by the authors of Ref.~\cite{Degrassi:2004mx}, 
there is no cut at $\mh = 0$ even in presence of massless fermions; 
indeed, the two-particle cut of the first diagram of \fig{noVKzerocut} is zero 
because of the helicity structure of the diagram.

\begin{figure}[ht]
    \includegraphics[bb=0 0 567 384,height=6.cm,width=7.cm]{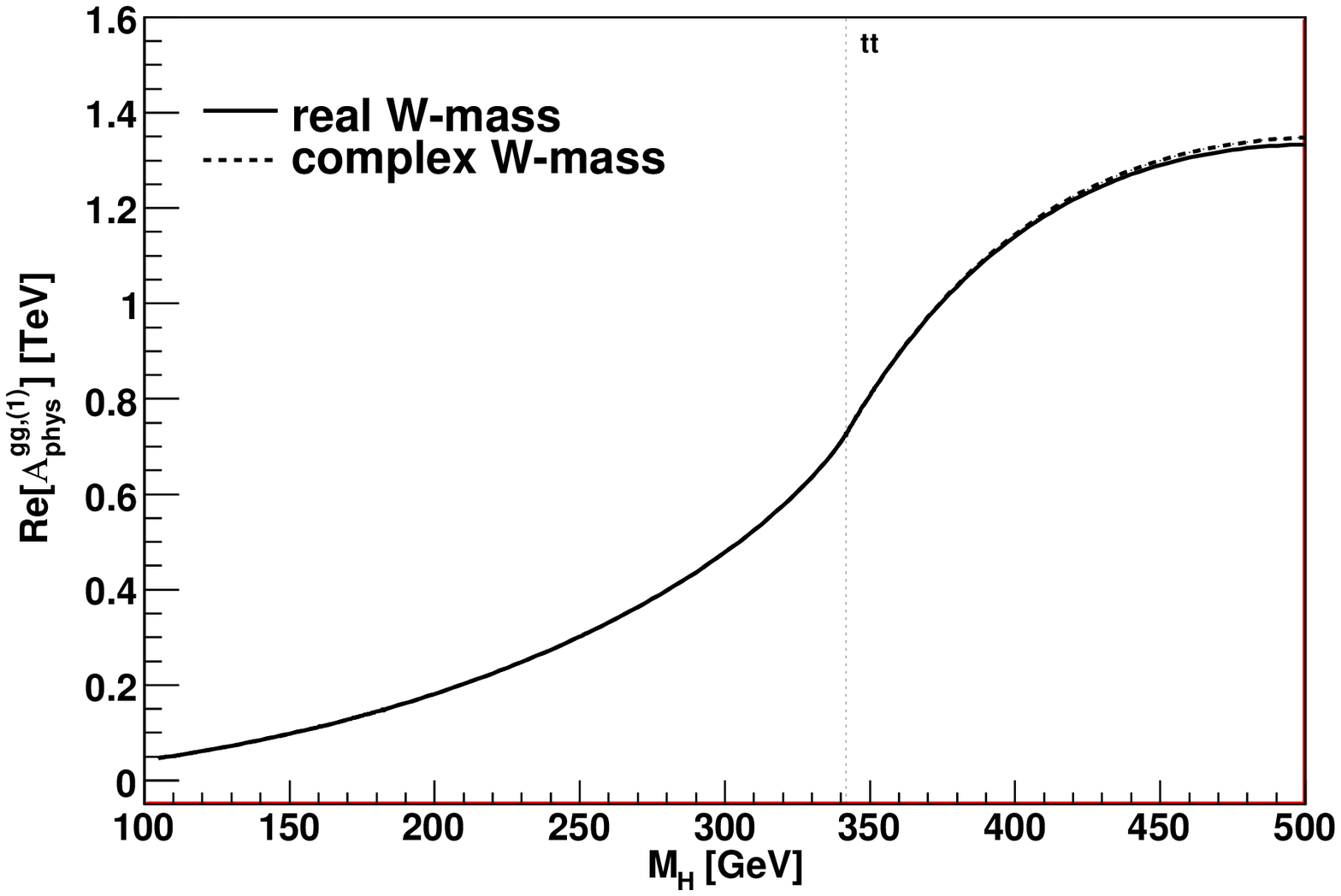}
    \hspace{1.cm}
    \includegraphics[bb=0 0 567 384,height=6.cm,width=7.cm]{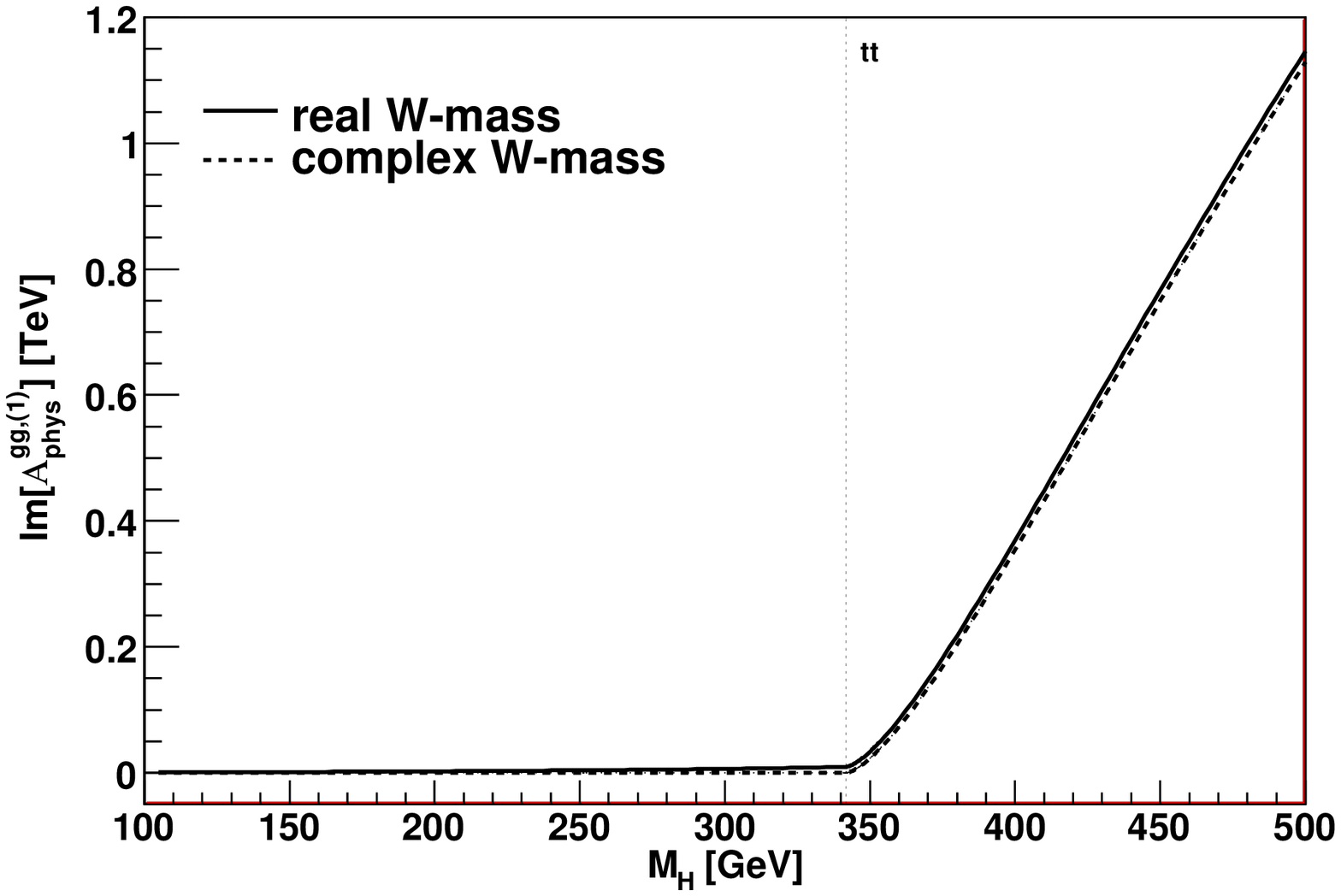}
    \vspace{-0.4cm}
    \caption[]{\label{OLlmh} Real and imaginary parts of the one-loop $H
      \to gg$ amplitude.}
  \end{figure}

The $\mh = \mw,\mz$ cuts are identified by the configuration shown in the second and
third diagrams of \fig{noVKzerocut}. Note that, for $H \to \gamma \gamma$,
the imaginary part is not exactly zero below the 
single-$W\,$threshold; this is due to the introduction of complex masses for 
vector bosons, as we will explain in detail in section~\ref{subsec:complexmass}. 
It is worth mentioning the behavior of the one-loop amplitude which is rapidly 
decreasing for small values of $\mh$, as shown in \fig{OLlmh} for $H \to g g$. 
Here, once again, we use a complex $W\,$ mass and, therefore, the imaginary part 
is different from zero even below the $t \overline{t}\,$threshold.
We have performed a dedicated 
analysis of the behavior of the amplitude around {\em single} thresholds which 
shows that regular behavior is a consequence of a delicate mechanism of numerical 
cancellation among several diagrams, e.g. $V^{\ssH}, V^{\ssG}$ doubly collinear 
and $V^{\ssE}$ simply collinear.

Having discussed the single vector-boson thresholds, we move to the double ones, 
identified for example by the cut of the fourth diagram of \fig{noVKzerocut}.
\subsection{Square-root singularities}\label{sqroots}
In this section we are interested in the problem of possible square-root 
singularities of the amplitude. When present, they are unphysical, although 
integrable; the solution to the apparent puzzle consists, as we will argue, in 
replacing real masses of unstable particles with their complex poles.

In order to illustrate the problem, let us consider the following integral,
related to a generalized one-loop two-point function:
  \bq
  I^n_{\alpha} = \intfx{x}\,x^n\,\Bigl( x^2 - x + \mu^2 - i\,0\Bigr)^{-\alpha},
  \label{iexam}
  \eq
where $n$ is a non-negative integer. When $\beta = 0$, with $\beta^2= 
1-4\,\mu^2$, the integration contour is pinched between the two 
singular points $x_{\pm}= (1\pm \beta)/2$, and the integral of \eqn{iexam} 
is singular, with a branch point of the two-particle cut $\mu^2 = 1/4$;
for $\alpha \ge 1$ we find
  \bq
  I^0_{\alpha} \sim
  \Bigl(-\,\frac{\beta}{2}\Bigr)^{1-2\,\alpha}\,
  B\left(\frac{1}{2},\alpha-\frac{1}{2}\right),\qquad \beta \to 0,
  \eq
where $B$ is the Euler beta function. To be more explicit, let us consider 
the UV decomposition of~section~\ref{conv} for the generalized one-loop two-point 
function with equal masses
  \bq
  B_0(k,l\,;\,s,m,m) = \sum_{i=-1}^1\,B_0(k,l\,;\,s,m,m\,;\,i)\,F^1_i(s),
  \eq
where $k$ and $l$ are the powers of the two propagators.
Concerning the finite parts for the $k=l=1$ and $k=2,l=1$ configurations, 
we obtain
\bqa
\label{deriv}
B_0(1,1\,;\,s,m,m\,;\,0) &=& 2 - \ln\frac{m^2}{s+i\, 0} - \beta\,L_{\beta},
\qquad B_0(2,1\,;\,s,m,m\,;\,0)=-\frac{1}{\beta}\,L_\beta,
\nl
L_{\beta} &=& \ln\frac{\beta+1}{\beta-1}, \qquad 
\beta^2= 1 - 4\,\frac{m^2}{s+i\,0}.
\eqa
In general, for $k+l > 2$, we have
\bq
B_0(k,l\,;\,s,m,m\,;\,0) \sim \beta^{5-2(k+l)}, \qquad \beta \to 0,
\eq
and we can easily conclude that the generalized one-loop two-point function
has an (unphysical) singularity at $s= 4\,m^2$ $\forall k,l\, | \,k+l > 2$.
For practical applications, one has to distinguish between the above- and 
below-threshold regions,
\bqa
\ba{lll}
\beta^2 = a^2 \ge 0, 
\;&\;
L_{\beta} = \ln\frac{1+a}{1-a} - i\,\pi \sim -i\,\pi, 
\;&\;
s \to 4\,m^2\mid_+,
\\
\beta^2 = - a^2 \le 0,
\;&\;
L_{\beta} = i\,\Bigl( \arctan\frac{2\,a}{1-a^2} + \pi\Bigr)
\sim i\,\pi, 
\;&\; s \to 4\,m^2\mid_-.
\ea
\eqa
Therefore, the generalized one-loop two point function with $k+l=3$ generates 
a square-root $1/ \beta$-divergent behavior, with $\beta^2 = 1-4\,\mws/\mhs$ or 
$\beta^2 = 1-4\,\mzs/\mhs$, associated with the normal threshold related to the 
two-particle cut (the leading Landau singularity).
\begin{figure}[ht]
\vspace{-0.5cm}
$$
\scalebox{0.7}{
\begin{picture}(220,70)(0,0)
 \SetWidth{1.4}
 \Line(0,0)(40,0)         
 \LongArrow(5,8)(25,8)          \Text(10,13)[cb]{\Large{$s$}}
 \Line(140,-35)(100,-35)  
 \Line(140,35)(100,35)    
 \Line(100,-35)(40,0)           \Text(70,-32)[cb]{\Large{$m$}}
 \Line(100,-35)(100,35)         
 \Line(100,35)(40,0)            \Text(70,26)[cb]{\Large{$m$}}
  \Text(190,-4)[cb]{\Large $\Longrightarrow$}
  \Text(190,-14)[cb]{\large contraction}
\end{picture}
}
\qquad
\scalebox{0.7}{
\begin{picture}(130,70)(0,0)
 \SetWidth{1.4}
 \Line(0,0)(35,0)         
 \LongArrow(5,8)(25,8)          \Text(10,13)[cb]{\Large{$s$}}
 \CArc(60,0)(25,0,360)          \Text(60,31)[cb]{\Large{$m$}}
                                \Text(60,-37)[cb]{\Large{$m$}}
 \Line(85,0)(120,35)
 \Line(85,0)(120,-35)
\end{picture}
}
$$
\vspace{0.3cm}
\caption[]{One-loop vertex with two equal masses and its sub-diagram
giving the sub-leading square-root singularity.}
\label{vertTH}
\end{figure}
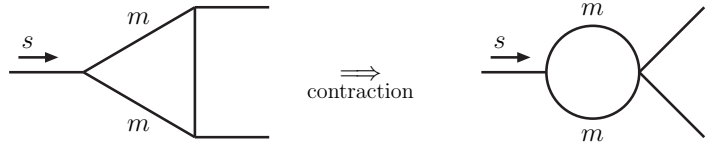

Let us now consider three-point functions, and investigate the behavior of the 
vertex of \fig{vertTH} around the normal threshold located at $s = 4\,m^2$. The 
leading Landau singularity of the vertex is the so-called anomalous threshold; 
the normal threshold shows up as a sub-leading singularity. 
Since the sub-leading singularity for a graph is the leading one for 
any of the contracted sub-graphs, we easily conclude that the singular 
behavior of the one-loop vertex of \fig{vertTH} around $s= 4\,m^2$ is due 
to the sub-graph where we shrink a line to a point; the singularity is a 
branch point in the complex $s\,$-plane. 

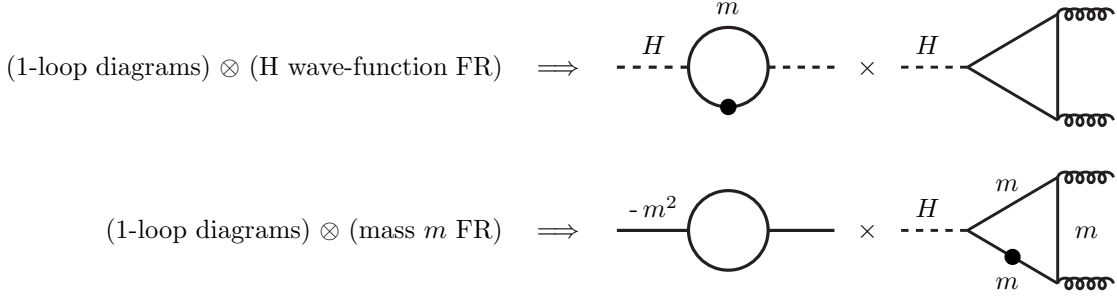
\begin{figure}[ht]
\vspace{-0.8cm}
\bqas
\mbox{(1-loop diagrams) $\otimes$ (H wave-function FR)}
\quad &\Longrightarrow& \quad
\raisebox{0.1cm}{
\begin{picture}(80,40)(0,0)
 \SetWidth{1.2}
 \DashLine(-2,0)(25,0){3}           \Text(12,5)[cb]{$H$}
 \DashLine(55,0)(80,0){3}
  \CArc(40,0)(15,0,360)              \Text(40,20)[cb]{$m$}
 \CCirc(40,-15){2.5}{1}{1}
\end{picture}
}
\,\,\times\,\,
\raisebox{0.1cm}{
\begin{picture}(80,40)(0,0)
 \SetWidth{1.1}
 \DashLine(0,0)(25,0){3}            \Text(10,5)[cb]{$H$}
 \Gluon(59,-20)(80,-20){2}{4}
 \Gluon(59,20)(80,20){2}{4}
 \Line(25,0)(59,20)
 \Line(59,20)(59,-20)
 \Line(59,-20)(25,0)
\end{picture}
}
\\ \\
\mbox{(1-loop diagrams) $\otimes$ (mass $m$ FR)}
\quad &\Longrightarrow& \quad
\raisebox{0.1cm}{
\begin{picture}(80,40)(0,0)
 \SetWidth{1.2}
 \Line(-2,0)(25,0)          \Text(12,5)[cb]{-$\,m^2$}
 \Line(55,0)(80,0)
 \CArc(40,0)(15,0,360)
\end{picture}
}
\,\,\times\,\,
\raisebox{0.1cm}{
\begin{picture}(80,40)(0,0)
 \SetWidth{1.1}
 \DashLine(0,0)(25,0){3}            \Text(10,5)[cb]{$H$}
 \Gluon(59,-20)(80,-20){2}{4}
 \Gluon(59,20)(80,20){2}{4}
 \Line(25,0)(59,20)                 \Text(40,15)[cb]{$m$}
 \Line(59,20)(59,-20)               \Text(70,-3)[cb]{$m$}
 \Line(59,-20)(25,0)                \Text(40,-22)[cb]{$m$}
 \CCirc(42,-10){2.5}{1}{1}
\end{picture}
}
\eqas
\vspace{0.2cm}
\caption[]{Singular $\beta^{-1}$ behavior at the normal $m$ threshold 
coming from WFR and mass finite renormalization.}
\label{hasCdI} 
\end{figure}

The same argument can be repeated for all diagrams with any number of 
external legs where we can cut {\em two and only two} lines with mass $m$; 
any normal threshold will be a sub-(sub-$\,\dots\,$) leading singularity, 
and a $1/\beta$ behavior will show up only if the reduced sub-graph, 
responsible for the singularity, can be reduced to the generalized one-loop 
two-point function of \eqn{deriv}, ${\dot B}_0(s,m,m) = B_0(2,1\,;\,s,m,m;0)$. 
A ${\dot B}_0(s,m,m)$ function is related to the derivative of a 
$B_0(1,1\,;\,s,m,m)$ function and therefore emerges in the computation 
whenever we include WFR factors for the external legs or we 
perform finite renormalization of the mass of an internal particle, 
as depicted in \fig{hasCdI}.
In the second case, generalized one-loop triangle functions appear,
and it is known that they can always be reduced to ${\dot B}_0$
functions using IBP identities (this is just another way to say 
that the normal threshold is a sub-leading singularity for a generalized
three-point function).

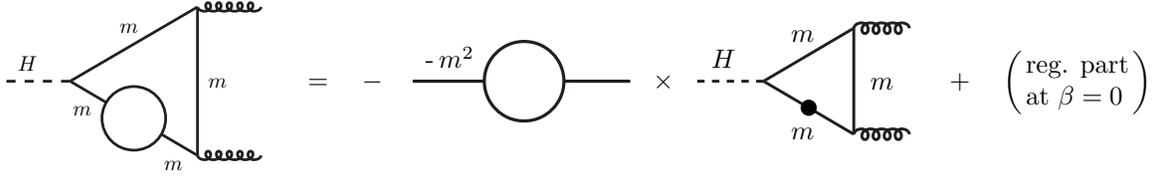
\begin{figure}[ht]
\vspace{-1cm}
$$
\raisebox{0.1cm}{\scalebox{0.8}{
\begin{picture}(140,70)(10,0)
 \SetWidth{1.3}
 \DashLine(10,0)(40,0){4}           \Text(20,5)[cb]{$H$}
 \Gluon(130,-35)(100,-35){2}{5}
 \Gluon(130,35)(100,35){2}{5}
 \CArc(70,-17.5)(15,0,360)   
 \Line(57,-10)(40,0)                \Text(46,-16)[cb]{$m$}
 \Line(100,-35)(83,-25)             \Text(89,-42)[cb]{$m$}
 \Line(100,35)(40,0)                \Text(68,23)[cb]{$m$}
 \Line(100,-35)(100,35)             \Text(110,-3)[cb]{$m$}
\end{picture}
}}
\!\!
=
\quad
-
\quad
\raisebox{0.1cm}{
\begin{picture}(80,40)(0,0)
 \SetWidth{1.2}
 \Line(-2,0)(25,0)                     \Text(12,5)[cb]{-$\,m^2$}
 \Line(55,0)(80,0)                   
 \CArc(40,0)(15,0,360)
\end{picture}
}
\,\,\times\,\,
\raisebox{0.1cm}{
\begin{picture}(80,40)(0,0)
 \SetWidth{1.1}
 \DashLine(0,0)(25,0){3}            \Text(10,5)[cb]{$H$}
 \Gluon(59,-20)(80,-20){2}{4}
 \Gluon(59,20)(80,20){2}{4}
 \Line(25,0)(59,20)                 \Text(40,15)[cb]{$m$}
 \Line(59,20)(59,-20)               \Text(70,-3)[cb]{$m$}
 \Line(59,-20)(25,0)                \Text(40,-22)[cb]{$m$}
 \CCirc(42,-10){2.5}{1}{1}
\end{picture}
}
\quad
+
\quad
\left(
\ba{l}
\hbox{reg. part}\\
\hbox{at } \beta=0
\ea
\right)
$$
\vspace{0.5cm}
\caption[]{Singular $\beta^{-1}$ behavior at the normal $m$ threshold 
coming from self-energy insertions ($V^{\ssM}$ topology).
The generalized one-loop triangle function in the right-hand side is 
related to the generalized one-loop self-energy of \eqn{deriv} 
through IBP identities.}
\label{hasCdII} 
\end{figure}

Concerning genuine two-loop diagrams, we observe that a ${\dot B}_0$ configuration 
can only arise if we have a self-energy insertion in a two-loop diagram.
A two-loop vertex containing a self-energy insertion, leading to a 
$1/\beta$-divergent behavior, is depicted in \fig{hasCdII}. 
For this diagram it is possible to find a representation where the 
singular part is completely written in terms of one-loop diagrams, 
as shown in the figure. 
The remainder can be cast in a form suited for numerical integration.

Note that the unphysical $1\slash \beta$ behavior, generated by the diagram
of \fig{hasCdII}, exactly cancels the one coming from mass finite 
renormalization. This is strictly true only in a complex 
renormalization scheme because, for on-shell real masses, only the real 
part of the self-energy in \fig{hasCdI} is taken and the cancellation 
does not take place for the corresponding imaginary part generated by 
the two-loop diagram of \fig{hasCdII}.
This is not surprising at all: self-energy insertions, signaling 
the presence of an unstable particle, should not be there. 
They are the consequence of a misleading organization of the perturbative 
expansion; Dyson-resummed propagators should be used and complex poles 
should replace real on-shell masses.
The remaining $1/\beta$ singularity is therefore coming only from the 
wave-function renormalization of the Higgs boson.
This unphysical behavior is strictly connected to the problem of 
defining a proper WFR for an unstable particle, as pointed out 
in~section~\ref{Hm&WFR}.

It is interesting to note that for the $t{\bar t}\,$ threshold in 
$H \to gg$ the ${\dot B}_0\,$-functions that are potentially dangerous 
always appear multiplied by $\beta^2$, as it happens for QCD corrections; 
the same is not true for pseudo-scalar Higgs decay, cfr. Fig.~4 
of~\cite{Harlander:2005rq}.

Finally, we observe that a more severe behavior associated with $\beta \to 0$ 
should not show up; for instance, we have verified that $1/\beta^2$ terms which 
appear as a consequence of the reduction procedure for the 
$H \to \gamma \gamma$ decay are of the form $F(\mhs)/ (\mhs-4\,\mws)$, 
with $F(4\,\mws) = 0$.
\subsection{Logarithmic singularities\label{sec:LogSing}}
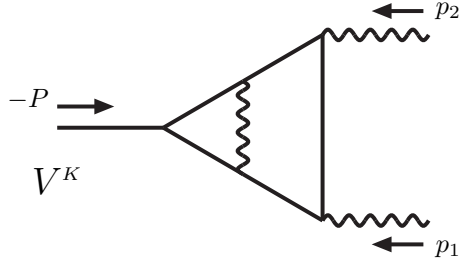
\begin{figure}[!ht]
\vspace{-0.5cm}
\begin{center}
\begin{picture}(150,70)(0,0)
 \SetWidth{1.5}
 \Line(0,0)(40,0)         
 \Photon(140,-35)(100,-35){2}{5}  
 \Photon(140,35)(100,35){2}{5}    
 \Photon(70,-17.5)(70,17.5){2}{5}
 \Line(70,-17.5)(40,0)
 \Line(70,17.5)(40,0) 
 \Line(100,-35)(70,-17.5)
 \Line(100,-35)(100,35)
 \Line(100,35)(70,17.5)
 \LongArrow(0,8)(20,8)          \Text(-11,7)[cb]{$-P$}
 \LongArrow(138,-44)(120,-44)   \Text(148,-49)[cb]{$p_1$}
 \LongArrow(138,44)(120,44)     \Text(148,41)[cb]{$p_2$}
 \Text(0,-25)[cb]{\Large $V^{\ssK}$}
\end{picture}
\end{center}
\vspace{1.3cm}
\caption[]{\label{TLvertbca} The irreducible scalar two-loop vertex
diagram of the $V^{\ssK}$ family showing up a logarithmic divergence. 
Solid lines represent a massive particle with mass $m$, whereas wavy 
lines correspond to massless particles.}
\end{figure}

Let us consider the scalar two-loop diagram of the $V^{\ssK}$ family
shown in \fig{TLvertbca}, and derive the corresponding integral in parametric 
space. We introduce the quadratic forms 
\bq
\chi(x) = \lpar x - \frac{1}{2}\rpar^2 - \frac{1}{4}\,\beta^2, \qquad
\xi(x,y) = x \,\lpar x  - 1 \rpar\,y^2 + \frac{1}{4}\,\lpar 1 - \beta^2\rpar,
\qquad \beta^2 = 1 - \frac{4\,m^2}{s},
\eq
where $s=-P^2>0$ and $m$ is the mass of the solid line (wavy lines correspond to
massless particles). 
We obtain
\bq
V^{\ssK} = \frac{2}{s^2}
           \int_0^1\,\frac{dx\,dy}{y\,\chi(x)}\,\left[
\li{2}{1 - \frac{y\,\chi(x)}{\chi(x y)}} -
\li{2}{1 - \frac{y\,\chi(x)}{\xi(x,y)}}\right].
\eq
Since we are interested in the behavior around $\beta \to 0$, we split 
$V^{\ssK}$ into a singular and regular part,
\bq
V^{\ssK} = V^{\ssK}_{\rm sing} + V^{\ssK}_{\rm reg} 
         = \frac{2}{s^2}\int_0^1\,\frac{dx\,dy}{y\,\chi(x)}\,\left[
\li{2}{1 - \frac{y\,\chi(x)}{\chi(x y)}} - \zeta(2) \right]
	 +V^{\ssK}_{\rm reg}.
\eq
The singular part $V^{\ssK}_{\rm sing}$ will be written
as~\cite{Ferroglia:2002mz} 
\bq
V^{\ssK}_{\rm sing} = \frac{2}{s^2} \! \intfx{t} \frac{\ln t}{1-t} I(t),
\quad
I(t) =\int_0^1 \!\! dx dy \Bigl[ (1\!-\!t) \chi(x y) + t y \chi(x)\Bigr]^{-1} =
\int_0^1\!\! dx dy \Bigl[ a\lpar x\! -\! X\rpar^2 + \lambda\Bigr]^{-1},
\label{VKsing}
\eq
where we have introduced the short-hand notations
\bq
a = \tau y, \qquad X = \frac{1}{2\,\tau}, \qquad
\lambda=  \frac{t\,(1-t)}{4\,\tau}\,\left[ (1-y)^2 -
\beta^2\,\lpar y + T\rpar\,\lpar y + \frac{1}{T}\rpar\right],
\eq
with $\tau = (1-t)\,y + t$ and $T = t/(1-t) > 0$. $I(t)$ can be split into 
two parts,
\bq
I(t)= B\lpar \frac{1}{2},\frac{1}{2}\rpar\,
\int_{y_{min}}^1\!\!\!dy\;a^{-1/2}\,\lambda^{-1/2} 
- \frac{1}{2}\,\sum_{i=1,2}\,
\int_0^1 dx\,dy\,
(-1)^i\,X_i\,x^{-1/2}\,\lpar a X^2_i + \lambda x\rpar^{-1}.
\label{singsplit}
\eq
Here $X_1 = - X$, $X_2 = 1 - X$ and $B(x,y)$ is the Euler beta function. 
The second term of \eqn{singsplit} is regular for $\beta=0$; the first term, 
instead, shows a singularity due to the fact that 
$\lambda \sim (1-y)^2$ for $\beta\to0$. However, we have a singular behavior 
only if $0 \le X \le 1$, which requires 
$y \ge y_{\rm min}=\max \{0\,,\,(t-1/2)/(t-1)\}$.
Being interested in the leading behavior for $\beta \to 0$, 
we can extend the integration domain in the first term to $[0,1]$ without
modifying the divergent behavior of the diagram.
The singular part is then given by
$$
I_{\rm sing}(t) = \frac{2\,\pi}{\sqrt{t (1-t)}}\!
\intsx{y}\,y^{-1/2}\,\Bigl[ (1-y)^2 - \beta^2\,
\lpar 1 + T\,y\rpar\,\lpar 1 + \frac{y}{T}\rpar\Bigr]^{-1/2}\!
= \frac{2\,\pi}{\sqrt{ t (1-t)}}\,J(t),
$$
\bqa
J(t) &=& \frac{1}{2 \pi i}\,\int_{-i\,\infty}^{+i\,\infty}\,\!\!\!\!ds\,
B\lpar s ,\frac{1}{2}-s\rpar\,\lpar - \beta^2 - i\,0\rpar^{s-1/2}\,
\intfx{y}\,y^{-1/2}\,(1-y)^{-2\,s}\,\lpar 1+T\,y\rpar^{s-1/2}\,
\lpar 1 + \frac{y}{T}\rpar^{s-1/2}
\nl
{}&=& \frac{1}{2 \pi i}\int_{-i\,\infty}^{+i\,\infty}\!\!\!\!\!\!\!ds\,
\frac{\egam{s}\egam{1/2-s}\egam{1-2 s}}{\egam{3/2-2 s}}
\lpar - \beta^2 - i\,0\rpar^{s-1/2}
F_1\lpar \frac{1}{2},\frac{1}{2}\!-\!s,
\frac{1}{2}\!-\!s,\frac{3}{2}\!-\!2s;- T,- \frac{1}{T}\rpar,
\eqa
where $0 < \Reb s < 1/2$ and $F_1$ denotes the first Appell function.
In order to obtain the expansion corresponding to $\beta \to 0$, we close the 
integration contour over the right-hand complex half-plane at 
infinity. The leading (double) pole is at $s = 1/2$; therefore, we obtain
\bq
J(t) = -\,\frac{1}{2}\,\ln \lpar - \beta^2 - i\,0 \rpar 
     + {\cal O}(1),\quad \beta\to0.
\eq
Inserting the result into \eqn{VKsing} and using 
$\intfx{t}\,t^{-1/2}\,\lpar 1 - t\rpar^{-3/2}\,\ln t = -\,2\,\pi$,
we get 
\bq
V^{\ssK}_{\rm sing} = \frac{4\,\pi^2}{s^2}\,\ln \lpar - \beta^2 - i\,0 \rpar 
     + {\cal O}(1),\quad \beta\to0.
\label{ourR}
\eq
If the massive loop in \fig{TLvertbca} is made of top quarks, the
contribution of the $V^{\ssK}$ integral to the amplitude behaves like
$\beta^2\,V^{\ssK}$ and, therefore, the logarithmic singularity is
$\beta^2\,$-protected at threshold; however, the same is not true for a
$W\,$-loop. Our result of \eqn{ourR} is confirmed by the evaluation of
$V^{\ssK}$ of Ref.~\cite{Anastasiou:2006hc} in terms of generalized 
log-sine functions. Starting from Eq.~(6.34) of
Ref.~\cite{Anastasiou:2006hc} and using the results of
Ref.~\cite{Kalmykov:2004xg} we expand around $\theta = \pi$, where $x =
e^{i\,\theta} = (\beta - 1)\,(\beta + 1)$, with $0 < \theta < \pi$.
This gives for the leading behavior of $V^{\ssK}$ below threshold  
$(\pi^2/2)\,\ln(\theta - \pi)$, where $\ln (-\beta^2) = \ln (\theta-\pi)^2 -
\ln 2$. The same behavior can also be extracted from the results of 
Ref.~\cite{Harlander:2005rq}.

Logarithmic singularities of the kind discussed in this section are a remnant 
of the one-loop Coulomb singularity of one-loop sub-diagrams. 
An alternative approach that automatically resums large Coulomb singularities at 
threshold has been pursued in Ref.~\cite{Melnikov:1994jb}.
The reader should be aware that in the pseudo-scalar decay the real and 
imaginary parts of the form factor may be significantly different from 
the lowest order perturbative ones\footnote{A discussion with M.~Spira 
on the last two points is gratefully acknowledged.}.

\subsection{Complex masses}
\label{subsec:complexmass}
In this section we set up and discuss our implementation of a consistent and 
gauge-invariant treatment of unstable particles in NNLO radiative corrections.

Our two-loop renormalization scheme has been described in details in section~\ref{Ampl} 
where counterterms have been introduced, different choices of IPSs considered and 
finite renormalization of lagrangian parameters discussed. In short, this 
represents the so-called
\bei
\item{\underline{RM - scheme}}
\eei
where masses are the real on-shell ones; it gives the extension of the generalized 
minimal subtraction scheme up to two loop level.
The analysis of section~\ref{NormTH} has shown the presence of pathological features and
the cure proceeds in two steps. 
Our first, pragmatical, solution to the problems induced by unstable internal 
particles has been presented in Ref.~\cite{Passarino:2007fp} (for an alternative
approach to the problem of unphysical threshold singularities connected to WFR
factors, see Ref.~\cite{Kniehl:2002wn}); the corresponding scheme
will be termed minimal complex mass scheme (hereafter MCM), the first emergency
kit.
\bei
\item{\underline{MCM - scheme}}
\eei
To evaluate the amplitude we start by removing the $\Reb$ label in those terms 
that, coming from finite renormalization, violate WSTIs. 
For instance, when we compute the doubly contracted WSTI for the full 
two-loop amplitude in $H \to \gamma \gamma$ we obtain the result of \eqn{para}:
pure two-loop contribution to the WSTI gives
$\Sigma_{\ssH}^{(1)}(-\mhs)$ while finite renormalization gives its 
real part $\Reb\,\Sigma_{\ssH}^{(1)}(-\mhs)$.
Therefore, the WSTI is violated above the $WW\,$ threshold,
as shown in section~\ref{sec:aaa}.

Furthermore, we decompose the amplitude for $H \to \gamma \gamma$ according to
\bq
{\cal A}_{\rm phys} = 
\Bigl(\sqrt{2}\,\gf\,\mws\Bigr)^{1/2}\frac{\alpha}{2\,\pi}\,A_{\rm
  phys},
\quad
A_{\rm phys} = 
  A^{(1)}_{\,\rm ex}
+ \,\frac{\gf\,\mws}{2\,\sqrt{2}\,\pi^2}\,
  \biggl[
    \frac{A^{(2)}_{\ssR}}{\beta}
  + A^{(2)}_{\ssL}\,\ln\lpar - \beta^2 - i\,0\rpar
  + A^{(2)}_{\rm rem}
  \biggr],
\label{decompo}
\eq
and prove that, as expected, $A^{(2)}_{\ssR}$, $A^{(2)}_{\ssL}$ and 
$A^{(2)}_{\rm rem}$ separately satisfy the WSTI. The latter fact 
allows us to minimally modify $A^{(2)}_{\ssR,\ssL}$ by working in the 
complex-mass scheme of Ref.~\cite{Denner:2005fg}: we include complex
masses in the gauge-invariant leading part of the two-loop amplitude as
well as in the one-loop part. 

The decomposition of \eqn{decompo} deserves a further comment. 
As we stressed in section~\ref{sqroots}, there are three sources of 
$1/\beta\,$ terms: a) pure two-loop diagrams of the $V_{\ssM}\,$ family, i.e. 
bubble insertions on the internal lines of the one-loop triangle; 
b) $W$-mass renormalization, i.e. on-shell $W$ self-energy $\,\times\,$ the 
mass-squared derivative of the one-loop $W\,$ triangle (the latter giving 
rise to $1/\beta$); c) Higgs wave-function renormalization factor $\,\times\,$ 
lowest order (the former giving rise to $1/\beta$).
One can easily prove that only c) survives in MCM and a,b), which are separately
singular, add up to a finite contribution ($\beta \to 0$); their divergence
is an artifact of expanding Dyson-resummed propagators in the on-shell approach.

The $\ln\beta\,$-dependent term originates from pure two-loop diagrams of the 
$V_{\ssK}\,$ family and it is a remnant of the one-loop Coulomb singularity
of one-loop sub-diagrams.

The amplitude for $H \to gg$ is different in some points, for instance we have no 
$V^{\ssK}$ diagram with a logarithmic behavior. Also the violation of WSTIs, 
described above, is specific to $H \to \gamma \gamma$ since we have no one-loop 
bosonic triangle. Further details have been presented in Ref.~\cite{Actis:2008uh}.

\begin{figure}[htb]
\includegraphics[bb=0 0 567 405,height=6.cm,width=7.cm]{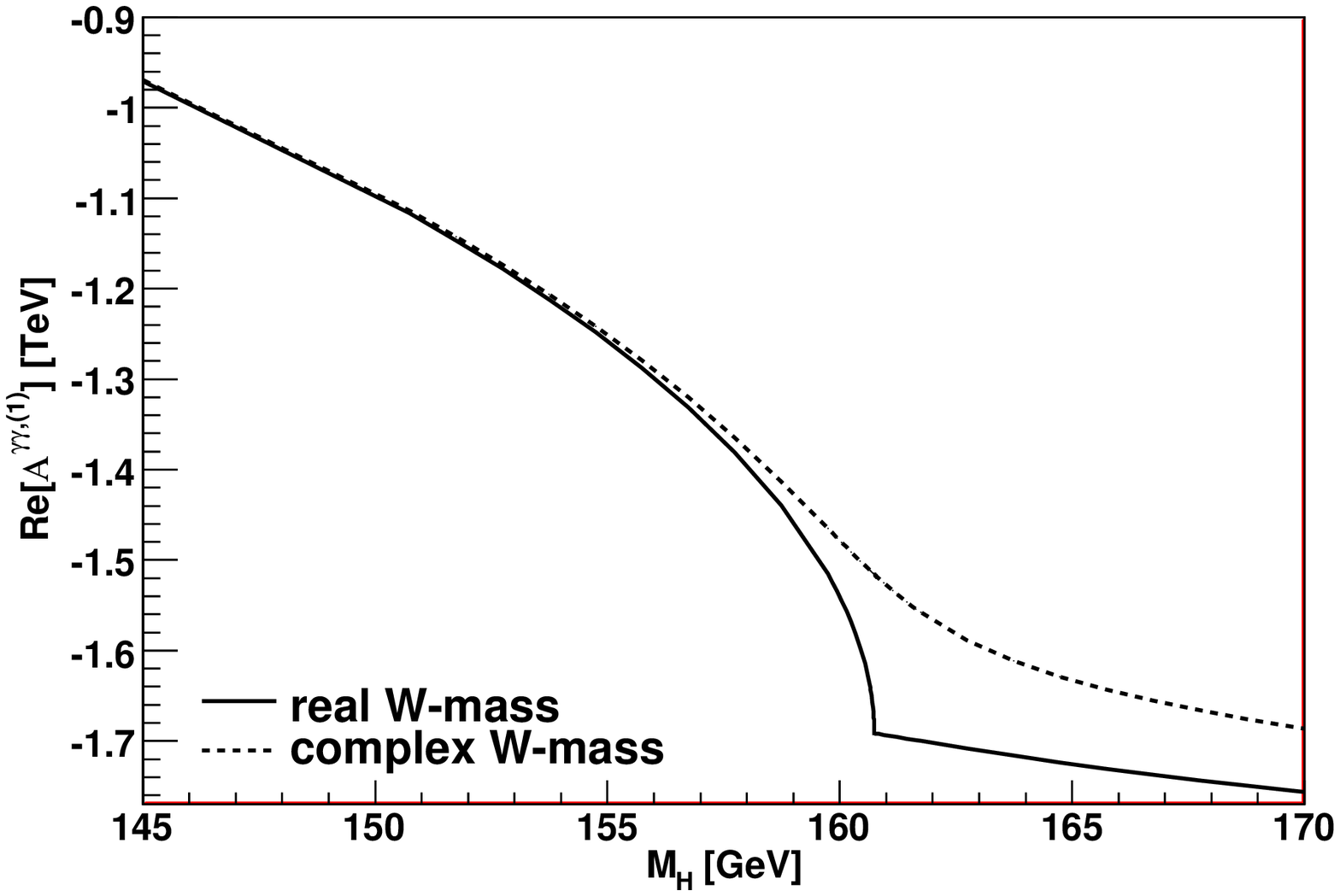}
\hspace{1.cm}
\includegraphics[bb=0 0 567 405,height=6.cm,width=7.cm]{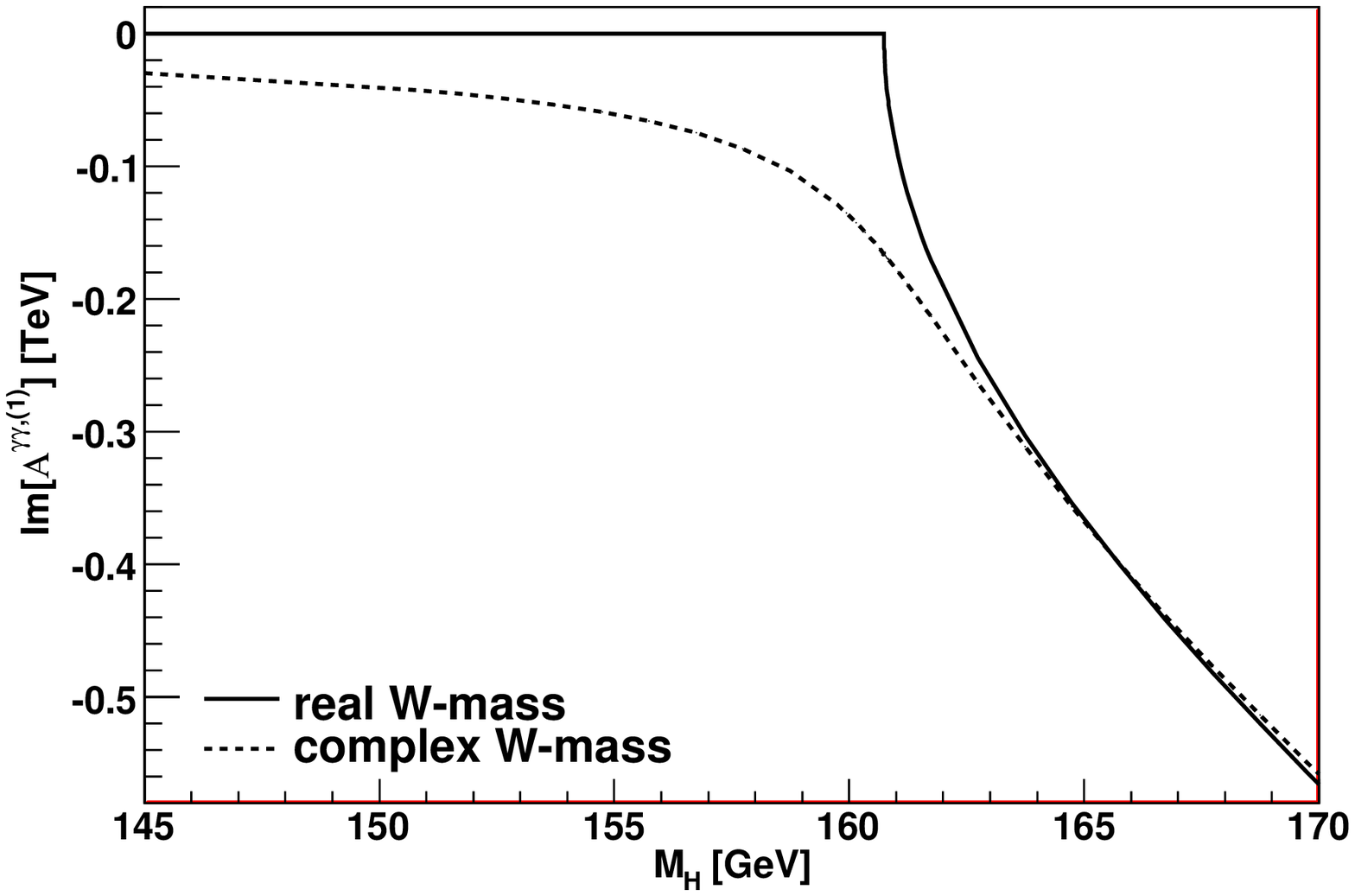}
\vspace{-0.4cm}
\caption[]{\label{OLFigure} Real and imaginary parts of the one-loop $H
\to \gamma \gamma$ amplitude with real and complex $W\,$-boson mass.
Note the sizable difference with \fig{OLlmh}.}
\end{figure}

The one-loop $H \to \gamma \gamma$ amplitude, with a complex $W\,$ mass, is 
shown in \fig{OLFigure} around the $WW\,$ threshold including a comparison
with the real-$W\,$-mass amplitude.
We have also analyzed the effect of (artificially) varying the imaginary part
of the $W\,$-boson complex mass (more details can be found in 
Ref.~\cite{Actis:2008uh}), showing that our {\em complex} result reproduces 
the {\em real} one (with a complex Higgs-mass renormalization condition) 
in the limit $\Gamma_{\ssW} \to 0$. Here, a comment is needed: we introduce
complex masses as poles on the second Riemann sheet, a fact that requires
a careful analytical continuation of loop integrals. 

Let us consider the decay $H \to \gamma \gamma$ in the RM scheme; on-shell 
renormalization spoils the complete cancellation between pure two-loop diagrams 
and $W$-mass renormalization mentioned before. As a consequence the RM scheme 
badly fails to approximate the complex mass results above the $WW\,$ threshold; 
here RM is not the smooth limit $\Gamma_{\ssW} \to 0$ of MCM and we understand the 
reason, RM is missing a cancellation which is instrumental in building a consistent
theory of unstable particles. If RM scheme is further modified by arbitrarily 
forcing this cancellation the corresponding result is the smooth limit just 
mentioned. Thus we have an additional argument to reject RM; at two-loop this
scheme is wholly inconsistent.

It is worth noting that cancellation between mass renormalization for some
internal line and bubble insertion in the same line, as far as divergent terms
are concerned, is a strict consequence of Dyson resummation with complex poles, 
see Eq.~(195) of Ref.~\cite{Actis:2006rc} and consequent discussion.

Finally, we mention the fact that MCM can be forced to agree with RM in the
limit $\Gamma_{\ssW} \to 0$ only if we adopt the unjustified approach
of continuing the $W$ self-energy in different Riemann sheets depending on
its origin, irreducible two-loop diagrams or finite renormalization.

In a nutshell, the MCM scheme has been designed to cure the unphysical infinities
of two-loop amplitudes, namely those points where the amplitude is artificially 
infinite; it does not deal with cusps associated with the crossing of 
normal-thresholds present in $A^{(2)}_{\rm rem}$, as described in detail in 
Ref.~\cite{Actis:2008uh}.
Note that this is not only an aesthetical issue but also a concrete problem in 
assessing the impact of electroweak NLO corrections on, say, Higgs production via 
gluon gluon fusion: observing the effect of NNLO QCD corrections with respect to 
NLO ones one is lead to understand possible sources of additional large 
corrections. Electroweak corrections, typically around the $WW\,$ threshold, can 
reach a $10\%$ in the MCM scheme due to a magnified cusped behavior.

\bei
\item{\underline{CM - scheme}}
\eei
Further to the last point, we have undertaken the task of introducing the 
(complete) complex-mass scheme (CM), based on Ref.~\cite{Denner:2005fg}, as 
explained for a two-loop calculation in Sect.~10 of~\cite{Actis:2006rc}. This 
means that all two-loop diagrams must be computed with complex $\mw, \mz$ masses,
whereas the top quark mass is the on-shell mass; note, however, that we keep the 
external Higgs boson on-shell and do not perform the ultimate step of introducing 
a (complex) pole residue and the associated partial width for the decaying Higgs boson.

For $H \to \gamma \gamma$ (in $R_{\xi}\,$ gauge) we have a bosonic triangle at 
one loop, which contains a factor $m_{\ssH}$ (from the $H-\phi-\phi$ vertex, see 
\fig{SteWiReno}), which needs to be renormalized through the replacement of the 
renormalized Higgs mass with the physical Higgs-boson mass. This fact introduces 
the real parts of $B_0\,$ functions, which lead to a violation of the WST 
identities above the $WW\,$ threshold, as described in \eqn{para} (note that 
wave-function renormalization factors never pose similar problem); also in the 
CM scheme the corresponding real label is removed, even if the external 
Higgs boson is assumed to be an on-shell particle (see comment above). 

As a final note one can say that CM scheme is the default for our results, the
other schemes being assigned to the role of benchmark.
\section{Extraction of collinear singularities} 
\label{EoCS}
In this section we discuss the problem of collinear singularities 
showing up in the calculation of a given (pseudo-)observable. 
Any method that aims to produce theoretical predictions 
for (pseudo-)observable quantities organizes the calculation of the 
corresponding $S$-matrix element into several building blocks, and the 
analytical structure of the total amplitude will not necessarily be 
the same of the single components. From this point of view, collinear 
singularities are a clear example: sometimes, a collinear-free amplitude 
is split into components which are separately divergent in the collinear 
regime. Therefore, singularities must be regularized and singular terms 
have to be extracted. 

It is worth noting that no numerical evaluation can be attempted before two
basic steps have been performed. On the one hand, all singular terms
(ultraviolet, infrared, collinear) of the amplitude have to be extracted,
and their cancellation or absorption into parton distribution functions
have to be explicitly checked. On the other hand, all enhanced terms have
to be isolated, such that numerical integration is only limited to 
{\em smooth} remainders; at this level there is no need to worry about the 
length of the remainders.  

Focusing our attention on electroweak processes, several methods aimed
to deal with collinear singularities have been developed and presented in 
the literature, as in Ref.~\cite{Denner:2001gw}, for one-loop leading 
logarithms in electroweak radiative corrections, and in
Ref.~\cite{Denner:2006jr}, for two-loop electroweak NLO logarithmic 
corrections to massless and massive fermionic processes.

Concerning electroweak corrections to the decay of the
Higgs boson into two photons or two gluons, collinear divergencies
are related to the coupling between photons or gluons with light fermions.
Therefore, there is no substantial difference between the two
processes and in the following, without loss of generality, we
will concentrate on the process $H \to \gamma\gamma$. 
On the contrary, the QCD corrections generate special types of
divergencies in the gluonic decay, because of the three-gluon
coupling.

A common approach to the problem of collinear divergencies is to consider all 
light fermions of the theory as massless states; dimensional regularization is 
then used to control the collinear behavior of single components of
the amplitude. 
In our approach, we prefer to keep the physical light-fermion masses 
to act as regulators, and to express the collinear behavior in terms of 
logarithms of these masses. 
After analytically checking that singular parts cancel in the total, 
we can safely get rid of the regularization parameter and include 
all collinear-free remainders into the total amplitude. 
These finite parts will be cast in a form which is functional to 
numerical integration. 

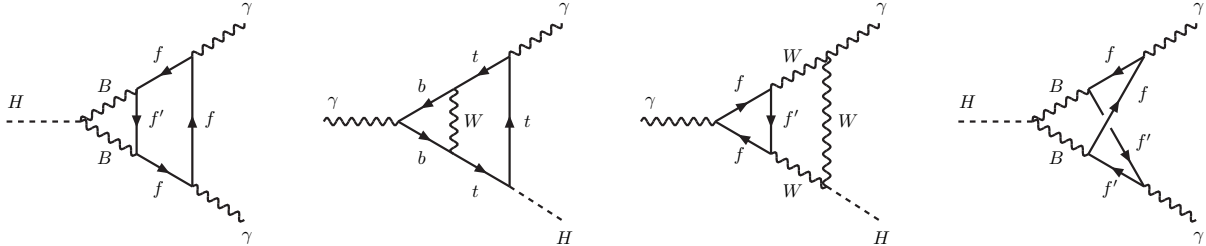
\begin{figure}[ht]
\begin{center}
\begin{tabular}{cccc}
\scalebox{0.7}{
\begin{picture}(150,75)(0,0)
 \SetWidth{1.2}
 \DashLine(0,0)(40,0){3}                \Text(5,7)[cb]{$H$}      
 \Photon(128,-53)(100,-35){2}{5}        \Text(130,-66)[cb]{$\gamma$}
 \Photon(128,53)(100,35){2}{5}          \Text(130,58)[cb]{$\gamma$} 
 \Photon(70,-17.5)(40,0){2}{5}          \Text(53,-23)[cb]{$B$}
 \Photon(70,17.5)(40,0){2}{5}           \Text(53,16)[cb]{$B$}
 \ArrowLine(70,17.5)(70,-17.5)          \Text(81,-3)[cb]{$f'$}
 \ArrowLine(70,-17.5)(100,-35)          \Text(82,-41)[cb]{$f$}
 \ArrowLine(100,-35)(100,35)            \Text(110,-3)[cb]{$f$}
 \ArrowLine(100,35)(70,17.5)            \Text(82,32)[cb]{$f$}
\end{picture}
}
&\!\!\!
\scalebox{0.7}{
\begin{picture}(150,75)(0,0)
 \SetWidth{1.2}
 \Photon(0,0)(40,0){2}{6}               \Text(5,7)[cb]{$\gamma$}      
 \DashLine(128,-53)(100,-35){3}         \Text(130,-66)[cb]{$H$}
 \Photon(128,53)(100,35){2}{5}          \Text(130,58)[cb]{$\gamma$} 
 \ArrowLine(40,0)(70,-17.5)             \Text(53,-23)[cb]{$b$}
 \ArrowLine(70,17.5)(40,0)              \Text(53,16)[cb]{$b$}
 \Photon(70,17.5)(70,-17.5){2}{5}       \Text(81,-3)[cb]{$W$}
 \ArrowLine(70,-17.5)(100,-35)          \Text(82,-41)[cb]{$t$}
 \ArrowLine(100,-35)(100,35)            \Text(110,-3)[cb]{$t$}
 \ArrowLine(100,35)(70,17.5)            \Text(82,32)[cb]{$t$}
\end{picture}
}
&\!\!\!
\scalebox{0.7}{
\begin{picture}(150,75)(0,0)
 \SetWidth{1.2}
 \Photon(0,0)(40,0){2}{6}               \Text(5,7)[cb]{$\gamma$}      
 \DashLine(128,-53)(100,-35){3}         \Text(130,-66)[cb]{$H$}
 \Photon(128,53)(100,35){2}{5}          \Text(130,58)[cb]{$\gamma$} 
 \ArrowLine(70,-17.5)(40,0)             \Text(53,-23)[cb]{$f$}
 \ArrowLine(40,0)(70,17.5)              \Text(53,16)[cb]{$f$}
 \ArrowLine(70,17.5)(70,-17.5)          \Text(81,-3)[cb]{$f'$}
 \Photon(70,-17.5)(100,-35){2}{5}       \Text(82,-41)[cb]{$W$}
 \Photon(100,-35)(100,35){2}{9}         \Text(112,-3)[cb]{$W$}
 \Photon(100,35)(70,17.5){2}{5}         \Text(82,32)[cb]{$W$}
\end{picture}
}
&\!\!\!
\scalebox{0.7}{
\begin{picture}(120,75)(0,0)
 \SetWidth{1.2}
 \DashLine(0,0)(40,0){3}                \Text(5,7)[cb]{$H$}      
 \Photon(128,-53)(100,-35){2}{5}        \Text(130,-66)[cb]{$\gamma$}
 \Photon(128,53)(100,35){2}{5}          \Text(130,58)[cb]{$\gamma$} 
 \ArrowLine(70,-17.5)(100,35)           \Text(100,10)[cb]{$f$}
 \ArrowLine(100,35)(70,17.5)            \Text(82,32)[cb]{$f$}
 \Photon(70,-17.5)(40,0){2}{5}          \Text(53,-23)[cb]{$B$}
 \Photon(70,17.5)(40,0){2}{5}           \Text(53,16)[cb]{$B$}
 \ArrowLine(82,-3.5)(100,-35)
 \Line(70,17.5)(78,3.5)                 \Text(100,-16)[cb]{$f'$}
 \ArrowLine(100,-35)(70,-17.5)          \Text(82,-41)[cb]{$f'$}
\end{picture}
}
\end{tabular}
\end{center}
\vspace{1cm}
\caption[]{
Complete list of diagrams with photons coupled to light fermions ($f$). 
In the figure $f'$ denotes any fermion (light or not, equal to $f$ or not).
The capital $B$ indicates a boson which can be either a $Z$ or a $W$. 
\label{fig:scdc0}}
\end{figure}

At two-loop level we encounter three different situations illustrated
in Fig.~\ref{fig:scdc0}:
1) one and only one of the two external photons is coupled to a light-fermion 
current (second, third diagram and fourth heavy-$f'$ diagram);
2) both photons are coupled to the same light-fermion current in a loop 
(first diagram); 
3) one photon is coupled to a light-fermion current in one loop, 
the other photon in the other loop (fourth diagram with light $f'$).

The treatment of the above-mentioned configurations simplifies when 
taking into account the reduction $\,\otimes\,$ symmetrization procedure 
described in section~\ref{sec:RedSym}. This aspect is considered in 
section~\ref{sec:collred} where we show that all diagrams of type 2) cancel.
Diagrams of type 1) and 3) are then discussed in section~\ref{sec:sc} 
and in section~\ref{sec:dc}, where we analytically extract the coefficients of the 
collinear logarithms. In section~\ref{sec:vmcoll} the special case of the $V^{\ada}$ 
configuration is considered; divergencies which are peculiar of the QCD 
corrections to $H\to g g$ are shortly reviewed in appendix~\ref{hereisnab}.
\subsection{Collinear behavior and tensor reduction}
\label{sec:collred} 
In this section we address a general question which has important
consequences in classifying those collinear configurations which are 
actually needed for the calculation. 
For the $H\to\gamma\gamma$ process, the diagrams where photons couple 
to light fermions belong to the $V^{\ssK}$ and $V^{\ssH}$ families with
the mass patterns depicted in \fig{fig:scdc0}.
Two naive approaches to the treatment of collinear singularities
would consist on the one hand in evaluating all the possible tensor integrals
associated to these families; on the other hand, to express them in terms
of some set of Master Integrals (hereafter MIs) using IBP identities.

The first option is not very convenient, since it requires to 
evaluate explicitly a large set of integrals. However, also a blind 
application of reduction through IBP identities has a drawback: the MIs 
representing the basis integrals cannot be a priori predicted (e.g. by 
using the standard Laporta algorithm~\cite{Laporta:1996mq}; constructive
approaches can be found in Ref.~\cite{Tarasov:2004ks}), and complicated 
collinear-divergent MIs could show up in the final answer. In this section 
we show that our approach, based on the reduction $\otimes$ symmetrization 
procedure of section~\ref{sec:RedSym}, represents the optimal solution: it identifies 
the smallest set of collinear-divergent integrals whose structure is simple 
enough to allow for an analytical extraction of all collinear singularities.

To introduce our argument, we start considering an $N\,$-point one-loop 
function with external momenta $p_1,\,\dots\,,p_{\ssN}$ and with scalar 
products in the numerator. 
After introducing $P^{\mu}_0 = 0$ and 
$P^{\mu}_{i} = p^{\mu}_1 + \,\cdots\, + p^{\mu}_i$, we consider the integral
\bq
S_{n\,;\,\ssN}\lpar f \rpar = 
\frac{\mu^{\ep}}{i\,\pi^2}\,\int\,d^nq\,
\frac{f\lpar q\,,\,\{p\}\rpar}{\prod_{i=0}^{N-1}\,[i]},
\qquad
[i] = \lpar q + P_{i} \rpar^2 + m^2_i,
\eq
and we perform a standard-reduction procedure to simpler functions; 
taking for instance four-point functions, it is a well-known fact that
\bq
S_{n\,;\,4}\lpar f \rpar = 
\sum_i\,b_i\,B_0(P^2_i) + \sum_{i,j}\,c_{ij}\,C_0(P^2_i,P^2_j) +
\sum_{i,j,k}\,d_{ijk}\,D_0(P^2_i,P^2_j,P^2_k) + R,
\eq
where $B_0$, $C_0$ and $D_0$ are scalar two-, three- and four-point 
functions and $R$ is the so-called rational term. Let us consider, 
in particular, the following example:
\bq
S_{n\,;\,4}\lpar q\cdot p_1 \rpar = 
\frac{\mu^{\ep}}{i\,\pi^2}\,\int\,d^nq\,
\frac{\spro{q}{p_1}}{\prod_{i=0}^{3}\,[i]} = 
-\,\sum_{i=1}^3\,D_{1i}\,H_{1i},
\eq
where the matrix $H$ is given by $H_{ij}= -\,\spro{p_i}{p_j}$,
$G = {\rm det}\,H$ is the Gram determinant associated with the four-point
function and $D_{1i}$ are standard form factors~\cite{Passarino:1979jh}. 
In standard reduction, one goes on expressing the $D_{1i}$ form factors 
in terms of $D_0$ and of three-point functions, with inverse powers of 
$G$. However, a more careful application of the method will make use of
\bq
\label{ATcoeff}
D_{1i} = -\,\frac{1}{2}\,H^{-1}_{ij}\,d_j,
\qquad
d_i = D^{(i+1)}_0 - D^{(i)}_0 - 2\,K_i\,D_0,
\qquad
K_i= \frac{1}{2}\,(P_i^2 - P_{i-1}^2 + m^2_i - m^2_{i-1} ),
\eq
where $D^{(i)}_0$ is the scalar triangle obtained by removing the propagator 
$i$ from the box. Therefore, we obtain
\bq
S_{n\,;\,4}\lpar q\cdot p_1 \rpar = 
\frac{1}{2}\,\sum_{i,j=1}^3\,H^{-1}_{ij}\,H_{1i}\,d_j = \frac{1}{2}\,d_1,
\eq
without explicit factors involving $G$. Furthermore, from \eqn{ATcoeff},
we see that the coefficient of $D_0$ in the reduction is 
$(m^2_0 - m^2_1 - p^2_1)/2$. Note that at the leading Landau singularity 
of the box, corresponding to the anomalous threshold~\cite{Goria:2008ny}, we must 
have
\bq
q^2 + m^2_0 = 0, 
\quad 
\lpar q+P_i\rpar^2 + m^2_i = 0
\quad \to \quad
\spro{q}{p_1}= \frac{1}{2}\,(m^2_0 - m^2_1 - p^2_1),
\eq
which is equal to the coefficient of the $D_0$ function. This is a general 
property: a careful application of standard reduction to an $N\,$-point 
function with any scalar product gives as coefficient for the scalar 
$N\,$-point integral the value of the scalar product at the anomalous 
threshold.

To summarize, in standard reduction for a $N\,$-point function each 
reducible scalar product in the numerator is replaced by a difference of 
propagators plus a $K\,$ factor, predicted by factorization properties at the 
anomalous threshold. The procedure can be continued and one finds 
$(N-1)-$point functions with reducible and also irreducible scalar 
products; for the latter inverse powers of Gram determinants will remain.

Imagine now that our $N\,$-point one-loop function is a sub-diagram 
(with loop momentum $q_2$) of a two-loop diagram (with momenta $q_1$, $q_2$). 
The numerator will contain, in general, reducible and irreducible scalar 
products.
If only reducible scalar products are present and if, after algebraic 
reduction $N \to N-1$ (as we said earlier, no inverse Gram determinants), the 
coefficients of the corresponding scalar, vector or tensor one-loop 
diagrams turn out to be zero, then the two-loop diagram will not appear 
in the final result and only its reduced child diagrams will do. 
In particular, if the original two-loop diagram is collinear divergent,
the singular behavior can be read off from its sub-diagrams, which is a 
simpler problem because one propagator less is involved.

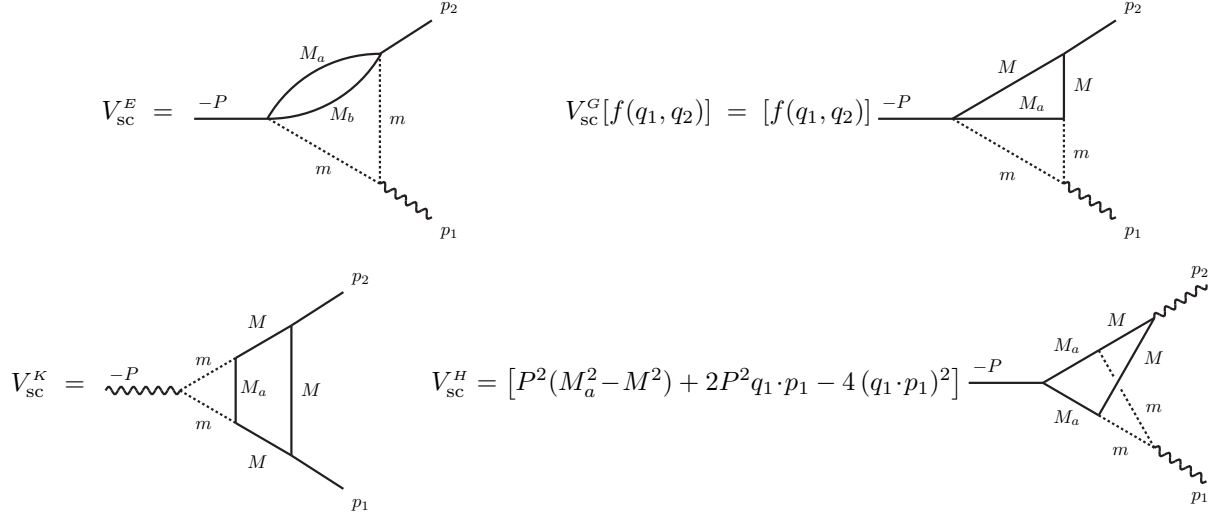
\begin{figure}[ht]
$$
V^{\aba}_{\rm sc}
\;=\;
\scalebox{0.7}{
\begin{picture}(140,75)(0,0)
 \SetWidth{1.2}
 \Line(0,0)(40,0)                    \Text(10,5)[cb]{$-P$}
 \Photon(128,-53)(100,-35){2}{5}     \Text(138,-65)[cb]{$p_1$}
 \Line(128,53)(100,35)               \Text(138,57)[cb]{$p_2$}
 \CArc(100,-35)(70,90,150)           \Text(64,31)[cb]{$M_a$}
 \CArc(40,70)(70,270,330)            \Text(80,-2)[cb]{$M_b$}
 \DashLine(100,-35)(40,0){1.4}       \Text(70,-30)[cb]{$m$}
 \DashLine(100,-35)(100,35){1.4}     \Text(110,-3)[cb]{$m$}
\end{picture}
}
\qquad\qquad
V^{\bba}_{\rm sc}[f(q_1,q_2)]
\;=\;
[f(q_1,q_2)]
\scalebox{0.7}{
\begin{picture}(140,75)(0,0)
 \SetWidth{1.2}
 \Line(0,0)(40,0)                    \Text(10,5)[cb]{$-P$}
 \Photon(128,-53)(100,-35){2}{5}     \Text(138,-65)[cb]{$p_1$}
 \Line(128,53)(100,35)               \Text(138,57)[cb]{$p_2$}
 \DashLine(100,-35)(40,0){1.4}       \Text(70,-33)[cb]{$m$}
 \DashLine(100,-35)(100,0){1.4}      \Text(110,-21)[cb]{$m$}
 \Line(100,0)(40,0)                  \Text(83,4)[cb]{$M_a$}
 \Line(100,0)(100,35)                \Text(111,15)[cb]{$M$}
 \Line(100,35)(40,0)                 \Text(70,25)[cb]{$M$}
\end{picture}
}
$$
\vspace{1cm}
$$
V^{\bca}_{\rm sc}
\;=\;
\scalebox{0.7}{
\begin{picture}(140,75)(0,0)
 \SetWidth{1.2}
 \Photon(0,0)(40,0){2}{6}            \Text(10,5)[cb]{$-P$}
 \Line(128,-53)(100,-35)             \Text(138,-65)[cb]{$p_1$}
 \Line(128,53)(100,35)               \Text(138,57)[cb]{$p_2$}
 \DashLine(70,-17.5)(40,0){1.4}      \Text(53,-21)[cb]{$m$}
 \DashLine(70,17.5)(40,0){1.4}       \Text(53,14)[cb]{$m$}
 \Line(70,-17.5)(70,17.5)            \Text(80,-3)[cb]{$M_a$}
 \Line(100,-35)(70,-17.5)            \Text(82,-42)[cb]{$M$}
 \Line(100,-35)(100,35)              \Text(111,-3)[cb]{$M$}
 \Line(100,35)(70,17.5)              \Text(82,34)[cb]{$M$}
\end{picture}
}
\qquad\;
V^{\bbb}_{\rm sc}
=
\left[ 
P^2(M_a^2\!-\!M^2) + 2P^2\spro{q_1\!}{\!p_1} - 4\,(\spro{q_1\!}{\!p_1})^2
\right]\!
\raisebox{0.1cm}{\scalebox{0.7}{
\begin{picture}(140,75)(0,0)
 \SetWidth{1.2}
 \Line(0,0)(40,0)                       \Text(10,5)[cb]{$-P$}    
 \Photon(128,-53)(100,-35){2}{5}        \Text(125,-65)[cb]{$p_1$}
 \Photon(128,53)(100,35){2}{5}          \Text(125,57)[cb]{$p_2$} 
 \Line(70,-17.5)(100,35)                \Text(100,10)[cb]{$M$}
 \Line(100,35)(70,17.5)                 \Text(80,31)[cb]{$M$}
 \Line(70,-17.5)(40,0)                  \Text(53,-23)[cb]{$M_a$}
 \Line(70,17.5)(40,0)                   \Text(53,16)[cb]{$M_a$}
 \DashLine(82,-3.5)(100,-35){1.4}
 \DashLine(70,17.5)(78,3.5){1.4}        \Text(100,-16)[cb]{$m$}
 \DashLine(100,-35)(70,-17.5){1.4}      \Text(82,-39)[cb]{$m$}
\end{picture}
}}
$$
\vspace{1cm}
\caption[]{
Definition of all MIs with one external massless particle coupled 
to a light particle for the process $H\to \gamma\gamma$.
The dot-lines indicate light particles of mass $m$, the wavy line a 
massless particle and the solid lines whatever particles. 
The mass $M$ is strictly heavy, while $M_a$ and $M_b$ can also be light.
}
\label{sc}
\end{figure}

This is what happens with the first diagram in \fig{fig:scdc0};
as described in appendix~\ref{proppo}, the physical content of the diagrams 
for the process $H\to\gamma\gamma$ can be extracted contracting the 
tensor in the amplitude by means of the projector $P^{\mu\nu}_{\ssD}=
D_{3\,;\,\mu\nu}$ of \eqn{PDproj}.
After the standard reduction $6{\rm legs} \to 5{\rm legs}$ is applied 
(see section~\ref{sec:RedSym}), when $m_{\!f}\to 0$ and for 
arbitrary $M_\ssB$ and $m_{\!f'}$, we obtain
\footnote{In this section we use low-case letters for denoting the masses
of {\it light} particles, which will be neglected after proving the
cancellation of collinear logarithms.}:
\bq
\raisebox{0.1cm}{\scalebox{0.8}{
\begin{picture}(140,75)(0,0)
 \SetWidth{1.2}
 \DashLine(0,0)(40,0){3}            
 \Photon(128,-53)(100,-35){2}{5}    
 \Photon(128,53)(100,35){2}{5}      
 \Photon(70,-17.5)(40,0){2}{5}      \Text(53,-23)[cb]{$B$}
 \Photon(70,17.5)(40,0){2}{5}       \Text(53,16)[cb]{$B$}
 \ArrowLine(70,17.5)(70,-17.5)      \Text(82,-3)[cb]{${\!f'}$}
 \ArrowLine(70,-17.5)(100,-35)      \Text(82,-39)[cb]{${\!f}$}
 \ArrowLine(100,-35)(100,35)        \Text(110,-3)[cb]{${\!f}$}
 \ArrowLine(100,35)(70,17.5)        \Text(82,32)[cb]{${\!f}$}
\LongArrow(4,8)(24,8)       \Text(9,13)[cb]{$-P$}
\LongArrow(118,-56)(104,-47)\Text(125,-65)[cb]{$p_1$}\Text(135,-53)[cb]{$\nu$}
\LongArrow(118,56)(104,47)  \Text(125,57)[cb]{$p_2$}\Text(135,45)[cb]{$\mu$}
\end{picture}
}}
\!\!\!\!\!\!\!\!\!\!\!\!
\otimes\,P^{\mu\nu}_{\ssD}\,\,
=
\,\,
C_{\!\ssK}^{\rm dc}\,( m_{\!f'}^2 - M_\ssB^2 \,+\, 2\,\spro{q_1}{p_1} )
\,
\raisebox{0.1cm}{\scalebox{0.8}{
\begin{picture}(140,75)(0,0)
 \SetWidth{1.2}
 \DashLine(0,0)(40,0){3}            
 \Photon(128,-53)(100,-35){2}{5}    
 \Photon(128,53)(100,35){2}{5}      
 \Line(70,-17.5)(40,0)                \Text(53,-23)[cb]{$M_B$}
 \Line(70,17.5)(40,0)                 \Text(53,16)[cb]{$M_B$}
 \Line(70,17.5)(70,-17.5)             \Text(82,-3)[cb]{$m_{\!f'}$}
 \DashLine(70,-17.5)(100,-35){1.4}    \Text(82,-39)[cb]{$m_{\!f}$}
 \DashLine(100,-35)(100,35){1.4}      \Text(110,-3)[cb]{$m_{\!f}$}
 \DashLine(100,35)(70,17.5){1.4}      \Text(82,31)[cb]{$m_{\!f}$}
\LongArrow(4,8)(24,8)       \Text(9,13)[cb]{$-P$}
\LongArrow(118,-56)(104,-47)\Text(125,-65)[cb]{$p_1$}
\LongArrow(118,56)(104,47)  \Text(125,57)[cb]{$p_2$}
\end{picture}
}}
\!\!\!\!\!\!\!\!\!\!\!
+\;
\left(\!
\ba{l}
\hbox{reduced}\\
\hbox{diagrams}
\ea
\!\!
\right),\!\!
\label{redVK1}
\vspace{2cm}
\eq
where the coefficient $C^{\rm dc}_\ssK$ is $\propto (n-4)\,P^2$.
Since the scalar and vector $V^{\ssK}$ are UV finite, we can take the
limit $n \to 4$; therefore, as we anticipated, six-propagator terms disappear 
from the projected $V^{\ssK}$ and only reduced diagrams with at most 
one photon coupled to light-fermion lines survive.
Concerning reduced diagrams, the collinear-divergent ones are always of the type 
$V^{\aba}$ and $V^{\bba}$, collected in~\fig{sc}.
Note that for the $V^{\aba}$ type only the scalar configuration survives, while 
for $V^{\bba}$ we have to deal with tensor integrals up to rank two.

The same procedure will then be applied to the second and third diagram of 
\fig{fig:scdc0}.
In these cases the coefficient of $V^{\ssK}$ does not vanish; in the end,
however, only scalar configurations survive,
\bqa
\raisebox{0.1cm}{\scalebox{0.7}{
\begin{picture}(140,75)(0,0)
 \SetWidth{1.2}
 \Photon(0,0)(40,0){2}{6}               
 \DashLine(128,-53)(100,-35){3}         
 \Photon(128,53)(100,35){2}{5}          
 \ArrowLine(40,0)(70,-17.5)             \Text(53,-23)[cb]{${\!b}$} 
 \ArrowLine(70,17.5)(40,0)              \Text(53,15)[cb]{${\!b}$}  
 \Photon(70,17.5)(70,-17.5){2}{5}       \Text(82,-3)[cb]{${\!_W}$} 
 \ArrowLine(70,-17.5)(100,-35)          \Text(82,-41)[cb]{${\!t}$}
 \ArrowLine(100,-35)(100,35)            \Text(112,-3)[cb]{${\!t}$}
 \ArrowLine(100,35)(70,17.5)            \Text(82,32)[cb]{${\!t}$} 
 \LongArrow(4,8)(24,8)       \Text(9,13)[cb]{$p_1$}\Text(-5,-8)[cb]{$\nu$}
 \LongArrow(118,-56)(104,-47)\Text(125,-65)[cb]{$-P$}
 \LongArrow(118,56)(104,47)  \Text(125,57)[cb]{$p_2$}\Text(135,45)[cb]{$\mu$}
\end{picture}
}}
\!\!\!\!\!\!\!\!\!
\otimes\,P^{\mu\nu}_{\ssD}\,\,
&\;=\;&
\,\,C_{\!\ssK\!,1}^{\rm sc}\;
\raisebox{0.1cm}{\scalebox{0.7}{
\begin{picture}(140,75)(0,0)
 \SetWidth{1.2}
 \Photon(0,0)(40,0){2}{5}              
 \DashLine(128,-53)(100,-35){3}    
 \Photon(128,53)(100,35){2}{5}      
 \DashLine(40,0)(70,-17.5){3}           \Text(53,-23)[cb]{$m_{\!b}$} 
 \DashLine(70,17.5)(40,0){3}            \Text(53,15)[cb]{$m_{\!b}$}  
 \Line(70,17.5)(70,-17.5)               \Text(82,-3)[cb]{$M_{\!_W}$} 
 \Line(70,-17.5)(100,-35)               \Text(82,-41)[cb]{$M_{\!t}$}
 \Line(100,-35)(100,35)                 \Text(112,-3)[cb]{$M_{\!t}$}
 \Line(100,35)(70,17.5)                 \Text(82,32)[cb]{$M_{\!t}$} 
\LongArrow(4,8)(24,8)       \Text(9,13)[cb]{$p_1$}
\LongArrow(118,-56)(104,-47)\Text(125,-65)[cb]{$-P$}
\LongArrow(118,56)(104,47)  \Text(125,57)[cb]{$p_2$}
\end{picture}
}}
+\quad
\left(\!
\ba{l}
\hbox{reduced}\\
\hbox{diagrams}
\ea
\!\!
\right),
\nl \nl \nl \nl
\raisebox{0.1cm}{\scalebox{0.7}{
\begin{picture}(140,75)(0,0)
 \SetWidth{1.2}
 \Photon(0,0)(40,0){2}{6}               
 \DashLine(128,-53)(100,-35){3}         
 \Photon(128,53)(100,35){2}{5}          
 \ArrowLine(70,-17.5)(40,0)             \Text(53,-23)[cb]{${\!f}$} 
 \ArrowLine(40,0)(70,17.5)              \Text(53,15)[cb]{${\!f}$}  
 \ArrowLine(70,17.5)(70,-17.5)          \Text(81,-3)[cb]{${\!f'}$} 
 \Photon(70,-17.5)(100,-35){2}{5}       \Text(82,-41)[cb]{${\!_W}$}
 \Photon(100,-35)(100,35){2}{9}         \Text(112,-3)[cb]{${\!_W}$}
 \Photon(100,35)(70,17.5){2}{5}         \Text(82,32)[cb]{${\!_W}$} 
 \LongArrow(4,8)(24,8)       \Text(9,13)[cb]{$p_1$}\Text(-5,-8)[cb]{$\nu$}
 \LongArrow(118,-56)(104,-47)\Text(125,-65)[cb]{$-P$}
 \LongArrow(118,56)(104,47)  \Text(125,57)[cb]{$p_2$}\Text(135,45)[cb]{$\mu$}
\end{picture}
}}
\!\!\!\!\!\!\!\!\!
\otimes\,P^{\mu\nu}_{\ssD}\,\,
&\;=\;&
\,\,C_{\!\ssK\!,2}^{\rm sc}\;
\raisebox{0.1cm}{\scalebox{0.7}{
\begin{picture}(140,75)(0,0)
 \SetWidth{1.2}
 \Photon(0,0)(40,0){2}{5}              
 \DashLine(128,-53)(100,-35){3}    
 \Photon(128,53)(100,35){2}{5}      
 \DashLine(70,-17.5)(40,0){1.4}       \Text(53,-23)[cb]{$m_{\!f}$} 
 \DashLine(70,17.5)(40,0){1.4}        \Text(53,15)[cb]{$m_{\!f}$}  
 \Line(70,17.5)(70,-17.5)             \Text(81,-3)[cb]{$m_{\!f'}$} 
 \Line(70,-17.5)(100,-35)             \Text(82,-41)[cb]{$M_{\!_W}$}
 \Line(100,-35)(100,35)               \Text(112,-3)[cb]{$M_{\!_W}$}
 \Line(100,35)(70,17.5)               \Text(82,32)[cb]{$M_{\!_W}$} 
\LongArrow(4,8)(24,8)       \Text(9,13)[cb]{$p_1$}
\LongArrow(118,-56)(104,-47)\Text(125,-65)[cb]{$-P$}
\LongArrow(118,56)(104,47)  \Text(125,57)[cb]{$p_2$}
\end{picture}
}}
+\quad
\left(\!
\ba{l}
\hbox{reduced}\\
\hbox{diagrams}
\ea
\!\!
\right).
\label{redVK2}
\eqa
\vspace{1cm}

\noindent
The coefficients $C_{\!\ssK\!,1}^{\rm sc}$ and $C_{\!\ssK\!,2}^{\rm sc}$
are independent from the loop momenta.
The collinear-divergent reduced diagrams belong to the families 
$V^{\aba}_{\rm sc}$ and $V^{\bba}_{\rm sc}$ depicted in~\fig{sc}; 
the new object to be computed is the scalar diagram $V^{\bca}_{\rm sc}$,
also defined in~\fig{sc}.

For the last diagram of \fig{fig:scdc0} we have to distinguish two cases, 
with $f'$ heavy (i.e. $f'$ is the top quark) or light.
In both cases, the coefficients of the tensor $\,V^{\ssH}$ do not vanish, but 
some universal structure in $q_1$ and $q_2$ can be identified, irrespective of
the coupling between the boson $B$ and the fermions $f$ and $f'$.
In the case of light $f'$, the reduced diagrams show also a doubly collinear
behavior of the $V^{\ssG}$ type, characterized by 
a universal structure in the loop momenta. Indeed we obtain
\bq
\raisebox{0.1cm}{\scalebox{0.6}{
\begin{picture}(140,75)(0,0)
 \SetWidth{1.2}
 \DashLine(0,0)(40,0){3}                
 \Photon(128,-53)(100,-35){2}{5}        
 \Photon(128,53)(100,35){2}{5}          
 \ArrowLine(70,-17.5)(100,35)           \Text(100,10)[cb]{${\!b}$}
 \ArrowLine(100,35)(70,17.5)            \Text(80,31)[cb]{${\!b}$}
 \Photon(70,-17.5)(40,0){2}{5}          \Text(53,-24)[cb]{${\!_W}$}
 \Photon(70,17.5)(40,0){2}{5}           \Text(53,17)[cb]{${\!_W}$}
 \ArrowLine(82,-3.5)(100,-35)
 \Line(70,17.5)(78,3.5)                 \Text(100,-16)[cb]{${\!t}$}
 \ArrowLine(100,-35)(70,-17.5)          \Text(82,-39)[cb]{${\!t}$}
 \LongArrow(4,8)(24,8)       \Text(9,13)[cb]{$-P$}
 \LongArrow(118,-56)(104,-47)\Text(125,-65)[cb]{$p_1$}\Text(135,-53)[cb]{$\nu$}
 \LongArrow(118,56)(104,47)  \Text(125,57)[cb]{$p_2$}\Text(135,45)[cb]{$\mu$}
\end{picture}
}}
\!\!\!\!\!\!\!\!\!\!\!\!\!\!\!
\otimes\,P^{\mu\nu}_{\ssD}
=
C_{\!\ssH}^{\rm sc}\,V^{\bbb}_{\rm sc}\;
+\;
{\rm r.d.,}
\qquad
\raisebox{0.1cm}{\scalebox{0.6}{
\begin{picture}(140,75)(0,0)
 \SetWidth{1.2}
 \DashLine(0,0)(40,0){3}                
 \Photon(128,-53)(100,-35){2}{5}        
 \Photon(128,53)(100,35){2}{5}          
 \ArrowLine(70,-17.5)(100,35)           \Text(100,10)[cb]{${\!f}$}
 \ArrowLine(100,35)(70,17.5)            \Text(80,31)[cb]{${\!f}$}
 \Photon(70,-17.5)(40,0){2}{5}          \Text(53,-24)[cb]{$B$}
 \Photon(70,17.5)(40,0){2}{5}           \Text(53,17)[cb]{$B$}
 \ArrowLine(82,-3.5)(100,-35)
 \Line(70,17.5)(78,3.5)                 \Text(100,-16)[cb]{${\!f'}$}
 \ArrowLine(100,-35)(70,-17.5)          \Text(82,-39)[cb]{${\!f'}$}
 \LongArrow(4,8)(24,8)       \Text(9,13)[cb]{$-P$}
 \LongArrow(118,-56)(104,-47)\Text(125,-65)[cb]{$p_1$}\Text(135,-53)[cb]{$\nu$}
 \LongArrow(118,56)(104,47)  \Text(125,57)[cb]{$p_2$}\Text(135,45)[cb]{$\mu$}
\end{picture}
}}
\!\!\!\!\!\!\!\!\!\!\!\!\!\!\!
\otimes\,P^{\mu\nu}_{\ssD}
=
C_{\!\ssH}^{\rm dc}\,V^{\bbb}_{\rm dc}\;
+\;
C_{\!\ssG}^{\rm dc}\,V^{\bba}_{\rm dc}\;
+\;
{\rm r.d.,}
\label{redVH}
\eq
\vspace{0.8cm}

\noindent
where the coefficients $C_{\!\ssH}^{\rm sc}$, $C_{\!\ssH}^{\rm dc}$ and 
$C_{\!\ssG}^{\rm dc}$ depend on the coupling between the boson $B$ and the 
fermions.
The diagram $V^{\bbb}_{\rm sc}$ is defined in~\fig{sc}, while 
$V^{\bbb}_{\rm dc}$ and $V^{\bba}_{\rm dc}$ are shown \fig{dc};
the reduced diagrams (r.d.) are again of the form $V^{\aba}_{\rm sc}$ 
and $V^{\bba}_{\rm sc}$ of~\fig{sc}.
\begin{figure}[!ht]
\vspace{-1cm}
{\small
$$
V^{\bbb}_{\rm dc}
=
\left[
P^2\!M^2 \!+\! 2P^2\spro{q_1\!\!}{\!p_1} \!-\! 4(\spro{q_1\!\!}{\!p_1})^2
\right]\!\!\!\!\!\!
\raisebox{0.1cm}{\scalebox{0.6}{
\begin{picture}(140,75)(0,0)
 \SetWidth{1.2}
 \Line(10,0)(40,0)                       \Text(22,5)[cb]{$-P$}    
 \Photon(128,-53)(100,-35){2}{5}        \Text(125,-65)[cb]{$p_1$}
 \Photon(128,53)(100,35){2}{5}          \Text(125,57)[cb]{$p_2$} 
 \DashLine(70,-17.5)(100,35){1.4}       \Text(100,10)[cb]{$m$}
 \DashLine(100,35)(70,17.5){1.4}        \Text(80,31)[cb]{$m$}
 \Line(70,-17.5)(40,0)                  \Text(53,-23)[cb]{$M$}
 \Line(70,17.5)(40,0)                   \Text(53,16)[cb]{$M$}
 \DashLine(82,-3.5)(100,-35){1.4}
 \DashLine(70,17.5)(78,3.5){1.4}        \Text(100,-16)[cb]{$m'$}
 \DashLine(100,-35)(70,-17.5){1.4}      \Text(82,-39)[cb]{$m'$}
\end{picture}
}}
\quad\!
V^{\bba}_{\rm dc}
=
\left[ 
M^2 \!-\! 4\,(M^2\!+\!P^2\!+\!\spro{q_1\!\!}{\!p_2})
             \frac{\spro{q_2\!\!}{\!p_1}}{P^2} 
\right]\!\!\!\!\!\!
\raisebox{0.1cm}{\scalebox{0.6}{
\begin{picture}(128,75)(0,0)
 \SetWidth{1.2}
 \Line(10,0)(40,0)                       \Text(22,5)[cb]{$-P$}        
 \Photon(128,-53)(100,-35){2}{5}        \Text(125,-65)[cb]{$p_1$}
 \Photon(128,53)(100,35){2}{5}          \Text(125,57)[cb]{$p_2$} 
 \DashLine(100,-35)(40,0){1.4}          \Text(70,-33)[cb]{$m'$}
 \DashLine(100,-35)(100,0){1.4}         \Text(110,-21)[cb]{$m'$}
 \Line(100,0)(40,0)                     \Text(83,4)[cb]{$M$}
 \DashLine(100,0)(100,35){1.4}          \Text(111,15)[cb]{$m$}
 \DashLine(100,35)(40,0){1.4}           \Text(70,25)[cb]{$m$}
\end{picture}
}}
$$
}
\vspace{0.3cm}
\caption[]{
Definition of all MIs with two external massless particles 
coupled to a light particle for the process $H\to \gamma\gamma$.
The dot-lines indicate light particles of mass $m$ or $m'$ ($m'$ can be 
equal to $m$ or not), the wavy line a massless particle and the solid lines 
a heavy particle of mass $M$.
}
\label{dc}
\end{figure}

Summarizing, we can say that the  reduction $\otimes$ symmetrization procedure 
allows us to identify the smallest sub-set of all diagrams with 
collinear divergencies, collected in~\fig{sc} and~\fig{dc}; they can be 
taken as MIs for the set of simply and doubly collinear 
configurations. The extraction of the corresponding collinear logarithms will 
be treated in the next two sections.
\subsection{Vertices with one photon coupled to light fermions}
\label{sec:sc}
In this section we extract the coefficients of the collinear logarithms 
for the MIs shown in \fig{sc} and compute the associated collinear-finite 
parts.
For the $V^{\aba}_{\rm sc}$ configuration (first diagram in \fig{sc}), 
we use the result derived in section~5.1 of Ref.~\cite{Ferroglia:2003yj}, 
and evaluate the limits $p_1^2 \to 0$ and $m_3=m_4= m \to 0$,
  \bqa \label{ve:res}
    V^{\aba}_{\rm sc}
    &=&
    - 2\,\bigg[ F^2_{-2}(s) - F^2_{-1}(s)
      \left( \ln\frac{m^2}{s} - \frac{1}{2} \right) \bigg]
    - \frac{1}{2}\,\ln^2\frac{m^2}{s}
    - \ln\frac{m^2}{s}\intsx{x}\,dz\,\ln\frac{\chiu{\aba}}{s}
    \nl
    &{}&
    + \intsx{x}\,dz\,\bigg[\frac{1}{2}\,\ln^2\frac{\chiu{\aba}}{s}
    + \ln\frac{z}{x(1\!-\!x)}\,\ln\frac{\chiu{\aba}}{s}
    - \li{2}{\frac{p_2^2\,xz\,(1\!-\!x)}{\chiu{\aba}}}\bigg]
    - \frac{1}{2}\,\zeta(2) + \frac{3}{2},
  \eqa
where $\chiu{\aba}= P^2\,x\,(1-x)\,(1-z) + p_2^2\,x\,z\,(1-x) + M_a^2\,(1-x) + 
M_b^2\,x$ and the two-loop UV factors $F_{-i}^2$, with $i=1,2$, have been 
defined in \eqn{UVfactors}. 
We readily identify $\chiu{\aba}$ with the polynomial
associated with the Feynman-parameter representation of the one-loop two-point 
function $B_0(1,1,(1\!-\!z)P^2+zp_2^2,M_a,M_b)$ ($x$ is the Feynman parameter).
In particular, the coefficient of the collinear logarithm can be written 
through the one-fold integral representation of a one-loop function,
\vspace{-0.7cm}

\bq
\label{ve}
V^{\aba}_{\rm sc}
\;=\;
\raisebox{0.1cm}{\scalebox{0.6}{
\begin{picture}(140,75)(0,0)
 \SetWidth{1.2}
 \Line(0,0)(40,0)                    \Text(10,5)[cb]{$-P$}
 \Photon(128,-53)(100,-35){2}{5}     \Text(138,-65)[cb]{$p_1$}
 \Line(128,53)(100,35)               \Text(138,57)[cb]{$p_2$}
 \CArc(100,-35)(70,90,150)           \Text(64,31)[cb]{$M_a$}
 \CArc(40,70)(70,270,330)            \Text(80,-2)[cb]{$M_b$}
 \DashLine(100,-35)(40,0){1.4}       \Text(70,-30)[cb]{$m$}
 \DashLine(100,-35)(100,35){1.4}     \Text(110,-3)[cb]{$m$}
\end{picture}
}}
\!\!
=\,
\ln\frac{m^2}{s}\,
\bigg[ 
1 \!-\! \frac{\ep}{2}\,\Delta_{\ssU\ssV}(s) 
\!-\! \frac{\ep}{4}\,\ln\frac{m^2}{s}
\bigg]\!
\intsx{z}\,
\raisebox{0.1cm}{\scalebox{0.6}{
\begin{picture}(100,50)(0,0)
 \SetWidth{1.2}
 \Line(0,0)(30,0)                     \Text(10,5)[cb]{$-P$}
 \Line(70,0)(100,0)                   \Text(90,5)[cb]{$p_2$}
 \CArc(50,0)(20,0,360)                \Text(52,24)[cb]{$M_a$}
                                      \Text(52,-33)[cb]{$M_b$}
\Photon(5,-20)(30,0){2}{5}            \Text(8,-33)[cb]{$(1\!-\!z)\,p_1$}
\Photon(70,0)(95,-20){2}{5}           \Text(97,-33)[cb]{$z\,p_1$}
\end{picture}
}}
\quad
+\,\,
V^{\aba}_{\rm sc,fin},
\vspace{1.8cm}
\eq
where $\Delta_{\ssU\ssV}$ can be read in \eqn{defMSB} and $V^{\aba}_{\rm sc,fin}$
denotes the collinear-free remainder (see \eqn{ve:res}). 
This simple result shows a feature that we will encounter also in more-complicated 
configurations: collinear singularities can be represented through objects with 
well-known analytical properties, and the cancellation of all collinear logarithms 
at the amplitude level can be analytically verified.

It will be shown in the following that the collinear behavior of the 
remaining three vertices of~\fig{sc} is also embedded in a one-loop 
integration.
The key observation is that all these configurations contain the product of 
two propagators of the same type:
  \bq \label{collprop}
    \frac{1}{(q^2+m^2)[(q+p)^2+m^2]},\qquad \hbox{with} \qquad
    \left\{
      \ba{l}
      p= p_1, \quad \quad q= q_1 \quad \hbox{ for } V^{\bba}_{\rm sc} \\
      p= P, \;\quad \quad q= q_1 \quad \hbox{ for } V^{\bca}_{\rm sc} \\
      p= - p_1, \quad \; q= q_2 \quad \hbox{ for } V^{\bbb}_{\rm sc} \\
      \ea
    \right. .
  \eq
In particular, let us consider the scalar configurations 
$V^{\bba}_{\rm sc}[1]$ and $V^{\bca}_{\rm sc}$ (scalar configurations of the 
second and third diagrams in \fig{sc}); we define 
$J_\ssN$ as the scalar sub-loop containing $q$, where $N$ denotes the 
number of $q\,$-dependent propagators in addition to those of \eqn{collprop}. 
For $V^{\bba}_{\rm sc}[1]$ and $V^{\bca}_{\rm sc}$ we have $N= 1$ and consider
\bq \label{J1start}
J_1= 
\frac{\mu^{4-n}}{i\,\pi^2}
\int\,d^nq\,\frac{1}{(q^2+m^2)[(q+p)^2+m^2][(q-q_2)^2+M_a^2]}.
\eq
Next, we introduce Feynman parameters $z$ and $y$, integrate over $q$ and 
set $n=4$, obtaining
\bq
J_1= \intsxy{z}{y}\,\frac{1}{V}, \qquad 
V= [\, A - y\,(q_2+p)^2 \,]\,y + m^2\,(1-y), \qquad
A= (q_2+p\,z)^2 + M_a^2.
\label{J1}
\eq
The result shows that the singularity for $m\ = 0$ is generated, in parametric 
space, at the point $y = 0$; 
we introduce $V_0= A\,y + m^2$, a simple polynomial having the same 
collinear properties of $V$, and add and subtract $1/ V_0$ at the integrand 
level, getting
\bq
J_1 = \int_0^1\,\frac{dz}{A}\,\ln \frac{A\,z - m^2}{s} 
+ \int_0^1 dz \int_0^z \frac{dy}{y}\,
  \bigg[ \frac{1}{A - y\,(q_2+p)^2} - \frac{1}{A} \bigg]
+ \ord{m^2},
\label{J1res}
\eq
where the first term is the collinear-divergent part of $J_1$. 
The complete expressions for $V^{\bba}_{\rm sc}[1]$ and 
$V^{\bca}_{\rm sc}$, and the related coefficients of the collinear logarithms,
are obtained inserting the result for $J_1$ in the $q_2$ integrals,
\bqa
V^{\bba}_{\rm sc}[1] &=&  
\frac{1}{i\pi^2}\ln\frac{m^2}{s}\intsx{z}\int\!
\frac{d^4q_2}
     {[(q_2+p_1\,z)^2 + M_a^2]\,[(q_2+p_1)^2 + M^2]\,[(q_2+P)^2 + M^2]}
+ V^{\bba}_{\rm sc,fin}[1],
\nl
V^{\bca}_{\rm sc} &=&  
\frac{1}{i\pi^2}\! \ln\frac{m^2}{s} \! \intsx{z}\int\!\!
\frac{d^4q_2}
     {[(q_2\!+\!P\,z)^2 \!+\! M_a^2]\,(q_2^2 \!+\! M^2)\,
      [(q_2\!+\!p_1)^2 \!+\! M^2]\,[(q_2\!+\!P)^2 \!+\! M^2]}
+ V^{\bca}_{\rm sc,fin},
\eqa
with the following diagrammatical correspondences,
\vspace{-0.6cm}

\bqa
V^{\bba}_{\rm sc}[1] \;=\;
\raisebox{0.1cm}{\scalebox{0.57}{
\begin{picture}(140,75)(0,0)
 \SetWidth{1.2}
 \Line(0,0)(40,0)                    \Text(10,5)[cb]{$-P$}
 \Photon(128,-53)(100,-35){2}{5}     \Text(138,-65)[cb]{$p_1$}
 \Line(128,53)(100,35)               \Text(138,57)[cb]{$p_2$}
 \DashLine(100,-35)(40,0){1.4}       \Text(70,-33)[cb]{$m$}
 \DashLine(100,-35)(100,0){1.4}      \Text(110,-21)[cb]{$m$}
 \Line(100,0)(40,0)                  \Text(83,4)[cb]{$M_a$}
 \Line(100,0)(100,35)                \Text(111,15)[cb]{$M$}
 \Line(100,35)(40,0)                 \Text(70,25)[cb]{$M$}
\end{picture}
}}
\!\!
&=&
\quad
\ln\frac{m^2}{s}\,
\intsx{z}\,
\raisebox{0.1cm}{\scalebox{0.57}{
\begin{picture}(140,75)(0,0)
 \SetWidth{1.2}
 \Line(0,0)(40,0)                    \Text(10,5)[cb]{$-P$}
 \Photon(10,-30)(40,0){2}{5}         \Text(16,-44)[cb]{$zp_1$}
 \Photon(128,-53)(100,-35){2}{5}     \Text(138,-65)[cb]{$(1\!-\!z)p_1$}
 \Line(128,53)(100,35)               \Text(138,57)[cb]{$p_2$}
 \Line(100,-35)(40,0)                \Text(70,-33)[cb]{$M_a$}
 \Line(100,-35)(100,35)              \Text(110,-3)[cb]{$M$}
 \Line(100,35)(40,0)                 \Text(70,25)[cb]{$M$}
\end{picture}
}}
\quad
+\,\,
V^{\bba}_{\rm sc,fin}[1],
\nl \nl \nl \nl
V^{\bca}_{\rm sc}
\;=\;
\raisebox{0.1cm}{\scalebox{0.57}{
\begin{picture}(140,75)(0,0)
 \SetWidth{1.2}
 \Photon(0,0)(40,0){2}{6}            \Text(10,5)[cb]{$-P$}
 \Line(128,-53)(100,-35)             \Text(138,-65)[cb]{$p_1$}
 \Line(128,53)(100,35)               \Text(138,57)[cb]{$p_2$}
 \DashLine(70,-17.5)(40,0){1.4}      \Text(53,-21)[cb]{$m$}
 \DashLine(70,17.5)(40,0){1.4}       \Text(53,14)[cb]{$m$}
 \Line(70,-17.5)(70,17.5)            \Text(80,-3)[cb]{$M_a$}
 \Line(100,-35)(70,-17.5)            \Text(82,-42)[cb]{$M$}
 \Line(100,-35)(100,35)              \Text(111,-3)[cb]{$M$}
 \Line(100,35)(70,17.5)              \Text(82,34)[cb]{$M$}
\end{picture}
}}
\!\!
&=&
\quad
\ln\frac{m^2}{s}\,
\intsx{z}\,
\raisebox{0.1cm}{\scalebox{0.57}{
\begin{picture}(140,75)(0,0)
 \SetWidth{1.2}
 \Photon(2,-53)(30,-35){2}{5}        \Text(-2,-65)[cb]{$-zP$}
 \Photon(2,53)(30,35){2}{5}          \Text(-2,57)[cb]{$-(1\!-\!z)P$}
 \Line(128,-53)(100,-35)             \Text(138,-65)[cb]{$p_1$}
 \Line(128,53)(100,35)               \Text(138,57)[cb]{$p_2$}
 \Line(30,-35)(30,35)                \Text(19,-3)[cb]{$M_a$}
 \Line(100,-35)(30,-35)              \Text(65,40)[cb]{$M$}
 \Line(100,-35)(100,35)              \Text(111,-3)[cb]{$M$}
 \Line(30,35)(100,35)                \Text(65,-46)[cb]{$M$}
\end{picture}
}}
\quad
+\,\,
V^{\bca}_{\rm sc,fin}.
\\\nl\nl\nn
\label{vg&vk}
\eqa
The computation of the collinear-finite parts $V^{\bba}_{\rm sc,fin}[1]$ and 
$V^{\bca}_{\rm sc,fin}$ will be described in more detail in
section~\ref{sec:collfinG} and section~\ref{sec:collfinK}.

The $V^{\bbb}_{\rm sc}$ configuration (fourth diagram in \fig{sc}) shows two 
additional $q$-dependent propagators; we consider the generalization of 
\eqn{J1start},
\bq
J_\ssN= 
\frac{\mu^{4-n}}{i\,\pi^2}
\int\,\frac{d^nq}{(q^2+m^2)[(q+p)^2+m^2]}\,
\prod_{i=1}^\ssN\,\frac{1}{(q+k_i)^2+m_i^2},
\label{JN}
\eq
and later we will specialize to the  $V^{\bbb}_{\rm sc}$ case,
setting $N=2$, $k_1=-q_2-p_1$, $k_2=-q_1+p_2$ and $m_1=m_2=M_a$.
We start introducing a Feynman parametrization,
\bqa \label{propN}
\prod_{i=1}^\ssN\,\frac{1}{(q+k_i)^2+m_i^2}&=& 
\Gamma(N)\int\!dS_{\ssN-1}(\{x\})\,\frac{1}{[(q+K)^2+\mu^2]^\ssN},
\nl
\mu^2= \sum_{i=1}^\ssN\,(x_{i-1}-x_i)\,(m_i^2+k_i^2) &-& K^2,
\qquad
K= \sum_{i=1}^\ssN\,(x_{i-1}-x_i)\,k_i,
\qquad
x_0= 1,\quad x_\ssN=0.
\eqa
Next, we combine the resulting three propagators of \eqn{JN}
with variables $z$ and $y$ and perform the $q$ integration.
Since the $V^{\bbb}_{\rm sc}$ configuration in \fig{sc} is UV finite, 
we can set $n=4$ and obtain
\bq
J_\ssN= 
\Gamma(N)\!\!\intsxy{z}{y}\!\int\!\!dS_{\ssN\!-\!1}(\{x\})
\frac{y^{\ssN\!-\!1}}{V^{\ssN}},
\qquad
V= [A \!-\! y (K\!-\!p)^2 ]\,y \!+\! m^2 (1\!-\!y),
\qquad
A= (K\!-\!p\,z)^2 \!+\! \mu^2.
\label{JNpar}
\eq
Subtracting and adding at the integrand level
$1/ V_0^{\ssN}$, with $V_0= A\,y + m^2$, we can extract 
the collinear logarithm through the $y$ integration,
\bq
\label{JNres}
J_\ssN\! =
- \Gamma(N)\!\!\int\!\!dS_{\ssN\!-\!1}(\{x\})\!\!\intsx{z}\,
\bigg\{
  \frac{1}{A^\ssN}
  \bigg( \ln\!\frac{m^2}{s} \!-\! \ln\!\frac{Az}{s} \!+\! 
\sum_{n=1}^{\ssN-1}\frac{1}{n} \bigg)
- \intsx{y}\,\frac{[A\!-\!y (K\!-\!p)^2]^{-\ssN}}{y}\bigg|_+
\bigg\}\!
+ {\cal O}(m^{\!2}),
\eq
where the $'+'$ distribution has been defined in \eqn{plusdist}.
The coefficient of the collinear logarithm can be further simplified if we 
move back to momentum space integrating over all Feynman variables $\{x\}$,
\bq
\Gamma(N)\int\!\!dS_{\ssN-1}(\{x\})\,\frac{1}{A^\ssN}=
\prod_{i=1}^\ssN\,\frac{1}{(k_i-p\,z)^2+m_i^2}.
\label{propback}
\eq
Formally, we have reconstructed the product of $N$ propagators on the 
right-hand side of \eqn{JN}, replacing the loop momentum $q$ by $-p\,z$. 
\eqn{JNres} has been derived under rather-general assumptions, and it can 
be used for extracting the collinear behavior of any two-loop scalar 
(respect to $q$) UV-finite diagram with one external massless particle 
coupled to one light particle;
its graphical representation reads
\vspace{-0.5cm}
\bq
\raisebox{0.1cm}{\scalebox{0.65}{
\begin{picture}(180,75)(0,0)
 \SetWidth{1.2}
 \Photon(0,0)(40,0){2}{5}              \Text(10,5)[cb]{$p$}
 \DashCArc(75,0)(35,60,300){1.4}       \Text(75,39)[cb]{$m$}
                                       \Text(75,-45)[cb]{$m$}
\LongArrowArc(75,0)(42,120,150)        \Text(40,34)[cb]{$q$}
\LongArrowArc(75,0)(42,210,240)        \Text(27,-34)[cb]{$q+p$}
 \GCirc(130,0){45}{0.8}                
 \Text(130,-3)[cb]{$q_a^{\mu_1}\!\!\dots q_a^{\mu_m}$}
\end{picture}
}}
\!
=
\,\,
\ln\frac{m^2}{s}\,
\intsx{z}\!\!\!\!\!
\raisebox{0.1cm}{\scalebox{0.65}{
\begin{picture}(90,75)(0,0)
 \SetWidth{1.2}
 \Photon(0,45)(30,26){2}{5}             \Text(-2,49)[cb]{$zp$}
 \Photon(0,-45)(30,-26){2}{5}           \Text(-2,-57)[cb]{$(1\!-\!z)p$}
 \GCirc(46,0){30}{0.8}                
 \Text(46,-3)[cb]{$q_a^{\mu_1}\!\!\dots q_a^{\mu_m}$}
\end{picture}
}}
\!\!\!\!
+\,\,
\hbox{coll. fin.},
\vspace{0.8cm}
\label{scalar}
\eq
where the bubble denotes a generic one-loop diagram involving a tensor 
structure $q_a^{\mu_1}\!\!\dots q_a^{\mu_m}$, $q_a$ is the loop momentum,
the wavy line represents the external massless particle ($p^2=0$) and the 
dotted one is the light particle with mass $m$.

The result of \eqn{scalar} can be immediately applied to the 
$V^{\bbb}_{\rm sc}$ configuration of \fig{sc}; note that we do not confine 
ourselves to the scalar configuration, but we consider the full 
$q_1$-dependent structure appearing in the coefficient of the diagram. 
We obtain
\bq
V^{\bbb}_{\rm sc}\;
=\;\;
\ln\frac{m^2}{s}
\intsx{z}\quad
[ P^2(M_a^2\!-\!M^2) + 2P^2\spro{q_1\!}{\!p_1} - 4\,(\spro{q_1\!}{\!p_1})^2 ]
\!\!\!\!\!\!\!\!\!\!
\raisebox{0.1cm}{\scalebox{0.65}{
\begin{picture}(140,75)(0,0)
 \SetWidth{1.2}
 \Photon(2,-53)(30,-35){2}{5}        \Text(-2,-65)[cb]{$(1\!-\!z)p_1$}
 \Line(2,53)(30,35)                  \Text(-2,57)[cb]{$-P$}
 \Line(128,-53)(100,-35)             \Text(138,-65)[cb]{$p_2$}
 \Photon(128,53)(100,35){2}{5}       \Text(138,57)[cb]{$zp_1$}
 \Line(30,-35)(30,35)                \Text(41,-3)[cb]{$M_a$}
 \Line(100,-35)(30,-35)              \Text(65,22)[cb]{$M_a$}
 \Line(100,-35)(100,35)              \Text(91,-3)[cb]{$M$}
 \Line(30,35)(100,35)                \Text(65,-29)[cb]{$M$}
 \LongArrow(108,-10)(108,10)         \Text(118,-3)[cb]{$q_1$}
\end{picture}
}}
+\quad
V^{\bbb}_{\rm sc,fin}.
\label{vh}
\vspace{1.2cm}
\eq
The procedure for dealing with the collinear-finite part $V^{\bbb}_{\rm sc,fin}$
will be described in section~\ref{sec:collfinH}.

Finally, let us discuss the extraction of the collinear logarithms for 
generic tensor and/or UV-divergent two-loop integrals. We introduce
\bq
J_\ssN^{\nu_1\cdots\nu_r}= 
\frac{\mu^{4-n}}{i\,\pi^2}
\int\,\frac{d^nq\,q^{\nu_1}\,\cdots\,q^{\nu_r}}{(q^2+m^2)[(q+p)^2+m^2]}\,
\prod_{i=1}^\ssN\,\frac{1}{(q+k_i)^2+m_i^2},
\label{JNtensor}
\eq
and use a Feynman parametrization analogous to the one used for deriving 
\eqn{JNpar}. In particular, before performing the $q$ integration, we have
\bqa \label{JNtensorp}
J_\ssN^{\nu_1\cdots\nu_r}&=& 
\frac{\mu^{4-n}\,\Gamma(N\!+\!2)}{i\,\pi^2}\!\!
\intsxy{z}{y}\!\int\!\!dS_{\ssN\!-\!1}(\{x\})\,y^{\ssN\!-\!1}\!\!
\int\!\!d^nq\,\prod_{i=1,r}\,q^{\nu_i}\,
{\cal V}^{-\ssN-2},
\nl
{\cal V}= q^2+2\,\spro{q}{\cal K}&+&{\cal M}^2,
\qquad\quad
{\cal K}= z\,p + y\,(K-p),
\qquad\quad
{\cal M}^2= (\mu^2+K^2)\,y + m^2\,(1-y),
\eqa
where $K$ and $\mu^2$ have been defined in \eqn{propN}. The $q$ integration is 
performed according to
\bqa
\int\!\!d^nq\,\prod_{i=1,r}\,q^{\nu_i}\,{\cal V}^{-\ssN-2}
&=&
\bigg(\!\! -\frac{1}{2} \bigg)^{\!\!r}\,\,
\frac{\Gamma(N+2-r)}{\Gamma(N+2)}\,
\prod_{i=1,r}\,\frac{\partial}{\partial{\cal K}^{\nu_i}}
\int\!\!d^nq\,{\cal V}^{-\ssN-2+r}
\nl
&=&
i\,\pi^{2-\ep/2}
\bigg(\!\! -\frac{1}{2} \bigg)^{\!\!r}\,\,
\frac{\Gamma(N-r+\ep/2)}{\Gamma(N+2)}\,
\prod_{i=1,r}\,\frac{\partial}{\partial{\cal K}^{\nu_i}}\,
V^{-\ssN+r-\ep/2},
\label{JNtensorpar}
\eqa
where $V={\cal M}^2 - {\cal K}^2=[\, A - y\,(K-p)^2 \,]\,y + m^2\,(1-y)$ is 
the same of \eqn{JNpar}.

The first derivative in \eqn{JNtensorpar} decreases the power 
of $V$ by one unit and generates an extra factor ${\cal K}$.
The second derivative acts on both ${\cal K}$ and $V$, producing two terms:
one where the power of $V$ does not change and the metric tensor 
appears; another one where the power of $V$ decreases by one unit and an 
additional factor ${\cal K}$ is generated.
After taking all $r$ derivatives, we find a term containing $V^{-\ssN-\ep/2}$: 
the one where all derivatives have acted on $V$, generating a factor 
${\cal K}^{\nu_1}\dots{\cal K}^{\nu_r}$. 
All other terms in the result will contain at least one power of the 
metric tensor, and for each of them the power of $V$ will be greater 
than $-N-\ep/2$, because at least one derivative has not acted on $V$.

We can show the following result: if the power of $V$ is greater than 
$-N-\ep/2$ then no collinear logarithm is generated. We cast the $y$ integral 
of \eqn{JNtensorp} as
\bq
\int_0^z\!dy\,\frac{y^\beta}{V^{\alpha}}= 
  \int_0^z\!dy\,\frac{y^\beta}{(A\,y+m^2)^{\alpha}}
+ \int_0^z\!dy\,y^\beta\,
  \left(\frac{1}{V^{\alpha}}-\frac{1}{V_0^{\alpha}}\right),
\label{Vint}
\eq
where we have added and subtracted in the integrand $V_0^{-\alpha}$, with 
$V_0= A\,y+m^2$.
From \eqn{JNtensorp}, \eqn{JNtensorpar} and the subsequent discussion we 
can argue that: 1) $\alpha\le N+\ep/2$ for all terms; 2) $\beta\ge N-1$,
since powers of $y$ in addition to the $y^{\ssN-1}$ term of \eqn{JNtensorp} 
can be embedded in the ${\cal K}$ factors.
Therefore, being $\beta-\alpha\ge-1-\ep/2$ in all cases, the second integral 
in \eqn{Vint} is always finite in the limit $m\to 0$: indeed
\bq \label{sim1}
\int_0^z\!dy\,y^\beta\,
\left(\frac{1}{V^{\alpha}}-\frac{1}{V_0^{\alpha}}\right) =
\int_0^z\!dy\,y^{\beta-\alpha}\,
\bigg\{\frac{1}{[A-y\,(K-p)^2]^{\alpha}}-\frac{1}{A^{\alpha}}\bigg\}
 + {\cal O}(m^2),
\qquad
\hbox{if} \quad \beta-\alpha > - 2.
\eq
The first term in \eqn{Vint}, instead, is collinear finite only if 
$\beta-\alpha > -1$; in this case indeed
\bq \label{sim2}
\int_0^z\!dy\,\frac{y^\beta}{(A\,y+m^2)^{\alpha}}= 
\frac{1}{(\beta-\alpha+1)A^\alpha} + {\cal O}(m^2).
\eq
Therefore, for all terms where the power of $V$ is greater or equal to 
$-N-\ep/2+1$, we have $\beta-\alpha\ge -\,\ep/2$, \eqn{sim1} and \eqn{sim2} 
simultaneously hold and no collinear logarithm is generated. 

As a result, we can extract the collinear behavior of $J_\ssN^{\nu_1\cdots\nu_r}$ 
replacing \eqn{JNtensorpar} in \eqn{JNtensorp}, and considering only the
term where all $r$ derivatives have acted on $V$; in this way we obtain
\bq
J_\ssN^{\nu_1\cdots\nu_r}= 
(-1)^r\,\Gamma\left(N+\frac{\ep}{2}\right)\!
\left(\frac{\mu^2}{\pi}\right)^{\!\!\ep/2}\!\!\!
\intsxy{z}{y}\!\int\!\!dS_{\ssN\!-\!1}(\{x\})\,y^{\ssN-1}\,
\prod_{i=1,r}\,{\cal K}^{\nu_i}\,V^{-\ssN-\ep/2}\, + \;\hbox{coll. fin.}
\label{JNtensorpp}
\eq
The expression for ${\cal K}$ given in \eqn{JNtensorp} contains a term 
proportional to $y$; it cannot produce a collinear logarithm in 
\eqn{JNtensorpp} because it increases the power of $y$ and leads to 
the above-mentioned condition $\beta-\alpha\ge -\,\ep/2$. 
Therefore, the collinear-divergent piece of $J_\ssN^{\nu_1\cdots\nu_r}$ 
is simply given by
\bq
J_\ssN^{\nu_1\cdots\nu_r}= 
\Gamma\left(N+\frac{\ep}{2}\right)\!
\left(\frac{\mu^2}{\pi}\right)^{\!\!\ep/2}\!\!\!
\intsxy{z}{y}\!\int\!\!dS_{\ssN\!-\!1}(\{x\})\,y^{\ssN-1}\,(-z)^r\,
\prod_{i=1,r}\,p^{\nu_i}\,V^{-\ssN-\ep/2}\, + \;\hbox{coll. fin.}
\label{JNtensorppp}
\eq
Now we add and subtract $V_0^{-\ssN-\ep/2}$, integrate in $y$ the added term and 
absorb in the collinear-finite remainder the subtracted one. 
The $y$ integration 
gives
\bq
\int_0^z\!\!dy\,\frac{y^{\ssN-1}}{(A\,y+m^2)^{\ssN+\ep/2}}= 
- \,\frac{\Gamma(N)\Gamma(1+\ep/2)}{\Gamma(N+\ep/2)}\,
\frac{1}{A^\ssN}\ln\left(\frac{m^2}{s}\right)\,
\sum_{k=0}^{\infty}\,\frac{1}{(k+1)!}\!
\left(\! -\frac{\ep}{2}\,\ln\frac{m^2}{s} \right)^k\, 
+ \;\hbox{coll. fin.}
\eq
Inserting the result in \eqn{JNtensorppp} and using \eqn{propback}, we 
finally extract the collinear logarithm for $J_\ssN^{\nu_1\cdots\nu_r}$,
\bq
J_\ssN^{\nu_1\cdots\nu_r}= 
\left(\frac{\mu^2}{s\,\pi}\right)^{\!\!\ep/2}\!\!\!\!\!
\Gamma\left(1\!+\!\frac{\ep}{2}\right)\!
\sum_{k=0}^{\infty}\,\frac{(\ep/2)^k}{(k\!+\!1)!}\!
\left(\! -\ln\frac{m^2}{s} \right)^{\!\!\!k+1}\!\!\!\!
\intsx{z}\,(-z)^r\!\prod_{j=1,r}\,p^{\nu_j}\!
\prod_{i=1}^\ssN\frac{1}{(k_i\!-\!pz)^2\!+\!m_i^2}\,
+ \;\hbox{coll. fin.}
\label{JNtensorres}
\eq
We are now able to compute the coefficient of the collinear 
logarithm for a two-loop diagram with a generic tensor structure in the 
configuration where only one external massless particle is coupled to 
a light internal one,
\vspace{-0.5cm}
\bq
\!
\raisebox{0.1cm}{\scalebox{0.65}{
\begin{picture}(180,75)(0,0)
 \SetWidth{1.2}
 \Photon(0,0)(40,0){2}{5}              \Text(10,5)[cb]{$p$}
 \DashCArc(75,0)(35,60,300){1.4}       \Text(75,39)[cb]{$m$}
                                       \Text(75,-45)[cb]{$m$}
\LongArrowArc(75,0)(42,120,150)        \Text(40,34)[cb]{$q$}
\LongArrowArc(75,0)(42,210,240)        \Text(27,-34)[cb]{$q+p$}
 \Text(65,-3)[cb]{$q^{\nu_1}\!\!\dots q^{\nu_r}$}
 \GCirc(130,0){45}{0.8}                
 \Text(130,-3)[cb]{$q_a^{\mu_1}\!\!\dots q_a^{\mu_m}$}
\end{picture}
}}
\!\!
=
\,
\ln\frac{m^2}{s}
\bigg[ 
1 
- \frac{\ep}{2}\Delta_{\ssU\!\ssV}(s) 
- \frac{\ep}{4}\ln\frac{m^2}{s} 
\bigg]\!
\intsx{z}\,(-z)^r\!\!\!\!\!\!
\raisebox{0.1cm}{\scalebox{0.65}{
\begin{picture}(90,75)(0,0)
 \SetWidth{1.2}
 \Photon(0,45)(30,26){2}{5}             \Text(-2,49)[cb]{$zp$}
 \Photon(0,-45)(30,-26){2}{5}           \Text(-2,-57)[cb]{$(1\!-\!z)p$}
 \GCirc(46,0){30}{0.8}                
 \Text(46,-3)[cb]{$q_a^{\mu_1}\!\!\dots q_a^{\mu_m}$}
\end{picture}
}}
\!\!\!\!
p^{\nu_1}\!\!\!\dots p^{\nu_r}
\,
+\;
\hbox{coll. fin.}\!\!\!
\vspace{0.8cm}
\label{tensor}
\eq
It is important to mention the striking correspondence with the special result 
derived for the simple configuration in \eqn{ve}.
\subsubsection{The collinear-finite part of $V^{\bba}_{\rm sc}$}
\label{sec:collfinG}
In order to compute the collinear-finite parts of the diagrams belonging to
the $V^{\bba}_{\rm sc}$ family, we slightly modify our approach.
We use a complete Feynman parametrization before extracting the collinear 
logarithms; they are then discarded and the collinear-finite remainder is 
written in a way that allows for direct numerical integration.
We follow the parametrization procedure described in section~7 of 
Ref.~\cite{Ferroglia:2003yj} for the scalar cases; the corresponding 
parametrization for tensor integrals can be found in section~9 of 
Ref.~\cite{Actis:2004bp}. 
Extracting the UV poles from these expressions and performing some changes of 
variables, we get 
\bqa
V^{\bba}_{\rm sc} &=& 
- \int_0^1\!\!\!dx_1\,dx_2\,dy_1\!\int_0^{y_1}\!\!\!\!\!dy_2\,
  (1\!-\!x_2)\,W_{\bba}^{-1},
\nl
V^{\bba}_{\rm sc}[q_i^\mu] &=& 
\int_0^1\!\!\!dx_1\,dx_2\,dy_1\!\int_0^{y_1}\!\!\!\!\!dy_2\,(1\!-\!x_2)\,
p_{\bba,i}^{\mu}\,W_{\bba}^{-1},
\quad
p_{\bba,i}^{\mu} = \sum_{h=1}^2\,a_{\bba,i}^h\,p_h^\mu,
\nl
V^{\bba}_{\rm sc}[q_i^\mu q_j^\nu] &=& 
- \int_0^1\!\!\!dx_1\,dx_2\,dy_1\!\int_0^{y_1}\!\!\!\!\!dy_2\,(1\!-\!x_2)\,
  p_{\bba,i}^{\mu}\,p_{\bba,j}^{\nu}\,W_{\bba}^{-1}
+ \Delta_{\bba,ij}^{\rm sc}\,\delta^{\mu\nu},
\label{vgsc:par1}
\eqa
where we have defined auxiliary quantities,
$\;\, \chi_\bba= p_2^2\,y_1\,(1\!-\!y_1) + M^2$ and
\bqa
W_\bba\! &=&
x_2\,(1\!-\!x_2)\,A_\bba + y_2\,[x_2\,M_a^2 + (1\!-\!x_2)\,m^2],
\qquad\quad\quad\,
A_\bba =
(1\!-\!y_1)\,[ P^2\,x_1\,y_2 + p_2^2\,(y_1\!-\!x_1\,y_2) ] + M^2,
\nl
a_{\bba,1}^1\! &=& 1 - x_1\,[1 - x_2\,(1\!-\!y_2)],
\qquad
a_{\bba,1}^2 = x_2\,(1\!-\!y_1),
\qquad
a_{\bba,2}^1 = 1 \!-\! x_1\,y_2,
\qquad
a_{\bba,2}^2 = 1\!-\!y_1,
\nl
\Delta_{\bba,11}^{\rm sc}\! &=& 
-\,\frac{1}{2}\Bigg\{\,
  F^2_{-2}(s)\,
+ F^2_{-1}(s)\,
  \bigg( \frac{3}{4} - \intsx{y_1}\,\ln\frac{\chi_\bba}{s} \bigg)
+ \frac{5}{16}
+ \frac{1}{4}\,\zeta(2)
\\
&&
\qquad
+ \,\frac{1}{2}\!\int_0^1\!\!\!\!dy_1\Bigg[\!
    \ln^{\!2}\!\frac{\chi_\bba}{s}
  - (\ln y_1 \!+\! 2)\ln\!\frac{\chi_\bba}{s}
  - 2\!\!\int_0^1\!\!\!\!dx_1 dx_2\!\!\int_0^{y_1}\!\!\!\!\!\!dy_2
    \bigg(\!
      x_2\ln\!\frac{W_\bba}{s}
    + \frac{1\!-\!x_2}{y_2}\ln\!\frac{W_\bba}{W_{\!\bba}|_{y_2\!=0}}
    \bigg)
  \Bigg]
\Bigg\}, 
%
%
\nl
\Delta_{\bba,12}^{\rm sc}\! 
&=& 
\frac{1}{2}\bigg[
- \frac{1}{2}\,F^2_{-1}(s) 
+ \frac{7}{8}
+ \int_0^1\!\!\!dx_1\,dx_2\,dy_1\!\int_0^{y_1}\!\!\!\!\!dy_2\,
  \ln\frac{W_\bba}{s}\bigg],
\nl
\Delta_{\bba,22}^{\rm sc}\! &=& 
- \,\frac{1}{2}\bigg[
    F^2_{\!-\!2}(s)\!
  + \!F^2_{\!-\!1}(s)\bigg( \frac{3}{4} - \ln\!\frac{m^{\!2}}{s} \!\bigg)\!
  + \!\frac{7}{16}\!
  + \!\frac{\zeta(2)}{4}
  + \!\frac{1}{2}\ln^{\!2}\!\frac{m^2}{s} 
  - \!\frac{3}{4}\!\ln\!\frac{m^2}{s}
  - \!\!\int_0^1\!\!\!\!\!dx_1dx_2dy_1\!\!\int_0^{y_1}\!\!\!\!\!\!\!dy_2
    \frac{1}{x_2}\ln\!\frac{W_\bba}{W_{\!\bba}|_{x_2\!=0}} \bigg].
\nonumber
\label{vgsc:par2}
\eqa
In order to get the collinear-finite parts, we have to extract and 
discard the collinear logarithms $\ln(m^2/s)$. Concerning the coefficients
of the metric tensor, only $\Delta_{\bba,22}^{\rm sc}$ 
has a collinear-divergent behavior;
$\Delta_{\bba,11}^{\rm sc}$ and $\Delta_{\bba,12}^{\rm sc}$ are
in a form suited for numerical integration, and there we simply set $m=0$.
For $\Delta_{\bba,22}^{\rm sc}$, some of the collinear logarithms are already 
explicit and we can discard them; the last term of $\Delta_{\bba,22}^{\rm sc}$ 
in \eqn{vgsc:par2}, instead, requires a further step.
Introducing $W_{\bba,0}= x_2\,A_\bba + x_2\,y_2\,M_a^2 + y_2\,m^2$ and
$W_{\bba,1}= A_\bba + y_2\,M_a^2$, we have
\bqa \label{collDOPP}
\int_0^1\!\!\frac{dx_2}{x_2}\,\!\ln\!\frac{W_\bba}{W_\bba|_{x_2=0}}
&=&
\int_0^1\!\!\frac{dx_2}{x_2}\,\bigg(
\ln\frac{W_\bba}{W_{\bba,0}} + \ln\frac{W_{\bba,0}}{W_\bba|_{x_2=0}} 
\bigg)\nn
\\
&=&
\int_0^1\!\!\frac{dx_2}{x_2}\, 
\ln\frac{(1\!-\!x_2)\,A_\bba + y_2\,M_a^2}{A_\bba + y_2\,M_3^2} 
- \li{2}{-\frac{W_{\bba,1}}{y_2\,m^2}}
+ {\cal O}(m^2)
\nl
&=&
  \frac{1}{2}\ln^2\!\frac{m^2}{s}
+ \ln\!\frac{m^2}{s}\ln\!\frac{y_2\,s}{W_{\bba,1}}
- \li{2\!}{\!\frac{A_\bba}{W_{\bba,1}}\!}
+ \frac{1}{2}\ln^2\!\frac{y_2\,s}{W_{\bba,1}}
+ \zeta(2)
+ {\cal O}(\!m^2).
\eqa
The integrals in \eqn{vgsc:par1} containing $W_\bba^{-1}$ show a 
collinear-divergent behavior for $m\to 0$ at the point $x_2=0$; 
therefore, terms where $W_\bba^{-1}$ is multiplied by $x_2$ are not 
singular and there we can simply set $m=0$. 
If no factor $x_2$ is present, instead, we write
\bqa
\int_0^1\!\!\!dx_2\,\frac{1}{W_\bba}
&=&
  \int_0^1\!\!\!dx_2\,\frac{1}{W_{\bba,0}} 
+ \int_0^1\!\!\!dx_2\,\Big( \frac{1}{W_\bba} - \frac{1}{W_{\bba,0}} \Big)
\nl
&=&
\frac{1}{A_\bba \!+\! y_2\,M_a^2}\,
\ln\frac{A_\bba \!+\! y_2\,(M_a^2+m^2)}{y_2\,m^2}
+ \int_0^1\!\!\frac{dx_2}{x_2}\,\bigg[
  \frac{1}{(1\!-\!x_2)\,A_\bba \!+\! y_2\,M_a^2} 
- \frac{1}{A_\bba \!+ y_2\,M_a^2} 
\bigg]
+ {\cal O}(m^2)
\nl
&=&
\frac{1}{W_{\bba,1}}\,\bigg(
- \ln\!\frac{m^2}{s}
+ \ln\!\frac{M_a^2}{s}
+ 2\,\ln\!\frac{W_{\bba,1}}{y_2\,M_a^2}
\bigg)
+ {\cal O}(m^2).
\label{Wg-1}
\eqa
Now all collinear logarithms have been extracted and can be discarded.
The finite remainder, however, can not be directly integrated numerically, 
because of the presence of denominators which can vanish inside the 
integration region.
On the other hand, these denominators are always linear in the variable $y_2$, 
being both $A_\bba$ and $W_\bba$ linear in $y_2$, and all terms generated 
from $W_\bba^{-1}$ have the following dependence on $y_2$:
\bq 
\frac{y_2^n}{a\,y_2+b},
\qquad
\frac{y_2^n}{a\,y_2+b}\,\ln\frac{a\,y_2+b}{c},
\qquad
y_2^n\,\frac{\ln y_2}{a\,y_2+b}.
\eq
For them we write
\bqa 
&&
\frac{\ln(a\,y_2+b)}{a\,y_2+b}=
\frac{1}{a}\,\partial_{y_2}\,\bigg[
  \frac{1}{2}\,\ln^2\left(1+\frac{a}{b}\,y_2 \right) 
+ \ln b\,\ln\left(1+\frac{a}{b}\,y_2 \right)
\bigg],
\nl
&&
\frac{\ln y_2}{a\,y_2+b}= 
\frac{1}{a}\,\partial_{y_2}\,\bigg[
\ln y_2\,\ln\left(1+\frac{a}{b}\,y_2 \right) + \li{2}{-\frac{a}{b}\,y_2}
\bigg],
\qquad
\frac{1}{a\,y_2+b} = 
\frac{1}{a}\,\partial_{y_2}\,\ln\left(1+\frac{a}{b}\,y_2 \right).
\label{eq:partial}
\eqa
The crucial point in \eqn{eq:partial} is that in the right-hand side 
the $1/a$ is always multiplied by a 
regulator function which goes to zero when the $a$ vanishes. 
The $-i0$ prescription associated to each mass ensures the validity of 
these expressions and can be used to produce the right imaginary parts of 
the logarithms.
After integrating by parts in $y_2$, we get for the collinear-finite parts
$$
V^{\bba}_{\rm sc,fin} = 
I^{\bba}_{\rm sc}[0,1,0,0,1,1],
\qquad
V^{\bba}_{\rm sc,fin}[q_i^\mu] = 
- \,\sum_{h=1}^2p_h^\mu\,
  I^{\bba}_{\rm sc}[b_{\bba,i}^h,c_{\bba,i}^h,d_{\bba,i}^h,
                    d_{\bba,i}^h,e_{\bba,i}^h,\bar{e}_{\bba,i}^h],
$$
\bq
V^{\bba}_{\rm sc,fin}[q_i^\mu q_j^\nu] =
  \Delta_{\bba,ij}^{\rm fin}\,\delta^{\mu\nu}
+ \sum_{h,k=1}^2\,p_h^\mu\,p_k^\nu\,
  I^{\bba}_{\rm sc}[b_{\bba,ij}^{hk},c_{\bba,ij}^{hk},d_{\bba,ij}^{hk},
                    \bar{d}_{\bba,ij}^{hk},e_{\bba,ij}^{hk},
                    \bar{e}_{\bba,ij}^{hk}],
\label{tableVG}
\eq
where the auxiliary function $I^{\bba}_{\rm sc}$ depends on six
parameters, symbolically denoted by $\{b\}_6$:
\bqa
I^{\bba}_{\rm sc}[\{b\}_6]
&=& 
\int_0^1\!\!\!dx_1dx_2dy_1dy_2\,\Bigg\{\,
  \frac{b_1}{\rho_{\bba}}\,
  \ln\left( 1 + \frac{y_1\,y_2\,\rho_{\bba}}{x_2\,\chi_\bba} \right)
+ \frac{b_2}{\rho_{\bba}}\,
  \ln\left(\! 1 + \frac{y_1\rho_{\bba}}{x_2\chi_\bba} \!\right)
\nl
&&
- \frac{b_3}{\eta_{\bba}}\bigg[\!
    \ln^2\!\left(\! 1 \!+\! \frac{y_1 y_2 \eta_{\bba}}{\chi_\bba} \!\right)
  + \bigg(\! 
    \ln\!\frac{M_a^2}{s} + 2\ln\!\frac{\chi_\bba}{y_1 y_2 M_a^2} 
    \!\bigg)\!
    \ln\!\left(\! 1 \!+\! \frac{y_1 y_2 \eta_{\bba}}{\chi_\bba} \!\right)
  \!\bigg]
+ \frac{2b_4}{\eta_{\bba}}
  \ln\!\left(\! 1 + \frac{y_1 y_2 \eta_{\bba}}{\chi_\bba} \!\right)
\nl
&&
- \frac{b_5}{\eta_{\bba}}\,\Bigg[
    \ln^2\!\left(\! 1 \!+\! \frac{y_1\,\eta_{\bba}}{\chi_\bba} \!\right)
  + \bigg(\! \ln\!\frac{M_a^2}{s} + 2\ln\!\frac{\chi_\bba}{y_1 M_a^2} \!\bigg)
    \ln\!\left(\! 1 \!+\! \frac{y_1\,\eta_{\bba}}{\chi_\bba} \!\right)
  \!\Bigg]
+ \frac{2b_6}{\eta_{\bba}}\,
  \li{2\!}{\!-\frac{y_1\eta_{\bba}}{\chi_\bba}\!}
\!\Bigg\}.
\label{Igsc}
\eqa
We recall that $\chi_\bba = p_2^2\,y_1\,(1\!-\!y_1) + M^2$; 
in addition, we have introduced the polynomials:
\bq
\rho_\bba= 
x_1 x_2 (1\!-\!y_1) (P^2\!-\!p_2^2) - x_2 M^2 + M_a^2,
\qquad
\eta_\bba= x_1\,(1-y_1)\,(P^2-p_2^2) - M^2 + M_a^2.
\eq
Concerning the coefficients of the metric tensor, we have defined 
$\Delta_{\bba,11}^{\rm fin}= \Delta_{\bba,11}^{\rm sc}|_{m=0}$, 
$\Delta_{\bba,12}^{\rm fin}= \Delta_{\bba,12}^{\rm sc}|_{m=0}$,
since they are collinear free; for $\Delta_{\bba,22}^{\rm fin}$,
we start from \eqn{vgsc:par2} and insert the result of \eqn{collDOPP},
\bq
\Delta_{\bba,22}^{\rm fin} = 
- \,\frac{1}{2}\bigg\{
    F^2_{-2}(s)
  + \frac{3}{4} F^2_{-1}(s)
  + \frac{7}{16} 
  - \frac{\zeta(2)}{4}
  + \int_0^1\!\!\!\!dx_1 dy_1\!\!\int_0^{y_1}\!\!\!\!\!\!dy_2\,
    \bigg[
      \li{2\!}{\!\frac{A_\bba}{A_\bba \!+\! y_2 M_a^2}\!}
    - \frac{1}{2}\ln^2\!\frac{A_\bba \!+\! y_2 M_a^2}{y_2\,s}
  \bigg]
\bigg\}.
\eq
Finally, we show the explicit expressions for the arguments of the 
vector integrals (second equation in \eqn{tableVG}),
\bqa
\ba{lllll}
b_{\bba,1}^{1} = - x_1\,x_2\,y_1,
\;&\;
b_{\bba,1}^{2} = 0,
\;&\;
b_{\bba,2}^{1}= x_1\,y_1,
\;&\;
b_{\bba,2}^{2}= 0,
\;&\; \\
c_{\bba,1}^{1}= 1 - x_1 - x_1\,x_2\,(1-y_1),
\;&\;
c_{\bba,1}^{2}= - x_2\,(1-y_1),
\;&\;
c_{\bba,2}^{1}= 1 - x_1\,y_1,
\;&\;
c_{\bba,2}^{2}= 1 - y_1,
\;&\; \\
d_{\bba,1}^{1}= 0,
\;&\;
d_{\bba,1}^{2}= 0,
\;&\;
d_{\bba,2}^{1}= x_1\,y_1,
\;&\;
d_{\bba,2}^{2}= 0,
\;&\; \\
e_{\bba,1}^{1}= 1 - x_1,
\;&\;
e_{\bba,1}^{2}= 0,
\;&\;
e_{\bba,2}^{1}= 1 - x_1\,y_1,
\;&\;
e_{\bba,2}^{2}= 1-y_1,
\;&\; \\
\bar{e}_{\bba,1}^{1}= 1 - x_1,
\;&\;
\bar{e}_{\bba,1}^{2}= 0,
\;&\;
\bar{e}_{\bba,2}^{1}= 1,
\;&\;
\bar{e}_{\bba,2}^{2}= 1-y_1,
\ea
\eqa
and for those of the tensor integrals (third equation in \eqn{tableVG}),
\bq
\ba{ll}
b_{\bba,11}^{11}= 
- 2\,x_1\,x_2\,y_1\,\{ 1 - x_1\,[ x_2 + y_1\,y_2 (1 - x_2)] \},
\, & \,
b_{\bba,11}^{12}= b_{\bba,11}^{21}= 
- x_1\,x_2\,y_1\,(1\!-\!x_2)\,(1\!-\!y_1),
\\
b_{\bba,11}^{22}= 0,
\, & \,
b_{\bba,12}^{11}= 
x_1\,y_1\,( 1 - x_1 - x_2 - x_1\,x_2 + 2\,x_1\,x_2\,y_1\,y_2 ),
\\
b_{\bba,12}^{12}= b_{\bba,12}^{21}= - x_1\,x_2\,y_1\,(1\!-\!y_1),
\, & \,
b_{\bba,12}^{22}= 0,
\\
b_{\bba,22}^{11}= 2\,x_1\,y_1\,( 1 - x_1\,y_1\,y_2 ),
\, & \,
b_{\bba,22}^{12}= b_{\bba,22}^{21}= x_1\,y_1\,(1\!-\!y_1),
\\
b_{\bba,22}^{22}= 0,
\, & \,
c_{\bba,11}^{11}= 
[ 1 - x_1 - x_1\,x_2\,(1\!-\!y_1) ]^2 - x_1^2\,x_2\,(1\!-\!y_1)^2,
\\
c_{\bba,11}^{12}= c_{\bba,11}^{21}= 
- x_2 (1\!-\!y_1) \{ 1 - x_1 [1- (1\!-\!x_2) (1\!-\!y_1)] \},
\, & \,
c_{\bba,11}^{22}= - x_2\,(1\!-\!x_2)\,(1\!-\!y_1)^2,
\\
c_{\bba,12}^{11}= 
(1\!-\!x_1\,y_1)\,[ 1 - x_1 - x_1\,x_2\,(1\!-\!y_1) ],
\, & \,
c_{\bba,12}^{12}= (1\!-\!y_1)\,[ 1 - x_1 - x_1\,x_2\,(1\!-\!y_1) ],
\\
c_{\bba,12}^{21}= - x_2\,(1\!-\!y_1)\,( 1 - x_1\,y_1 ),
\, & \,
c_{\bba,12}^{22}= - x_2\,(1\!-\!y_1)^2,
\\
c_{\bba,22}^{11}= ( 1 - x_1\,y_1 )^2,
\, & \,
c_{\bba,22}^{12}= c_{\bba,22}^{21}= (1\!-\!y_1)\,( 1 - x_1\,y_1 ),
\\
c_{\bba,22}^{22}= (1\!-\!y_1)^2,
\, & \,
d_{\bba,11}^{ij}= \bar{d}_{\bba,11}^{ij}= 0,
\\
d_{\bba,12}^{11}= \bar{d}_{\bba,12}^{11}= x_1\,y_1\,(1\!-\!x_1),
\, & \,
d_{\bba,12}^{12}= d_{\bba,12}^{21}= d_{\bba,12}^{22}= 
\bar{d}_{\bba,12}^{12}= \bar{d}_{\bba,12}^{21}= \bar{d}_{\bba,12}^{22}= 0,
\\
d_{\bba,22}^{11}= 2\,x_1\,y_1\,( 1 - x_1\,y_1\,y_2 ),
\, & \,
\bar{d}_{\bba,22}^{11}= x_1\,y_1\,( 2 - x_1\,y_1\,y_2 ),
\\
d_{\bba,22}^{12}= d_{\bba,22}^{21}= 
\bar{d}_{\bba,22}^{12}= d_{\bba,22}^{21}= x_1\,y_1\,(1\!-\!y_1),
\, & \,
d_{\bba,22}^{22}= \bar{d}_{\bba,22}^{22}= 0,
\\
e_{\bba,11}^{11}= \bar{e}_{\bba,11}^{11}= (1-x_1)^2,
\, & \,
e_{\bba,11}^{12}= e_{\bba,11}^{21}= e_{\bba,11}^{22}= 
\bar{e}_{\bba,11}^{12}= \bar{e}_{\bba,11}^{21}= \bar{e}_{\bba,11}^{22}= 0,
\\
e_{\bba,12}^{11}= (1\!-\!x_1)\,(1\!-\!x_1\,y_1),
\, & \,
\bar{e}_{\bba,12}^{11}= 1\!-\!x_1,
\\
e_{\bba,12}^{12}= \bar{e}_{\bba,12}^{12}= (1\!-\!x_1)\,(1\!-\!y_1),
\, & \,
e_{\bba,12}^{2j}= \bar{e}_{\bba,12}^{2j}= 0,
\\
e_{\bba,22}^{11}= (1\!-\!x_1\,y_1)^2,
\, & \,
\bar{e}_{\bba,22}^{11}= 1,
\\
e_{\bba,22}^{12}= e_{\bba,22}^{21}= (1\!-\!y_1)\,(1\!-\!x_1\,y_1),
\, & \,
\bar{e}_{\bba,22}^{12}= \bar{e}_{\bba,22}^{21}= 1\!-\!y_1,
\\
e_{\bba,22}^{22}= \bar{e}_{\bba,22}^{22}= (1\!-\!y_1)^2.
\, & \,
\ea
\eq
\subsubsection{The collinear-finite part of $V^{\bca}_{\rm sc}$}
\label{sec:collfinK} 
In this section we consider the $V^{\bca}_{\rm sc}$ configuration, using
the Feynman parametrization of Ref.~\cite{Passarino:2006gv}. For the needed 
UV-finite scalar configuration we set $\ep=0$ in Eq.~(159) of 
Ref.~\cite{Passarino:2006gv}, perform the $y_3$ integration and make the 
appropriate change of variables in order to simplify the integral. In the 
collinear configuration we obtain
\bq
V^{\bca}_{\rm sc}=
- \int_0^1\!\!\!dx_1\,dx_2\,dy_1\!\int_0^{y_1}\!\!\!\!\!dy_2\,
x_1\,(1-x_2)\,y_2\,
\sum_{i=a,b}\, \frac{1}{W_{\bca,i}\,U_{\bca,i}},
\eq
where we have defined
\bqa
\ba{ll}
W_{\bca,a} =
x_2\,(1\!-\!x_2)\,A_{\bca,a} + y_2\,[ m^2\,(1\!-\!x_2) + M_a^2\,x_2 ],
\quad & \quad
U_{\bca,a} = A_{\bca,a} + y_2\,M^2,
\\
A_{\bca,a} =
(1\!-\!y_1)\,[ p_1^2\,(y_1\!-\!x_1\,y_2) + 
p_2^2\,x_1\,y_2 ] + (1\!-\!y_2)\,M^2,
\quad & \quad
\\
W_{\bca,b} = W_{\bca,a}(p_1\leftrightarrow p_2),
\quad & \quad
U_{\bca,b}= U_{\bca,b}(p_1\leftrightarrow p_2).
\ea
\eqa
The integrand behaves as $1/x_2$ for $m=0$, revealing the presence of a 
collinear divergence. 
In order to extract the collinear logarithm, we introduce the polynomial
$W_{\bca,a0}= x_2\,A_{\bca,a} + y_2\,(m^2 + M_a^2\,x_2)$ (similarly for 
$W_{\bca,b}$). 
Then, we perform the integration over $x_2$,
\bqa
\int_0^1\!\!\!dx_2\,\frac{1\!-\!x_2}{W_{\bca,a}} 
&=&
  \int_0^1\!\!\!dx_2\,\Bigl[ \frac{1}{W_{\bca,a0}} 
+ \Big( \frac{1}{W_{\bca,a}} - \frac{1}{W_{\bca,a0}} \Big)\bigg|_{m=0}
- \frac{x_2}{W_{\bca,a}}\bigg|_{m=0} \Bigr] + \ord{m^2}
\nl
&=&
\frac{1}{A_{\bca\!,a} \!+\! y_2 M_a^2}\,\bigg( \!\!
- \!\ln\!\frac{m^2}{s}\!
+ \ln\!\frac{M_a^2}{s}
+ 2 \ln\!\frac{A_{\bca\!,a} \!+\! y_2 M_a^2}{y_2 M_a^2}
\bigg)\!
- \! \frac{1}{A_{\bca\!,a}}\ln\!\frac{A_{\bca\!,a} \!+\! y_2\,M_a^2}{y_2 M_a^2}
\! +\! {\cal O}(m^2).
\qquad
\eqa
All the polynomials involved in the computation are linear in $y_2$; therefore, 
an additional analytical integration is possible. 
First, we perform two splittings,
\bq
\frac{y_2}{U_{\bca,a}(A_{\bca\!,a} \!+\! y_2 M_a^2)} =
\frac{1}{M^2-M_a^2}\,
\bigg( \frac{1}{A_{\bca\!,a} \!+\! y_2 M_a^2} - \frac{1}{U_{\bca,a}} \bigg),
\qquad
\frac{y_2}{U_{\bca,a}A_{\bca\!,a}} =
\frac{1}{M^2}\,\bigg( \frac{1}{A_{\bca\!,a}} - \frac{1}{U_{\bca,a}} \bigg).
\eq
Secondly, we carry on the $y_2$ integration by means of the relations 
introduced in \eqn{eq:partial} and of the identity
\bq
\frac{\ln(cy_2\!+\!b)}{ay_2\!+\!b}= 
\frac{1}{a}\partial_{y_2}\bigg[\,
  \frac{1}{2}\ln^2\!\left(\! 1 \!+\! \frac{a}{b}y_2 \!\right) 
+ \ln b\,\ln\!\left(\! 1 \!+\! \frac{a}{b}y_2 \!\right) 
+ \li{2\!}{\!-\frac{a}{b}y_2\!}
+ \li{2\!}{\frac{\!(a\!-\!c)y_2}{a y_2\!+\!b}\!}
- \li{2\!}{\!-\frac{c}{b}y_2\!}
\!\bigg].
\eq
Note that for $a\to 0$ the content of the squared bracket vanishes; 
the complete result reads
\bqa
V^{\bca}_{\rm sc,fin}
&=&
\int_0^1\!\!\!\!dx_1dy_1\,\frac{x_1}{M^2}\,\Bigg\{\!
  \frac{1}{\rho_{\bca}}\,
  L_{\bca}(y_1\rho_{\bca},y_1M_a^2,y_1M_a^2;0)
- \frac{2\,M^2}{(M^2\!-\!M_a^2)\eta_{\bca}}\,
  L_{\bca}(y_1\eta_{\bca},0,y_1M_a^2;1)
\nl
&&
- \frac{1}{(p_1^2\!-\!p_2^2)y_1}\,\bigg[
    L_{\bca}(y_1\omega_{\bca},\mu_{\bca},(1\!-\!y_1)M_a^2;0)
  - \frac{2\,M^2}{M^2\!-\!M_a^2}L_{\bca}
(y_1\omega_{\bca},\mu_{\bca},(1\!-\!y_1)M_a^2;1)
  \bigg]
\!\Bigg\},
\eqa
where we have introduced
\bq
L_{\bca}(A,B,C;\alpha)= 
    \li{2\!}{\!\frac{-B}{\chi_\bca\!+\!A}\!}
  - \li{2\!}{\!\frac{A\!+\!B}{-\chi_\bca}\!}
  + \frac{1}{2}
    \ln^2\!\left(\! 1 \!+\! \frac{A}{\chi_\bca} \!\right)
  + \bigg(\! 
    \ln\!\frac{\chi_\bca}{C} + \frac{\alpha}{2}\ln\!\frac{M_a^2}{s} 
    \!\bigg)
    \ln\!\left(\!1 \!+\! \frac{A}{\chi_\bca} \!\right).
\eq
Here $\chiu{\bca}= y_1\,(1\!-\!y_1)\,p_2^2 + M^2$ and the polynomials appearing
in the arguments of $L_{\bca}$ are given by
\bq
\eta_{\bca}= \omega_{\bca} + M_a^2 - M^2,
\qquad
\rho_{\bca}= \omega_{\bca} - M^2,
\qquad
\omega_{\bca}= x_1\,(1\!-\!y_1)(p_1^2\!-\!p_2^2),
\qquad
\mu_{\bca}= (1\!-\!y_1)(M^2\!-\!M_a^2).
\eq
\subsubsection{The collinear-finite part of $V^{\bbb}_{\rm sc}$ 
\label{sec:collfinH}} 
As previously done for $V^{\bba}_{\rm sc}$ and $V^{\bca}_{\rm sc}$, we specify here
the parametrization of the diagram, following the procedure outlined 
in section~10.4 of Ref.~\cite{Ferroglia:2003yj}.
Setting the UV regulator $\ep$ to zero and performing some trivial change 
of variables, we can cast the collinear configuration of $V^{\bbb}_{\rm sc}$ as
\bq
V^{\bbb}_{\rm sc} = 
- P^2\,\int dC_5\lpar x,y,\{z\}\rpar\,
  x\,y\,(1\!-\!x)\,(1\!-\!y)\,\frac{a_{\bbb\!}}{W_{\bbb}^{2}},
\label{vhsc:par1}
\eq
where we have defined the following quantities: 
\bqa \label{VHdefs}
\ba{ll}
W_\bbb = 
y\,(1\!-\!y)\,A_\bbb + y\,B_\bbb + (1\!-\!y)\,x\,(1\!-\!x)\,m^2,
\quad & \quad
A_\bbb = x\,(1\!-\!x)(1\!-\!z_2\!-\!z_3)(1\!-\!z_1\!-\!z_2)\,P^2,
\\
B_\bbb = x\,\chi_\bbb \!+\! (1\!-\!x)\,M^2,
\quad & \quad
\chi_\bbb = P^2\,z_2\,(1\!-\!z_2) + M_a^2,
\\
a_{\bbb} =
M_a^2 - M^2 - P^2\,{\cal Z}_{\bbb}\,(1-{\cal Z}_{\bbb}),
\quad & \quad
{\cal Z}_{\bbb}= z_1 + x\,(1\!-\!y)(1\!-\!z_1\!-\!z_2).
\ea
\eqa
Since the collinear singularity for $m\to 0$ shows up at the point $y=0$, 
we extract the collinear logarithm according to
$$
\int_0^1\!\!\!\!dy\frac{y(1\!-\!y)}{W_\bbb^2}
\!=\!
\int_0^1\!\!\!\!dy\,y\bigg[\!
  \frac{1}{W_{\!\!\bbb\!,0}^2}
+ \bigg(\!
  \frac{1\!-\!y}{W_{\!\!\bbb}^2} - \frac{1}{W_{\!\!\bbb\!,0}^2}
  \!\bigg) 
\bigg]
\!=\!
  \frac{1}{{\cal W}_{\!\bbb\!,0}^2}\!\bigg[
        \ln\!\frac{{\cal W}_{\!\bbb\!,0}}{x(1\!-\!x)s} 
  - \ln\!\frac{m^2}{s}
  - 1 
  \bigg]
+ \!\int_0^1\!\!\frac{dy}{y}\!\bigg(\!
  \frac{1\!-\!y}{{\cal W}_{\!\bbb}^2} - \frac{1}{{\cal W}_{\!\bbb\!,0}^2} 
  \!\bigg)
+ {\cal O}(m^{\!2}),
$$
\bq
\int_0^1\!dy\,\frac{y^{n+2}(1\!-\!y)}{W_\bbb^2}
=
  \int_0^1\!\!\!dy\,\frac{y^n(1\!-\!y)}{{\cal W}_{\!\bbb}^2}
+ {\cal O}(m^2),
\qquad
n \ge 0,
\label{eq:WH}
\eq
where we have introduced 
$W_{\!\!\bbb\!,0} \!= y\,{\cal W}_{\bbb\!,0} + x(1\!-\!x)\,m^2$, 
$\;{\cal W}_{\!\bbb} \!= (1\!-\!y)A_\bbb + B_{\!\bbb}$ and
${\cal W}_{\!\bbb\!,0} \!= A_\bbb \!+\! B_{\!\bbb}$.
We notice that all polynomials involved in the computation are combinations 
of the quantities $A_\bbb$ and $B_\bbb$, both linear in $z_1$;
therefore, for the terms involving ${\cal W}_{\!\bbb\!,0}^{-2}$ or 
${\cal W}_{\!\bbb}^{-2}$, we perform an integration by 
parts in $z_1$ using the following relations:
\bqa
\int_0^1\!\!\!dz_1\,z_1^n\,\frac{\ln(a z_1+b)}{(a z_1+b)^2}
&=&
\int_0^1\!\!\!dz_1\,\bigg[
  \frac{z_1^n}{(a z_1+b)^2}
 + n\,z_1^{n-1}\,\frac{\ln(a z_1+b)}{a(a z_1+b)}
\bigg]
- \,\frac{\ln(a+b)}{a(a+b)}
+ \delta_{n,0}\,\frac{\ln b}{a b},
\nl
\int_0^1\!\!\!dz_1\,\frac{z_1^n}{(a z_1+b)^2}
&=&
- \frac{n}{b}\,\int_0^1\!\!\!dz_1\,\frac{z_1^n}{a z_1+b}
+ \frac{1}{b(a+b)}.
\label{intbyparts}
\eqa
Let now be ${\cal P}$ and ${\cal Q}$ two generic polynomials in the 
Feynman variables. 
The result after this step can be formally written as a sum of terms of the 
form $1/ {\cal Q}{\cal P}$ or $\ln{\cal P}/ {\cal Q} {\cal P}$. 
These polynomials are directly related to $W_H$ and $W_{H,0}$ and therefore 
are linear in $z_3$, as one can easily prove by direct computation. 
Therefore, we can use the following relations, with $\Delta= a d - b c$:
\bqa
&&
\frac{1}{(a z_3+b)(c z_3+d)}=
\frac{1}{\Delta}\,\partial_{z_3}\,
\Big[ \ln(a z_3+b) - \ln(c z_3+d) \Big],
\nl
&&
\frac{\ln(c z_3\!+\!d)}{(a z_3\!+\!b)(c z_3\!+\!d)}=
\frac{1}{\Delta}\,\partial_{z_3}
\bigg[
  \ln\frac{\Delta}{a}\ln\frac{a z_3\!+\!b}{c z_3\!+\!d}
- \li{2}{\frac{\Delta}{a(c z_3\!+\!d)}}
- \ln\frac{\Delta}{a(c z_3\!+\!d)} 
  \ln\left(\! 1 \!-\! \frac{\Delta}{a(c z_3\!+\!d)} \!\right)
\! \bigg].
\label{primitives}
\eqa
After performing the integration over $z_3$, the remaining integrands can be 
cast as
\bq \label{casi}
\frac{1}{\omega_{\!\bbb}}
\ln^n\left(\! 1 \!+\! \frac{\omega_{\!\bbb}}{\cal P} \!\right)
\left[\; 1 \;,\; \ln\frac{\cal Q}{s} \;\right],
\qquad
\frac{1}{\omega_{\!\bbb}}\li{2}{\frac{\omega_{\!\bbb}}{\cal P}},
\qquad
\frac{1}{\omega_{\!\bbb}}\ln\frac{\cal P}{s}\,
\left[ 1 \;,\; \ln x \;,\; \ln(1-x) \;,\; \ln\frac{\omega_{\!\bbb}}{s} \right],
\eq
with $n=1,2$, where we have defined 
$\omega_{\!\bbb}= x\,\chi_\bbb \!+\! (1\!-\!x)\,M^2$.
The first two cases in \eqn{casi} can be numerically integrated; the last, instead, requires further 
manipulations. In particular, for the terms with $1/y$, coming from the last 
integral of the first relation contained in \eqn{eq:WH}, we have always the
combination
\bq
\!\frac{1}{\omega_{\!\bbb}}\!\int_0^1\!\!\!\,\frac{dy}{y}
\ln\!\frac{{\cal P}'(1\!-\!y)\!-\!\omega_{\!\bbb}}
          {{\cal P}'\!\!-\!\omega_{\!\bbb}}
=
-\frac{1}{\omega_{\!\bbb}}\li{2\!}{\!\frac{{\cal P}'}{{\cal P}'\!\!-\!\omega_{\!\bbb}}\!}
\!=
\!\frac{1}{\omega_{\!\bbb}}\bigg[
  \li{2\!}{\!\frac{\omega_{\!\bbb}}{{\cal P}'\!\!-\!\omega_{\!\bbb}}\!}
+ \ln\!\frac{\omega_{\!\bbb}}{{\cal P}'\!\!-\!\omega_{\!\bbb}}
  \ln\!
  \left(\! 
  1 \!-\! \frac{\omega_{\!\bbb}}{{\cal P}'\!\!-\!\omega_{\!\bbb}} 
  \!\right)
- \zeta(2)
\bigg]\!.
\eq
In the last expression, the terms with $\zeta(2)$ cancel out and the others 
are suited for numerical integration.
For the other terms of the type $\omega_{\!\bbb}^{-1}\ln({\cal P}/s)$, we 
explicitly write the $i\,0$ Feynman prescription and get
\bq
\frac{1}{\omega_{\!\bbb}}\ln\frac{\cal P}{s}=
\frac{1}{\omega_{\!\bbb}}
\ln\frac{ ({\cal P}\!+\!\omega_{\!\bbb}) - \omega_{\!\bbb} - i\,0}{s}
=
\frac{1}{\omega_{\!\bbb}}\bigg[
  \ln\!\bigg(\! 
  1 \!-\! \frac{\omega_{\!\bbb}}{{\cal P}\!+\!\omega_{\!\bbb}- i\,0}
  \!\bigg)
+ \ln\bigg( \frac{{\cal P}\!+\!\omega_{\!\bbb}}{s} - i\,0 \!\bigg)
\bigg],
\eq
where the first logarithm regulates the zero of the denominator 
$\omega_{\!\bbb}$, while the second is always of the following form:
\bq
\ln\!\left( \frac{{\cal P}\!+\!\omega_{\!\bbb}}{s} - i\,0 \!\right) =
\ln\!\left( x\,(1\!-\!x)\,\frac{{\cal P}'}{s} - i\,0 \right)= 
  \ln x + \ln(1\!-\!x) 
+ \ln\!\left( \frac{{\cal P}'}{s} - i\,0 \!\right).
\eq
Here ${\cal P}'$ does not depend on $x$.
As a consequence, for the terms containing this logarithm, we can integrate 
by parts in $x$, since $\omega_{\!\bbb}$ is linear in $x$, using
the three relations of \eqn{eq:partial} (replacing $y_2$ with $x$).

Collecting all pieces, we derive the final result for $V^{\ssH}_{\rm sc,fin}$,
\bqa \label{fullvh}
V^{\ssH}_{\rm sc,fin} 
&=&
\sum_{n=1}^{3}\int dC_4\lpar x,y,\{z\}\rpar\,\Bigg\{\!
\frac{2b_{\bbb\!,n}}{\omega_{\!\bbb}}\,\bigg[
  2\,\li{2\!\!}{\frac{\omega_{\!\bbb}}{\omega_{\!n}}\!}
+ \bigg(\!
  \ln\!\frac{x\,(1\!-\!x)\omega_{\!\bbb}}{s}
  - 2\ln\frac{\omega_{\!n}}{s}
  \!\bigg)
  \ln\!\left(\! 1 \!-\! \frac{\omega_{\!\bbb}}{\omega_{\!n}} \!\right)
\!\bigg]
\nl
&&
+ \frac{c_{\bbb\!,n}}{\omega_{\!\bbb}}
  \ln\!\left(\! 1 \!-\! \frac{\omega_{\!\bbb}}{\omega_{\!n\!,y}} \!\right)
\!\!\Bigg\}
+ \int_0^1\!\!\!dx\,dz_1\,dz_2\,
  \frac{1}{\bar{\omega}_{\!\bbb}}\,\Bigg\{
    \Theta_{\!1}\,\li{2}{-\frac{\bar{\omega}_{\!\bbb}}{M^2}}
  + \bigg( \Theta_{\!2}\ln\frac{M^2}{s} + \Theta_{\!3} \bigg)
    \ln\bigg(\! 1 \!+\! \frac{\bar{\omega}_{\!\bbb}}{M^2} \!\bigg)
\nl
&&
  + \,\Theta_{\!4}\bigg[ 
      \frac{1}{2}\ln\bigg(\! 1 \!+\! \frac{x\,\bar{\omega}_{\!\bbb}}{M^2} \!\bigg)
    + \ln\frac{M^2}{x\,s} - 2 
    \bigg]\!
    \ln\bigg(\! 1 \!+\! \frac{x\,\bar{\omega}_{\!\bbb}}{M^2} \!\bigg)
  + ( \Theta_{\!5} + \Theta_{\!6}\ln x )\,
    \ln\left(\! 1 \!-\! \frac{x\,\bar{\omega}_{\!\bbb}}{\chi_\bbb} \!\right)
  \Bigg\},
\eqa
where, in addition to $\omega_{\!\bbb}$ introduced after \eqn{casi}, we have 
defined auxiliary quantities in term of $\chi_\bbb$ of \eqn{VHdefs},
\bqa
\ba{lll}
\bar{\omega}_{\!\bbb}= \chi_\bbb \!-\! M^2,
\;&\;
\omega_{n} = \omega_{\!\bbb} + P^2 x\,(1\!-\!x)\,\sigma_n,
\;&\;
\omega_{n\!,y} = \omega_{\!\bbb} + P^2 x\,(1\!-\!x)\,(1\!-\!y)\,\sigma_n,
\\
\sigma_{1} = (1\!-\!z_2)^2,
\;&\;
\sigma_{2} = -z_2\,(1\!-\!z_2),
\;&\;
\sigma_{3} = (1\!-\!z_2)(1\!-\!z_1\!-\!z_2),
\ea
\eqa
\bq \label{THETA}
\Theta_{\!n}= 
  2\,d_{\!\bbb\!,n}\,\ln\left( \frac{P^2}{s}(1\!-\!z_1\!-\!z_2)-i\,0 \right)
+ \,i\,\pi\,e_{\!\bbb\!,n}\,{\rm sign}(P^2).
\eq
The coefficients appearing in \eqn{fullvh} and \eqn{THETA} read as
\bqa
\ba{ll}
b_{\bbb\!,1}= M^2 - M_a^2 + \frac{P^2}{4}( 4 z_2^2 - 6 z_2 + 1 ),
\;&\;
b_{\bbb\!,2}= M_a^2 - M^2 - \frac{P^2}{4}
[ (4 z_2^2 - 6 z_2 + 3)x - 2 ],
\\
b_{\bbb\!,3}= P^2(1\!-\!x)[ 1 - 2z_2 - 2(1\!-\!x)(1\!-\!z_1\!-\!z_2) ],
\;&\;
c_{\bbb\!,1}= -2 P^2 x\,(1\!-\!z_2)[ 1 - x\,(1\!-\!z_2)(2\!-\!y) ],
\\
c_{\bbb\!,2}= P^2 x\,[ 1 - 
\frac{x}{2}\,(2\!-\!y)(4 z_2^2 - 6 z_2 + 3) ],
\;&\;
c_{\bbb\!,3}= -2 P^2 x\,
[ 1 - 2z_2 - 2\,(2\!-\!2 x\!+\!x\,y)(1\!-\!z_1\!-\!z_2) ],
\\
d_{\bbb\!,1}= P^2( 1 - 2 z_1 ),
\;&\;
d_{\bbb\!,2}= 0,
\\
d_{\bbb\!,3}= P^2( z_1 - z_2 ),
\;&\;
d_{\bbb\!,4}= P^2[ 1 - 2z_2 - 4\,(1\!-\!x)(1\!-\!z_1\!-\!z_2) ],
\\
d_{\bbb\!,5}= - P^2[ 1 - 2z_2 - 2x\,(1\!-\!z_1\!-\!z_2) ],
\;&\;
d_{\bbb\!,6}= - P^2[ 1 - 2z_2 - 4x\,(1\!-\!z_1\!-\!z_2) ],
\\
e_{\bbb\!,1}= 4\,(M^2 - M_a^2) + \frac{P^2}{2}( 4 z_2^2 - 6 z_2 + 1 ),
\;&\;
e_{\bbb\!,2}= 2\,(M_a^2 - M^2) - \frac{P^2}{2}( 4 z_2^2 - 6 z_2 + 1 ),
\\
e_{\bbb\!,3}= \frac{P^2}{4}( 12 z_2^2 - 18 z_2 + 5 ),
\;&\;
e_{\bbb\!,4}= - P^2[ 1 - x\,( 4 z_2^2 - 6 z_2 + 3 ) ],
\\
e_{\bbb\!,5}= P^2[ x - 2\,(2\!-\!x)(1\!-\!z_2)(1\!-\!2z_2) ],
\;&\;
e_{\bbb\!,6}= P^2[ 1 - (1\!-\!x)( 4 z_2^2 - 6 z_2 + 3 ) ].
\ea
\eqa
\subsection{Vertices with two photons coupled to light fermions\label{sec:dc}}
In this section we extract the coefficients of the collinear logarithms
for the MIs illustrated in \fig{dc}, where both photons are attached
to light-fermion lines.
In general the two light masses $m$ and $m'$ are different 
and both vanishing small: therefore, we need to extract both collinear 
logarithms, $L=\ln(m^2/M^2)$ and $L'=\ln(m'^2/M^2)$. 
\subsubsection{The Master Integral $V^{\bba}_{\rm dc}$}
We apply to $V^{\bba}_{\rm dc}$ the parametrization introduced 
for $V^{\bbb}$ in section~10.4 of Ref.~\cite{Ferroglia:2003yj}.
We first combine the $q_1$ and $q_2$ propagators with $z_1$ and $z_2$ 
Feynman parameters; the resulting expression contains the 
product of three propagators, the first in $q_1$, the second in $q_1-q_2$ and 
the last in $q_2$,
\bqa
V^{\bba}_{\rm dc}
&=&
\frac{1}{P^2}\,
\frac{\mu^{2\ep}}{\pi^4}\int\!d^nq_1d^nq_2\int_0^1\!\!dz_1\,dz_2\,
\frac{P^2\,M^2 - 4\,(M^2+P^2+\spro{q_1}{p_2})\,\spro{q_2}{p_1}}
     {[12]^2\,[3]\,[45]^2},
\nl
{[12]} &= & q_1^2 + 2\,z_1\,\spro{q_1\!}{\!p_1} + m'^2,
\quad
{[3]} = (q_1-q_2)^2 + M^2,
\quad
{[45]}= q_2^2 + 2\,\spro{q_2\!}{\!(p_1\!+\!z_2\,p_2)} + z_2\,P^2 + m^2.
\eqa
The terms $[12]$ and $[3]$ are then combined with a variable $x$ and the $q_1$ 
integration is performed generating a new $q_2$ propagator; 
the resulting two $q_2$-dependent propagators are then combined with a 
new variable $y$, so that the integration in $q_2$ can be carried on. 
After extracting the UV simple pole in $\ep$, the result simply reads
\bqa
V^{\bba}_{\rm dc}
&=&
  \frac{1}{2}\,F^2_{-1}(M^2)
- \frac{7}{8}
- \int dC_4\lpar x,y,\{z\}\rpar\,(1\!-\!y)\,
  \left[\,
  \ln\frac{U_{\bba}}{M^2} + (1\!-\!x)\;\frac{c_{\bba}}{U_{\bba}} 
  \,\right],
\nl&&
c_{\bba}= 
M^2\,\big[1+2\,z_2\,(1\!-\!y)\big] 
+ P^2\,z_2\,(1\!-\!y)\,\big\{1+z_1\,[1-x\,(1\!-\!y)]\big\},
\nl&&
U_{\bba}= 
  z_1\,z_2\,x\,y\,(1\!-\!x)\,(1\!-\!y)\,P^2 + x\,y\,M^2
+ x\,(1\!-\!x)\,(1\!-\!y)\,m^2 + y\,(1\!-\!x)\,m'^2\,.
\eqa
Only the term containing $U_\ssG^{-1}$ is singular for a vanishing fermion 
mass; indeed, for $m\to 0$ ($m'\to 0$), a term $y^{-1}$($x^{-1}$) can be 
factorized out of $U_\ssG^{-1}$, leading to a divergent integral.
Therefore, if $U_\ssG^{-1}$ is multiplied by the product $x y$, we can simply 
set the fermion masses to zero, while if $U_\ssG^{-1}$ is multiplied by  
$x$ or $y$ we have to perform a subtraction,
\bq
\frac{x}{U_\ssG}= 
      x\,\bigg( \frac{1}{U_{\ssG}} - \frac{1}{U_{\ssG,y}} \bigg) 
\,+\, \frac{x}{U_{\ssG,y}},
\qquad\quad
\frac{y}{U_\ssG}= 
      y\,\bigg( \frac{1}{U_{\ssG}} - \frac{1}{U_{\ssG,x}} \bigg) 
\,+\, \frac{y}{U_{\ssG,x}},
\label{ssubVG}
\eq
where the subscripts $x$ and $y$ indicate that inside $U_\ssG$ we have set
$x^2=0$ and $y^2=0$.
The first bracket in both equations is collinear free while the last term 
of the first (second) formula generates the collinear logarithm $L$ ($L'$), 
after an explicit integration in $y$ ($x$). 
For terms where the coefficient of $U_\ssG^{-1}$ contains neither $x$ 
nor $y$ a double subtraction is needed,
\bq
\frac{1}{U_\ssG}= 
  \bigg(   \frac{1}{U_{\ssG}} - \frac{1}{U_{\ssG,x}}
         - \frac{1}{U_{\ssG,y}} + \frac{1}{U_{\ssG,xy}} \bigg)
\,+\, \bigg( \frac{1}{U_{\ssG,x}} - \frac{1}{U_{\ssG,xy}} \bigg) 
\,+\, \bigg( \frac{1}{U_{\ssG,y}} - \frac{1}{U_{\ssG,xy}} \bigg) 
\,+\, \frac{1}{U_{\ssG,xy}}, \label{dsubVG}
\eq
with $x^2=y^2=0$ in the polynomial with the subscript $xy$. 
The first bracket of \eqn{dsubVG} contains no singularity; in the second 
(third) bracket, we integrate in $x$ ($y$) generating the collinear logarithm 
$L'$ ($L$). Finally, in the last term, both integrations in $x$ and $y$ are 
performed  and the product $L\,L'$ appears.
Having extracted the collinear logarithms, we can now set the fermion 
masses to zero, obtaining
\bq
V^{\bba}_{\rm dc}=
  \frac{1}{2}\,F^2_{-1}(M^2)
+ \frac{3}{8}
+ \int dC_4\lpar x,y,\{z\}\rpar\,
  \big(\, 
        {\cal I}_\ssG\,L\,L'
  \,+\, {\cal I}_\ssG^y\,L
  \,+\, {\cal I}_\ssG^x\,L'
  \,+\, {\cal I}_\ssG^{xy}\,
  \,\big),
\eq
where we have introduced
\bqa
{\cal I}_{\ssG}
&=&
\,-\, \alpha_{\ssG,1}\,\xi_{\ssG,0}^{-1},
\qquad\quad
{\cal I}_{\ssG}^u =
\alpha_{\ssG,1}\,\bigg( 
  \frac{\xi_{\ssG,u}^{-1}}{u} \bmid_+
+ \xi_{\ssG,0}^{-1}\,\ln\frac{\xi_{\ssG,0}}{M^2}
\bigg)
- \alpha_{\ssG,u}\,\xi_{\ssG,u}^{-1},
\qquad
u= x,y,
\nl
{\cal I}_{\ssG}^{xy}
&=&
\,-\,\alpha_{\ssG,1}\,\bigg\{\;
       \frac{\xi_{\ssG}^{-1}}{x\,y} \bmid_{++}
     + \sum_{u=x,y}\!
       \frac{\xi_{\ssG,u}^{-1}[\ln(\xi_{\ssG,u}/M^2)-1]}{u}\bmid_+
     + \xi_{\ssG,0}^{-1}\,
       \Big[ \ln^2\frac{\xi_{\ssG,0}}{M^2} + \zeta(2) \Big]
     - \xi_{\ssG,x}^{-1}\,\frac{\ln(1-x)}{x}
\;\bigg\}
\nl
&&
\,+\,\sum_{u=x,y}\!\alpha_{\ssG,u}\,\xi_{\ssG,u}^{-1}
     \Big( 2\,\ln\frac{\xi_{\ssG,u}}{M^2} - 1 \Big)
\,-\,\alpha_{\ssG,x}\,\xi_{\ssG,x}^{-1}\,\ln(1-x)
\,-\, \alpha_{\ssG}\,\xi_{\ssG}^{-1}
\,-\,(1-y)\,\ln\frac{\xi_{\ssG}}{M^2}.
\label{vgdc1}
\eqa
Here the $'+'$ and $'++'$ distributions have been defined in \eqn{plusdist} 
and \eqn{plusplusdist}; 
the $\alpha_\ssG$ coefficients and the $\xi_\ssG$ functions are given by
$$
\alpha_{\ssG} \!=
  P^2 z_2
  \big\{ (2\!-\!y)[1+z_1(2\!-\!x)] + z_1(1\!-\!x)(1\!-\!y)^2 \big\}
+ M^2\big[ 1 + 2z_2(2\!-\!y) \big], 
\quad 
\xi_{\ssG} \!=
P^2 z_1 z_2(1\!-\!x)(1\!-\!y) + M^2\!\!,
$$
\vspace{-1cm}
\bqa
\alpha_{\ssG,y}&=& \alpha_{\ssG}(x\!=\!1),
\qquad\qquad
\alpha_{\ssG,x}= \alpha_\ssG(y\!=\!1),
\qquad\qquad
\alpha_{\ssG,1}= \alpha_\ssG(x\!=\!1\!=\!y),
\nl
\xi_{\ssG,y}&=& \xi_\ssG(x\!=\!0),
\qquad\qquad\,
\xi_{\ssG,x}= \xi_\ssG(y\!=\!0),
\qquad\qquad\;\,
\xi_{\ssG,0}= \xi_\ssG(x\!=\!0\!=\!y).
\eqa
Since the $\xi_\ssG$ functions are linear in the four integration variables,
all integrations can be analytically performed.
Introducing $\omega= -P^2/M^2$ and $l_{\omega} = \ln(1-\omega)$, the result is
\bqa
V^{\bba}_{\rm dc}
&=&
  \frac{1}{2}\,F^2_{-1}(M^2)
- \frac{43}{8}
- \frac{1}{\omega^2}\,\bigg\{
    L\,L'\,
    \big[ 2\,\omega + (1\!-\!\omega)\,(2\!-\!\omega)\,l_{\omega} \big]
  + L\,\big[
      2\,\omega\,(1\!+\!\omega)
    + 2\,(1\!-\!\omega)\,l_{\omega}
    - \,(1\!-\!\omega)\,(2\!-\!\omega)\,l^2_{\omega} 
\nl
&&
    + \omega\,(1\!-\!\omega)\,\li{2}{\omega}
    \big]
  + L'\,\big[
      6\,\omega 
    + (1\!-\!\omega)\,(4\!-\!\omega)\,l_{\omega} 
    - \,(1\!-\!\omega)\,(2\!-\!\omega)\,l^2_{\omega} 
    + (\omega^2\!-\!2\,\omega\!+\!2)\,\li{2}{\omega}
    \big]
\nl
&&
  + \bigg[ \frac{19}{2} + 4\,\zeta(2) \bigg]\,\omega
  + \bigg[ \frac{3}{2}\,(5\!+\!\omega) + 2\,(2\!-\!\omega)\,\zeta(2)\, \bigg]
    (1\!-\!\omega)\,l_{\omega}
  - \,(1\!-\!\omega)\,(6\!-\!\omega)\,l^2_{\omega} 
  + \frac{4}{3}\,(1\!-\!\omega)\,(2\!-\!\omega)\,l^3_{\omega} 
\nl
&&
  - \,(\omega^2\!-\!4\,\omega\!+\!2)\,\li{2}{\omega}
  + 2\,(1\!-\!\omega)\,(2\!-\!\omega)\,l_{\omega} \,\li{2}{\omega}
  - 2\,\omega\,\li{3}{\omega}
  - 2\,(3\,\omega\!-\!2)\,S_{1,2}\,(\omega)
  \bigg\},
\eqa
where $\li{n}{\omega}$ and $S_{1,2}(\omega)$ are Nielsen poly-logarithms.
\subsubsection{The Master Integral $V^{\bbb}_{\rm dc}$ \label{sec:vhdc}}
We parametrize $V^{\bbb}_{\rm dc}$ following section~10.4 of
Ref.~\cite{Ferroglia:2003yj}.
After combining the $q_1$, $q_1-q_2$ and $q_2$ propagators with Feynman parameters $z_1$, $z_2$ 
and $z_3$, we obtain
\bqa
V^{\bbb}_{\rm dc}
&=&
\frac{\mu^{2\ep}}{\pi^4}\int\!d^nq_1d^nq_2\int dC_3\lpar\{z\}\rpar\,
\frac{P^2\,M^2 \,+\, 2\,P^2\,\spro{q_1}{p_1} - 4\,(\spro{q_1}{p_1})^2}
     {[12]^2\,[34]^2\,[56]^2},\qquad
{[12]}= q_1^2 - 2\,z_1\,\spro{q_1}{p_2} + m^2,
\nl
{[34]} &=& (q_1-q_2)^2 + 2\,\spro{(q_1-q_2)}{(p_1-z_2\,P)} + M^2,
\quad
{[56]}= q_2^2 - 2\,(1\!-\!z_3)\,\spro{q_2}{p_1} + m'^2.
\eqa
Afterwards, $[12]$ and $[34]$ are combined with an additional variable 
$x$; we integrate in $q_1$ and obtain a new $q_2$ propagator which is 
combined with $[56]$ using a variable $y$. Finally, the $q_2$ integration is 
performed. The MI $V^{\bbb}_{\rm dc}$ is UV finite; after setting
$\epsilon=0$, we obtain
\bqa
V^{\bbb}_{\rm dc}
&=&
- \int dC_5\lpar x,y,\{z\}\rpar\,\,
  x\,y\,(1\!-\!x)\,(1\!-\!y)\;\frac{c_{\bbb}}{U_{\bbb}^{2}},
\nl
c_{\bbb}&=& 
  P^2\,M^2 
+ P^4\,[z_1-x\,(1\!-\!y)\,(z_1\!-\!z_2)]\,[1-z_1+x\,(1\!-\!y)\,(z_1\!-\!z_2)],
\nl
U_{\bbb}&=& 
  x\,y\,(1\!-\!x)\,(1\!-\!y)\,(z_1\!-\!z_2)(1\!-\!z_2\!-\!z_3)\,P^2 
+ x\,y\,\chi
+ y\,(1\!-\!x)\,m^2 + x\,(1\!-\!x)\,(1\!-\!y)\,m'^2\,,
\eqa
where we have introduced $\chi= z_2\,(1\!-\!z_2)\,P^2+M^2$.

The collinear-singular behavior corresponding
to $m\to 0$ ($m'\to 0$) shows up at $x=0$ ($y=0$).
In fact, in this limit, the global $xy$ factor combines with the 
$1/x^2$ ($1/y^2$) term coming from $U_\ssH^{-2}$ and produces a global 
$x^{-1}$ ($y^{-1}$) divergent integral.
Subtraction terms are evidently not needed when the coefficient of 
$U_\ssH^{-2}$ contains the product $x^2 y^2$; for terms where 
$U_\ssH^{-2}$ is multiplied by either $x^2 y$, $x y^2$ or $x y$, we apply the 
following subtraction formulae:
\bqa
x^2\,y\,(1\!-\!x)\,(1\!-\!y)\,U_\ssH^{-2}
&=&
      x^2\,y\,(1\!-\!x)\,[ (1\!-\!y)\,U_{\ssH}^{-2} - U_{\ssH\!,y}^{-2} ] 
\,+\, x^2\,y\,(1\!-\!x)\,U_{\ssH\!,y}^{-2},
\label{ssubVHy}
\\
x\,y^2\,(1\!-\!x)\,(1\!-\!y)\,U_\ssH^{-2}
&=&
      x\,y^2\,(1\!-\!y)\,[ (1\!-\!x)\,U_{\ssH}^{-2} - U_{\ssH\!,x}^{-2} ] 
\,+\, x\,y^2\,(1\!-\!y)\,U_{\ssH\!,x}^{-2},
\label{ssubVHx}
\\
x\,y\,(1\!-\!x)(1\!-\!y)\,U_\ssH^{-2}
&=&
  x\,y\,[   (1\!-\!x)(1\!-\!y)\,U_{\ssH}^{-2} - (1\!-\!y)\,U_{\ssH\!,x}^{-2}
          - (1\!-\!x)\,U_{\ssH\!,y}^{-2} + U_{\ssH\!,xy}^{-2} ]
\nl
&+&
  x\,y\,[ (1\!-\!y)\,U_{\ssH\!,x}^{-2} - U_{\ssH\!,xy}^{-2} ] 
+ x\,y\,[ (1\!-\!x)\,U_{\ssH\!,y}^{-2} - U_{\ssH\!,xy}^{-2} ] 
+ x\,y\,U_{\ssH\!,xy}^{-2}.
\label{dsubVH}
\eqa
The subscripts $x$, $y$ and $xy$ for $U_\ssH$ indicate that we
have set $x^2=0$, $y^2=0$ and $x^2=y^2=0$ respectively. 

The first squared brackets in all three Eqs.(\ref{ssubVHy}-\ref{dsubVH}) 
are collinear free.
In the second (third) squared bracket of \eqn{dsubVH} and in the last term of 
\eqn{ssubVHx} (\eqn{ssubVHy}), the integration in $x$ ($y$) generates 
the collinear logarithms $L$ ($L'$).
In the last term of \eqn{dsubVH} we perform both integrations in $x$ and $y$ 
obtaining the product $L\,L'$.
At this point, the collinear-divergent behavior is explicit and we can set 
$m=m'=0$ inside all polynomials, getting
\bq
V^{\ssH}_{\rm dc}=
\int dC_5\lpar x,y,\{z\}\rpar\,
\big(\, 
      {\cal I}_\ssH\,L\,L'
\,+\, {\cal I}_\ssH^y\,L
\,+\, {\cal I}_\ssH^x\,L'
\,\big)
+ V^{\ssH}_{\rm dc,fin},
\label{calII}
\eq
\bq
{\cal I}_{\ssH}=
\,-\, \beta_{\ssH}\,\xi_{\ssH\!,0}^{-2},
\qquad\quad
{\cal I}_{\ssH}^u =
\beta_{\ssH}\bigg[ 
  \frac{(1\!-\!u)\,\xi_{\ssH\!,u}^{-2}}{u} \bmid_+\!
+ \xi_{\ssH\!,0}^{-2}\,\Big(\! \ln\frac{\xi_{\ssH\!,0}}{M^2} - 1 \!\Big)
\bigg]
+ \alpha_{\ssH\!,u}\,\xi_{\ssH\!,u}^{-2},
\qquad\;
u= x,y.
\label{calIII}
\eq
Here $V^{\ssH}_{\rm dc,fin}$ reads
\bqa
V^{\ssH}_{\rm dc,fin}\!\!
&=&
\!\!\int\!\! dC_5(x,y,\!\{\!z\!\})\Bigg\{\!
  \alpha_{\ssH}(1\!-\!y)\xi_{\ssH}^{-2}
- \alpha_{\ssH\!,x}\!\Bigg[
    \frac{(1\!\!-\!\!y)\xi_{\ssH}^{-2}}{y}\bmid_+\!
  + \xi_{\ssH\!,x}^{-2}
    \bigg(\!\! \ln\!\frac{\xi_{\ssH\!,x}}{(1\!\!-\!\!x)M^2} \!-\! 1 \!\bigg)
  \!\Bigg]
- \beta_{\ssH}\!\Bigg[
    \frac{(1\!\!-\!\!x)(1\!\!-\!\!y)\xi_{\ssH}^{-2}}{xy} \bmid_{++}\;
\nl
&&\!\!\!\!\!\!\!
  + \!\sum_{u=x,y}\!\!
    \frac{(1\!-\!u)\xi_{\ssH\!,u}^{-2}(\,\ln(\xi_{\ssH\!,u}/M^2)\!-\!1 )}{u}
    \bmid_+\!
  + \xi_{\ssH\!,0}^{-2}\bigg(\!\!
      \ln^{\!2}\!\frac{\xi_{\ssH\!,0}}{M^2}
    - 2\,\ln\!\frac{\xi_{\ssH\!,0}}{M^2}
    + \zeta(2) 
    \!\bigg)
  - (1\!-\!x)\xi_{\ssH\!,x}^{-2}\frac{\ln(1\!-\!x)}{x}
\Bigg]
\!\Bigg\}\!,
\label{vhdcfin1}
\eqa
where the $\alpha_\ssH$ and $\beta_\ssH$ coefficients and the $\xi_\ssH$ 
functions are given by
\bqa
\alpha_{\ssH}&=&
  P^4\,(1\!-\!x)\,(1\!-\!z_1\!-\!z_2)\,
  \big\{ (1\!-\!z_1\!-\!z_2)\,[1-x\,(2\!-\!y)] - z_1 + z_2 \big\},
\qquad
\beta_{\ssH}= P^2\,\big[ P^2\,z_1\,(1\!-\!z_1) + M^2 \big],
\nl
\xi_{\ssH}&=&
  P^2\,\big[ (1\!-\!z_1)\,(1\!-\!z_2)\,(1\!-\!z_3) + z_1\,z_2\,z_3 \big]\,
       (1\!-\!x)\,(1\!-\!y) 
+ P^2\,z_2\,(1\!-\!z_2)\,(x+y-xy)+ M^2,
\nl
\alpha_{\ssH\!,y} &=&  \alpha_{\ssH}(x\!=\!1)= 0,\quad
\alpha_{\ssH\!,x}\!=\! \alpha_\ssH(y\!=\!1),
\quad
\xi_{\ssH\!,y}\!=\! \xi_\ssH(x\!=\!0),
\quad
\xi_{\ssH\!,x}\!=\! \xi_\ssH(y\!=\!0),
\quad
\xi_{\ssH\!,0}\!=\! \xi_\ssH(x\!=\!0\!=\!y).
\eqa
We consider now the coefficients ${\cal I}_{\ssH}$, ${\cal I}_{\ssH}^x$
and ${\cal I}_{\ssH}^y$, written in \eqn{calIII}, of the collinear logarithms of
\eqn{calII}, and express them in terms of one-loop functions.
After changing $y\to x$ in ${\cal I}_{\ssH}^y$, we observe that the 
coefficients of $L$ and $L'$ are integrals in $x$, $z_1$, $z_2$ and $z_3$.
Then, we use the following trick to get rid of $\ln(\xi_{\ssH\!,0}/M^2)$,
\bq
\xi_{\ssH\!,0}^{-2}\,\Bigg(\! \ln\frac{\xi_{\ssH\!,0}}{M^2} - 1 \!\Bigg)=
\xi_{\ssH\!,0}^{-2}\,\ln\frac{P^2\,z_2\,(1\!-\!z_2)+M^2}{M^2}
\,+\, \intsx{x}\frac{(1\!-\!x)\,\xi_{\ssH\!,x}^{-2}}{x} \bmid_+ .
\label{specialtrick}
\eq
Since $\beta_\ssH$ and $\alpha_{\ssH\!,x}$ do not depend on $z_3$ and 
$\xi_{\ssH\!,x}$ is linear in $z_3$, the $z_3$ integration can be easily 
performed, leading to
\bqa
V^{\ssH}_{\rm dc}&=&
\int_0^1\!\!dx\,dz_1\,dz_2\,
\big(\, 
      {\cal J}_\ssH^{f\!f'}\,L\,L'
\,+\, {\cal J}_\ssH^{f}\,L
\,+\, {\cal J}_\ssH^{f'}\,L'
\,\big)
\,+\,V^{\ssH}_{\rm dc,fin},
\qquad
{\cal J}_{\ssH}^{f\!f'}=
- \beta_{\ssH} \eta_{\ssH\!,0}^{-1} \rho_{\ssH\!,0}^{-1},
\nl
{\cal J}_{\ssH}^{f}
\!&=&
\beta_{\ssH}\bigg[ 
  2\frac{(1\!-\!x)\eta_{\ssH\!,x}^{-1}\rho_{\ssH\!,x}^{-1}}{x} \bmid_+\!\!\!
+ \eta_{\ssH\!,0}^{-1} \rho_{\ssH\!,0}^{-1}
  \ln\frac{P^2\,z_2\,(1\!-\!z_2)+M^2}{M^2}
\bigg],\quad
{\cal J}_{\ssH}^{f'}
=
{\cal J}_{\ssH}^{f} + 
\alpha_{\ssH\!,x} \eta_{\ssH\!,x}^{-1} \rho_{\ssH\!,x}^{-1},
\eqa
where we have defined
\bqa
\ba{ll}
\eta_{\ssH\!,x}= 
P^2 z_1\,z_2\,(1\!-\!x) + P^2\,z_2\,(1\!-\!z_2)\,x+ M^2,
\quad & \quad
\eta_{\ssH\!,0}= P^2 z_1\,z_2 + M^2,
\\
\rho_{\ssH\!,x}=
P^2(1\!-\!z_1)\,(1\!-\!z_2)\,(1\!-\!x) + P^2\,z_2\,(1\!-\!z_2)\,x+ M^2,
\quad & \quad
\rho_{\ssH\!,0}= P^2(1\!-\!z_1)\,(1\!-\!z_2) + M^2.
\ea
\eqa
At this point, we perform some partial fractioning,
\bqa
\beta_{\ssH}\,(1\!-\!x)\,\eta_{\ssH\!,x}^{-1}\,\rho_{\ssH\!,x}^{-1}
&=&
P^2\,\big\{
(z_1-x\,z_2)\,
\eta_{\ssH\!,x}^{-1} + [1-z_1-x\,(1\!-\!z_2)]\,\rho_{\ssH\!,x}^{-1} 
\big\},
\nl
\beta_{\ssH}\eta_{\ssH\!,0}^{-1}\,\rho_{\ssH\!,0}^{-1}
&=&
P^2\,\big[
z_1\,\eta_{\ssH\!,0}^{-1} + (1\!-\!z_1)\,\rho_{\ssH\!,0}^{-1}
\big],
\nl
\alpha_{\ssH\!,x}\,\eta_{\ssH\!,x}^{-1}\,\rho_{\ssH\!,x}^{-1}
&=&
P^2\,\big[ (1\!-\!z_1\!-\!z_2)\,(1-x) - z_1 + z_2 \big]\,
(\eta_{\ssH\!,x}^{-1} - \rho_{\ssH\!,x}^{-1}).
\eqa
Next, we change $z_1\to 1-z_1$, $z_2\to 1-z_2$ in all terms containing 
$\rho_{\ssH\!,x}$ or $\rho_{\ssH\!,0}$, which will then become $\eta_{\ssH\!,x}$ or 
$\eta_{\ssH\!,0}$, and obtain
\bqa
{\cal J}_{\ssH}^{f}
&=&
  4\,P^2\,z_1\,\frac{\eta_{\ssH\!,x}^{-1}}{x} \bmid_+\!
- 4\,P^2\,z_2\,\eta_{\ssH\!,x}^{-1}
+ 2\,P^2\,z_1\,\eta_{\ssH\!,0}^{-1}\,
  \ln\frac{P^2\,z_2\,(1\!-\!z_2)+M^2}{M^2},
\nl
{\cal J}_{\ssH}^{f'}
&=&
{\cal J}_{\ssH}^{f} 
+ 2\,P^2\,\big[ (1\!-\!z_1\!-\!z_2)\,(1-x) - 
z_1 + z_2 \big]\,\eta_{\ssH\!,x}^{-1},
\qquad\quad
{\cal J}_{\ssH}^{f\!f'}= - 2\,P^2\,z_1\,\eta_{\ssH\!,0}^{-1}.
\label{JH}
\eqa
Finally, we can get rid of the logarithm in \eqn{JH} by using a trick 
similar to the one introduced in \eqn{specialtrick},
\bq
\eta_{\ssH\!,0}^{-1}\,\ln\frac{P^2\,z_2\,(1\!-\!z_2)+M^2}{M^2}=
- \intsx{x}\,\frac{\eta_{\ssH\!,x}^{-1}}{x}\bmid_+ 
+ \,\eta_{\ssH\!,0}^{-1}\,\ln\frac{\eta_{\ssH\!,0}}{M^2}.
\eq
For $z_2\,\eta_{\ssH\!,x}^{-1}$ in the first line of \eqn{JH} we use instead
\bq
2\,P^2\,z_2\,\eta_{\ssH\!,x}^{-1}=
  P^2\,(1\!-\!z_1)\,\eta_{\ssH\!,x}^{-1}
+ P^2\,z_1\,\frac{\eta_{\ssH\!,x}^{-1}}{x}\bmid_+
- \frac{1}{x}\,\partial_{z_2}\,\ln\frac{\eta_{\ssH\!,x}}{\eta_{\ssH\!,0}}.
\eq
In this way the terms containing the $'+'$ distribution cancel in \eqn{JH}.
In all terms which contain $\eta_{\ssH\!,x}^{-1}$, we first change 
$z_2\to z_2/x$ and then $z_1\to z$, $z_2\to x_2$ and $x\to 1-x_1+x_2$.
In the rest we are able to integrate in all variables $x$, $z_1$ and $z_2$, 
obtaining
\bqa
\!\!\!\!\!\!\!\!\!\!\!\!
V^{\ssH}_{\rm dc}
&=&
     2\,\bigg( 1 - \frac{1\!+\!\omega}{\omega}\,l_{\omega} \bigg)\,
     L\,L'
\,+\,2\,\bigg[
       1 
     + \frac{1\!+\!\omega}{\omega}\,l_{\omega}\,
       \big( l_{\omega} - 1 \big)
     + \li{2}{\omega}
     \bigg]\,
     ( L + L' )
\nl
&&
- \,2\,P^2\!\!\int_0^1\!\!\!\!dz\,dx_1\!\!\int_0^{x_1}\!\!\!\!\!\!\!dx_2\,
  \big\{ 
    (1\!-\!z)\,L
   + \big[ 1 \!-\! x_2\,z \!-\! x_1\,(1\!-\!z) \big]\,L' 
  \big\}\,
  \kappa_\ssH^{-1}\,
+ \,V^{\ssH}_{\rm dc,fin},
\eqa
where we have introduced $\omega= -P^2/M^2$, $l_{\omega} = \ln(1-\omega)$ 
and 
$\kappa_\ssH= P^2\,x_2\,[1\!-\!x_2\,z\!-\!x_1\,(1\!-\!z)] 
+ M^2\,(1\!-\!x_1\!+\!x_2)$;
the latter can be easily recognized as the polynomial of a one-loop triangle 
with $z$-dependent momenta.
Therefore, we can express all coefficients of the collinear logarithms in 
terms of integrals of one-loop functions as for the single-collinear case 
treated in section~\ref{sec:sc},
\bqa
\!\!\!\!\!\!\!\!\!\!\!\!
V^{\ssH}_{\rm dc}
&=&
     2\,\bigg( 1 - \frac{1\!+\!\omega}{\omega}\,l_{\omega} \bigg)\,
     L\,L'
\,+\,2\,\bigg[ 
       1 
     + \frac{1\!+\!\omega}{\omega}\,l_{\omega}\,
       \big( l_{\omega} - 1 \big)
     + \li{2}{\omega}
     \bigg]\,
     ( L + L' )
\nl
&&
- \,2\intsx{z}\,
  \big[ (1\!-\!z)P^2\,L + ( P^2 + 2\,\spro{q}{p_2} )\,L' \big]
\raisebox{0.1cm}{\scalebox{0.6}{
\begin{picture}(140,75)(0,0)
 \SetWidth{1.2}
 \Line(0,0)(40,0)                    \Text(10,5)[cb]{$-P$}
 \Photon(128,-53)(100,-35){2}{5}     \Text(138,-67)[cb]{$(1\!-\!z)p_1$}
 \Photon(128,53)(100,35){2}{5}       \Text(138,57)[cb]{$p_2+zp_1$}
 \Line(100,-35)(40,0)                \Text(70,-34)[cb]{$M$}
 \Photon(100,-35)(100,35){2}{9}      \Text(110,-3)[cb]{$0$}
 \Line(100,35)(40,0)                 \Text(70,26)[cb]{$M$}
\end{picture}
}}
\qquad+\quad V^{\ssH}_{\rm dc,fin}.
\label{vhdc}
\eqa
\vspace{1cm}

\noindent
In \eqn{vhdc} the triangle of loop momentum $q$ is to be intended in the 
graphical representation given in appendix~\ref{app:diagrams}.

For the collinear-finite part $V^\ssH_{\rm dc,fin}$ we do not need an 
analytical expression, 
since it will be enough to put it in a form suited for numerical evaluation.
The situation is similar to the one for $V^\ssH_{\rm sc,fin}$, computed in 
section~\ref{sec:collfinH};
all terms of \eqn{vhdcfin1} have the following structure, with 
respect to $z_1$ and $z_3$:
\bq
f(n,m)= z_1^n\,\frac{\ln^m\xi}{\xi^2},
\qquad\quad
\xi= a\,z_1\,z_3+b\,z_1+c\,z_3+d,
\qquad\quad
n,m= 0,1,2.
\eq
Therefore, we can integrate by parts in $z_1$ after inserting
\bq
\frac{\ln^2\xi}{\xi^2}= 
-\frac{2}{a z_3\!+\!b} \partial_{z_1} \bigg[\frac{1}{\xi}
\bigg( 1\! +\! \ln\xi \!+\!\frac{1}{2}\ln^2\xi\bigg)\bigg],
\quad
\frac{\ln\xi}{\xi^2}= 
- \frac{1}{a z_3\!+\!b}\,\partial_{z_1} \bigg[\frac{1}{\xi}
\bigg( 1 \!+\! \ln\xi \bigg)\bigg],
\quad
\frac{1}{\xi^2}= - \frac{1}{a z_3\!+\!b}\partial_{z_1}\frac{1}{\xi},
\eq
obtaining the result
\bqa
V^{\ssH}_{\rm dc,fin}\! 
&=\!&
\int dC_5\lpar x,y,\{z\}\rpar\,\Bigg\{\!
  (1\!-\!x)\bigg[\frac{1\!-\!y}{\xi_{\ssA}}
  \bigg( \frac{\alpha_1}{\xi_{\ssB}}\! -\! \frac{\alpha_z}{\xi} \bigg)\!
+\! \bigg( \frac{1\!-\!y}{y \xi_{\ssA}}
  \bigg( 
  \frac{z_1\,\alpha_{x z}}{\xi} \!-\! \frac{\alpha_{x 1}}{\xi_{\ssB}} 
  \bigg)
  \bigg)\bigg|_+ \!\!\!\!
-\!
  \frac{\ln(1\!-\!x)}{\xi_{x \ssA}}
  \bigg( 
    \frac{z_1 \alpha_{x z}}{\xi_{x}}\! 
  -\! \frac{\alpha_{x 1}}{\xi_{x \ssB}} 
  \bigg)\bigg]
\nl
&+& \frac{1}{\xi_{2}}\,\bigg(
    \frac{\alpha_{x 1}}{\xi_{x \ssB}}\ln\frac{\xi_{x \ssB}}{M^2}
  - \frac{\alpha_{x 0}}{\xi_{x \ssA}}\ln\frac{\xi_{x \ssA}}{M^2}
  - \frac{\alpha_{x z}}{\xi_{x}}\ln\frac{\xi_{x}}{M^2}
  \bigg)
+
  \bigg[ 
  \frac{(1\!-\!x)(1\!-\!y)}{x\,y\,\xi_{\ssA}}\,
  \bigg( \frac{z_1\,\beta_z}{\xi} - \frac{\beta_0}{\xi_{\ssB}} \bigg)
  \bigg]\bigg|_{++}
\nl
&+& \frac{1}{\xi_{2}}\,\sum_{u=x,y}\!
  \bigg[
  \frac{1}{u}\,\bigg(
    \frac{\beta_0}{\xi_{u \ssB}}\ln\frac{\xi_{u \ssB}}{M^2}
  - \frac{\beta_0}{\xi_{u \ssA}}\ln\frac{\xi_{u \ssA}}{M^2}
  - \frac{\beta_z}{\xi_{u}}\ln\frac{\xi_{u}}{M^2}
  \bigg)
  \bigg]\bigg|_+
+\!
  \frac{1}{\xi_{2}}\!\bigg(
    \frac{\beta_0}{\xi_{0 \ssB}}\ln^2\!\frac{\xi_{0 \ssB}}{M^2}
  - \frac{\beta_0}{\xi_{0 \ssA}}\ln^2\!\frac{\xi_{0 \ssA}}{M^2}
\nl
  &-& \frac{\beta_z}{\xi_{0}}\ln^2\!\frac{\xi_{0}}{M^2}
  \!\bigg)\!
+ \frac{\zeta(2)}{\xi_{0 \ssA}}
  \bigg( \frac{z_1\beta_z}{\xi_{0}} - \frac{\beta_0}{\xi_{0 \ssB}} \!\bigg)
- \frac{1\!-\!x}{x \xi_{x \ssA}}\ln(1\!-\!x)
  \bigg( \frac{z_1\beta_z}{\xi_{x}} - \frac{\beta_0}{\xi_{x \ssB}} \!\bigg)
\!\Bigg\},
\eqa
where we have used the following short-hand notations for the $\xi$ functions,
\bqa
\ba{llll}
\xi= \xi_\ssH,
\;&\;
\xi_x= \xi_{\ssH\!,x}\,,
\;&\;
\xi_y= \xi_{\ssH\!,y}\,,
\;&\;
\xi_0= \xi_{\ssH\!,0}\,,
\qquad\quad
\xi_{2}= P^2\,(z_2 + z_3 - 1),
\\
\xi_{\ssA}= \xi(z_1=0),
\;&\;
\xi_{x \ssA}= \xi_{x}(z_1=0),
\;&\;
\xi_{y \ssA}= \xi_{y}(z_1=0),
\;&\;
\xi_{0 \ssA}= \xi_{0}(z_1=0),
\\
\xi_{\ssB}= \xi(z_1=1),
\;&\;
\xi_{x \ssB}= \xi_{x}(z_1=1),
\;&\;
\xi_{y \ssB}= \xi_{y}(z_1=1),
\;&\;
\xi_{0 \ssB}= \xi_{0}(z_1=1),
\ea
\eqa
and for the $\alpha$ and $\beta$ coefficients,
\bqa
\ba{ll}
\alpha_z=
P^4\,z_1\,
\big\{ 1 - 2\,(1\!-\!z_1\!-\!z_2)\,[1-x\,(2\!-\!y)] - 2\,(1\!-\!z_1) \big\},
\;&\;
\alpha_1=
  P^4\,z_2\,\big[ 1 - x\,z_2\,(2\!-\!y) \big],
\\
\alpha_{x z}= 
P^4\,\big[ 1 - 2\,(1\!-\!x)\,(1\!-\!z_1\!-\!z_2) - 2\,(1\!-\!z_1) \big],
\;&\;
\alpha_{x 1}= P^4\,z_2\,(1 - x\,z_2),
\\
\alpha_{x 0}= P^4\,(1\!-\!z_2)\,[1 - x\,(1\!-\!z_2)],
\;&\;
\beta_z= P^4\,(1-2\,z_1),
\qquad
\beta_0= P^2\,M^2.
\ea
\eqa
The result contains terms having the following structures in $z_3$:
$1/ {\cal Q}{\cal P}$ and $1/ {\cal Q}{\cal P}\ln^n{\cal P}$, with $n=1,2$,
where ${\cal Q}$ and ${\cal P}$ are linear in $z_3$. The integration in $z_3$ can 
be performed using \eqn{primitives} for the first two cases
and the following relation for the last one, with $\Delta=ad-bc$:
\bqa
\frac{\ln^2(c z_3\!+\!d)}{(a z_3\!+\!b)(c z_3\!+\!d)}
&=&
\frac{1}{\Delta}\,\partial_{x_3}
\Bigg\{
  \ln^2\frac{\Delta}{a}
  \Big[ \ln(a z_3\!+\!b) - \ln(c z_3\!+\!d) \Big]
- 2\li{3\!}{\frac{\Delta}{a(c z_3\!+\!d)}}
\nl
&&
- 2\ln(c z_3\!+\!d)\,\li{2\!}{\frac{\Delta}{a(c z_3\!+\!d)}}
- \left[ \ln^2\frac{\Delta}{a} - \ln^2(c z_3\!+\!d) \right]
  \ln\!\left(\! 1 \!-\! \frac{\Delta}{a(c z_3\!+\!d)} \!\right)
\Bigg\}.
\eqa
We also integrate analytically the $'+'$ distributions and use the 
properties of the dilogarithms to cast the result in one of the 
following forms:
\bq
\frac{\cal F}{\chi}
\ln^n\!\left(\! 1 \!-\! \frac{\chi}{\cal P} \!\right),
\;\,
n= 1,2;
\quad
\frac{\cal F}{\chi}\li{2\!}{\frac{\chi}{\cal P}};
\quad
\frac{\cal F}{\chi}\li{3\!}{\frac{\chi}{\cal P}};
\quad
\frac{\cal F}{\chi}S_{1\!,2\!}\left( \frac{\chi}{\cal P} \right);
\quad
\frac{1}{\chi}\ln\frac{\cal P}{M^2}\ln^k\!\!\frac{\chi}{M^2},
\;\,
k= 0,1,2,
\eq
where ${\cal F}$ is a polynomial or a logarithm of polynomials.
The first four forms are ready for numerical  integration, 
since the zeros of $\chi$ are compensated by a regulator function.
For the last case, we observe that the polynomial ${\cal P}$ is always 
of a special form:
making explicit the $i\,0$ Feynman prescription we have
\bq
\frac{1}{\chi}\ln\frac{\cal P}{M^2} =
\frac{1}{\chi}\bigg[
  \ln\!\left(\! 
  1 \!-\! \frac{\chi}{{\cal P}\!+\!\chi- i\,0}
  \!\right)
+ \ln\left( \frac{{\cal P}\!+\!\chi}{M^2} - i\,0 \!\right)
\bigg],
\qquad
{\cal P}\!+\!\chi= \pm \,a\,P^2\,(1\!-\!z_1\!-\!z_2)^n,
\quad
n= 0,1,
\label{eq:trick2}
\eq
where $a$ is the product of non-negative factors $(1\!-\!x)$, $(1\!-\!y)$, 
$z_2$ and $(1\!-\!z_2)$ which can be extracted from the logarithm.
Our procedure has generated a regulator function for the zeros of 
$\chi$ and we obtain a simple remainder which can be integrated 
in one variable (in our case $z_1$).

Let us consider now terms without a regulator function: thanks to the 
special form of $V^{\ssH}_{\rm dc}$, the integration generates a factor 
$\chi$ in the numerator, which cancels the $\chi$ in the denominator.
Note that this cancellation was anyway expected: if $V^{\ssH}_{\rm dc}$ 
is a master integral, then it must have the right threshold properties 
of the original diagram.
From the beginning (see section~\ref{NormTH}), we do not expect threshold 
singularities in $V^{\ssH}$ and a non-regulated $1/\chi$ factor, 
having a $1/\sqrt{1-4M^2/s}$ threshold behavior, is not allowed.
The final expression for $V^{\ssH}_{\rm dc}$ reads then
\bqa
V^{\ssH}_{\rm dc,fin} \!
&=&
\!\int\!\! dC_4(x,y,\{z\})\frac{P^2}{\chi}\sum_{i=1}^{3}\Bigg\{
  \frac{\beta_{a,i}}{2}\ln\!\left(\! 1 \!-\! \frac{\chi}{\xi_{a,i}} \!\right)
+ \frac{\beta_{b,i}}{2}\bigg[
    2\,\li{2\!}{\!\frac{\chi}{\xi_{b,i}}\!}
  - \left(\! 
    2\ln\!\frac{\xi_{b,i}}{M^2} \!-\! \ln\!\frac{(1\!-\!x)\chi}{M^2} 
    \right)
    \ln\!\left(\! 1 \!-\! \frac{\chi}{\xi_{b,i}} \!\right)
  \!\bigg]
\nl
&-&\!
  2\beta_{c,i}\bigg[
    2S_{1,2\!}\left(\! \frac{\chi}{\xi_{c,i}} \!\right)
  - 6\li{3\!}{\!\frac{\chi}{\xi_{c,i}}\!}
  + 2\bigg(\!\!
      \ln\!\Big(\! 1 \!-\! \frac{\chi}{\xi_{c,i}} \!\Big) 
    - 3\ln\!\frac{\xi_{c,i}}{M^2}
    + \ln\!\frac{\chi}{M^2}
    \!\bigg)\li{2\!}{\!\frac{\chi}{\xi_{c,i}}\!}
  - \ln\!\frac{\xi_{c,i}}{\chi}
    \ln^2\!\left(\! 1 \!-\! \frac{\chi}{\xi_{c,i}} \!\right)
\nl
&+&\!
    \bigg( 3\ln\!\frac{\xi_{c,i}}{M^2} \!-\! 2\ln\!\frac{\chi}{M^2} \bigg)
    \ln\!\frac{\xi_{c,i}}{M^2}
    \ln\!\left(\! 1 \!-\! \frac{\chi}{\xi_{c,i}} \!\right)
  \!\bigg]
\Bigg\}
- 2\,i\pi\,{\rm sign}(P^2)\int_0^1\!\!\!dz_2
\bigg[\! \ln^2\!\!\frac{\chi}{M^2} - 2\,\zeta(2) \bigg],
\!\!\!\label{vhdcfin}
\eqa
where the polynomial in the denominator is $\chi= z_2\,(1-z_2)\,P^2 + M^2$. 
The $\xi$ functions and the $\beta$ coefficients are given by
\bqa
\ba{ll}
\xi_{a,1}\!\!=\! 
  z_2[ 1 \!-\! z_2 \!-\! (1\!-\!x)(1\!-\!y)(1\!-\!z_1\!-\!z_2) ]P^2 \!+\! M^2\!\!,
\;&\;
\xi_{b,1}\!\!=\! z_2[z_1\!-\!x\,(1\!-\!z_1\!-\!z_2) ]P^2 \!+\! M^2\!\!,
\\
\xi_{c,1}\!\!=\! z_1z_2P^2 \!+\! M^2\!\!,
\;&\;
\xi_{a,2}\!= z_2[ 1 \!-\! z_2\,(x\!+\!y\!-\!x\,y) ]P^2 + M^2,
\\
\xi_{b,2}\!= z_2(1\!-\!x\,z_2)P^2 + M^2,
\;&\;
\xi_{c,2}\!= z_2P^2 + M^2,
\\
\xi_{a,3}\!= z_2(1\!-\!z_2)[1\!-\!y\,(1\!-\!x)]P^2 \!+\! M^2,
\;&\;
\xi_{b,3}\!= xz_2(1\!-\!z_2)P^2 + M^2,
\\
\xi_{c,3}\!= M^2,
\ea
\eqa
\bqa
\ba{lll}
\beta_{a,1}\!= 4\,[ 1 \!-\! 2\,( 2 \!-\! 2x \!+\! xy)(1\!-\!z_1\!-\!z_2) ],
\;&\;
\beta_{b,1}\!= 
4 [1 \!-\! 2 (1\!-\!z_1) \!-\! 2 (1\!-\!x)(1\!-\!z_1\!-\!z_2)],
\;&\;
\beta_{c,1}= 1 \!-\! 2\,z_1,
\\
\beta_{a,2}\!= - 4 z_2 ( 1 \!-\! 2xz_2 \!+\! xyz_2 ),
\;&\;
\beta_{b,2}\!= - 4\,z_2\,(1\!-\!xz_2),
\;&\;
\beta_{c,2}= \! -\! M^2/P^2\!,
\\
\beta_{a,3}\!= 2 - x\,( 1 \!-\! 2z_2 \!+\! 4z_2^2)(1\!+\!y),
\;&\;
\beta_{b,3}\!= 2 - x + 2\,x\,z_2\,(1\!-\!2z_2),
\;&\;
\beta_{c,3}= M^2/P^2.
\ea
\eqa
\subsection{The special case of the $V^{\ada}$ family\label{sec:vmcoll}}
\begin{figure}[ht]
\begin{center}
\begin{tabular}{cc}
\begin{picture}(140,75)(0,0)
 \SetWidth{1.2}
 \Line(0,0)(40,0)                    \Text(10,5)[cb]{$-P$}
 \Photon(128,-53)(100,-35){2}{5}     \Text(138,-65)[cb]{$p_1$}
 \Photon(128,53)(100,35){2}{5}       \Text(138,57)[cb]{$p_2$}
 \DashLine(57,-10)(40,0){1.4}        \Text(46,-16)[cb]{$m$}
 \CArc(70,-17.5)(15,0,360)           \Text(80,-1)[cb]{$M_1$}
                                     \Text(58,-41)[cb]{$M_2$}
 \DashLine(100,-35)(83,-25){1.4}     \Text(89,-42)[cb]{$m$}
 \DashLine(100,-35)(100,35){1.4}     \Text(110,-3)[cb]{$m$}
 \DashLine(100,35)(40,0){1.4}        \Text(68,23)[cb]{$m$}
 \Text(15,-50)[cb]{$V^{\ada}_a$}
\end{picture}
&\qquad\qquad\qquad
\begin{picture}(140,75)(0,0)
 \SetWidth{1.2}
 \Photon(0,0)(40,0){2}{6}            \Text(10,5)[cb]{$-P$}
 \Photon(128,-53)(100,-35){2}{5}     \Text(138,-65)[cb]{$p_1$}
 \Line(128,53)(100,35)               \Text(138,57)[cb]{$p_2$}
 \DashLine(57,-10)(40,0){1.4}        \Text(46,-16)[cb]{$m$}
 \CArc(70,-17.5)(15,0,360)           \Text(80,-1)[cb]{$M_1$}
                                     \Text(58,-41)[cb]{$M_2$}
 \DashLine(100,-35)(83,-25){1.4}     \Text(89,-42)[cb]{$m$}
 \DashLine(100,-35)(100,35){1.4}     \Text(110,-3)[cb]{$m$}
 \DashLine(100,35)(40,0){1.4}        \Text(68,23)[cb]{$m$}
 \Text(15,-50)[cb]{$V^{\ada}_b$}
\end{picture}
\end{tabular}
\end{center}
\vspace{2cm}
\caption[]{The $V^{\ada}$ family, characterized by a $1/m^2$ behavior in
the $m\to 0$ limit. Dot-lines have a small mass $m$; wavy 
lines are massless.} 
\label{vm}
\end{figure}
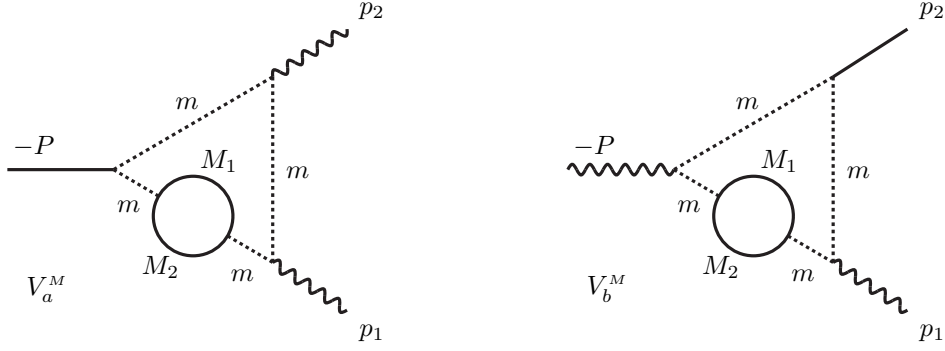

The computation of the amplitude for $H \to \gamma \gamma$ has been performed
keeping the light-fermion masses during the whole generation$\,\otimes\,$
simplification procedure; they have been set to zero just after the
extraction of the collinear logarithms.  
Therefore, at intermediate steps, we have diagrams multiplied by a small
mass which, in general, will disappear when the fermion mass goes to
zero. 
However, some of these diagrams belong to the $V^{\ada}$ family, and
develop a $1/m^2$ behavior. They are shown in \fig{vm} and they would in
general lead to a residual contribution when their
coefficient is proportional to $m^2$.
It is worth noting that, concerning the $H \to \gamma \gamma$ amplitude, only
the scalar $V^{\ada}$ function remains after the reduction$\,\otimes\,$
symmetrization procedure; in fact, all tensor structures have been reduced to 
simpler topologies. 

We parametrize the $V^{\ada}_i$ ($i=a,b$) integrals shown in Fig.~\ref{vm} following
section~8 of Ref.~\cite{Ferroglia:2003yj} and obtain
\bqa
&&V^{\ada}_i
=
-\left(\frac{\mu^{2}}{\pi}\right)^\ep\*\Gamma\left(2+\ep\right)\*
\int_{0}^{1}\!dx\* dy\*\int_{0}^{y}\!dz_1\*\int_{0}^{z_1}\!dz_2\*
[x\*(1-x)]^{-\tfrac{\ep}{2}}\*
(1-y)^{\tfrac{\ep}{2}-1}\*(y-z_1)\*
\chi_i(x,y,z_1,z_2)^{-2-\ep},
\nl
&&
\chiu{a}= m_x^2\*(1\!-\!y)\!+\!P^2\*z_2\*(1\!-\!z_1)\!+\!m^2\*y, \quad
\chiu{b}= m_x^2\*(1\!-\!y)\!+\!p_2^2\*z_2\*(z_1\!-\!z_2)\!+\!m^2\*y,
\eqa
with $m_x^2=[M_1^2\*(1-x)+M_2^2\*x]/ [x\*(1-x)]$.
In order to extract the aforementioned $1/m^2$ behavior, we apply a
Bernstein-Sato-Tkachov algorithm~\cite{Tkachov:1996wh} on $\chi_i$,
\bqa
\!\!\!\chiu{a}(x,y,z_1,z_2)^{-2-\ep}&=&\frac{1}{m^2}\*
  \left\{1+\frac{1}{1+\ep}\*\left[
  (y-1)\*\partial_y
 +\frac{1}{2}\*(z_1-1)\*\partial_{z_1}
 +\frac{1}{2}\*z_2\*\partial_{z_2}
 \right]\right\}\*
 \chiu{a}(x,y,z_1,z_2)^{-1-\ep},\nl
\!\!\!\chiu{b}(x,y,z_1,z_2)^{-2-\ep}&=&\frac{1}{m^2}\*
  \left\{1+\frac{1}{1+\ep}\*\left[
  (y-1)\*\partial_y
 +\frac{1}{2}\*z_1\*\partial_{z_1}
 +\frac{1}{2}\*z_2\*\partial_{z_2}
 \right]\right\}
 \chiu{b}(x,y,z_1,z_2)^{-1-\ep}.
\label{bstder}
\eqa
After integration by parts the poles are extracted. 
Analyzing the behavior at small $m$, we find that, in addition to the $1/m^2$
behavior, also $\ln m^2$ terms are present.
The structure of the remaining integrals is simple enough to allow for a
direct integration in all variables. The result yields
\bqa
\label{vma}
m^2
\times
\raisebox{0.1cm}{\scalebox{0.65}{
\begin{picture}(140,75)(0,0)
 \SetWidth{1.2}
 \Line(0,0)(40,0)                    \Text(10,5)[cb]{$-P$}
 \Photon(128,-53)(100,-35){2}{5}     \Text(138,-65)[cb]{$p_1$}
 \Photon(128,53)(100,35){2}{5}       \Text(138,57)[cb]{$p_2$}
 \DashLine(57,-10)(40,0){1.4}        \Text(46,-16)[cb]{$m$}
 \CArc(70,-17.5)(15,0,360)           \Text(80,-1)[cb]{$M_1$}
                                     \Text(58,-41)[cb]{$M_2$}
 \DashLine(100,-35)(83,-25){1.4}     \Text(89,-42)[cb]{$m$}
 \DashLine(100,-35)(100,35){1.4}     \Text(110,-3)[cb]{$m$}
 \DashLine(100,35)(40,0){1.4}        \Text(68,23)[cb]{$m$}
\end{picture}
}}
\!\!
&=&
-\frac{1}{P^2}\*\bigg[
2\*F^2_{-1}(s)
-\ln\left(-\frac{m^2}{P^2}\right)+\hat{\rho}(-P^2,M_1^2,M_2^2)+1 
\bigg] \nn \\
&&\qquad+\, {\cal O}(\ep)+{\cal O}(m^2),\;\; s=-P^2,
\eqa
\bqa
\label{vmb}
m^2
\times
\raisebox{0.1cm}{\scalebox{0.65}{
\begin{picture}(140,75)(0,0)
 \SetWidth{1.2}
 \Photon(0,0)(40,0){2}{6}            \Text(10,5)[cb]{$-P$}
 \Photon(128,-53)(100,-35){2}{5}     \Text(138,-65)[cb]{$p_1$}
 \Line(128,53)(100,35)               \Text(138,57)[cb]{$p_2$}
 \DashLine(57,-10)(40,0){1.4}        \Text(46,-16)[cb]{$m$}
 \CArc(70,-17.5)(15,0,360)           \Text(80,-1)[cb]{$M_1$}
                                     \Text(58,-41)[cb]{$M_2$}
 \DashLine(100,-35)(83,-25){1.4}     \Text(89,-42)[cb]{$m$}
 \DashLine(100,-35)(100,35){1.4}     \Text(110,-3)[cb]{$m$}
 \DashLine(100,35)(40,0){1.4}        \Text(68,23)[cb]{$m$}
\end{picture}
}}
\!\!
&=&
-\frac{1}{p_2^2}\*
\bigg\{ 
-2\*F^2_{-1}(s)\*
\left[\ln\left(-\frac{m^2}{p_2^2}\right)+2+i\*\pi\right] 
+\ln^2\left(-\frac{m^2}{p_2^2}\right)\nn\\
&&\qquad
+ \ln\left(-\frac{m^2}{p_2^2}\right)\*
\left[
 1 + i\,\pi
-\hat{\rho}(-p_2^2,M_1^2,M_2^2)
\right]-2-i\*\pi+\frac{\pi^2}{6}\nn\\
&&\qquad-\left(2+i\*\pi\right)\*
\hat{\rho}(-p_2^2,M_1^2,M_2^2)
\bigg\}
+\, {\cal O}(\ep)+{\cal O}(m^2),\;\;\;s=-p_2^2,
\eqa
and with $\hat{\rho}(x,y,z)=[y\ln(y/x)-z\ln(z/x)] / (z-y)$.
Therefore, before taking the massless limit in the terms of the amplitude
generated by the $V^{\ada}$ family, we introduce the results of 
\eqn{vma} and \eqn{vmb}. 

It turns out that all terms coming from these kinds of $V^{\ada}$ diagrams, where
the Higgs boson and the two photons are coupled to light fermions, are
either zero or cancel analytically one another. As expected, they lead to a
total vanishing contribution; we would have obtained the same result treating
them as massless diagrams from the very beginning.
Note that diagrams with leptons, up and down quarks of the first and second 
generations and bottom quarks cancel separately.

\vspace{0.5cm}
Summarizing, in this section we have discussed an algorithm which allows to 
express the coefficients of all collinear logarithms in terms of one-loop 
integrals. 
In order to verify their cancellation, we then reduce all one-loop diagrams
to scalar integrals with propagators raised to a canonical power one; this
last step requires the standard reduction of Ref.~\cite{Passarino:1979jh} 
and the IBP identities of Ref.~\cite{Tkachov:1981wb}. Each collinear-divergent 
diagram contributes to the total result with its collinear-free part which is 
manipulated as described above and then computed numerically.
\section{Evaluation of massive diagrams} 
\label{EMD}
For the evaluation of massive two-loop diagrams, we follow the methods
of Refs.~\cite{Passarino:2001wv,Passarino:2001jd} for self-energies 
and of Refs.~\cite{Ferroglia:2003yj,Passarino:2006gv,Actis:2004bp} for 
vertices. 
For the three vertex families $V^{\ssG}$, $V^{\ssK}$ and $V^{\ssH}$, we 
modify our approach, since two external massless particles appear and
simpler results can be derived. Since the massive diagrams do not lead
to collinear logarithms, the techniques described in this 
section can also be applied in the case of external light particles
(setting their squared momenta to zero).

\vspace{0.5cm}
\underline{\emph{$V^{\ssG}$ family}}.
Let us consider the scalar, vector and tensor integrals of the $V^{\ssG}$ 
family, in the configuration where collinear divergences are absent and 
one of the two external squared momenta $p_1^2$ and $p_2^2$ vanishes. 
Setting without loss of generality 
$p_2^2=0$, we follow the parametrizations described in section~7 of 
Ref.~\cite{Ferroglia:2003yj} and in section~9 of Ref.~\cite{Actis:2004bp}, 
in close analogy with the procedure already described in section~\ref{sec:collfinG}. 
The extraction of the UV poles is trivial; after an appropriate change of 
variables, we can write the $V^{\ssG}$ functions in parametric space as
  \bqa \label{YG}
  V^{\bba}|_{p_2^2=0} &=& 
  - \int_0^1\!\!\!dx_1\,dx_2\,dy_1\!\int_0^{y_1}\!\!\!\!\!dy_2\,
  (1\!-\!x_2)\,Y_{\bba}^{-1},
  \nl
  V^{\bba}[q_i^\mu]_{p_2^2=0} &=& 
  \int_0^1\!\!\!dx_1\,dx_2\,dy_1\!\int_0^{y_1}\!\!\!\!\!dy_2\,(1\!-\!x_2)\,
  p_{\bba,i}^{\mu}\,Y_{\bba}^{-1}, \quad
  p_{\bba,i}^{\mu} = \sum_{h=1}^2 a_{\bba,i}^h\,p_h^\mu,
  \nl
  V^{\bba}[q_i^\mu q_j^\nu]_{p_2^2=0}  &=& 
  - \int_0^1\!\!\!dx_1\,dx_2\,dy_1\!\int_0^{y_1}\!\!\!\!\!dy_2\,(1\!-\!x_2)\,
  p_{\bba,i}^{\mu}\,p_{\bba,i}^{\nu}\,Y_{\bba}^{-1}
  + \Delta_{\bba,ij}\,\delta^{\mu\nu},
  \eqa
where $Y_{\bba}$ reads
  \bqa \label{defYG1}
  &&
  Y_{\bba}= a_{\ssG}\,y_1 + b_{\ssG}, \qquad\quad
  a_{\ssG}= (1\!-\!x_2)\,\sigma_{\ssG},\qquad\quad
  \sigma_{\ssG}= x_2\,[ x_1\,y_2\,( p_1^2 - P^2 ) + m_4^2 - m_5^2 ],
  \nl
  &&
  b_{\ssG}=
  x_1\,x_2\,y_2(1\!-\!x_2)[ P^2 \!-\! p_1^2(1\!-\!x_1\!+\!y_2\!-\!x_2\,y_2) ]
  + x_2(1\!-\!x_2)\,(  m_5^2 \!-\! m_4^2y_2 )
  + y_2(1\!-\!x_2)\chiu{\ssG,x}
  + m_3^2 x_2\,y_2,
  \nl
  &&
  \chiu{\ssG,x}= p_1^2 x_1(1\!-\!x_1) +  m_1^2x_1 + m_2^2(1\!-\!x_1),
  \eqa
the coefficients of the tensor structures involving the external momenta are 
given by
  \bq \label{defYG2}
  a_{\bba,1}^1 = 1 - x_1\,[1 - x_2\,(1\!-\!y_2)], \qquad
  a_{\bba,1}^2 = x_2\,(1\!-\!y_1), \qquad
  a_{\bba,2}^1 = 1 \!-\! x_1\,y_2, \qquad
  a_{\bba,2}^2 = 1\!-\!y_1,
  \eq
and the coefficients of the Kronecker delta are
  \bqa \label{KronVG}
  \Delta_{\bba,11}
  &=&
  - \frac{1}{2}\,F_{-2}^{2}(s) 
  + \frac{1}{2}\,F_{-1}^{2}(s)\,
  \bigg( \intsx{y_1}\,\ln\frac{m_{45}^2}{s} - \frac{3}{4} \bigg)
  - \frac{5}{32} - \frac{1}{8}\,\zeta(2)
  - \frac{1}{4}\!\int_0^1\!\!\!dy_1\ln^2\frac{m_{45}^2}{s}
  \nl
  &&
  + \frac{1}{4}\!\int_0^1\!\!\!dy_1(\ln y_1 + 2)\ln\frac{m_{45}^2}{s}
  + \frac{1}{2}\!\int_0^1\!\!\!dx_1 dx_2 dy_1\!\int_0^{y_1}\!\!\!\!\!\!dy_2\,
  \bigg(
  x_2\ln\frac{Y_\bba}{s}
  + \frac{1\!-\!x_2}{y_2}\,\ln\frac{Y_\bba}{Y_\bba|_{y_2=0}}
  \bigg),
  \nl
  \Delta_{\bba,12}
  &=&
  - \frac{1}{4}\,F_{-1}^{2}(s)
  + \frac{7}{16}
  + \frac{1}{2}\,\int_0^1\!\!\!dx_1\,dx_2\,dy_1\!\int_0^{y_1}\!\!\!\!\!dy_2\,
  \ln\frac{Y_\bba}{s},
  \nl
  \Delta_{\bba,22}
  &=&
  - \frac{1}{2}\,F_{-2}^{2}(s) 
  + \frac{1}{2}\,F_{-1}^{2}(s)\,
  \bigg( \intsx{x_1}\ln\frac{\chiu{\ssG,x}}{s} - \frac{3}{4} \bigg)
  - \frac{7}{32} - \frac{1}{8}\,\zeta(2)
  - \frac{1}{4}\,\intsx{x_1}\ln^2\frac{\chiu{\ssG,x}}{s} 
  \nl
  &&
  + \frac{3}{8}\,\intsx{x_1}\ln\frac{\chiu{\ssG,x}}{s}
  + \frac{1}{2}\,\int_0^1\!\!\!dx_1\,dx_2\,dy_1\!\int_0^{y_1}\!\!\!\!\!dy_2\,
  \frac{1}{x_2}\,\ln\frac{Y_\bba}{Y_\bba|_{x_2=0}},
  \eqa
with $m_{45}^2= m_4^2\,y_1 + m_5^2(1\!-\!y_1)$ and $s=-P^2$. 

The terms proportional to $\delta^{\mu\nu}$ in \eqn{KronVG} can be directly 
integrated numerically; for those terms in \eqn{YG} containing $Y_\ssG^{-1}$ we 
observe that the polynomial $Y_\ssG$ is linear in $y_1$ and integrate by parts,
  \bq
  \int_{y_2}^1\!\!\!dy_1\,\frac{y_1^n}{Y_\ssG} =
  \frac{1}{a_{\ssG}}\,\Bigg[
  \ln\left( 1 + \frac{a_{\ssG}}{b_{\ssG}} \right)
  - y_2^n\ln\left( 1 + \frac{a_{\ssG}y_2}{b_{\ssG}} \right)
  - n\int_{y_2}^1\!\!\!dy_1\,y_1^{n-1}
  \ln\left( 1 + \frac{a_{\ssG}y_1}{b_{\ssG}} \right)
  \Bigg].
  \eq
Therefore, the expressions of \eqn{YG} can be written in terms of simpler
integral representations,
  \bqa \label{VGsmooth}
  V^{\bba}|_{p_2^2=0}
  &=& 
  - \int_0^1\!\!\!dx_1\,dx_2\,dy_2\,
  \frac{1}{\sigma_{\ssG}}\bigg[
    \ln\left( 1 + \frac{a_{\ssG}}{b_{\ssG}} \right)
  - \ln\left( 1 + \frac{a_{\ssG}\,y_2}{b_{\ssG}} \right)
  \bigg],
  \nl
  V^{\bba}[q_i^\mu]_{p_2^2=0}
  &=& 
  \int_0^1\!\!\!\!dx_1dx_2dy_1dy_2\Bigg\{\!
  \frac{\alpha_{\bba,i}^1}{\sigma_{\ssG}}p_1^\mu\bigg[\!
  \ln\!\left(\! 1 \!+\! \frac{a_{\ssG}}{b_{\ssG}} \!\right)
  - \ln\!\left(\! 1 \!+\! \frac{a_{\ssG}y_2}{b_{\ssG}} \!\right)
  \!\bigg]
  \nl
  &&
  + \,\alpha_{\bba,i}^2p_2^\mu\bigg[
    \frac{y_1}{\bar{a}_{\ssG}}
    \ln\!\left(\! 1 \!+\! \frac{\bar{a}_{\ssG}y_1}{\bar{b}_{\ssG}} \!\right)
  -  \frac{1\!-\!y_2}{\sigma_{\ssG}}
    \ln\!\left(\! 1 \!+\! \frac{a_{\ssG}y_2}{b_{\ssG}} \!\right)
  \!\bigg]
  \!\Bigg\},
  \nl
  V^{\bba}[q_i^\mu q_j^\nu]_{p_2^2=0}
  &=& 
  \Delta_{\bba,ij}\,\delta^{\mu\nu}
  - \int_0^1\!\!\!dx_1dx_2dy_1dy_2\,\Bigg\{\!
    \alpha_{\bba,i}^1\,\alpha_{\bba,j}^1\,p_1^\mu\,p_1^\nu\,\bigg[\!
      \ln\!\left(\! 1 \!+\! \frac{a_{\ssG}}{b_{\ssG}} \!\right)
    - \ln\!\left(\! 1 \!+\! \frac{a_{\ssG}y_2}{b_{\ssG}} \!\right)
    \!\bigg]
    \nl
    &&
  + \,(   \alpha_{\bba,i}^1\,\alpha_{\bba,j}^2\,p_1^\mu\,p_2^\nu\,
        + \alpha_{\bba,i}^2\,\alpha_{\bba,j}^1\,p_2^\mu\,p_1^\nu\,)
    \bigg[
      \frac{y_1}{\bar{a}_{\ssG}}
      \ln\!\left(\! 1 \!+\! \frac{\bar{a}_{\ssG}y_1}{\bar{b}_{\ssG}} \!\right)
    - \frac{1\!-\!y_2}{\sigma_{\ssG}}
      \ln\!\left(\! 1 \!+\! \frac{a_{\ssG}y_2}{b_{\ssG}} \!\right)
    \!\bigg]
    \nl
    &&
  + \,\alpha_{\bba,i}^2\,\alpha_{\bba,j}^2\,p_2^\mu\,p_2^\nu\,\bigg[
      \frac{2\,y_1(1\!-\!y_1)}{\bar{a}_{\ssG}}
      \ln\!\left(\! 1 \!+\! \frac{\bar{a}_{\ssG}y_1}{\bar{b}_{\ssG}} \!\right)
    - \frac{(1\!-\!y_2)^2}{\sigma_{\ssG}}
      \ln\!\left(\! 1 \!+\! \frac{a_{\ssG}y_2}{b_{\ssG}} \!\right)
    \!\bigg]
  \Bigg\},
  \eqa
where we have introduced short-hand notations related to \eqn{defYG1} and 
\eqn{defYG2},
  \bq
  \bar{a}_{\ssG} \!= a_{\ssG}|_{y_2\to y_1y_2},
  \qquad
  \bar{b}_{\ssG} \!= b_{\ssG}|_{y_2\to y_1y_2},
  \qquad
  \alpha_{\bba,1}^1 \!= a_{\bba,1}^1,
  \qquad
  \alpha_{\bba,1}^2 \!= x_2,
  \qquad
  \alpha_{\bba,2}^1 \!= a_{\bba,2}^1,
  \qquad
  \alpha_{\bba,2}^2 \!= 1.
  \eq
Looking at \eqn{VGsmooth} we see that the zeros of $\sigma_{\ssG}$ in the 
denominator are smoothly compensated by logarithms that vanish for 
$\sigma_{\ssG}=0$ ($a_{\ssG}$ is proportional to $\sigma_{\ssG}$). However, in 
the process under consideration we have $m_4=m_5$, and we can encounter some 
numerical instability; in this case, $\sigma_{\ssG}$ is proportional to $x_2\,y_2$ 
and $b_{\ssG}$ vanishes for $x_2=y_2=0$.
Since the zero of $b_{\ssG}$ is of lower order with respect to the zero of 
$\sigma_{\ssG}$, the instability can be cured by a sector decomposition of the 
unit square~\cite{Binoth:2000ps},
  \bq
  \int_0^1\!\!\!\!dx_2dy_2\,f(x_2,y_2)= 
  \bigg(
  \int_0^1\!\!\!\!\!dx_2\!\!\int_0^{x_2}\!\!\!\!\!\!\!dy_2
  + \int_0^1\!\!\!\!\!dy_2\!\!\int_0^{y_2}\!\!\!\!\!\!\!dx_2
  \bigg)f(x_2,y_2)= 
  \int_0^1\!\!\!\!dx_2dy_2\,\Big[ x_2 f(x_2,x_2y_2) + y_2 f(x_2y_2,y_2) \Big].
  \eq

\vspace{0.5cm}
\underline{\emph{$V^{\ssK}$ family}}.
We analyze here the $V^{\ssK}$ configurations with $P^2=0$ and $p_1^2=p_2^2=0$,
focusing on the scalar diagram, since it is the  only one which survives after 
the reduction procedure described in section~\ref{sec:ScalarReduct}.
Note that the method explained here can be applied, with a straightforward 
generalization, also to tensor integrals. We set $\ep=0$ in Eq.~(159) of 
Ref.~\cite{Passarino:2006gv}, perform the $y_3$ integration and make some change 
of variables simplifying the integral,
\bq \label{bracks}
V^{\bca} =
\int_0^1\!\!\!dx_1\,dx_2\,dy_1\!\int_0^{y_1}\!\!\!\!\!dy_2\;x_1\,x_2
\sum_{i=a,b}
\frac{1}{{\cal B}_{\!\bca\!,i}}\,\bigg[
  \frac{x_2\,(1\!-\!x_2)}
       {x_2(1\!\!-\!\!x_2){\cal A}_{\bca\!,i} + y_2{\cal B}_{\!\bca\!,i}}
- \frac{1}{{\cal A}_{\bca\!,i}}
\bigg],
\eq
  \bqa
  {\cal A}_{\bca,a}
  &=&
  P^2 x_1 y_2 (y_1\!-\!x_1\,y_2)
  + (y_1\!-\!x_1\,y_2)[ p_1^2(1\!-\!y_1) + m_4^2 ]
  + x_1\,y_2[ p_2^2(1\!-\!y_1) + m_6^2]
  + (1\!-\!y_1)\,m_5^2,
  \nl
  {\cal B}_{\bca,a}
  &=&
  P^2 x_2^2 x_1(1\!-\!x_1) 
  + x_2[ m_1^2(1\!-\!x_1) + m_2^2 x_1 ]
  + (1\!-\!x_2) m_3^2
  - x_2(1\!-\!x_2)[ m_4^2(1\!-\!x_1) + m_6^2 x_1 ],
  \nl
  {\cal A}_{\bca,b} &=&
  {\cal A}_{\bca,a}
  (m_1\leftrightarrow m_2,m_4\leftrightarrow m_6,p_1\leftrightarrow p_2),
  \qquad
  {\cal B}_{\bca,b}  =
  {\cal B}_{\bca,a}
  (m_1\leftrightarrow m_2,m_4\leftrightarrow m_6,p_1\leftrightarrow p_2).
  \eqa
Note that the zeros of ${\cal B}_{\bca,a}$ and of ${\cal B}_{\bca,b}$ in 
\eqn{bracks} are compensated by a corresponding zero in the squared bracket. 
The general procedure described in Ref.~\cite{Ferroglia:2003yj} writes the 
integral in $y_1$ and $y_2$ as a one-loop $C$ function; however, after setting 
$P^2=0$ or $p_1^2=p_2^2=0$, we observe that ${\cal A}_{\bca,a}$ and 
${\cal A}_{\bca,b}$ are linear in $y_2$ and $y_1$, respectively, thus allowing 
for an integration by parts where we use the last relation of \eqn{eq:partial}.
For the case $P^2=0$ we obtain
$$
V^{\!\bca}|_{P^2=0}\! =
\!\int_0^1\!\!\!\!\!dx_1 dx_2 dy_1\frac{x_2}{{\cal B}_{\!\bca\!,a}}\!
\Bigg[\!
  \frac{x_1 c_\bca}
       {x_1 c_\bca a_{\!\bca\!,a} \!+\! {\cal B}_{\!\bca\!,a}}\!
  \ln\!\left(\! 
       1 \!+\! \frac{x_1 c_\bca a_{\!\bca\!,a}\!+\!{\cal B}_{\!\bca\!,a}}
                    {c_\bca b_{\!\bca\!,a}}\,y_1 
       \!\right)
\!-\! \frac{1}{a_{\!\bca\!,a}}\!
       \ln\!\left(\! 1 \!+\! 
                     \frac{x_1 y_1 a_{\!\bca\!,a}}{b_{\!\bca\!,a}} \!\right)
\!\!\Bigg]
+\,
\left\{\!
  {
    \ba{l}
    m_1 \!\!\leftrightarrow\! m_2 \\
    m_4 \!\!\leftrightarrow\! m_6 \\
     p_1    \leftrightarrow\! p_2
    \ea
  }
  \!\!\right\}
$$
\bq
a_{\bca\!,a}= 
(p_2^2\!-\!p_1^2)(1\!-\!y_1) + m_6^2 - m_4^2,
\quad\;
b_{\bca\!,a}= p_1^2y_1(1\!-\!y_1) + m_4^2y_1 + m_5^2(1\!-\!y_1),
\quad\;
c_\bca= x_2(1\!-\!x_2).
\eq
A similar result holds for the configuration where $p_1^2= p_2^2= 0$,
  \bq \label{finvk}
  \!
  V^{\bca}|_{p_1^2=p_2^2=0}
  =
  \int_0^1\!\!\!\!dx_1 dx_2 dy_2\,
  \frac{x_1 x_2}{\alpha_{\!\bca\!,a}{\cal B}_{\!\bca\!,a}}\,
  \bigg[
  \ln\!\left(\!
      1 + \frac{\gamma_\bca{\cal B}_{\!\bca\!,a}}
               {\alpha_{\!\bca\!,a}\!+\!\beta_{\!\bca\!,a}}
     \!\right)
   - \ln\!\left(\!
      1 + \frac{\gamma_\bca{\cal B}_{\!\bca\!,a}}
               {\alpha_{\!\bca\!,a}y_2\!+\!\beta_{\!\bca\!,a}}
     \!\right)
   \!\bigg]
   \;+\;
   \left\{
     \ba{l}
     m_1\leftrightarrow m_2 \\
     m_4\leftrightarrow m_6
     \ea
   \right\},
   \eq
  \bq
  \alpha_{\bca,a}= P^2 x_1 y_2 + m_4^2 - m_5^2,
  \qquad
  \beta_{\bca,a}= - P^2 x_1^2 y_2^2 + x_1\,y_2 (m_6^2-m_4^2) + m_5^2,
  \qquad
  \gamma_\bca= \frac{y_2}{x_2(1\!-\!x_2)}.
  \eq
There are special configurations where the result can be further simplified 
integrating in one additional variable. This step becomes necessary to achieve 
numerical stability in all cases where special relations between masses 
move some zeros of the denominator to the border of the integration domain, 
as in \eqn{finvk} when $m_3=0$, $m_1=m_4$ and $m_2=m_6$;
in this case we can factorize $x_2^2$ from ${\cal B}_{\!\bca\!,i}$ and 
carry on the integration in $x_2$,
  \bq
  \int_0^1\frac{dx_2}{x_2}\bigg[
  \ln\!\left(\! 1 \!+\! \frac{A x_2}{1\!-\!x_2} \!\right) 
  - \ln\!\left(\! 1 \!+\! \frac{B x_2}{1\!-\!x_2} \!\right) 
  \bigg]
  =
  - \li{2\!}{1\!-\!A} + \li{2\!}{1\!-\!B}
  .
  \eq

\vspace{0.5cm}
\underline{\emph{$V^{\ssH}$ family}}.
For evaluating diagrams of the $V^{\ssH}$ family without collinear 
singularities, we rely on the methods developed in Ref.~\cite{Hollik:2005ns} 
where the requirement is that two external squared momenta vanish, as in the 
process $H\to \gamma\gamma$; this condition 
allows for a double integration by parts in parametric space. However, 
the method described in Ref.~\cite{Hollik:2005ns} needs special care when 
applied in the region above threshold (a case not covered in 
Ref.~\cite{Hollik:2005ns}). 
Following the same parametrization procedure described in section~\ref{sec:vhdc}, 
we set $p_1^2=p_2^2=0$ and obtain for diagrams up to rank two
  \bqa \label{deffa}
  V^{\bbb}|_{p_1^2=p_2^2=0} &=& 
  - \int dC_5\lpar x,y,\{z\}\rpar\,
  x\,y\,(1\!-\!x)\,(1\!-\!y)\,Y_{\bbb}^{-2},
  \nl
  V^{\bbb}[q_i^\mu]_{p_1^2=p_2^2=0} &=& 
  - \int dC_5\lpar x,y,\{z\}\rpar\,
  x\,y\,(1\!-\!x)\,(1\!-\!y)\,Y_{\bbb}^{-2}\,
  p_{\bbb,i}^{\mu}, \quad
  p_{\bbb,i}^{\mu} = \sum_{h=1}^2 a_{\bbb\!,i}^h\,p_h^\mu,
  \nl
  V^{\bbb}[q_i^\mu q_j^\nu]_{p_1^2=p_2^2=0} &=& 
  - \!\int dC_5\lpar x,y,\{z\}\rpar\,\frac{1\!-\!y}{Y_{\bbb}}\,\bigg[
    x\,y\,(1\!-\!x)\,Y_{\bbb}^{-1}
    p_{\bbb,i}^{\mu}\,p_{\bbb,j}^{\nu}
  + \frac{1}{2}  a_{\bbb\!,ij}^{0}\,\delta^{\mu\nu}
  \bigg],
  \eqa
where $Y_{\bbb}$ is given by
\bqa
&&
\!\!\!\!
Y_{\!\bbb}\!= 
xy(1\!-\!x)(1\!-\!y)\chiu{\bbb\!,0}
+ y(1\!-\!x)\chiu{\bbb\!,1}
+ xy\chiu{\bbb\!,2}
+ x(1\!-\!x)(1\!-\!y)\chiu{\bbb\!,3},
\quad\!
\chiu{\!\bbb\!,0}\!=
(1\!-\!z_2\!-\!z_3)(1\!-\!z_1\!-\!z_2)P^2\!\!\!,
\nl
&&
\!\!\!\!
\chiu{\!\bbb\!,1}\!= m_1^2(1\!-\!z_1) + m_2^2z_1,
\qquad
\chiu{\!\bbb\!,2}\!= P^2\,z_2\,(1\!-\!z_2) + m_3^2\,(1\!-\!z_2) + m_4^2\,z_2,
\qquad
\chiu{\!\bbb\!,3}\!= m_5^2(1\!-\!z_3) \!+\! m_6^2z_3,
\qquad\quad
\eqa
and the coefficients of the tensor structures read as
  \bqa
  a_{\bbb\!,1}^1 &=& - x\,(1\!-\!y)(1\!-\!z_2\!-\!z_3), 
  \qquad
  a_{\bbb\!,1}^2  =  1 \!-\! z_1 \!-\! x\,(1\!-\!y)(1\!-\!z_2-\!z_1), 
  \qquad
  a_{\bbb\!,2}^1  =  z_3 \!+\! y\,(1\!-\!z_2\!-\!z_3), 
  \nl
  a_{\bbb\!,2}^2 &=& y\,(1\!-\!z_1\!-\!z_2), 
  \qquad\quad\quad\quad
  a_{\bbb\!,11}^{0} = x\,[1 - x\,(1\!-\!y)],
  \qquad
  a_{\bbb\!,12}^{0} = x\,y,
  \qquad
  a_{\bbb\!,22}^{0} = y.
  \eqa
Focusing on the configurations which appear for the $H\to\gamma\gamma$ decay, 
we set $m_2=m_1$, $m_4=m_3$, $m_6=m_5$ and use the symmetries of the result;
vector integrals can be simply written in terms of the scalar one,
  \bq \label{Hvecs}
  V_{\!s}^{\bbb}[q_1^\mu] = \frac{1}{2}\,V_{\!s}^{\bbb}p_2^\mu,
  \qquad\qquad
  V_{\!s}^{\bbb}[q_2^\mu] = \frac{1}{2}\,V_{\!s}^{\bbb}p_1^\mu,
  \eq
where the subscript $s$ denotes the configuration $p_1^2=p_2^2=0$, $m_2=m_1$, 
$m_4=m_3$, $m_6=m_5$. Furthermore, the expression for $Y_\bbb$ shows a simple
dependence on $z_1$ and $z_3$,
  \bqa
  Y_\bbb &=& 
  \alpha_{\ssH}\,[ z_1\,z_3 - (1\!-\!z_2)(z_1+z_3) + (1\!-\!z_2)^2 ] 
  + \beta_{\ssH},
  \qquad
  \alpha_{\ssH}= x\,y\,(1\!-\!x)(1\!-\!y)P^2,\nl
  \beta_{\ssH}&=& 
  y\,(1\!-\!x)m_1^2
  + x\,y\,\chiu{\bbb\!,2}
  + x\,(1\!-\!x)(1\!-\!y)m_5^2,
  \qquad\quad
  \chiu{\bbb\!,2}= P^2\,z_2\,(1\!-\!z_2) + m_3^2.
  \eqa
This structure corresponds to the collinear case $V^{\bbb}_{\rm sc}$ 
discussed in section~\ref{sec:collfinH}; integrating by parts in $z_1$ and $z_3$ 
according to \eqn{intbyparts} and \eqn{primitives}, we obtain ($n,k\neq 0$)
\bqa \label{longInts}
\int_0^1\!\!\!\!\!dz_1 dz_3\,\frac{z_1^n z_3^k}{Y_\bbb^2} 
&=&
\frac{n\,\!k}{\alpha_{\ssH}\beta_{\ssH}}\!
\int_0^1\!\!\!\!dz_1dz_3\,z_1^{n-1}z_3^{k-1}\bigg[
\ln\frac{\alpha_{\ssH}(1\!-\!z_2\!-\!z_3)(1\!-\!z_1\!-\!z_2)\!+\!\beta_{\ssH}}
        {-\alpha_{\ssH}\,z_2(1\!-\!z_1\!-\!z_2)\!+\!\beta_{\ssH}}
        + \ln\frac{\alpha_{\ssH}\,z_2^2\!+\!\beta_{\ssH}}
        {-\alpha_{\ssH}\,z_2(1\!-\!z_3\!-\!z_2)\!+\!\beta_{\ssH}}
        \bigg],
        \nl
        \int_0^1\!\!\!dz_1dz_3\,\frac{z_1^n}{Y_\bbb^2} 
&=&
\frac{n}{\alpha_{\ssH}\beta_{\ssH}}\!\int_0^1\!\!\!dz_1\,z_1^{n-1}\bigg[
\ln\frac{\alpha_{\ssH}(1\!-\!z_2)(1\!-\!z_1\!-\!z_2)\!+\!\beta_{\ssH}}
        {-\alpha_{\ssH}\,z_2(1\!-\!z_1\!-\!z_2)\!+\!\beta_{\ssH}}
        + \ln\frac{\alpha_{\ssH}\,z_2^2\!+\!\beta_{\ssH}}
        {-\alpha_{\ssH}\,z_2(1\!-\!z_2)\!+\!\beta_{\ssH}}
\bigg],
\nl
\int_0^1\!\!\!dz_1dz_3\,\frac{z_3^k}{Y_\bbb^2} 
&=&
\frac{k}{\alpha_{\ssH}\beta_{\ssH}}\!\int_0^1\!\!\!dz_3\,z_3^{k-1}\bigg[
\ln\frac{\alpha_{\ssH}(1\!-\!z_2)(1\!-\!z_3\!-\!z_2)\!+\!\beta_{\ssH}}
        {-\alpha_{\ssH}\,z_2(1\!-\!z_3\!-\!z_2)\!+\!\beta_{\ssH}}
+ \ln\frac{\alpha_{\ssH}\,z_2^2\!+\!\beta_{\ssH}}
        {-\alpha_{\ssH}\,z_2(1\!-\!z_2)\!+\!\beta_{\ssH}}
\bigg],
\nl
\int_0^1\!\!\!dz_1dz_3\,\frac{1}{Y_\bbb^2} 
&=&
\frac{1}{\alpha_{\ssH}\beta_{\ssH}}\bigg[
\ln\frac{\alpha_{\ssH}(1\!-\!z_2)^2\!+\!\beta_{\ssH}}
        {-\alpha_{\ssH}\,z_2(1\!-\!z_2)\!+\!\beta_{\ssH}}
        + \ln\frac{\alpha_{\ssH}\,z_2^2\!+\!\beta_{\ssH}}
        {-\alpha_{\ssH}\,z_2(1\!-\!z_2)\!+\!\beta_{\ssH}}
\bigg],
\nl
\int_0^1\!\!\!dz_3\,\frac{1}{Y_\bbb} 
&=&
- \frac{1}{\alpha_{\ssH}(1\!-\!z_1\!-\!z_2)}
\ln\left( 1 - \frac{\alpha_{\ssH}(1\!-\!z_1\!-\!z_2)}
                   {\alpha_{\ssH}(1\!-\!z_2)(1\!-\!z_1\!-\!z_2)+\beta_{\ssH}}
   \right),
\eqa
where in the last equation, which is related to the tensor form factors 
proportional to $\delta^{\mu \nu}$, we can carry on numerical integration.
For the other four expressions in \eqn{longInts} we have to pay attention to the 
zeros of $\beta_{\ssH}$ (the factor $\alpha_{\ssH}^{-1}$ cancels with the 
numerator in \eqn{deffa}).
It can be easily proved that the real part of the expression contained in 
the squared bracket vanishes in the limit $\beta_{\ssH}\to 0$; however, an 
imaginary part can in general survive. Below threshold, for $s=-P^2<4\,m_3^2$, 
$\beta_{\ssH}$ does not change sign in the integration region; therefore, for 
the real part of the diagram below threshold (as done in~\cite{Hollik:2005ns}) 
it is enough to take the real part of the integrand and then integrate 
numerically. For $H\to\gamma\gamma$, we are interested also in the 
behavior above threshold, where we have to take into account the imaginary 
part of the squared bracket.
Following the approach of section~\ref{sec:collfinH}, we make explicit the $i\, 0$ 
prescription for masses and perform the following symbolic decomposition:
  \bq
  \frac{1}{\beta_{\ssH}}
  \ln\frac{{\cal Q}_1 \!+\! \beta_{\ssH}}{{\cal Q}_2 \!+\! \beta_{\ssH}}=
  \frac{1}{\beta_{\ssH}}\ln\frac{{\cal Q}_1 \!+\! \beta_{\ssH} \!-\! i\,0}
                       {{\cal Q}_2 \!+\! \beta_{\ssH} \!-\! i\,0}=
  \frac{1}{\beta_{\ssH}}\bigg[
  \ln\!\left(\! 
  1 \!+\! \frac{\beta_{\ssH}}{{\cal Q}_1 \!-\! i\,0}
  \!\right)
  - \ln\!\left(\! 
  1 \!+\! \frac{\beta_{\ssH}}{{\cal Q}_2 \!-\! i\,0}
  \!\right)
  + \ln\frac{{\cal Q}_1 - i\,0}{{\cal Q}_2 - i\,0}
  \bigg].
  \label{thisisneeded}
  \eq
The first two terms are regular when $\beta_{\ssH}\to0$; for the last term
of \eqn{thisisneeded} we notice that $\beta_{\ssH}$ is linearly dependent on $y$ 
and in the argument of the logarithm a factor $y$ can be simplified between 
${\cal Q}_1$ and ${\cal Q}_2$. Therefore, we can integrate by parts in $y$ getting
  \bqa
  V_{\!s}^{\ssH}
  &=&
  \frac{1}{P^2}\int_0^1\!\!\!\!dx\,dy\,dz_2\,\Bigg\{\!
  \frac{2}{\beta_{\ssH}}\,\bigg[
  \ln\!\left(\! 1 \!-\! \frac{\beta_{\ssH}}{\beta_{\!1}} \!\right)
  - \ln\!\left(\! 1 \!-\! \frac{\beta_{\ssH}}{\beta_{\!2}} \!\right)
  \!\bigg]
  + i\,\pi\,{\rm sign}(P^2)\frac{2}{\bar{\beta}_{\ssH}}
  \ln\!\left(\! 1 \!+\! \frac{\bar{\beta}_{\ssH}}{\beta_0} \!\right)
  \!\!\Bigg\},
  \nl
  V_{\!s}^{\ssH}[q_i^\mu q_j^\nu]
  &=&
  \frac{1}{P^2}
  \int dC_5\lpar x,y,\{z\}\rpar\,\Bigg\{\,
  \delta^{\mu\nu}
  \frac{a_{\bbb\!,ij}^{0}\,(1\!-\!y)}{2\alpha_{\!\ssH}(1\!-\!z_1\!-\!z_2)}
  \ln\!\left(\! 1 \!+\! \frac{\alpha_{\!\ssH}(1\!-\!z_1\!-\!z_2)}
                             {\beta_3} \!\right)
                           \nl
  &&
  +  \sum_{h,k=1}^{2}\,p_h^\mu\,p_k^\nu\,\Bigg[
  \sum_{n=1}^{4}\,
  \frac{\alpha_{\bbb\!,ij}^{n\!,hk}}{\beta_{\ssH}}
  \ln\!\left(\! 1 \!-\! \frac{\beta_{\ssH}}{\beta_{\!n}} \!\right)
  - i\,\pi\,{\rm sign}(P^2)\frac{1}{\bar{\beta}_{\ssH}}\sum_{l=1}^{2}
  \Theta_{\bbb\!,ij}^{l\!,hk}\!
  \ln\!\left(\! 1 \!+\! \frac{u_l\,\bar{\beta}_{\ssH}}{\beta_0} \!\right)
  \!\Bigg]
  \Bigg\},
  \eqa
with $u_1=1$ and $u_2=y$. Here we have introduced short-hand notations for
the $\beta$ factors,
  \bqa
  &&
  \beta_{\ssH}= \bar{\beta}_{\ssH}\,y + \beta_0,
  \qquad
  \bar{\beta}_{\ssH}= 
  (1\!-\!x)m_1^2
  + x\,\chiu{\bbb\!,2}
  - x\,(1\!-\!x)m_5^2,
  \qquad\quad\,\,\,
  \beta_0= x\,(1\!-\!x)m_5^2,
  \qquad
  \beta_{n} \!= \beta_{\!\ssH} \!+\! \alpha_{\!\ssH}\,\sigma_n,
  \nl
  &&
  \sigma_{1} = (1\!-\!z_2)^2,
  \quad\quad\,\,\,\,\,\,
  \sigma_{2} = - z_2(1\!-\!z_2),
  \quad
  \sigma_{3} = (1\!-\!z_2)(1\!-\!z_1\!-\!z_2),
  \quad
  \sigma_{4} = (1\!-\!z_3\!-\!z_2)(1\!-\!z_1\!-\!z_2),
  \eqa
for the $\Theta\,$-functions and for the $\alpha$, $\beta$, $\gamma$ and $\delta$ 
coefficients
  \bq
  \Theta_{\bbb\!,ij}^{l\!,hk}= 
  \beta_{\bbb\!,ij}^{l\!,hk} 
  + \gamma_{\bbb\!,ij}^{l\!,hk}\,\Theta(z_1\!+\!z_2\!-\!1)
  + \delta_{\bbb\!,ij}^{l\!,hk}\,
  \Theta(z_1\!+\!z_2\!-\!1)\Theta(z_3\!+\!z_2\!-\!1),
  \quad
  l= 1,2,
  \eq
\bqa
\ba{ll}
\alpha_{\bbb\!,11}^{1\!,11}= 2x^2(1\!-\!y)^2(1\!-\!z_2)^2, 
\;&\;
\alpha_{\bbb\!,11}^{2\!,11}= 
-\frac{x^2}{2}(1\!-\!y)^2( 4z_2^2\! -\! 6z_2\! +\! 3 ), 
\\
\alpha_{\bbb\!,11}^{3\!,11}= - 4\,x^2(1\!-\!y)^2(1\!-\!z_1\!-\!z_2), 
\;&\;
\alpha_{\bbb\!,11}^{4\!,11}= 0,
\\
\alpha_{\bbb\!,11}^{1\!,12}\!= \alpha_{\bbb\!,11}^{1\!,21}\!= 
- \!x (1\!-\!y) (1\!-\!z_2) [ 1\! - \!2x(1\!-\!y)(1\!-\!z_2) ],
\;&\; 
\alpha_{\bbb\!,11}^{2\!,12}\!= \alpha_{\bbb\!,11}^{2\!,21}\!= 
- \frac{x}{2} \! (1\!-\!y)[ 1\! +\! x(1\!-\!y)( \!4z_2^2 \!-\! 6 
z_2 \!+\! 1 ) ], 
\\
\alpha_{\bbb\!,11}^{3\!,12}\!= \alpha_{\bbb\!,11}^{3\!,21}\!= 
x(1\!-\!y)\{ 1 + 2(1\!-\!z_2)[1 - 2x(1\!-\!y)] \}, 
\;&\;
\alpha_{\bbb\!,11}^{4\!,12}= \alpha_{\bbb\!,11}^{4\!,21}= 
- x(1\!-\!y)[ 1 - x(1\!-\!y) ], 
\\
\alpha_{\bbb\!,11}^{1\!,22}= 
1 \!-\! 2x(1\!-\!y)(1\!-\!z_2)[1 \!-\! x(1\!-\!y)(1\!-\!z_2) ], 
\;&\;
\alpha_{\bbb\!,11}^{2\!,22}= 
- [ 1 \!-\! x(1\!-\!y) + \frac{x^2}{2} (1\!-\!y)^2
( 4z_2^2 \!-\! 6z_2 \!+\! 3 ) ], 
\\
\alpha_{\bbb\!,11}^{3\!,22}\!= 
- 2\,[ 1 \!-\! x(1\!-\!y) ]\,[ 1 \!-\! 2z_1\! -\! 
2x(1\!-\!y)(1\!-\!z_1\!-\!z_2) ], 
\;&\;
\alpha_{\bbb\!,11}^{4\!,22}= 0,
\ea
\nl
&&
\eqa
\bqa
\ba{ll}
\beta_{\bbb\!,11}^{1\!,11}= 0,
\;&\;
\beta_{\bbb\!,11}^{2\!,11}= 2x^2(1\!-\!y)(1\!-\!z_2)(1-2z_2), 
\\
\gamma_{\bbb\!,11}^{1\!,11}= 0,
\;&\;
\gamma_{\bbb\!,11}^{2\!,11}= 2x^2(1\!-\!y)( 5 - 4z_1 - 4z_2 ), 
\\
\delta_{\bbb\!,11}^{1\!,11}= \delta_{\bbb\!,11}^{2\!,11}= 
\beta_{\bbb\!,12}^{1\!,11}= \beta_{\bbb\!,21}^{1\!,11}= 0,
\;&\;
\beta_{\bbb\!,12}^{2\!,11}= \beta_{\bbb\!,21}^{2\!,11}= 
\frac{x}{2}[ 1\! +\! 2x(1\!-\!y)( 4z_2^2\! -\! 6z_2\! +\! 1 ) ], 
\\
\gamma_{\bbb\!,12}^{1\!,11}= \gamma_{\bbb\!,21}^{1\!,11}= 0,
\;&\;
\gamma_{\bbb\!,12}^{2\!,11}= \gamma_{\bbb\!,21}^{2\!,11}= 
- x(1-2z_2) [ 1 \!-\! 4x(1\!-\!y) ], 
\\
\delta_{\bbb\!,12}^{1\!,11}= \delta_{\bbb\!,21}^{1\!,11}= 0,
\;&\;
\delta_{\bbb\!,12}^{2\!,11}= \delta_{\bbb\!,21}^{2\!,11}= 
- 2x [ 1 \!-\! 2x(1\!-\!y) ], 
\\
\beta_{\bbb\!,22}^{1\!,11}= \frac{1}{2}, 
\;&\;
\beta_{\bbb\!,22}^{2\!,11}= x[ 1 - 2xz_2(1\!-\!y)(3-2z_2) ], 
\\
\gamma_{\bbb\!,22}^{1\!,11}= 3 - 4z_1, 
\;&\;
\gamma_{\bbb\!,22}^{2\!,11}= 
- 2x[ 5 \!-\! 4z_1 \!-\! 2z_2 \!-\! x(1\!-\!y)( 7 \!-\! 
4z_1 \!-\! 4z_2) ], 
\\
\delta_{\bbb\!,22}^{1\!,11}= 0,
\;&\;
\delta_{\bbb\!,22}^{2\!,11}= 0.
\ea
\nl
&&
\eqa
Note that we have presented here only the coefficients of
$V_{\!s}^{\ssH}[q_1^\mu q_1^\nu]$, the surviving configuration after the
reduction~$\,\otimes\,$~symmetrization procedure. Other tensor structures 
can be obtained with the help of the same procedure. 
\section{Numerical results \label{hicsunt}}
In this section we discuss the numerical results for the two-loop corrections to 
the decay widths of the processes $H\to gg $ and $H\to \gamma \gamma$, given by
\bq
\label{eq:DecayWidth}
\Gamma(H\to gg )= 
\frac{\alpha_s^2(\mu_R^2)\*G_F\*\mw^2}{4\*\sqrt{2}\*\pi^3\*\mh}\left|A^{gg}_{\rm{phys}}\right|^2
\quad\mbox{and}\quad
\Gamma(H\to\gamma\gamma)=
\frac{\alpha^2\*G_F\*\mw^2}{32\*\sqrt{2}\*\pi^3\*\mh}\left|A^{\gamma\gamma}_{\rm{phys}}\right|^2,
\eq
with $ A^{gg}_{\rm phys}\,
\frac{\alpha_s(\mu_R^2)}{2\,\pi}\Bigl(\sqrt{2}\,\gf\,\mws\Bigr)^{1/2}\!\!={\cal
A}$ from \eqn{AMP:ampSplit4Glu} and $ A^{\gamma\gamma}_{\rm
phys}\,\frac{\alpha}{2\,\pi}\Bigl(\sqrt{2}\,\gf\,\mws\Bigr)^{1/2}\!\!={\cal
A}$ from \eqn{AMP:ampSplit3bb}. 

One important production mechanism of the Standard Model Higgs boson at the LHC is 
the gluon-fusion channel, $pp \to gg +X \to H + X$. Its partonic cross section 
$\sigma$, to LO in QCD, can by related to $A^{gg}_{\rm phys}$ via
\bq
\sigma( gg \to H)=
\frac{\alpha_s^2(\mu_R^2)\*G_F\*\mw^2}{32\*\sqrt{2}\*\pi\*\mh^4}
\left|A^{gg}_{\rm{phys}}\right|^2.
\eq

The relative correction $\delta$, induced by the higher order corrections, is 
given by $\Gamma=\Gamma_0\*(1+\delta)$($\sigma=\sigma_0\*(1+\delta)$), where 
$\Gamma_0$ ($\sigma_0$) is the lowest order quantity. 
For the decay $H\to \gamma\gamma$ we split the relative correction in
$\delta=\delta_{\EW}+\delta_{\QCD}$, to distinguish the contributions
arising from electroweak and QCD corrections. For the process $H\to
 gg $($ gg \to H$) we consider only two-loop electroweak corrections;
interplay between electroweak and QCD corrections has been discussed
in Ref.~\cite{Actis:2008ug} where an estimate of the remaining theoretical uncertainty 
is also presented.

As discussed in section~\ref{sqroots} and section~\ref{sec:LogSing}, 
threshold singularities appear in both reactions if we restrict the calculation
to the RM scheme (see section~\ref{subsec:complexmass}). For the case 
$H\to \gamma\gamma$ we consider a 
phenomenologically relevant Higgs-mass range from $100\,$GeV to $170\,$GeV, which 
contains the $WW\,$ threshold. For the gluon-gluon case we consider a Higgs-mass
range from $100\,$GeV to $500\,$GeV. Here we cross not only the $WW\,$ threshold, 
but also the $ZZ\,$ and $\overline{t} t\,$ thresholds. 
The $WW\,$ and $ZZ\,$ thresholds have a different behavior as compared to
the $\overline{t}t\,$ threshold. Indeed, the amplitude around the $\overline{t}t\,$ 
threshold contains a potentially dangerous two-point function, as explained in 
section~\ref{NormTH}, but in contrast to the $WW\,$ and $ZZ\,$ cases this singular 
function (for $\mhs = 4\,M_t^2$) is here protected by a multiplicative factor
$\beta^2=1-4\*M_t^2/s$. In order to cure these singularities we
have introduced complex masses~\cite{Denner:2005fg} as described in
section~\ref{subsec:complexmass}. We stress the fact that at the single $W\,$ and 
$Z$ thresholds no special enhancement occurs.

All light-fermion masses have been set to zero in the collinear-free amplitude 
and we have defined the $W$-and $Z$-boson {\em experimental} complex poles by
\bq
s_{j} = \mu_{j}\,\lpar \mu_{j} - i\,\gamma_{j}\rpar,
\quad 
\mu^2_{j} = M^2_j - \Gamma^2_{j},
\quad
\gamma_{j} = \Gamma_{j}\,\lpar 1 - \Gamma^2_{j}/(2\,M^2_j)\rpar,
\label{replacement}
\eq
with $j=\{W,Z\}$. As input parameters for the numerical evaluation we have used the
following values~\cite{Amsler:2008zz,:2007bxa}:
\[
\begin{array}{llll}
\mw = 80.398\,\GeV,  \;\; & \;\; 
\mz = 91.1876\,\GeV, \;\; & \;\;
M_t = 170.9\,\GeV,   \;\; & \;\;
\Gamma_{\ssW} = 2.093\,\GeV,\\
\gf = 1.16637\,\times\,10^{-5}\,\GeV^{-2},  \;\; & \;\; 
\alpha(0) = 1/137.0359911,                  \;\; & \;\; 
\alpha_{\ssS}\lpar \mz\rpar= 0.118,         \;\; & \;\; 
\Gamma_{\ssZ} = 2.4952\,\GeV.
\end{array}
\]

\begin{figure}[!htb]
\begin{center}
\includegraphics[bb=0 0 567 384,width=12.cm]{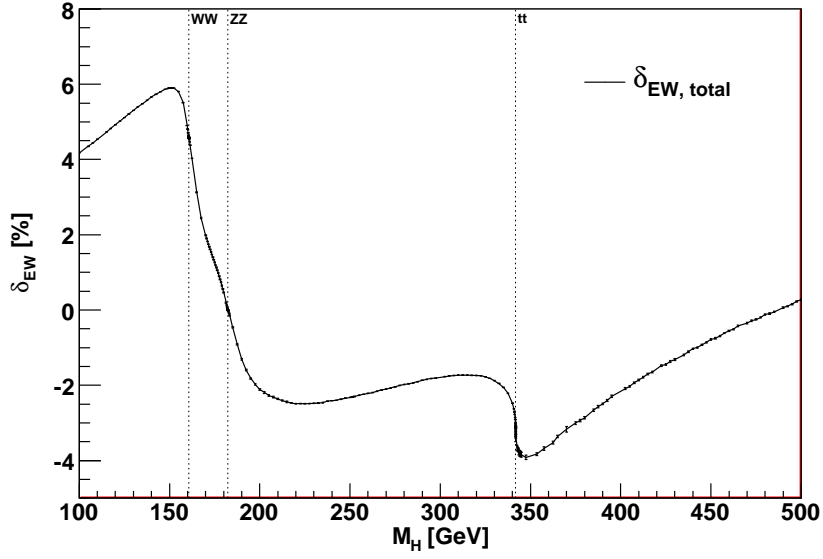}
\caption[]{\label{fig:dEWHgluglu} The two-loop electroweak percentage
  corrections for the decay width $\Gamma(H\to
   gg )$ and the total partonic cross section
  $\sigma( gg \to H)$. The solid line denotes the total electroweak 
  correction, including also top quarks (first + second + third generation).  
  The vertical dash-lines  indicate the location of electroweak thresholds.}
\end{center}
\end{figure}

For $\Gamma(H\to gg)$ and $\sigma( gg \to H)$ the behavior of $\delta_{\EW}$ 
as a function of $\mh$ is given in \fig{fig:dEWHgluglu}. For the gluon-gluon case, 
treated in the CM scheme, we observe a smooth behavior in the full range of 
$\mh$ where all cusps present in the MCM scheme have disappeared; we may 
conclude that results in the CM scheme nicely interpolate those of the
RM scheme around thresholds. Therefore, all pathological aspects 
associated to the crossing of thresholds have disappeared. It is also worth 
mentioning that the CM scheme greatly improves stability in the numerical 
evaluation of all master integrals, especially those in the $V^{\ssH}$ family.

In the shown Higgs-mass range the percentage correction varies between about 
+6\% and -4\%. 

It is important to consider this result in more details: around the $WW\,$ 
threshold we find a maximum for the total electroweak percentage correction, and 
only a light shoulder of the curve around the $ZZ\,$ threshold. Both 
characteristics are less suppressed if one considers only light fermions
with real masses, shown in \fig{Compaone}. Let us 
define light-fermion correction as those coming from the first and second 
generation of light (massless) quarks and the one from bottom quarks only; each of 
these three classes of diagrams constitutes a gauge-invariant subset.

It is worth noting that below the $WW\,$ threshold the contributions from top 
quarks are small but they become significant above the $ZZ\,$ threshold and 
even more important around the $\overline{t}t\,$ threshold, where the curve 
exhibits a minimum, which is absent for light fermions, see \fig{Compaone}. In 
this region the top-quark contribution leads to a sizable effect of about -4\%.

The results for $\Gamma(H \to gg)$  have been used to derive NLO electroweak corrections
to Higgs production at hadron colliders in Ref.~\cite{Actis:2008ug}.
 
\begin{figure}[!hbt]
\begin{center}
\includegraphics[bb=0 0 567 384,width=12.cm]{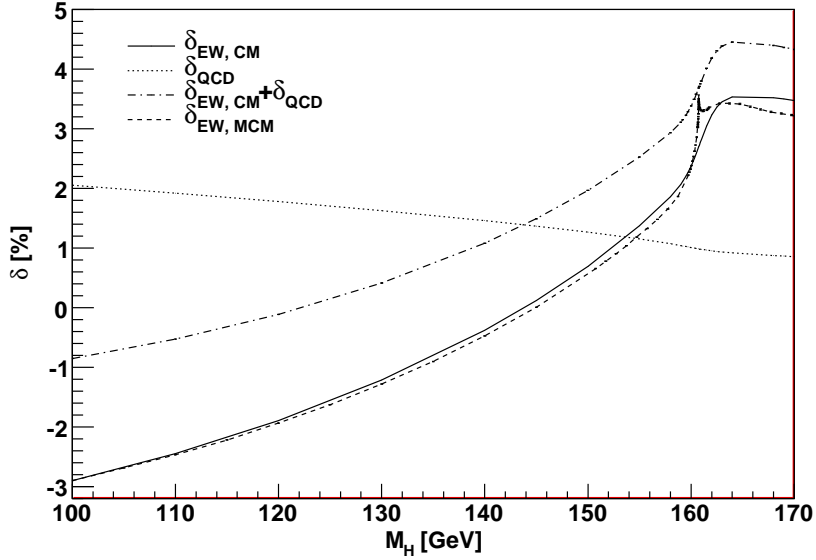}
\caption[]{\label{dEWHgamgam} The two-loop electroweak and QCD percentage
  corrections for the $H\to \gamma\gamma$ decay.}
\end{center}
\end{figure}

\begin{figure}[!hbt]
\begin{center}
\includegraphics[bb=0 0 567 384,width=12.cm]{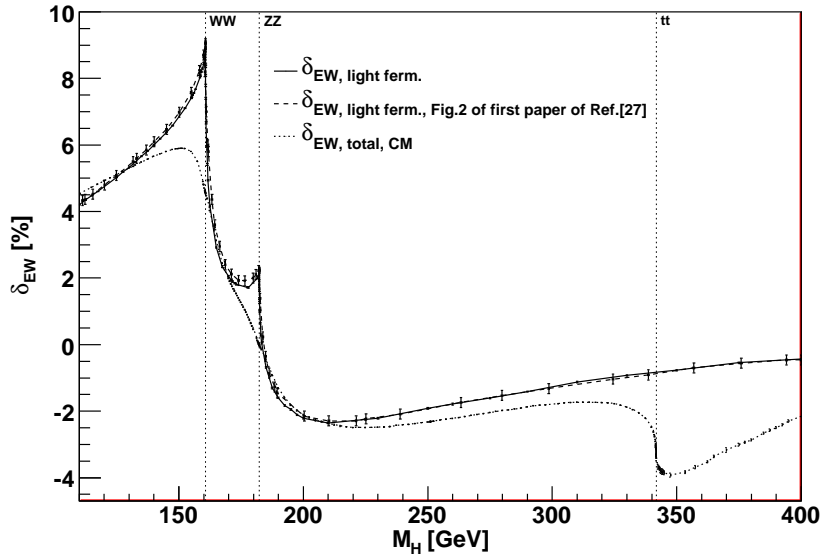}
\caption[]{\label{Compaone} The two-loop electroweak percentage
  corrections for $H\to gg$ showing a comparison: total in CM
  scheme, light fermions only in RM scheme, light fermions only obtained
  from Ref.~\cite{Aglietti:2004nj} (see main text for the error bars).}
\end{center}
\end{figure}

The numerical result for the percentage correction to the partial width
$\Gamma(H\to \gamma\gamma)$ has been presented in Ref.\cite{Passarino:2007fp} in 
the minimal complex-mass (MCM) setup and is shown in \fig{dEWHgamgam},
as well as the extension to the full complex-mass (CM) scheme.
The QCD corrections (dotted curve) as well as the CM electroweak
corrections (solid-line) are shown separately. If it is true that below the
$WW\,$ threshold $\delta_{\QCD}$ and $\delta_{\EW}$ almost compensate
and lead to a small total correction (dash-dot line), the new results tell us
that, above the $WW\,$ threshold, both contributions are positive and lead to a
sizable overall effect of approximately $4\%$. In the considered Higgs-boson mass
range the total correction varies between $-1\%<\delta_{\rm tot}<4\%$.

To summarize, the electroweak correction around the $WW\,$ threshold shown in
\fig{dEWHgamgam} (dash-line) has been produced by minimally modifying the 
two-loop amplitude through the usage of a complex $\mw$ mass. We have extended 
our calculation in order to treat the whole amplitude completely with complex 
$W\,$ and $Z\,$ masses. This allows us to study the remaining cusp and the 
influence of the complex mass on the electroweak correction, which is shown 
through the solid line.

We have compared our numerical results in the region below the $WW$
threshold with those of Ref.~\cite{Degrassi:2004mx}. Eqs.(8) and (11) of
Ref.~\cite{Degrassi:2004mx} contain typos%
\footnote{Thanks to G. Degrassi and F. Maltoni for prompt confirmation.}
; once we correct, very good agreement between our calculations, up to
$130-140\,$GeV, is found. To be more precise, we split the total
contribution into light fermions and top (namely diagrams
with at least one top line); for the light-fermion contribution the
agreement is always very good because the result of
Ref.~\cite{Aglietti:2004nj} is exact, whereas, for the top part, the
comparison gives a check of how good is the expansion of
Ref.~\cite{Degrassi:2004mx}.  We confirm what is expected, for light
Higgs-boson masses (say, around $115\,$GeV) the agreement between our results is
again very good and starts deteriorating for heavier values (say,
starting from $140\,$GeV).
In \fig{Compaone} we have summarized our findings for $\delta_{\EW}$ in
gluon-gluon fusion; here we also include the light-fermion part of the corrections
compared with the ones of Ref.~\cite{Aglietti:2004nj}. 
The result of Ref.~\cite{Aglietti:2004nj} is known completely in terms of harmonic
poly-logarithms. For simplicity we used the tool
EasyNData~\cite{Uwer:2007rs} to read out the result
from Fig.~2 of the first paper of Ref.~\cite{Aglietti:2004nj}, which is shown
in \fig{Compaone} as a dash-line. The error bars originate from our
estimation of how good we can read out the data from the plot.
Once again, the relevant news are the behavior above $WW$ threshold induced by 
top quarks and the around-threshold behavior in the CM scheme.
\begin{figure}[!hbt]
\begin{center}
\includegraphics[bb=0 0 567 384,width=12.cm]{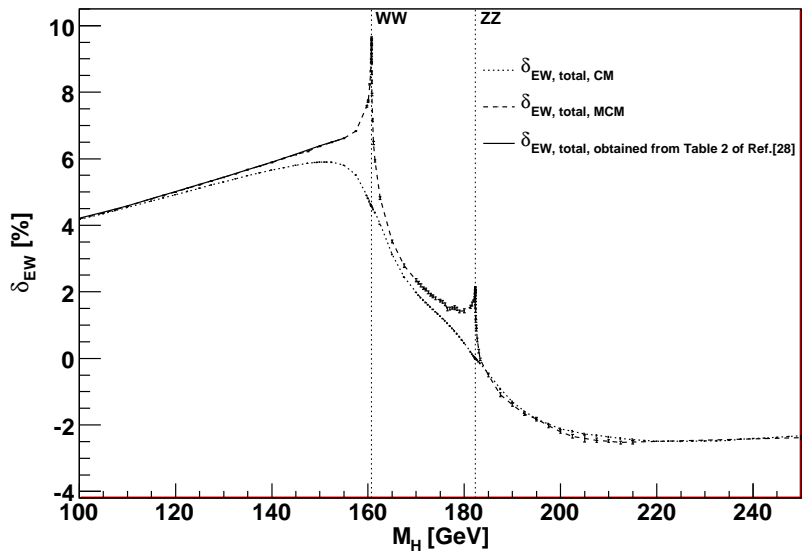}
\caption[]{\label{Compatwo} An analogous plot to \fig{Compaone} for a
narrower range of the Higgs-boson mass. Here we compare our results in
the CM and MCM schemes with the results (below 155 GeV) of
Ref.~\cite{Degrassi:2004mx}}
\end{center}
\end{figure}

Differences become even more striking when we consider the total
corrections in a narrower range in the Higgs-boson mass, as done in
\fig{Compatwo}.
Here we concentrate our attention
around the region containing both $WW$ and $ZZ$ thresholds and compare our
final result (CM scheme) with the one obtained from Tab.~2 of 
Ref.~\cite{Degrassi:2004mx}.
 
The very good agreement when we use the MCM scheme and the subsequent extension to 
CM scheme make us confident of the goodness of the result.
\section{Conclusions}
\label{sec:conc}
In this paper we have provided all technical details for a stand-alone numerical 
calculation of the full two-loop corrections to the decay widths
$H \to \gamma \gamma$ and $H \to gg$, where $H$ stands for the Standard 
Model Higgs boson. The techniques introduced in this context, however, have a 
much wider range of application, i.e. they are general enough to be used for 
all kinematical configurations of $1 \to 2$ processes at the two-loop level.

The generation of diagrams for any given process has always been performed
with the FORM~\cite{Vermaseren:2000nd} program {$\GS$}~\cite{GraphShot} 
which performs simplifications and symbolic manipulations of the
loop integrals, like performing traces, removing reducible scalar products and 
symmetrizing the integrals. 

The amplitudes contain both QCD and electroweak contributions and our
strategy has been to present complete calculations where all enhanced terms 
are extracted analytically and their cancellation shown, whenever it occurs; 
the remainder of the amplitude is a multi-dimensional integral in parametric 
space with an integrand which, after suitable algebraic manipulations, is 
transformed into a smooth function therefore avoiding well-known problems of 
numerical instabilities, even in those regions of parametric space which are
notoriously plagued by normal threshold singularities; the latter are always
cured by the introduction of complex masses, solving at the same time a
conceptual problem, internal unstable particles, and a practical one, avoiding
numerical instabilities associated with the behavior around singularities lying 
on the real axis of the external Mandelstam invariant.

The strategy for the rest of the calculations consists in regularizing 
the collinear singularities and in extracting the singular pieces from all 
singular terms. After checking -- analytically -- that singular parts cancel 
in the total (if it applies), one can safely take rid of the regularization 
parameter and include all collinear-free remainders into the total amplitude. 

As far as numerical results are concerned we can safely state that the methods
developed in this paper produce extremely accurate predictions for any value 
of the Higgs-boson mass, including the full dependence on the $W$-, $Z$- and 
Higgs-boson masses and on the top-quark mass. A consistent and gauge-invariant 
treatment of unstable particles allows to produce precise results around the 
physical thresholds: both a minimal and a complete version of this scheme are 
studied. It is found that the use of the complex mass scheme smoothens the 
threshold singularities and that a complete implementation of this scheme is
needed in order to get the NLO electroweak corrections under control. 

When applied to the gluon fusion channel, 
$pp \to gg +X \to H + X$, our results show that the electroweak scaling factor 
which should multiply the QCD-corrected cross-section is between 
$-4\%$ and $+6\%$ in the range $100\,\GeV < \mh < 500\,\GeV$, without incongruent 
effects around the physical electroweak thresholds.
Finally we observe that around the $\overline{t}t\,$ threshold the top-quark 
contribution to $H \to gg$ leads to a sizable effect of about $-4\%$. 
\Acknowledgments
We gratefully acknowledge important discussions with Giuseppe Degrassi,
Ansgar Denner, Stefan Dittmaier, Robert Harlander, Philipp Kant, Massimiliano
Grazzini, Fabio Maltoni, Michael Spira and Peter Zerwas.

The authors thank the Galileo Galilei Institute for Theoretical Physics
for the hospitality and the INFN for partial support during the
completion of this work.

This work has been supported by MIUR under contract 2001023713$\_$006, 
by the European Community's Marie Curie Research Training Network {\it Tools and 
Precision Calculations for Physics Discoveries at Colliders} under contract 
MRTN-CT-2006-035505, by the U.S. Department of Energy under contract No. 
DE-AC02-98CH10886 and by the Deutsche Forschungsgemeinschaft through 
Sonderforschungsbereich/Transregio 9 {\it Computergest\"utzte Theoretische 
Teilchenphysik}. Feynman diagrams have been drawn with the packages 
{\sc Axodraw}~\cite{Vermaseren:1994je} and {\sc Jaxo\-draw}~\cite{Binosi:2003yf}.
\clearpage
\appendix
\section{Self-energies, vertices and tadpoles\label{app:diagrams}}
\label{app:topos}
In this section we collect our conventions for the diagrams involved in the 
paper.
\begin{figure}[!ht]
\vspace{-0.9cm}
\bqa
{[f(q)]}\;
\raisebox{0.1cm}{\scalebox{0.7}{
\begin{picture}(130,75)(0,0)
 \SetWidth{1.2}
 \Line(0,0)(35,0)         
 \LongArrow(5,8)(25,8)          \Text(10,13)[cb]{$P$}
 \CArc(60,0)(25,0,360)          \Text(60,29)[cb]{$1$}\Text(60,-35)[cb]{$2$}
 \Line(85,0)(120,0)         
 \Text(10,-30)[cb]{\Large $B$}
\end{picture}
}}
\!\!\!\!\!
&=&
\quad
\frac{\mu^{\ep}}{i\,\pi^2}\int\!d^nq\,\frac{f(q)}{[1][2]},
\quad \hbox{with} \quad
\left\{
\ba{l}
{[1]} = q^2+m^2_1 \\	
{[2]} = (q+P)^2+m^2_2
\ea
\right.
\nl
{[f(q)]}\;
\raisebox{0.1cm}{\scalebox{0.7}{
\begin{picture}(130,75)(0,0)
 \SetWidth{1.2}
 \Line(0,0)(40,0)         
 \LongArrow(5,8)(25,8)          \Text(10,13)[cb]{$-P$}
 \Line(128,-53)(100,-35)  
 \LongArrow(118,-56)(104,-47)   \Text(125,-65)[cb]{$p_1$}
 \Line(128,53)(100,35)    
 \LongArrow(118,56)(104,47)     \Text(125,57)[cb]{$p_2$}
 \Line(100,-35)(40,0)           \Text(70,-30)[cb]{$1$}
 \Line(100,-35)(100,35)         \Text(107,-3)[cb]{$2$}
 \Line(100,35)(40,0)           \Text(70,25)[cb]{$3$}
 \Text(0,-40)[cb]{\Large $C$}
\end{picture}
}}
&=&
\quad
\frac{\mu^{\ep}}{i\,\pi^2}\int\!d^nq\,\frac{f(q)}{[1][2][3]},
\quad \hbox{with} \quad
\left\{
\ba{l}
{[1]} = q^2+m^2_1 \\
{[2]} = (q+p_1)^2+m^2_2 \\
{[3]} = (q+P)^2+m^2_3
\ea
\right.
\nn
\eqa
\vspace{0.6cm}
\caption[]{The one-loop self-energy and vertex. 
$f$ is a generic polynomial in the loop momentum $q$.
The dimension of the space-time is $n=4-\ep$ and $\mu$ is the 
renormalization scale.} 
\label{OL}
\end{figure}
\begin{figure}[!ht]
\vspace{-0.9cm}
\bqa
{[f(q_1,q_2)]}\;
\raisebox{0.1cm}{\scalebox{0.7}{
\begin{picture}(130,75)(0,0)
 \SetWidth{1.2}
 \Line(0,0)(35,0)         
 \LongArrow(5,8)(25,8)          \Text(10,13)[cb]{$P$}
 \CArc(60,0)(25,0,360)          \Text(60,29)[cb]{$1$}\Text(60,-35)[cb]{$3$}
 \Line(35,0)(85,0)              \Text(60,3)[cb]{$2$}
 \Line(85,0)(120,0)         
 \Text(10,-30)[cb]{\Large $S^{\ssA}$}
\end{picture}
}}
\!\!\!\!\!
&=&
\quad
\frac{\mu^{2\ep}}{\pi^4}\int\!d^nq_1d^nq_2\,\frac{f(q_1,q_2)}{[1][2][3]},
\quad \hbox{with} \quad
\left\{
\ba{l}
{[1]} = q^2_1+m^2_1 \\
{[2]} = (q_1-q_2+P)^2+m^2_2 \\
{[3]} = q_2^2+m^2_3
\ea
\right.
\nl
{[f(q_1,q_2)]}\;
\raisebox{0.1cm}{\scalebox{0.7}{
\begin{picture}(130,75)(0,0)
 \SetWidth{1.2}
 \Line(0,0)(35,0)         
 \LongArrow(5,8)(25,8)          \Text(10,13)[cb]{$P$}
 \CArc(60,0)(25,0,360)          \Text(40,22)[cb]{$1$}
                                \Text(80,22)[cb]{$3$}
                                \Text(60,-35)[cb]{$4$}
 \CArc(35,25)(25,-90,0)         \Text(60,0)[cb]{$2$}
 \Line(85,0)(120,0)         
 \Text(10,-30)[cb]{\Large $S^{\ssC}$}
\end{picture}
}}
\!\!\!\!\!
&=&
\quad
\frac{\mu^{2\ep}}{\pi^4}\int\!d^nq_1d^nq_2\,\frac{f(q_1,q_2)}{[1][2][3][4]},
\quad \hbox{with} \quad
\left\{
\ba{l}
{[1]} = q^2_1+m^2_1 \\
{[2]} = (q_1-q_2)^2+m^2_2 \\
{[3]} = q_2^2+m^2_3 \\
{[4]} = (q_2+P)^2+m^2_4
\ea
\right.
\nl
{[f(q_1,q_2)]}\;
\raisebox{0.1cm}{\scalebox{0.7}{
\begin{picture}(130,75)(0,0)
 \SetWidth{1.2}
 \Line(0,0)(35,0)         
 \LongArrow(5,8)(25,8)          \Text(10,13)[cb]{$P$}
 \CArc(60,0)(25,0,360)          \Text(60,29)[cb]{$1$}
                                \Text(88,11)[cb]{$3$}
                                \Text(60,-35)[cb]{$4$}
                                \Text(31,11)[cb]{$5$}
 \CArc(60,35)(25,-135,-45)      \Text(60,0)[cb]{$2$}
 \Line(85,0)(120,0)         
 \Text(10,-30)[cb]{\Large $S^{\ssE}$}
\end{picture}
}}
\!\!\!\!\!
&=&
\quad
\frac{\mu^{2\ep}}{\pi^4}\int\!d^nq_1d^nq_2\,\frac{f(q_1,q_2)}{[1][2][3][4][5]},
\quad \hbox{with} \quad
\left\{
\ba{l}
{[1]} = q^2_1+m^2_1 \\
{[2]} = (q_1-q_2)^2+m^2_2 \\
{[3]} = q_2^2+m^2_3 \\
{[4]} = (q_2+P)^2+m^2_4 \\
{[5]} = q_2^2+m^2_5
\ea
\right.
\nl
{[f(q_1,q_2)]}\;
\raisebox{0.1cm}{\scalebox{0.7}{
\begin{picture}(130,75)(0,0)
 \SetWidth{1.2}
 \Line(0,0)(35,0)         
 \LongArrow(5,8)(25,8)          \Text(10,13)[cb]{$P$}
 \CArc(60,0)(25,0,360)          \Text(34,15)[cb]{$1$}
                                \Text(86,15)[cb]{$4$}
                                \Text(33,-21)[cb]{$2$}
                                \Text(87,-21)[cb]{$5$}
 \Line(60,25)(60,-25)           \Text(67,-3)[cb]{$3$}
 \Line(85,0)(120,0)         
 \Text(10,-30)[cb]{\Large $S^{\ssD}$}
\end{picture}
}}
\!\!\!\!\!
&=&
\quad
\frac{\mu^{2\ep}}{\pi^4}\int\!d^nq_1d^nq_2\,\frac{f(q_1,q_2)}{[1][2][3][4][5]},
\quad \hbox{with} \quad
\left\{
\ba{l}
{[1]} = q^2_1+m^2_1 \\
{[2]} = (q_1+P)^2+m^2_2 \\
{[3]} = (q_1-q_2)^2+m^2_3 \\
{[4]} = q_2^2+m^2_4 \\
{[5]} = (q_2+P)^2+m^2_5
\ea
\right.
\nn
\eqa
\vspace{-0.6cm}
\caption[]{The irreducible two-loop self-energies diagrams. 
$f$ is a generic polynomial in the loop momenta $q_1$ and $q_2$.
The dimension of the space-time is $n=4-\ep$ and $\mu$ is the 
renormalization scale.} 
\label{TLselfenrgies}
\end{figure}
\begin{figure}[!ht]
\vspace{-0.9cm}
\bqa
{[f(q_1,q_2)]}\;
\raisebox{0.1cm}{\scalebox{0.7}{
\begin{picture}(130,75)(0,0)
 \SetWidth{1.2}
 \Line(0,0)(40,0)         
 \LongArrow(5,8)(25,8)          \Text(10,13)[cb]{$-P$}
 \Line(128,-53)(100,-35)  
 \LongArrow(128,-43)(114,-34)   \Text(138,-50)[cb]{$p_1$}
 \Line(128,53)(100,35)    
 \LongArrow(128,43)(114,34)     \Text(138,42)[cb]{$p_2$}
 \CArc(100,-35)(70,90,150)      \Text(64,31)[cb]{$1$}
 \CArc(40,70)(70,270,330)       \Text(80,-2)[cb]{$2$}
 \Line(100,-35)(40,0)           \Text(70,-30)[cb]{$4$}
 \Line(100,-35)(100,35)         \Text(107,-3)[cb]{$3$}
 \Text(0,-40)[cb]{\Large $V^{\aba}$}
\end{picture}
}}
\!\!\!\!\!
&=&
\quad
\frac{\mu^{2\ep}}{\pi^4}\int\!d^nq_1d^nq_2\,\frac{f(q_1,q_2)}{[1][2][3][4]},
\quad \hbox{with} \quad
\left\{
\ba{l}
{[1]} = q^2_1+m^2_1 \\
{[2]} = (q_1-q_2)^2+m^2_2 \\
{[3]} = (q_2+p_2)^2+m^2_3 \\
{[4]} = (q_2+P)^2+m^2_4
\ea
\right.
\nl
{[f(q_1,q_2)]}\;
\raisebox{0.1cm}{\scalebox{0.7}{
\begin{picture}(130,75)(0,0)
 \SetWidth{1.2}
 \Line(0,0)(42,0)         
 \LongArrow(5,8)(25,8)          \Text(10,13)[cb]{$-P$}
 \Line(128,-53)(100,-35)  
 \LongArrow(128,-43)(114,-34)   \Text(138,-50)[cb]{$p_1$}
 \Line(128,53)(100,35)    
 \LongArrow(128,43)(114,34)     \Text(138,42)[cb]{$p_2$}
 \CArc(55,-9)(15,0,360)         \Text(75,-3)[cb]{$1$}\Text(38,-27)[cb]{$2$}
 \Line(100,-35)(67,-15.75)      \Text(80,-37)[cb]{$3$}
 \Line(100,-35)(100,35)         \Text(107,-3)[cb]{$4$}
 \Line(100,35)(45,3)            \Text(68,23)[cb]{$5$}
 \Text(0,-40)[cb]{\Large $V^{\aca}$}
\end{picture}
}}
\!\!\!\!\!
&=&
\quad
\frac{\mu^{2\ep}}{\pi^4}\int\!d^nq_1d^nq_2\,\frac{f(q_1,q_2)}{[1][2][3][4][5]},
\quad \hbox{with} \quad
\left\{
\ba{l}
{[1]} = q^2_1+m^2_1 \\
{[2]} = (q_1-q_2)^2+m^2_2 \\
{[3]} = q_2^2+m^2_3 \\
{[4]} = (q_2+p_1)^2+m^2_4 \\
{[5]} = (q_2+P)^2+m^2_5 
\ea
\right.
\nn\\[-0.2cm]
{[f(q_1,q_2)]}\;
\raisebox{0.1cm}{\scalebox{0.7}{
\begin{picture}(130,75)(0,0)
 \SetWidth{1.2}
 \Line(0,0)(42,0)         
 \LongArrow(5,8)(25,8)          \Text(10,13)[cb]{$-P$}
 \Line(128,-53)(100,-35)  
 \LongArrow(128,-43)(114,-34)   \Text(138,-50)[cb]{$p_1$}
 \Line(128,53)(100,35)    
 \LongArrow(128,43)(114,34)     \Text(138,42)[cb]{$p_2$}
 \Line(57,-10)(40,0)            \Text(46,-16)[cb]{$6$}
 \CArc(70,-17.5)(15,0,360)      \Text(80,0)[cb]{$1$}\Text(58,-40)[cb]{$2$}
 \Line(100,-35)(83,-25)         \Text(89,-42)[cb]{$3$}
 \Line(100,-35)(100,35)         \Text(107,-3)[cb]{$4$}
 \Line(100,35)(40,0)            \Text(68,23)[cb]{$5$}
 \Text(0,-40)[cb]{\Large $V^{\ada}$}
\end{picture}
}}
\!\!\!\!\!
&=&
\quad
\frac{\mu^{2\ep}}{\pi^4}\int\!d^nq_1d^nq_2\,\frac{f(q_1,q_2)}{[1][2][3][4][5]},
\quad \hbox{with} \quad
\left\{
\ba{l}
{[1]} = q^2_1+m^2_1 \\
{[2]} = (q_1-q_2)^2+m^2_2 \\
{[3]} = q_2^2+m^2_3 \\
{[4]} = (q_2+p_1)^2+m^2_4 \\
{[5]} = (q_2+P)^2+m^2_5 \\
{[6]} = q_2^2+m^2_6 
\ea
\right.
\nn\\[-0.4cm]
{[f(q_1,q_2)]}\;
\raisebox{0.1cm}{\scalebox{0.7}{
\begin{picture}(130,75)(0,0)
 \SetWidth{1.2}
 \Line(0,0)(42,0)         
 \LongArrow(5,8)(25,8)          \Text(10,13)[cb]{$-P$}
 \Line(128,-53)(100,-35)  
 \LongArrow(128,-43)(114,-34)   \Text(138,-50)[cb]{$p_1$}
 \Line(128,53)(100,35)    
 \LongArrow(128,43)(114,34)     \Text(138,42)[cb]{$p_2$}
 \Line(100,-35)(40,0)           \Text(70,-33)[cb]{$1$}
 \Line(100,0)(100,35)           \Text(107,-21)[cb]{$2$}
 \Line(100,0)(40,0)             \Text(83,5)[cb]{$3$}
 \Line(100,-35)(100,0)          \Text(107,15)[cb]{$4$}
 \Line(100,35)(40,0)            \Text(70,25)[cb]{$5$}
 \Text(0,-40)[cb]{\Large $V^{\bba}$}
\end{picture}
}}
\!\!\!\!\!
&=&
\quad
\frac{\mu^{2\ep}}{\pi^4}\int\!d^nq_1d^nq_2\,\frac{f(q_1,q_2)}{[1][2][3][4][5]},
\quad \hbox{with} \quad
\left\{
\ba{l}
{[1]} = q^2_1+m^2_1 \\
{[2]} = (q_1+p_1)^2+m^2_2 \\
{[3]} = (q_1-q_2)^2+m^2_3 \\
{[4]} = (q_2+p_1)^2+m^2_4 \\
{[5]} = (q_2+P)^2+m^2_5 
\ea
\right.
\nn\\[-0.2cm]
{[f(q_1,q_2)]}\;
\raisebox{0.1cm}{\scalebox{0.7}{
\begin{picture}(130,75)(0,0)
 \SetWidth{1.2}
 \Line(0,0)(42,0)         
 \LongArrow(5,8)(25,8)          \Text(10,13)[cb]{$-P$}
 \Line(128,-53)(100,-35)  
 \LongArrow(128,-43)(114,-34)   \Text(138,-50)[cb]{$p_1$}
 \Line(128,53)(100,35)    
 \LongArrow(128,43)(114,34)     \Text(138,42)[cb]{$p_2$}
 \Line(70,-17.5)(40,0)             \Text(53,-21)[cb]{$1$}
 \Line(70,17.5)(40,0)              \Text(53,14)[cb]{$2$}
 \Line(70,-17.5)(70,17.5)          \Text(77,-3)[cb]{$3$}
 \Line(100,-35)(70,-17.5)          \Text(82,-39)[cb]{$4$}
 \Line(100,-35)(100,35)            \Text(107,-3)[cb]{$5$}
 \Line(100,35)(70,17.5)            \Text(82,31)[cb]{$6$}
 \Text(0,-40)[cb]{\Large $V^{\bca}$}
\end{picture}
}}
\!\!\!\!\!
&=&
\quad
\frac{\mu^{2\ep}}{\pi^4}\int\!d^nq_1d^nq_2\,
\frac{f(q_1,q_2)}{[1][2][3][4][5][6]},
\quad \hbox{with} \quad
\left\{
\ba{l}
{[1]} = q^2_1+m^2_1 \\
{[2]} = (q_1+P)^2+m^2_2 \\
{[3]} = (q_1-q_2)^2+m^2_3 \\
{[4]} = q^2_2+m^2_4 \\
{[5]} = (q_2+p_1)^2+m^2_5 \\
{[6]} = (q_2+P)^2+m^2_6 
\ea
\right.
\nn\\[-0.4cm]
{[f(q_1,q_2)]}\;
\raisebox{0.1cm}{\scalebox{0.7}{
\begin{picture}(130,75)(0,0)
 \SetWidth{1.2}
 \Line(0,0)(42,0)         
 \LongArrow(5,8)(25,8)          \Text(10,13)[cb]{$-P$}
 \Line(128,-53)(100,-35)  
 \LongArrow(128,-43)(114,-34)   \Text(138,-50)[cb]{$p_1$}
 \Line(128,53)(100,35)    
 \LongArrow(128,43)(114,34)     \Text(138,42)[cb]{$p_2$}
 \Line(70,-17.5)(100,35)                            \Text(97,10)[cb]{$2$}
 \Line(100,35)(70,17.5)                             \Text(82,31)[cb]{$1$}
 \Line(70,-17.5)(40,0)                              \Text(53,-21)[cb]{$4$}
 \Line(70,17.5)(40,0)                               \Text(53,14)[cb]{$3$}
 \Line(100,-35)(82,-3.5)\Line(78,3.5)(70,17.5)      \Text(97,-16)[cb]{$6$}
 \Line(100,-35)(70,-17.5)                           \Text(82,-39)[cb]{$5$}
 \Text(0,-40)[cb]{\Large $V^{\bbb}$}
\end{picture}
}}
\!\!\!\!\!
&=&
\quad
\frac{\mu^{2\ep}}{\pi^4}\int\!d^nq_1d^nq_2\,
\frac{f(q_1,q_2)}{[1][2][3][4][5][6]},
\quad \hbox{with} \quad
\left\{
\ba{l}
{[1]} = q_1^2+m^2_1 \\
{[2]} = (q_1-p_2)^2+m^2_2 \\
{[3]} = (q_1-q_2+p_1)^2+m^2_3 \\
{[4]} = (q_1-q_2-p_2)^2+m^2_4 \\
{[5]} = q_2^2+m^2_5 \\
{[6]} = (q_2-p_1)^2+m^2_6 
\ea
\right.
\nn
\eqa
\vspace{-0.6cm}
\caption[]{The irreducible two-loop vertex diagrams. 
$f$ is a generic polynomial in the loop momenta $q_1$ and $q_2$.
The dimension of the space-time is $n=4-\ep$ and $\mu$ is the 
renormalization scale.} 
\label{TLvertices}
\end{figure}
\begin{figure}[!ht]
\vspace{-0.9cm}
\bqa
\raisebox{0.1cm}{\scalebox{0.7}{
\begin{picture}(130,75)(0,0)
 \SetWidth{1.2}
 \CArc(60,0)(25,0,360)          \Text(60,15)[cb]{$1$}
 \Text(0,-3)[cb]{\Large $A$}
\end{picture}
}}
\!\!\!\!\!
&=&
\quad
\frac{\mu^{\ep}}{i\pi^2}\int\!d^nq_1\,\frac{1}{[1]},
\quad \hbox{with} \quad
\ba{l}
{[1]} = q^2_1+m^2_1 
\ea
\nonumber\\[-0.9cm]
\raisebox{0.1cm}{\scalebox{0.7}{
\begin{picture}(130,75)(0,0)
 \SetWidth{1.2}
 \CArc(60,0)(25,0,360)          \Text(30,-3)[cb]{$1$}
                                \Text(65,-3)[cb]{$2$}
 \Line(60,25)(60,-25)           \Text(90,-3)[cb]{$3$}
 \Text(0,-3)[cb]{\Large $T^{\ssA}$}
\end{picture}
}}
\!\!\!\!\!
&=&
\quad
\frac{\mu^{2\ep}}{\pi^4}\int\!d^nq_1d^nq_2\,\frac{1}{[1][2][3]},
\quad \hbox{with} \quad
\left\{
\ba{l}
{[1]} = q^2_1+m^2_1 \\
{[2]} = (q_1-q_2)^2+m^2_2 \\
{[3]} = q_2^2+m^2_3 
\ea
\right.
\nn
\eqa
\vspace{-0.6cm}
\caption[]{The irreducible one- and two-loop vacuum diagrams. 
The dimension of the space-time is $n=4-\ep$ and $\mu$ is the 
renormalization scale.} 
\label{OTLvacuums}
\end{figure}
\section{Properties of projectors \label{proppo}}
In this appendix we briefly summarize a general approach based on the work of
Ref.~\cite{Binoth:2002qh}. Amplitudes for two-loop $1 \to 2$ processes are 
decomposed into form factors which have to be extracted with proper projection 
operators. Let us consider tensor, one-loop, $N$-point functions in $n$ dimensions
($N \le 5, n= 4-\ep$)
\bq
S^{\mu\,\dots\,\nu}_{n\,\ssN} = \frac{\mu^{\ep}}{i\,\pi^2}\,\int\,d^nq\,
\frac{q^{\mu}\,\cdots\,q^{\nu}}{\prod_{i=0,N-1}\,(i)},
\quad
(i) = \lpar q + p_1 + \,\cdots\,+ p_i\rpar^2 + m^2_i,
\eq
where $p_0 = 0$. We select $N-1$ independent vectors, $r_i,\;i=1,
\,\dots\,,N-1$ and introduce the following notations:
\bq
G^{\ssN}_{ij} = G_{\ssN\,;\,ij} = 2\,\spro{r_i}{r_j}, \quad
{\cal G}^{\ssN} = {\rm det}\,G^{\ssN}.
\eq
For definiteness we will choose $r_i = p_i$. Next we introduce
\bq
D_{N\,;\,\mu\nu} = \frac{1}{n-N+1}\,\Bigl[ \delta_{\mu\nu} - 2\,r^t_{\mu}\,
G^{-1}_{\ssN}\,r_{\nu}\Bigr],
\qquad
R^{\mu}_{\ssN} = 2\,G^{-1}_{\ssN}\,r^{\mu}.
\label{PDproj}
\eq
They satisfy the following properties:
\bq
\delta_{\mu\nu}\,D^{\mu\nu}_{\ssN} = 1, \qquad
D^{\mu\alpha}_{\ssN}\,D_{\ssN\,;\,\alpha}^{\nu} = 
\frac{1}{n-N+1}\,D^{\mu\nu}_{\ssN},
\eq
\bq
D_{\ssN\,;\,\mu\nu}\,R^{\ssN\,;\,\nu}_i = 
D_{\ssN\,;\,\mu\nu}\,r^{\nu}_i = 0,
\quad
\spro{R^{\ssN}_i}{r_j} = \delta_{ij}, 
\quad
\spro{R^{\ssN}_i}{R^{\ssN}_j} = 2\,G^{-1}_{\ssN\,;\,ij}.
\eq
Let us consider now the action of these projectors on tensor integrals: we 
consider first
\bq
S^{\mu\nu}_{n\,\ssN} = \frac{\mu^{\ep}}{i\,\pi^2}\,\int\,d^nq\,
\frac{q^{\mu}\,q^{\nu}}{\prod_{i=0,N}\,(i)} =
\frac{\mu^{\ep}}{i\,\pi^2}\,\egam{N}\,\int\,d^nq\,\dsimp{N-1}\,
\frac{q^{\mu}\,q^{\nu}}{\lpar \spro{q}{q} + 2\,\spro{P}{q} + M^2\rpar^N},
\eq
\bq
P^{\mu} = \sum_{i=1}^{N-1}\,x_i\,p^{\mu}_i,
\quad
M^2 = \sum_{i=0}^{N-1}\,\lpar x_i - x_{i+1}\rpar\,
\Bigl[\lpar p_1+\,\cdots\,p_i\rpar^2 + m^2_i\Bigr],
\eq
with $x_0 = 1$ and $x_{\ssN} = 0$. Introducing
$\chi\lpar\{x\}\rpar = M^2 - \spro{P}{P}$, we obtain
\bq
S^{\mu\nu}_{n\,\ssN} = \lpar \frac{\mu^2}{\pi}\rpar^{\ep/2}\,
\dsimp{N-1}\,\chi^{2-N-\ep/2}\,\Bigl[
\egam{N-2+\frac{\ep}{2}}\,P^{\mu}\,P^{\nu} +
\frac{1}{2}\,\egam{N-3+\frac{\ep}{2}}\,\chi\,\delta^{\mu\nu}
\Bigr].
\eq
When the integral is projected with $D$ we have
\bq
P\,D_{\ssN}\,P = \sum_{i,j=1}^{N-1}\,x_i\,x_j\,p_i\,D_{\ssN}\,p_j = 0,
\qquad
D_{\ssN\,;\,\mu\nu}\,S^{\mu\nu}_{n\,\ssN} = \frac{1}{2}\,S_{n+2\,\ssN}.
\eq
For scalar diagrams we define the fundamental parametric representation,
\bqa
S_{n\,\ssN} &=& \lpar \frac{\mu^2}{\pi}\rpar^{2-n/2}\,
\egam{N-\frac{n}{2}}\,\dsimp{N-1}\,\chi^{n/2-N},
\nl
\chi &=& x^t\,H_{\ssN-1}\,x + 2\,K^t_{\ssN-1}\,x + L_{\ssN-1}
= (x - X)^t\,H_{\ssN-1}\,(x - X) + B_{\ssN}.
\eqa
Furthermore, we define new scalar objects,
\bq
S_{n\,\ssN}\lpar i,\,\cdots\,,j\rpar = \lpar \frac{\mu^2}{\pi}\rpar^{2-n/2}\,
\egam{N-\frac{n}{2}}\,\dsimp{N-1}\,x_i\,\cdots\,x_j\,\chi^{n/2-N}.
\eq
\bq
S_{n\,l}\lpar i,\,\cdots\,,j\rpar = \lpar \frac{\mu^2}{\pi}\rpar^{2-n/2}\,
\egam{l-\frac{n}{2}}\,(l)_n\lpar i,\,\cdots\,,j\rpar,
\eq
where $(2)= B, (3)= C$ etc. We obtain
\bq
R^{\ssN}_{i\mu} R^{\ssN}_{j\nu}\,S^{\mu\nu}_{n\,\ssN} =
2\,G^{-1}_{\ssN\,;\,ij}\,S_{n\,\ssN}(0) + \frac{1}{2}\,S_{n+2\,\ssN}(i,j).
\eq
Next we consider integrals with three momenta in the numerator
\bq
S^{\mu\nu\alpha}_{n\,\ssN} = -\,\lpar \frac{\mu^2}{\pi}\rpar^{2-n/2}\,
\dsimp{N-1}\,\chi^{n/2-N}\,\Bigl[
\egam{N-\frac{n}{2}}\,P^{\mu}\,P^{\nu}\,P^{\alpha} 
+ \frac{1}{2}\,\egam{N-1-\frac{n}{2}}\,\chi\,\delta^{\{\mu\nu}\,P^{\alpha\}} 
\Bigr],
\eq
where fully symmetrized tensors have been introduced. There are two kind of 
projections,
\bqa
D^{\ssN}_{\mu\nu}\,R^{\ssN}_{i\alpha}\,S^{\mu\nu\alpha}_{n\,\ssN} &=&
-\,\frac{1}{2}\,S_{n+2\,\ssN}(i),
\nl
R^{\ssN}_{i\mu} R^{\ssN}_{j\nu} R^{\ssN}_{k\alpha}\,
S^{\mu\nu\alpha}_{n\,\ssN} &=&
-\,S_{n\,\ssN}(ijk) 
-\,G^{-1}_{\ssN\,;\,ij}\,S_{n+2\,\ssN}(k)
-\,G^{-1}_{\ssN\,;\,ik}\,S_{n+2\,\ssN}(j)
-\,G^{-1}_{\ssN\,;\,jk}\,S_{n+2\,\ssN}(i).
\eqa
With four momenta in the numerator we have
\bqa
S^{\mu\nu\alpha\beta}_{n\,\ssN} &=& \lpar \frac{\mu^2}{\pi}\rpar^{2-n/2}\,
\dsimp{N-1}\,\chi^{n/2-N}\,\Bigl[
\egam{N-\frac{n}{2}}\,P^{\mu}\,P^{\nu}\,P^{\alpha}\,P^{\beta}
\nl
{}&+& \frac{1}{2}\,\egam{N-1-\frac{n}{2}}\,\chi\,
          \delta^{\{\mu\nu}\,P^{\beta}\,P^{\alpha\}}
+ \frac{1}{4}\,\egam{N-2-\frac{n}{2}}\,\chi^2\,
          + \delta^{\{\mu\nu}\,\delta^{\beta\alpha\}}\Bigr],
\eqa
where fully symmetrized tensors have been introduced.
Projections to be considered are:
\bq
D^{\ssN}_{\mu\nu}\,D^{\ssN}_{\alpha\beta}\,S^{\mu\nu\alpha\beta}_{n\,\ssN} =
\frac{1}{4}\,\frac{n-N+3}{n-N+1}\,S_{n+4\,\ssN}(0),
\eq
\bq
D^{\ssN}_{\mu\nu}\,R^{\ssN}_{i\alpha}\,R^{\ssN}_{j\beta}\,
S^{\mu\nu\alpha\beta}_{n\,\ssN} = \frac{1}{2}\,S_{n+2\,\ssN}(ij),
\qquad
R^{\ssN}_{i\mu}\,R^{\ssN}_{j\nu}\,R^{\ssN}_{k\alpha}\,R^{\ssN}_{l\beta}\,
S^{\mu\nu\alpha\beta}_{n\,\ssN} = S_{n\,\ssN}(ijkl).
\eq
\section{Non-abelian soft and collinear diagrams in $H \to gg$ 
\label{hereisnab}}
In this appendix we present a short summary of our procedure for extracting
soft/collinear divergent parts of the two-loop diagrams that contribute to the 
process $H \to gg$. Mass-singular, virtual, configurations appear when a)
a massless, internal, line is connected to two on-shell, external lines 
({\em soft}); b) a massless, external, line is connected to two massless,
internal lines ({\em collinear}).
Furthermore, we have a soft singularity whenever we attach an extra massless
line to an on-shell external line; if the latter is also massless an additional
collinear divergence arises. These QCD-like configurations have been
extensively discussed in the literature, e.g.~see Ref.~\cite{Spira:1995rr},
and in this section we briefly illustrate our approach to the problem. 

It is worth noting that in the QCD sector of the corrections to $H \to gg$ we 
select the regulator of collinear divergences to be the space-time dimension and
not the masses of the light quarks. On the contrary we select the latter 
in all situations where collinear singularities cancel in the total.
In the following we adopt the convention that a letter $i$ inside a one-loop
diagram denotes the scalar product $\spro{q}{p_i}$, where $q$ is the corresponding
loop momentum.

In the soft/collinear decomposition of a two-loop diagram we could 
highlight the nature of the singularity by introducing factors
$F^{\ssI\ssR}_{i} = F^{2}_{i}$ and write a soft/collinear decomposition for an 
arbitrary two-loop vertex,
\bq
V^{\ssI}_{\rm lab} = 
\sum_{i=0,2}\,V^{\ssI}_{{\rm lab}\,;\,i}\,F^{\ssI\ssR}_{-i}(\mhs), 
\label{scdec}
\eq
where the superscript $I$ denotes the family whereas the subscript {\em lab}
denotes the rank of the tensor ($0$ for scalar, etc.) as given in 
Ref.~\cite{Actis:2004bp}.
For those integrals where also ultraviolet divergences are present we need to
extend the decomposition of \eqn{scdec} to include higher orders of $\ep\,$ 
poles by introducing
\bq
F^2_{-3}(x) = -\frac{1}{\ep^3} + \frac{\DUV(x)}{\ep^2}
- \frac{1}{2}\,\frac{\DUV^2(x)}{\ep}
+ \frac{1}{3}\,\DUV^3(x).
\eq
Alternatively, we can factorize UV and IR/collinear divergences. Suppose that
we are considering a two-loop diagram with a single UV pole; remembering that
each loop can contribute at most one soft and collinear $1/\ep^2$ term the
highest possible pole in our integral is $1/\ep^5$. Consider the case
where the highest pole is $1/\ep^3$, then we first give the UV decomposition
and further decompose the coefficients with new IR factors:
\bq
V^{\ssI}_{\rm lab} = 
\sum_{i=0,3}\,V^{\ssI}_{{\rm lab}\,;\,i}\,F^2_{-i}(\mhs) = 
\sum_{i=0,1}\,V^{\ssI\,;\,\ssU\ssV}_{{\rm lab}\,;\,i}\,F^2_{-i}(\mhs),
\eq
\bq
V^{\ssI\,;\,\ssU\ssV}_{{\rm lab}\,;\,1} =
V^{\ssI}_{{\rm lab}\,;\,3}\,{\overline F}^{2\,;\,\ssI\ssR}_{-2}(\mhs),
\qquad
V^{\ssI\,;\,\ssU\ssV}_{{\rm lab}\,;\,0} =
\sum_{i=0,2}\,V^{\ssI}_{{\rm lab}\,;\,i}\,F^{2\,;\,\ssI\ssR}_{-i}(\mhs),
\eq 
\bq
F^{2\,;\,\ssI\ssR}_i(x) = F^{\ssI\ssR}_i(x),
\qquad 
{\overline F}^{2\,;\,\ssI\ssR}_{-2}(x) = -\,\frac{1}{\ep^2} + \frac{1}{6}\,
\DUV^2(x).
\eq
In the following we give a sample of our results; the full list can be found 
in~\cite{longNAB}.
\vspace{-0.3cm}
\begin{figure}[ht]
\begin{center}
\begin{picture}(130,75)(0,0)
 \SetWidth{1.2}
 \DashLine(0,0)(42,0){6}         \Text(10,7)[cb]{$H$}
 \Gluon(128,-53)(100,-35){2}{6}  \Text(132,-50)[cb]{$g$}
 \Gluon(128,53)(100,35){2}{6}    \Text(132,42)[cb]{$g$}
 \ArrowLine(40,0)(70,-17.5)         
 \ArrowLine(70,17.5)(40,0)       \Text(50,20)[cb]{$m$}
 \ArrowLine(70,-17.5)(70,17.5)       
 \Gluon(100,-35)(70,-17.5){2}{6}
 \Gluon(100,-35)(100,35){2}{10}
 \Gluon(100,35)(70,17.5){2}{6}
 \Text(30,-40)[cb]{\Large $V^{\bca;a}$}
\end{picture}
\qquad\quad
\begin{picture}(130,75)(0,0)
 \SetWidth{1.2}
 \DashLine(0,0)(42,0){6}         \Text(10,7)[cb]{$H$}
 \Gluon(128,-53)(100,-35){2}{6}  \Text(132,-50)[cb]{$g$}
 \Gluon(128,53)(100,35){2}{6}    \Text(132,42)[cb]{$g$}
 \ArrowLine(40,0)(70,-17.5)          
 \ArrowLine(70,-17.5)(100,0)         
 \Gluon(100,-35)(70,-17.5){2}{6}
 \Gluon(100,-35)(100,0){2}{6}
 \ArrowLine(100,0)(100,35)
 \ArrowLine(100,35)(40,0)
 \Text(30,-40)[cb]{\Large $V^{\bca;b}$}
\end{picture}
\qquad\quad
\begin{picture}(130,75)(0,0)
 \SetWidth{1.2}
 \DashLine(0,0)(42,0){6}         \Text(10,7)[cb]{$H$}
 \Gluon(128,-53)(100,-35){2}{6}  \Text(132,-50)[cb]{$g$}
 \Gluon(128,53)(100,35){2}{6}    \Text(132,42)[cb]{$g$}
 \ArrowLine(70,-17.5)(100,35)        
 \ArrowLine(100,35)(70,17.5)                             
 \ArrowLine(40,0)(70,-17.5)
 \ArrowLine(70,17.5)(40,0)
 \Gluon(100,-35)(82,-3.5){2}{6}\Gluon(78,3.5)(70,17.5){2}{6}
 \Gluon(100,-35)(70,-17.5){2}{6}
 \Text(30,-40)[cb]{\Large $V^{\bbb}$}
\end{picture}
\end{center}
\vspace{1.8cm}
\caption[]{The $V^{\ssK; a}$, $V^{\ssK;b}$ and $V^{\ssH}$ non-abelian 
soft/collinear configurations. Solid lines stand for a common mass $m$.}
\label{VKa}
\end{figure}
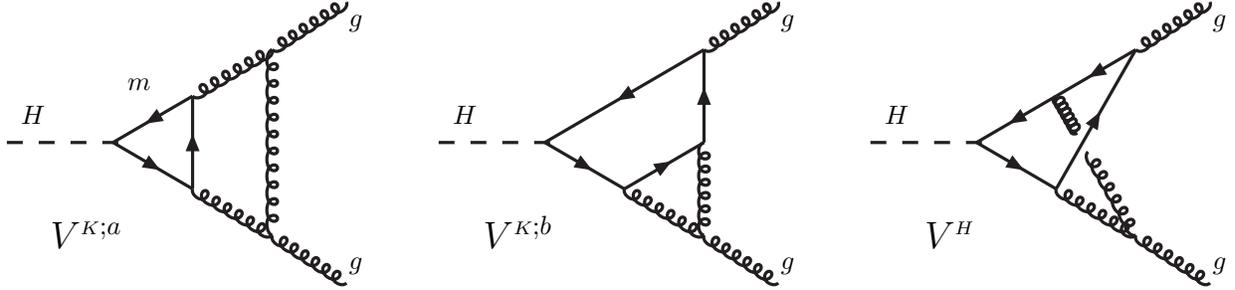

For the $V^{\ssK;a}$ configuration we introduce a quadratic form,
$\chi(x) = x(x-1) + m^2/\mhs$, where $m$ is the mass in the triangle of 
\fig{VKa}, and obtain
\bqa
V^{\ssK\,;\,a}_{-2} \!&=&\! 4\,\intfx{x}\,\frac{x}{\chi},
\nl
V^{\ssK\,;\,a}_{-1} \!&=&\! 2 \int_0^1\!\! \frac{dx}{\chi}\,\bigg\{\!
\!\int_0^{x} \!\! dy \ln\lpar \chi \!-\! y (x\!-\!y) \rpar
- \frac{1}{1\!-\!x} \ln\lpar \! 1 \!-\! \frac{x (1\!-\!x)}{\chi} \! \rpar
+ x \Bigl[\ln\lpar \chi \!-\! x (1\!-\!x)\rpar \!-\! 3 \ln\chi \!+ i\,\pi
\Bigr]\!\bigg\}.
\eqa

For the $V^{\ssK;b}$ configuration, we 
introduce a new quadratic form, $\xi(x,y)= \chi(x) + y^2 m^2/\mhs $,
and derive the following result:
\bq
V^{\ssK\,;\,b}_{-2} = 0,
\qquad
V^{\ssK\,;\,b}_{-1} = \intfxy{x}{y}\,\frac{\ln(1-y)}{\xi(x,y)}.
\eq
Integrating over $y$ we obtain
\bq
V^{\ssK\,;\,b}_{-1} = \frac{\mh}{2 m}\,\intfx{x}\,
\lpar -\,\chi\rpar^{-1/2}\,\bigg[
\text{Li}_2\lpar\frac{1}{y_+}\rpar -
\text{Li}_2\lpar\frac{1}{y_-}\rpar 
- \text{Li}_2\lpar\frac{1-x}{y_+}\rpar +
\text{Li}_2\lpar\frac{1-x}{y_-}\rpar\bigg],
\eq
with $y_{\pm} = 1 \pm \mh/ (m\,\sqrt{-\,\chi}\, )$.

Finally, for the non-planar $V^{\ssH}$ configuration of \fig{VKa}, we obtain
the following result:
\bq
V^{\ssH}_{-2} = 0,
\qquad
V^{\ssH}_{-1} = 4\,\int_0^1\,\frac{dx}{x}\,
\ln^2\Bigl( 1 - \frac{\mhs}{m^2}\,x (1-x)\Bigr),
\eq
and carry on the $x$ integration,
\bqa
V^{\ssH}_{-1} &=& 8\,\bigg\{ 
S_{1,2}\lpar \frac{1}{x_-}\rpar + S_{1,2}\lpar \frac{1}{x_+}\rpar +
S_{1,2}(x_-) + S_{1,2}(x_+) 
+\ln\frac{x_-}{x_+}\,\bigg[ \li{2}{\frac{x_- - x_+}{x_-}}
\nl
{} &-&
\li{2}{\frac{x_- - x_+}{x_- (1-x_+)}}\bigg] 
{} - S_{1,2}\lpar\frac{x_- - x_+}{x_-}\rpar +
S_{1,2}\lpar\frac{x_- - x_+}{x_- (1-x_+)}\rpar -
S_{1,2}\lpar\frac{x_+ - x_-}{x_+ - 1}\rpar\bigg\},
\eqa
with $x_{\pm} = 1/2 \pm 1/2 \sqrt{1 - 4\,m^2/ \mhs}$.
Here $\li{2}{z}$ and $S_{1,2}(z)$ are Nielsen poly-logarithms.
\section{Loop integrals with complex masses and momenta \label{allcmplx}}
Consider the logarithm of a complex number $z= z_{\ssR} + i\,z_{\ssI}$. 
Let ${\tilde z} = z_{\ssR} - i\,0$, we define
\bq
\ln_{(2)} \lpar z\,;\,{\tilde z} \rpar = 
\ln z - 2\,i\,\pi\,\theta \lpar - z_{\ssR} \rpar,
\eq
which satisfies
\bq
\lim_{z_{\ssI} \to 0}\,\ln_{(2)} \lpar z\,;\,{\tilde z} \rpar =   
\ln \lpar {\tilde z}\rpar.
\eq
When computing the amplitude in the CM scheme we will encounter, after reduction,
one-loop two-point functions where both masses and the external invariant are
complex. Let
\bq
\chi(x) = s_{\ssP}\,x^2 + \lpar M^2_2 - M^2_1 - s_{\ssP} \rpar\,x + M^2_1,
\eq
\bq
s_{\ssP} = M^2 - i\,\Gamma\,M, \qquad
M^2_i = m^2_i - i\,\gamma_i\,m_i.
\eq
The correct definition of the $B_0$ sfunction is as follows:
\bq
B_0 = \frac{1}{{\bar\ep}} - \intfx{x}\,\ln_{(2)}\lpar \chi\,;\,{\tilde\chi}\rpar,
\quad
{\tilde\chi} = \chi\bmid_{\Gamma,\gamma_i = 0} - i\,0,
\label{safe}
\eq
with the only restriction that $\chi(x)$ does not cross the negative real axis
for $x \in [0,1]$ (in which case the definition itself of a two-point function 
becomes dubious).  The best way of understanding the second Riemann sheet is to 
consider complex $p^2$ and zero masses; here $\chi = p^2\,x\,(1-x)$, therefore 
$\Reb\,\chi(x) < 0$ but $\Imb\,\chi(x) > 0$ which is the opposite of the 
$-\,i\,0$ prescription. Thus $\ln$ gives the wrong answer and $\ln_{(2)}$ the 
correct one based on the fact that we require a continuous limit {\em widths} 
$\,\to\,0$. Therefore, to compute $B_0$ one should start with the definition of the
corresponding function with masses and $p^2$ real and continue to complex
masses and $p^2$ on the correct Riemann sheet, i.e. with $\ln_{(2)}$ instead
of $\ln$. 
Note that in this function $y \pm i\,0$ is always treated as $y \pm i$. Of course,
one could use directly the logarithm of a complex number to compute $\ln x$.

When only the internal masses are complex there is no problem at all; indeed
we find
\bq
\Imb\,\chi = - \gamma_1\,m_1\,(1-x) - \gamma_2\,m_2\,x < 0,
\qquad
\Imb\,{\tilde\chi} = - 0 < 0.
\eq
This fact is also true for arbitrary two-loop diagrams as long as only internal
masses are continued into the complex plane.
There are special cases of complex $p^2$, complex masses.
In the actual calculation we have at most two scales: consider this situation 
where we are free to set $p^2= - M^2_2$ and obtain
\bq
\chi(x) = M^2_2\,x^2 + M^2_1\,(1 - x),
\eq
which means that $\Reb\,\chi(x) > 0$ for $x \in [0,1]$ and, therefore, we
can use {\em standard} results, i.e. $\ln_{(2)} \to \ln$.
Furthermore, when $M_1= 0$ we obtain
\bq
\chi(x) = p^2\,(1 - x) + M^2_2.
\eq
If $p^2= -M^2 + i\,\Gamma\,M$ then $\Reb\,\chi(x) < 0$ requires $M > m_2$
which is never satisfied in our case ($\mw < M_t, \mh$). Also here
$\ln_{(2)} \to \ln$. 
Therefore, as far as the calculation reported in this paper is concerned, the only 
relevant case is complex $p^2$ and zero internal masses.
\clearpage

\end{document}